# FORMULA TO DETERMINE THE COUNTRIES EQUILIBRIUM EXCHANGE RATE WITH THE DOLLAR AND PROPOSAL FOR A SECOND BRETTON WOODS CONFERENCE

Walter H. Bruckman
Department of Social Sciences
University of Puerto Rico
Cayey Campus

Abstrac

This paper presents the way in which can be determined the exchange rates that simultaneously balance the trade balances of all countries that trade with each other within a common market. A mathematical synthesis between the theory of comparative advantages of Ricardo and Mill and the New Theory on International Trade of Paul Krugman is also presented in this paper. This mathematical synthesis shows that these theories are complementary.

Also presented in this paper is a proposal to organize a common market of the American hemisphere. This economic alliance would allow to establish a political alliance for the common defense of the entire American hemisphere.

The formula we developed in this paper to determine the exchange rates of the countries solves the problem that Mexico, Canada, Europe, Japan and China currently experience with the United States in relation to the deficits and surpluses in the trade balance of the countries and their consequent impediment so that stable growth of international trade can be achieved.

## INTRODUCTION

The theory of comparative advantages of David Ricardo was later complemented by John Stuart Mill. The theory of comparative advantages of Ricardo and Mill is the strongest argument that exists to justify trade between two countries. The theory of comparative advantages explains how a country benefits from its international trade with another country (Ricardo's contribution) by increasing productivity and, consequently, the joint production of both countries and exchanging the surplus. The theory of comparative advantages also explains under what conditions two countries can optimize this benefit (Mill's contribution). However, the theory of comparative advantages does not explain under what conditions a country can increase joint productivity and optimize the benefit of its international trade in the case of more than two countries. That is, Ricardo and Mill's theory does not explain under what conditions a country can increase joint productivity and optimize the benefit of its international trade with multiple countries, which is the practical reality faced by all countries. Consequently, the application of the theory of comparative advantages as a theoretical instrument to guide the work of international trade in countries is limited. In addition, the theory, while analyzing the reason why a country A should keep its trade balance in equilibrium with another country B (Mill's contribution) as a way to optimize the benefit of international trade between the two countries, does not analyze the role which plays the equilibrium of the other balances within the balance of payments on the benefit that can be obtained from international trade.

In this paper we intend to complete the theory comparative advantages on international trade of Ricardo and Mill's by explaining under what conditions a country can optimize the benefit of its international trade with more than one country, that is, with multiple countries. Specifically, explain how to achieve simultaneous equilibrium in the trade balances of all countries that trade with each other. In addition, will be analyzed the effect of balances and imbalances on the financial balance, the capital balance and the current account balance on the benefit that a country can obtain from its international trade with another country and with multiple countries.

This complementation of the model of comparative advantages will allow to delineate the criteria under which the institutions in charge of regulating international trade between countries should be established and organized, such as the International Monetary Fund and the World Bank. Consequently, this complementation of the comparative advantages model constitutes a base on which to convene a new Bretton Woods conference.

## THE ESTABLISHMENT OF A COMMON MARKET OF THE AMERICAN HEMISPHERE

This theory has an immediate application as an instrument to remedy the current US trade war with China. One way in which the United States can reduce the loss of the benefits of trade with China is to move the lost trade with China to the American hemisphere by establishing a common market that involves all countries from Canada to Argentina and Chile, including Central America and the Caribbean. This common market would lay the foundations on which a common defense political alliance can be organized.

What are the obstacles that this initiative would have? The international trade treaties that have been established in America, Europe and Asia, are far from working perfectly. The problem they face is that they cannot avoid the deficits and surpluses in the trade balances of the countries that make up the treaties, nor the deficits and surpluses of the other balances within the balance of payments. Consequently, economic instability arises in international trade and recessions, both in countries with deficits and surpluses.

These deficits and surpluses mean that in the long term a stable economic development of international trade cannot be generated, which eventually leads to trade wars that lead to military wars such as the First and Second World War. After 73 years of the end of the Second World War, the countries of the world are again in serious disputes due to the deficits and surpluses that are generated from international trade. The tariff wars of the United States with China, Mexico, Canada, Europe and Japan, among others, give testimony to this problem. In the absence of economic principles that can explain how simultaneous equilibrium can be achieved in the trade balances of all countries that trade with each other, the tendency is for all countries to try to maintain a surplus in their trade balance through to devalue its currency against the dollar for being the country with the strongest and most accepted currency. At the beginning of each commercial cycle, the country with the most accepted currency (yesterday, England and today the United States) benefits from seigniorage by providing financing to other countries. However, in the long term the indebtedness of the deficit country becomes unsustainable because it affects its GDP and increases the threat of hyperinflation. It affects the GDP of debtor countries to the extent that this debt returns as foreign investment and produces profits and interests. In addition, because the exchange rates of countries that trade with each other are not independent of each other, but instead are interdependent, it happens that the efforts of some countries to keep their currencies devalued and maintain a surplus in their balances trade, cause appreciations in the currencies of other countries and their consequent deficits in the trade balances of those other countries with which their international trade is intertwined.

## WHAT HAS PRODUCED THE DEFICITS AND SURPLUSES IN THE TRADE BALANCES OF THE COUNTRIES?

What has produced the deficits and surpluses in the trade balances of the countries that have signed trade agreements in America, Europe and Asia, has its origin in the incorrect interpretation of the theory on international trade of David Ricardo and John Stuart Mill known as Theory of Comparative Advantages. This incorrect interpretation of the Theory of Comparative Advantages is what emerges from the proposals presented at the Bretton Woods conference and that culminated in the creation of the International Monetary Fund (IMF) and the World Bank (WB). This incorrect interpretation of the Theory of Comparative Advantages is also what emerges from the creation of the euro as a single currency in the European common market.

The incorrect interpretation of the Theory of Comparative Advantages led the countries in Bretton Woods to establish measures to regulate international trade that produced results contrary to the objective pursued and which was the stable growth of international trade. Three measures that were taken at Bretton Woods were 1. keep the exchange rate of the dollar with gold fixed at a rate of $35 an ounce and keep the exchange rates of the other currencies with the dollar and with gold as fixed as possible, 2. keep the equilibrium of payments balanced and 3. maintain the free mobility of capital.

Among the three objectives that the countries in Bretton Woods agreed to commit to, the most damaging was to keep exchange rates as fixed as possible because that prevents countries from achieving equilibrium in the trade balance through changes in the exchange rate. Exchange rates are the prices of the currencies in terms of the other currencies. The equilibrium exchange rates are those that equalize exports with imports and, consequently, that balance the trade balance. In consecuense, when it is established in Breton Woods that United State should fix its exchange rate with gold and the other countries fix their exchange rate with the dollar, keeping its balance of payments in equilibrium instead of the trade balance and maintaining the free mobility of capital so that could keep the balance of payment in equilibrium instead of the balance of trade, international trade was condemned to remain in instability permanently. These three measures produced a result that was contrary to the purpose for which the Bretton Woods conference was held and which was to achieve the development of stable international trade.

The rule that each country should keep in equilibrium its balance of payments instead of its trade balance and that was proposed by both the United States and England at Bretton Woods, brought as a consequence the permanent imbalance in trade balances. That is to say, it brought as a consequence the imbalance in the long term and therefore the instability in the growth of international trade. This norm, together with the norm of maintaining a fixed exchange rate of the dollar with gold and making the other countries keep their exchange rate fixed with the dollar, prevented, not only achieving equilibrium in the countries' trade balances, but also prevented countries from fully developing international trade based on comparative advantages, limiting it to trade based on absolute advantages.

The fixed exchange rate of the dollar with gold prevailed during the years 1944 to 1971. This norm or gold standard limited international trade based on comparative advantages by generating an effect equivalent to that of establishing a single currency for all countries that signed the Bretton Woods treaty. Such an agreement made gold, and therefore the dollar, a common currency for all signatories to the treaty. This rule had the same effect as it does today the rule of creating the euro as the single currency in the European Union. The euro as the single currency in the European Union does not work because it creates a permanent deficit in the trade balances of the least developed countries within the European Union (Greece, Portugal, Spain, etc.) with respect to the most developed ones (Germany, France, Italy) or, what is the same, creates a permanent surplus in the most developed countries within the European Union with respect to the least developed ones.

## Why the Euro as a Single Currency Doesn't Work?

In a perfectly competitive market, there is only one way a country can export goods to another country and that is by producing goods at lower prices than other countries. In common markets where each country has its own currency, all the country that wants to reduce its prices abroad has to do in order to export more is to increase its exchange rate. However, when the countries within a common market have a single currency, the situation is different. When there is a single currency between several countries, as in the case of the euro in the European Union or when there is a single currency between several regions, provinces or states within the same country, as in the case of the dollar in the United States, only there are two ways in which that country, region, or province can produce at lower prices to compete with the prices of the others: 1. reducing wages in that country, region, or province, 2. increasing the productivity of labor from that country, region or province. Since we should not expect that in the short term the productivity of labor will change unless technological changes occur, the only variable to be able to make the prices of a country, region or province lower and be competitive in the other countries, regions or provinces is by reducing the salary in that country, region or province. When there is a single currency between several countries or between several regions, provinces or states within the same country, it happens that as wages decrease in that country, region or province, their prices will decrease and they will be more competitive, which will allow to increase its exports to the other countries, regions or provinces.

In the European Union there is only one currency and the most developed countries (Germany, France, Italy) generate a continuous surplus in their respective trade balances with the least developed countries within the European Union (Greece, Portugal, Spain, etc.) or, what is the same, the least developed countries generate a continuous deficit in their trade balance with respect to the most developed countries within the European Union. These surpluses are due to the fact that the most developed countries have absolute advantages in the production of many goods, while the least developed countries do not have many goods where they have absolute advantages. Consequently, unless there are increases in productivity or decreases in wages in less developed countries, trade based on absolute advantages does not tend towards equilibrium in trade balances.

However, despite what has been said previously, it can be affirmed that any country B that has an absolute disadvantage in the production of all goods with respect to another country A, will have comparative advantages in the production of multiple goods. The only way this would not be so is when country B has exactly the same absolute disadvantage in the production of all goods and, as we know, that is highly unlikely. Consequently, country B will have comparative advantages with country A, even though country B has an absolute disadvantage in the production of all goods with respect to country A. Therefore, to the extent that there are decreases in the salary of the less developed country B, trade based on comparative advantages with country A will develop and move towards the equilibrium in the trade balance.

The statement in the previous paragraph is correct if we start from the premise of international trade in which there are only two countries A and B. When we start from the premise of international trade between multiple countries we can state that trade based on comparative advantages will tend towards equilibrium in the trade balance of each country with all the other countries considered together to the extent that there are decreases in the wages of each less developed country. That is, as we will demonstrate later in this paper, when we consider trade between multiple countries, this propensity towards the equilibrium generated by comparative advantages does not necessarily occur with respect to the trade balance between two countries, but with respect to trade balance of each country with all other countries considered together.

With respect to the European Union, we can conclude that, if we assume that labor productivity does not change in the short and medium term, then to the extent that wages decrease in less developed countries, it will happen that, in those goods where those least developed countries have absolute disadvantages, but nevertheless have greater comparative advantages, they will begin to have lower prices and may be exported to more developed countries. This will be so, despite the fact that developed countries have absolute advantages in the production of these goods. It is expected that the more wages decrease in the less developed countries, the more goods where those countries have the greatest comparative advantages will be added to the list of those that reach lower prices and can be exported to the more developed countries. Consequently, this dynamic will lead to equilibrium in the trade balance of each country with all the other countries considered as a whole.

As can be seen, it is the comparative advantages and not the absolute ones that make possible the equilibrium in the trade balances of the countries because they establish a trade dynamic that tends to equate exports with imports. Consequently, comparative advantages are what make it possible for each country to obtain maximum productivity and maximum joint production from its workforce, compatible with balance in its trade balance and, therefore, compatible with economic stability.

Trade based on the exploitation of comparative advantages is the one that is generated when free floating of currencies and free determination of exchange rates are allowed in the currency markets. Countries that have a common currency do not have flexible exchange rates. However, as we have discussed, trade based on the exploitation of comparative advantages in the case of countries that have a common currency it is possible as long as wages are used as an adjustment variable to balance countries' trade balances instead of Exchange rates. The problem with the above reasoning is that we cannot expect workers to accept reductions in their wages as adjustment variables to replace the exchange rate. On the contrary, we can expect that when wages are reduced strikes and protests will take place against the politicians elected, as well as emigration. Furthermore, we can expect emigration to occur when the free mobility of the labor factor is established, as is the case in the European Union. Emigration reduces the supply of labor and that puts pressure on wages not to drop.

None of this happens if each country keeps its own currency and the adjustment variable of the trade balance is the exchange rate instead of wages or emigration. It can be seen that, when each country has its own currency, an increase in the exchange rate that generates a depreciation of the currency in the less industrially developed countries (Greece, Portugal, Spain, etc.) will have the same effect as a reduction in wages and prices. However, workers will not protest or strike when the rate of change is increased.

Finally, there is a third variable that Germany and France can use and have used as an adjustment variable instead of the exchange rate to correct the imbalances in the trade balances of countries with recessions within the European Union and that is the indebtedness of countries with recession. The indebtedness serves in the short and medium term to remedy the imbalances in the trade balances (deficits) of Greece, Portugal and Spain and maintain the expansion of aggregate demand in, Germany, France and Italy, but in the long term it generates a problem that culminates in the collapse of recessionary economies.

In summary, we can conclude that, the ruro as a single currency does not work because to the extent that wages don't be downward flexible to that extent it allows trade based on absolute advantages, but it does not allow or limit trade based on comparative advantages. Trade based on absolute advantages does not tend to generate equilibrium in trade balances, while trade based on comparative advantages tends to generate equilibrium in trade balances. Consequently, the most efficient way to promote equilibrium in the countries' trade balances and, consequently, the stable development of international trade, is not by establishing a single currency for all countries where countries can only export goods where they have absolute advantages, but by establishing that all countries maintain their own currency and allow adjustments in currency exchange rates to enable countries to export goods where they have comparative advantages.

The above reasoning is correct insofar as the goods are homogeneous (perfect competition). When we assume non-homogeneous goods (monopolistic competition) the situation is more complex because, in addition to changes in prices, the tastes and preferences of consumers will be determining consumer demand for different goods. Under this circumstance, the productivity of labor and with it the comparative advantages will not be the only factor or variable that determines the demand and with it the competitiveness of the goods and exports of the countries. Consequently, under these circumstances, countries will export those goods whose prices in relation to the tastes and preferences of consumers are perceived as lower. However, as we will have an opportunity to demonstrate later in this paper, despite this complexity, the market under monopolistic competition will also lead to the same result as the market under perfect competition in terms of promoting equilibrium in the trade balance when wages go down or exchange rates are reduced.

## Why Does the Classic Gold Standard Do Not Work?

In the classical gold standard described by Hume, each country is responsible for the liquidity of its currency by setting its exchange rate with gold. The exchange rate of each country's national currency with gold is supposed to be kept fixed and deficits in the trade balance should be settled with gold transfers from the deficit country to the creditor country. Furthermore, the deficit country is supposed to reduce the amount of national currency in circulation in the same proportion as its gold reserves are reduced. Under such circumstances, it is assumed that the prices in gold and in national currency of the goods of each country should decrease in the same proportion as the gold reserves and the national currency in circulation decrease. In other words, prices in gold and national currency are supposed to be reduced through a deflationary process. The same reasoning is supposed to apply in the opposite direction when the country has a surplus in the trade balance. Under such circumstances, it is assumed that the prices in gold and in national currency of the goods of each country should increase in the same proportion as the gold reserves and the national currency in circulation increase. Consequently, as the quantity of gold increases or decreases and, consequently, the quantity of national currency in circulation within each country increases or decreases, the prices of each country must increase or decrease through the process of inflation or deflation, according to be the case. Under such circumstances, the adjustment variable to maintain equilibrium in trade balances is not the exchange rates but the inflation-deflation rates of prices and wages in each country.

Note that the establishment of the classical gold standard, as described by Hume, is equivalent to the establishment of a single currency for all the countries that use this system. In this system, the exchange rates of the national currencies with respect to gold are fixed and, therefore, the exchange rates of each currency with respect to the other currencies will be fixed as well. Consequently, since exchange rates are fixed, they cannot regulate equilibrium in trade balances. The exchange rates of each currency with respect to the other currencies will be fixed because by maintaining the exchange rates of the national currencies fixed with respect to gold, this means that the exchange rates between the national currencies have to be fixed for maintain the equivalence between the values of each currency with respect to gold. Otherwise there would be speculation in the currency markets and that speculation would maintain the equivalence between the values of the currencies with respect to the gold. Consequently, under the classical gold standard system, the adjustment variable to maintain equilibrium in the trade balances of the countries is not the exchange rate between the currencies that must be kept fixed as long as the exchange rates of each currency are fixed with respect to gold. The adjustment variable is the inflation-deflation rate. The inflation or deflation rate will have to increase or decrease the prices of the goods and with this make them less or more competitive abroad, as the case may be. The increase or decrease in the prices of the goods of each country expressed in gold and in national currency will be what allows a country to increase or decrease its imports and exports.

This system has two defects for which it does not work. On the one hand, it starts from the false premise that wages and prices are perfectly flexible to the downside and, on the other hand, it starts from the false premise that the governments of the countries will faithfully fulfill their commitment to adjust their money supply to their gold reserves.

The first false premise of the system exposed by Hume is to assume that wages and, consequently, prices are perfectly flexible downwards, which is not true since, as we have pointed out in the case of the European Union, workers tend to resist, protest and strike in the face of wage cuts. The downward assumption of flexible wages was one of the criticisms that Keynes made against classical economists. Consequently, those who advocate the gold standard today are unaware that we should not presume perfect downward wage flexibility. The classical gold standard system as exposed by Hume is an inefficient system because it does not describe reality by presuming perfect downward wage flexibility. To the extent that this assumption is not fulfilled, it is a system that generates the same effect as that of creating a single currency for all countries, in the same way as the euro in the European Union. In other words, this system allows trade based on absolute advantages, which, as we have pointed out in the case of the euro as the single currency of the European Union, does not lead to the equilibrium in trade balances. As we have pointed out, it is trade based on comparative advantages that tends towards equilibrium in the countries' trade balances. Therefore, to the extent that countries' wages and prices are not flexible downward, this system limits, hinders or prevents trade based on comparative advantages, which as we have pointed out is the one that leads to equilibrium in countries' trade balances.

The second false implicit premise from which Hume's model starts is to assume that the politicians of each country will have to sacrifice their personal ambitions regarding winning the next elections and that they will not employ an expansionary monetary policy that is favorable to those purposes. We know that this is false. Consequently, to the extent that governments expand their money supply beyond their gold reserves, at this rate they will be exporting their inflation and instability to other countries in the international market.

## Why the Bretton Woods Gold Standard Do Not Work?

In the case of the Bretton Woods gold standard, the United States had to maintain a relationship between its money supply and its gold reserves, but the other countries were not obliged to maintain a proportional relationship between the money supply and the gold existing in their central banks, since they could supplement their gold reserves with their dollar reserves that had to be backed by gold in the United States.

Consequently, the Bretton Woods gold standard has the same flaws as the classic gold standard. On the one hand, to the extent that wages and prices are not flexible downward, the system only allows trade based on absolute advantages. As we have pointed out, trade based on absolute advantages does not tend towards equilibrium in trade balances. It is trade based on comparative advantages that tends toward the equilibrium in the trade balance. On the other hand, to the extent that the politicians of the countries do not sacrifice their ambitions to win the next elections and refrain from increasing the money supply in some countries, to that extent the system will allow the export of inflation and instability from some countries to others.

However, in the Bretton Woods gold standard, the adjustment variable to maintain equilibrium in the trade balance does not depend solely on the inflation-deflation rate of each individual country, as in the case of the classic gold standard, but also it implicitly depends on the indebtedness of the countries as an adjustment variable to keep the balance of payments in equilibrium instead of the trade balance. Remember that, as we have already pointed out, in the case of the European Union, the indebtedness of the least industrially developed countries (Greece, Portugal, Spain, etc.) is the solution that the European common market has implicitly provided to the problem of deficits in the trade balances of the countries within this common market. As we have pointed out, this alternative action works in the short and medium term, but in the long term it generates collapse.

In the Bretton Woods gold standard, for indebtedness to function as an adjustment variable, it is necessary to establish the requirement that each country maintain equilibrium in the balance of payments instead of the trade balance. For the country with the most accepted currency (United States) this is not a very serious problem since it can expand its monetary base to finance its growing debt to the extent that other countries continue to accept it. In the case of the other countries, the situation is different because they will have to be open to the free flow of capital (dollars) to maintain equilibrium in their balance of payments and that is why this was the third measure agreed at Bretton Woods.

The problem with borrowing as a solution to imbalances in trade balances is that, as we have pointed out in the case of the euro in the European Union, it works in the short and medium term, but in the long term it leads to collapse. It will work as long as surplus countries are willing to continue accumulating dollars as a way to hoard wealth while the United States keeps expanding its monetary base and as long as deficit countries can continue to borrow as long as the United States lends it dollars that it prints. Recall what happened from 1944 to 1971. As the United States began to expand its monetary base to finance the Vietnam War beginning in 1960, to that extent inflation in the United States increased and, as expected, the Inflation in all other countries also increased. Consequently, the other countries stopped trusting and demanded the conversion of their dollars for gold causing the Bretton Woods system to collapse in 1971.

As we will see later, the solution to these problems described above lies in maintaining equilibrium in trade balances instead of equilibrium in balance of payments and using exchange rates as an adjustment variable instead of the reduction in wages or indebtedness to balance the balance of payments. With regard to the inflation problem, the solution lies in establishing that each country maintains an annual inflation limit and has to compensate the countries that have reserves of their currency for an annual amount equivalent to their annual inflation rate. At the end of this article, we will discuss how this measure can be implemented through the creation of a supranational institution in charge of implementing it.

In conclusion, in systems based on the gold standard, to the extent that wages and prices are not flexible downward, trade based on comparative advantage cannot be developed and that is the trade that tends to generate the equilibrium in trade balances. On the other hand, trade based on absolute advantages generated by the establishment of a common currency for all countries does not tend to maintain equilibrium in the trade balances of the countries. Consequently, trade based on the classical gold standard or the Bretton Woods gold standard, as well as trade based on a common currency, contrary to what its exponents propose, does not generate stability in the development of international trade. From this result we must conclude that all international trade based on either of these two gold standards, as well as international trade based on a common currency, should be discarded. In addition, we must conclude that the best way to regulate international trade is by determining the exchange rates between national currencies that keep in equilibrium the trade balances of all countries that trade with each other, and that is the purpose of this writing.

Evidence that Floating Exchange Rates Generate More Stable Growth in International Trade than Fixed Exchange Rates Established Under the Gold Standard

In the Bretton Woods gold standard, fixed exchange rates were maintained from 1944 to 1971. From 1971 this system was terminated, and in 1973 the floating of currencies allowed. Chart A shows how the exports of the United States from 1947 to 1971 fluctuate around 5%. On the contrary, from 1973 its growth was increasing until reaching almost triple, bordering on 13.5%. This is because, as the currencies floated against the dollar from 1971, the resulting exchange rates allowed the maximum development of trade based on comparative advantages, as well as based on absolute advantages.

Grafica A

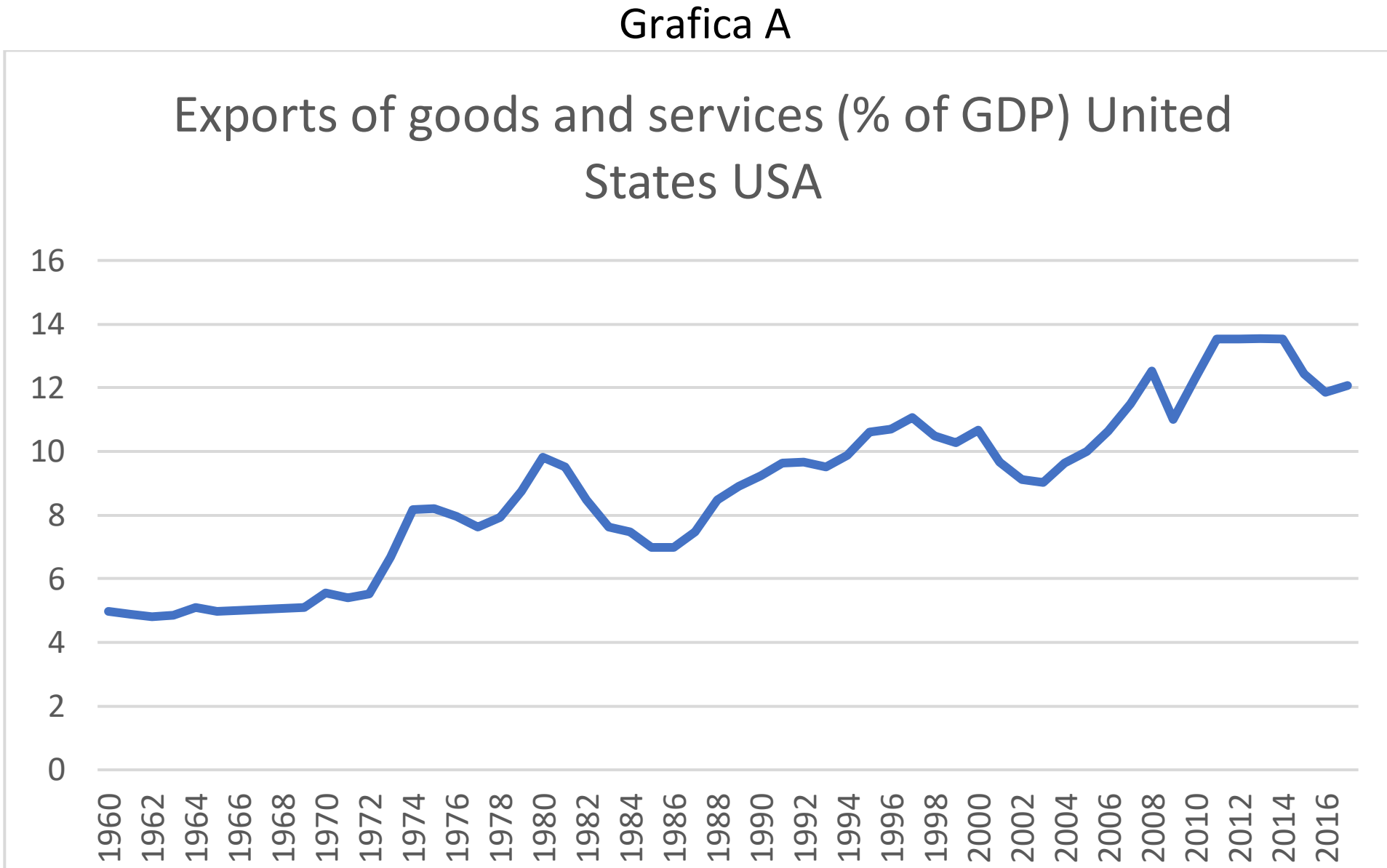

When the countries of the world agreed at Bretton Woods to set the value of the dollar with gold and the value of their national currencies with gold and the dollar, they made the same mistake that countries in the European Common Market committed by establishing a single currency with the euro. As we have pointed out, this situation prevailed from 1944 to 1971 and favored the development of trade that was based on absolute advantages, which, as we have explained, tends towards imbalance in trade balances, limiting the development of trade based on comparative advantages, which is the one that makes growth in trade possible and the one that tends towards equilibrium in trade balances.

During the years 1944 to 1971, countries that did not have wages low enough to offset by their lower labor productivity in the production of different goods (absolute disadvantage), could not increase their exports. Since, as we have noted, in all countries there is a natural resistance of workers to accept reductions in their wages, the result is that, to the extent that wages were not low enough to offset lower labor productivity, the International trade tends to be limited to exports of goods where there are absolute advantages and to reduce trade in goods where there are comparative advantages. In this way, the stable development of international trade is impeded by hindering the propensity towards keep in equilibrium trade balances. The level of imbalances in trade balances will depend, on the one hand, on the level of inflexibility of wages and, on the other hand, on the level of tariffs that are established to try to reduce the imbalances in trade balances that generate the inflexibility of wages.

From 1971, international trade grew rapidly as a consequence of the fact that with the elimination of the fixed exchange rate of the dollar with gold, international trade based on comparative advantages began to develop and grow. However, the problem of how to achieve balance in the trade balances remained unresolved, which is the question we intend to solve in this paper.

## Emigration as an Adjustment Variable to Correct Imbalances (Deficits) in Trade Balances

In the case of Europe, an additional problem is added. The incorrect interpretation of the theory of comparative advantages has led them to make the mistake of establishing, not only a common currency, but also the free mobility of the labor factor between the countries that make up the European Common Market. Consequently, apart from reductions in wages as an adjustment variable to correct imbalances (deficits) in trade balances, emigration is also introduced as an adjustment variable to correct imbalances (deficits) in trade balances. Therefore, Ireland, Germany, France and Italy, being the countries with the highest productivity, become the recipient countries of the unemployed workforce in Greece, Portugal, Spain, etc.

What happens in the European Union is the same thing that happens between different regions of the same country with different levels of economic development. In one country, people migrate from regions with low levels of industrialization to the most industrialized regions of the country. Because it is the same country, the nation as a whole does not lose economic or political power. For example, in the United States, because within the same country the mobility of the workforce is not impeded, the less industrialized central states lose population every year that moves to more industrialized states such as California or New York.

When in a common market between countries a single currency is established and the free mobility of labor, international trade becomes from an international trade based on comparative advantages to, an international trade based on absolute advantages.

Productivity in Greece, Portugal, Spain, etc., has been unable to compete with productivity in Germany, France and Italy. Consequently, the creation of a common currency has brought the recessions that occurred in Greece, Portugal, Spain in 2011 and the free mobility of the labor factor has generated internal emigration within the common market and the consequent growth of economic and political power from Germany, France and Italy at the cost of the loss of the economic and political power of Greece, Portugal and Spain.

## The Fundamental Problem of the Exchange Rates and of the Adjustment Variables

The fundamental problem of exchange rates as an adjustment variable to balance the trade balances of the countries is to determine what is the equilibrium exchange rate of each country with all other countries. If the equilibrium exchange rates of the countries are not independent of each other, as we shall demonstrate in this writing, then it is not possible to know what the equilibrium exchange rate of one country is with another country without simultaneously determining the equilibrium exchange rates of all countries with the other countries. Consequently, no country A can accuse another B of keeping its currency devalued below its equilibrium level with respect to country A because no one can know what is the equilibrium exchange rate of one country with another, that is, which It is the level of equilibrium surplus or deficit of one country A with another B, without determining in turn the equilibrium exchange rates of all other countries with respect to all other countries.

When international trade is based on a gold standard system or on a single currency system, the substitution of the exchange rates of the countries by the level of the wages of the countries as adjustment variables to balance the trade balances generates the same fundamental problem than the exchange rate.

When international trade is based on a gold standard system or on a single currency system, what we do is substitute the exchange rates of the countries for the level of wages of the countries as adjustment variables to balance trade balances of the countries. Consequently, the nature of the problem has not changed and remains the same. The fundamental problem of wages as an adjustment variable to balance trade balances is to determine what the equilibrium wage of each country is with respect to all other countries. If the equilibrium wage levels of the countries are not independent of each other, then one cannot know what the equilibrium wage level of one country B is with another country A without simultaneously determining the equilibrium wage levels of all the other countries with respect to all other countries. Consequently, under these circumstances no country A can accuse another B of keeping its wage level below its equilibrium level with respect to country A because no one can know what the equilibrium wage level of one country is with another. That is, no country can know what is the level of equilibrium surplus or deficit in the trade balance of a country A with another B without in turn determining the equilibrium wage level of all other countries with all other countries. Consequently, neither the creation of a gold standard, nor the creation of a common currency for all countries that trade with each other, solves the central problem of international trade, which is to achieve stable growth by maintaining the balance of the trade balances of all countries.

Aside from wages, countries that share a single currency can use any other variable to serve as substitute for the exchange rate. However, in addition to wages, any other variable that is used as an adjustment variable to substitute the exchange rate, such as the reduction of taxes on companies or the emigration of the unemployed population, will generate the same fundamental problem as generates the exchange rate. That is to say, the problem of determining the set of exchange rates that balances the trade balances of all countries is replaced by the problem of determining the set of business tax rates or the emigration rate that balances the trade balances from all countries.

The European Union recently sued Apple in Ireland for failing to pay 13 billion in taxes. But this claim is invalid because if a single currency is kept, then neither Germany nor France can justify that Ireland has no right to modify its level of wages to workers or its level of taxes on companies or its level of emigration as variables of adjustment to avoid deficits and promote surpluses in its trade balance. They cannot prohibit that variable from being used by Ireland as an adjustment variable, as that would condemn them to maintain a permanent deficit in their trade balance with Germany and France, as happened with Greece, Spain and Portugal in 2012.

If Ireland had its own currency independent of the Euro all it would have to do to export more to Germany or France would be to increase the exchange rate with the Euro. Since Ireland does not have its own currency but the same currency as the rest of the European Union countries, the Euro, all it can do to export more to Germany or France is to lower its salaries or reduce taxes on companies. That has the same effect as increasing its exchange rate. If the level of corporate taxes goes down, we should expect exports to increase for two reasons. In the case of companies operating in a perfectly competitive market, lower taxes allow lower prices. In the case of companies that operate in a monopolistically competitive market, lower taxes allow more companies to establish themselves in that country and export their products to other countries in the European Union where each monopolistic company has a protected niche in the tastes and consumer preferences.

Of what depends whether Ireland uses as an adjustment variable reducing corporate taxes instead of lowering wages? We have explained that it is impossible for politicians to lower the wages of workers as an adjustment variable to balance the trade balance without committing political suicide before the voters of their country and that is why wages are not flexible downwards. However, politicians can lower taxes on companies without having to commit suicide before the electorate. Consequently, this is what Ireland, Cyprus and other members of the European Union with a deficit in their trade balances do.

However, despite the fact that Ireland, Cyprus and the other countries with deficits in their trade balances have the perfect right to use wages or taxes as an adjustment variable, the fundamental problem remains unsolved since the question that would have to be answered is: What is the level of equilibrium corporate taxes for Ireland, Cyprus and the other countries? Again, the same problem arises. The level of business equilibrium tax in Ireland is not independent of the level of business equilibrium tax in all other countries within the European Union. A level of business tax close to zero, such as that maintained by Ireland, may be too low and be equivalent to maintaining an exchange rate above its equilibrium level, generating a permanent surplus in Ireland with respect to Germany or France. Consequently, what Germany or France have a right to claim from Ireland is that their corporate tax level is unfairly below the equilibrium tax level. In other words, we see that the underlying problem regarding the balance of trade balances remains unsolved.

The same would occur if emigration is used as an adjustment variable to maintain the balance of trade balances. In other words, the equilibrium level of emigration of the workforce from Ireland to Germany or France is not independent of the equilibrium level of emigration of the other countries with deficits in their trade balance within the European Union and the level of immigration from balance of the other countries with surpluses in their trade balances. In order to balance the trade balances of all the countries within the European Union using migration as an adjustment variable, it would be necessary to determine the equilibrium levels of emigration or immigration of all the countries within the European Union.

In conclusion, the European Union does not optimize its international trade by establishing a single currency, in the same way that it would not optimize its international trade if it established a system based on the gold standard. This is because to the extent that wages are not perfectly flexible downward, these systems establish international trade based on absolute advantages that do not tend towards equilibrium in the trade balances of the countries. As we have already pointed out, trade that tends towards equilibrium in trade balances is the trade based on comparative advantages and this is achieved when each country has its own currency and the exchange rate is used as an adjustment variable.

If it were possible that wages, business taxes and migration were perfectly flexible downwards or upwards, the only thing that is achieved with an international trading system based on a single currency it is a system based on wages or taxes or emigration as adjustment variables that will have the same problems of imbalances in trade balances that countries that use exchange rates as an adjustment variable. To the extent that these variables are not perfectly flexible downward or upward, this system will have adjustment variables that are much more rigid and difficult to modify than when each country conserves its own currency and exchange rates are used as an adjustment variable.

## THE SET OF EXCHANGE RATES THAT BALANCE THE TRADE BALANCES OF ALL COUNTRIES THAT TRADE WITH EACH OTHER

As can be seen, the way in which countries can benefit from the exploitation of comparative advantages as described by David Ricardo and John Stuart Mill, is by maintaining the condition that each country has its own currency and that this currency can vary its value with respect to the other currencies so that the equilibrium in the countries trade balance can be generated by establishing the exchange rates that generate that equilibrium.

The following explains the way in which can be determined the set of exchange rates that balance the trade balances of all countries that trade with each other within a common market. This explanation or formula allows to establish what are the causes of the deficits and surpluses in the trade balances of the countries of America, Europe and Asia that have signed trade agreements. The formula that we are going to develop is presented within the context of a proposal on how the American Hemisphere Common Market can be organized to avoid this problem of deficits and surpluses in the trade balance of countries and its consequent impediment to achieve a stable growth of international trade.

> Let us call a ***set of universal equilibrium exchange rates*** the set of exchange rates that balance the trade balances of all countries that trade with each other.

The questions that follow are: Is there a way to balance the trade balances of all countries that trade with each other? Or to be more precise, is there a set of universal equilibrium exchange rates that balance the trade balances of all countries that trade with each other? In the case that such a set exist, is there a way to determine these exchange rates that balance the trade balances of all countries that trade with each other in a common market?

The problem of how to achieve the development of an international trade that is stable among $n$ countries that trade with each other, is reduced to the problem of how to determine if it exists a set of universal equilibrium exchange rates that balance the trade balances of all $n$ countries that trade with each other and, if it exists, how that set of universal equilibrium exchange rates can be determined. In what follows we will answer these questions.

### THE EQUIVALENCE BETWEEN THE VALUES OF THE CURRENCIES

The set of universal equilibrium exchange rates that balance the trade balances of all $n$ countries that trade with each other, has to comply with a restriction and is to maintain the equivalence between the values of the currencies. How do we define the equivalence between the values of the currencies?

If we have a common market of $n$ countries, the number of exchange rates that determine trade between those $n$ countries will be equal to the number of permutations between $n$ countries taken from 2 in 2. That is, $nP_r = \frac{n!}{(n-r)!} = nP_2 = \frac{n!}{(n-2)!}$. We say that this set of $nP_2$ exchange rates is equivalent set when, of all the $n$ currencies of the $n$ countries, we have n-1 currencies that define their value (exchange rate) in relation to the most accepted currency and all the remaining [$nP_2$ - (n-1)] exchange rates are proportionate to the value of each currency expressed in terms of the most accepted currency.

Let's see an example. Assume three countries A, B and C. Suppose further that the currency of country A is the most widely accepted currency and suppose that the currencies of countries B and C are of lesser acceptance. Finally, suppose that the currencies of countries B and C define their value in relation to the most widely accepted currency of country A. Those two values are exchange rates:

TCab = Units of the currency of country A for each unit of the currency of country B
TCac = Units of the currency of country A for each unit of the currency of country C

From the previous premises, we say that the TCbc exchange rate between countries B and C is equivalent, if it maintains proportion with the TCab and TCac exchange rates of these two currencies with the currency of country A of greater acceptance.

If the exchange rate of these two currencies TCbc did not maintain a proportion with the exchange rates TCab and TCac of those two currencies with respect to the currency A of greater acceptance, then one of the two currencies B or C would be overvalued or undervalued with respect to the other in relation to currency A of greater acceptance. Consequently, if currency B were overvalued relative to currency C, speculators in the foreign exchange market could make a profit when, instead of directly buying currency A of greater acceptance with overvalued currency B, they indirectly buy currency A through to buy with overvalued currency B undervalued currency C and with it buy currency A.

Let us define the set of exchange rates that maintain the equivalence between the values of the currencies, as ***the set of equivalent exchange rates***.

As we have indicated, all set of exchange rates has to comply with a restriction and is to maintain the equivalence between the values of the currencies. This means that the set of $_nP_2$ universal equilibrium exchange rates used to balance the trade balances of all n countries that trade with each other must also be a set of equivalent exchange rates.

Theoretically there must be an indeterminate number of equivalent exchange rate sets. One for each different set of n-1 exchange rates that define the value of the currencies of the n-1 countries with respect to the currency of country A of most acceptance. The problem of how to achieve stable international trade can be reduced to the problem of how to ensure that the set of universal equilibrium exchange rates that balance the trade balances of all countries that trade with each other complies with the restriction of being a set of equivalent exchange rates.

The greater the number of countries that trade with each other in a common market, the more complex the task is to determine the set of universal equilibrium exchange rates that balance the trade balances of all countries and that are also equivalent. For example, in the case that concerns us of the common market of the American hemisphere, suppose that a group of n = 22 countries of the American hemisphere decide to establish a common market. There are $_{22}P_2 = \frac{22!}{(22-2)!} =$ 462 possible exchange rates among the n = 22 countries. These possible rates of exchange are given by the number of permutations of 22 countries taken from 2 in 2. What is involved is to determine what is the particular set of 462 universal equilibrium exchange rates that balance the trade balances of the 22 countries and also maintain the equivalence between the values of the currencies.

## THE DETERMINATION OF THE UNIVERSAL EQUILIBRIUM EXCHANGE RATES THAT ARE EQUIVALENT AND THE EQUILIBRIUM IN THE CAPITAL, FINANCIAL AND INCOME BALANCES

One way to determine the 462 universal equilibrium exchange rates that balance the trade balances of the 22 countries of the American hemisphere and that are also equivalent would be to float the 22 currencies in the 231 currency markets that would arise from the 462 exchange rates possible. But that flotation of the currencies would have to be done with several restrictions. The currencies must be floated while maintaining the necessary controls so that financial balances, capital balances and income balances within the countries' current account balances are kept in equilibrium, so that imbalances in the flows of Capitals do not affect the determination of exchange rates that balance the trade balances of all countries. This implies that the capital, financial and income balances must be in equilibrium independent of the balance of the trade balance and not jointly within the balance of payments. This would prevent capital mobility and, specifically, individual imbalances in these balances within the balance of payments from affecting the determination of the universal equilibrium exchange rates that balance the trade balances of all countries.

It is expected that this flotation of all currencies, while maintaining the restrictions on the aforementioned components of the balance of payments of the countries, generates the set of universal equilibrium exchange rates that balance the trade balances of all the countries that trade with each other. In addition, it is expected that, since such set of exchange rates is generated by the free interaction between the supply and demand of the currencies of all countries, the resulting exchange rates (universal equilibrium exchange rates) are also those that maintain the equivalence between the values of the currencies, apart from being those that equilibrium the trade balances of all countries.

Note that it is said that the financial and capital balance of each country should be in balance by individual and not jointly. Otherwise, they will produce an imbalance in the balance of income within the current account balance that would affect the determination of that country's exchange rate with the other countries. The reason is as follows. It is expected that the return on interest of a level of loans in the financial balance will be lower than the return on earnings of the same level of investments in the balance of capital. Therefore, equal capital but of the opposite sign in the financial and capital balances of a country, will have to produce different levels of performance that will be reflected in the balance of income within the current account balance of that country. Consequently, a deficit in the financial balance that is canceled with a surplus in the balance of capital, generates different returns that will not cancel each other in the balance of income within the current account balance and that, therefore, will affect the determination of the equilibrium exchange rate of that country within a system of floating exchange rates.

> Let us define the set of universal equilibrium exchange rates that maintain the equivalence between the values of the currencies, as the set of ***equivalent universal equilibrium exchange rates***.

In Section PROPOSAL FOR A SECOND BRETTON WOODS CONFERENCE more ahead we will present a way in which the equivalent universal equilibrium exchange rates can be generated without it being necessary to maintain equilibrium in the financial and capital balances.

### HOW TO SIMPLIFY THE DETERMINATION OF THE EQUIVALENT UNIVERSAL EQUILIBRIUM EXCHANGE RATES

We have said that, if all n currencies are floated, we can expect that, through the free interaction of the supply and demand functions of the currencies in the currency market, that set of equivalent universal equilibrium exchange rates will be generated that balance all trade balances. However, we have found another way to determine, with a considerably smaller effort, the set of equivalent universal equilibrium exchange rates. For this we have divided the set of equivalent universal equilibrium exchange rates into three subsets and we have made the last two subsets functions of the first. Let's see.

In order to simplify the demonstration, suppose a common market between n = 3 countries A, B and C. The possible exchange rates that should be determined are 6:

TCab, TCac, TCba, TCca, TCbc, TCcb

However, if we establish the restriction that all exchange rates must be equivalent, in reality only the first 2 exchange rates should be determined because the other 4 would be determined by the corresponding mathematical formulas that maintain the equivalence in the value of the currency. That is, the first 2 exchange rates TCab and TCac would be the only 2 that would have to be determined, because in order to maintain the restriction that the exchange rates are equivalent, the next 2 exchange rates TCba and TCca, which are the counterpart of the first two exchange rates would be determined by the reciprocal of the first two exchange rates. The formulas would be TCba = $\frac{1}{\text{TCab}}$ and TCca = $\frac{1}{\text{TCac}}$. Similarly, if we maintain the restriction that exchange rates have to be equivalent, the last 2 exchange rates, would be determined by the formulas TCbc = $\frac{\text{TCac}}{\text{TCab}}$ and TCcb = $\frac{\text{TCab}}{\text{TCac}}$.

Let us define or call ***base currency or reference currency*** the currency with the highest degree of acceptance among all countries that trade with each other in a common market and, consequently, with which the other countries have an interest in establishing the value of their currencies.

If we establish the currency of country A as the base currency, then:

Let us define or call the first 2 exchange rates TCab and TCac that are established with respect to the base or reference currency of country A and that can be determined freely, ***primary exchange rates***.

Let us define or call the 2 exchange rates TCba = $\frac{1}{\text{TCab}}$ y TCca = $\frac{1}{\text{TCac}}$ which are the counterpart of the primary exchange rates, ***counter-primary exchange rates***.

Let's define or call the remaining 2 exchange rates TCbc = $\frac{TCac}{TCab}$ y TCcb = $\frac{TCab}{TCac}$, ***non-primary exchange rates***.

We can conclude that the set of primary exchange rates plus the set of counter-primary exchange rates plus the set of non-primary exchange rates are equivalent. Therefore, when we determine the set of counter-primary exchange rates and the set of non-primary exchange rates using the algebraic formulas that define equivalence, any combination of primary exchange rates TCab and TCac that generates the equilibrium in the trade balances of all the countries that trade with each other (in the case of our example the three countries A, B and C) will also be generating the set of equivalent universal equilibrium exchange rates. That is, the set of equivalent universal equilibrium exchange rates would be equal to the sum of the set of primary exchange rates that generates the balance in the trade balances of all countries, plus the counter-primary exchange rates plus the non-primary exchange rates derived from the primary exchange rates.

As you can see, for the set of equilibrium primary exchange rates that in turn balances all the trade balances of all n countries comply with the restriction of maintaining equivalence between the values of currencies, counter-primary and non-primary exchange rates must be calculated using the formulas we explain paragraphs back.

Let us define the set of n-1 equilibrium primary exchange rates that in turn balances all the trade balances of all n countries that trade with each other and that also maintains the equivalence between the values of the currencies as the ***set of equivalent universal equilibrium Primary exchange rates***.

If we apply the above reasoning to the case of the 22 countries of the American hemisphere and assume that the currency of country A (United States) is the base currency or reference currency because it is the one with the highest degree of acceptance, then we would have 21 primary exchange rates to be determined freely by floating the currencies with the base currency. The remaining 21 counter-primary exchange rates and the 420 non-primary exchange rates would be determined by the formulas described before. That is to say, one would not have to worry about determining by floating the currencies 441 exchange rates, because they would be automatically determined by the corresponding formulas that define them, when determining the 21 primary exchange rates.

## THE IMPORTANCE OF THE REGULATION OF CAPITAL FLOWS IN THE DETERMINATION OF EXCHANGE RATES

As we point out, to determine the 21 equivalent universal equilibrium primary exchange rates that generates the equilibrium in the trade balances of the 22 countries of the American hemisphere, you can proceed to maintain the free flotation of the 21 currencies with the base or reference currency (the dollar) until the equilibrium in the trade balances of the 22 countries is obtained. As we have discussed, for the free flotation of the 21 currencies with respect to the base currency (the dollar) to generate the equivalent universal equilibrium primary exchange rates, it is necessary that, during the flotation process of the 21 currencies with respect to the dollar, countries maintain their financial balance, capital balance and income balance, within the current account balance in equilibrium, independent of the equilibrium in the trade balance. In other words, the equilibrium in these balances must be independent of the equilibrium in the balance of payments that includes the trade balance. In the end, the balance of payments would be in equilibrium, but not globally, but because its components would be in equilibrium and, therefore, the sum of those components would be in equilibrium. If there is a set of equivalence universal equilibrium primary exchange rates that balances the trade balances of all countries and if the equivalent exchange rates are obtained through the indicated algebraic formulas, it is expected that the free floating of the 21 currencies with the base currency (the dollar) generates that set of equivalent universal equilibrium primary exchange rates, provided that the other balances within the balance of payment (capital balance, financial balance and income balance) are kept in equilibrium.

Note that the previous measure of keeping capital mobility under control in order to obtain the equivalent universal equilibrium primary exchange rates is contrary to the current IMF policy of maintaining free international capital mobility with the objective of maintaining the equilibrium in the balance of payments of the countries as a means of counteracting the deficits or surpluses in the trade balances.

The 21 primary exchange rates must remain permanently floating so that they can incorporate in the equilibrium of the trade balances of the 21 countries the changes in the monetary mass or inflation of each country, the changes in the population, the changes in labor productivity in different industrial sectors, the changes in consumer tastes and preferences, etc.

Contrary to these complexities that we have described in order to maintain balance in the trade balances of all 22 countries in the American hemisphere that trade with each other, the IMF proposal to maintain the equilibrium in the balance of payments of all 22 countries that trade with each other is relatively simple. It only requires that each country make the necessary loans with the IMF to equilibrate its balance of payments. Even simpler is that, unlike exchange rates whose simultaneous universal equilibrium values are interdependent, in the case of financial transactions that each of the 22 countries can make with the IMF, these are transactions that are totally independent each other. The only problem with this IMF solution is that this relatively easy equilibrium to be achieved in the balance of payments of all countries that trade with each other, does not serve at all to avoid the instability of international trade. It is as if we wanted to cure a sore on the left arm of a patient by applying an acne remedy on the patient's face that has nothing to do with the sore on the left arm but is easy to apply. That is why we point out at the beginning of this paper that the solutions provided by the Bretton Woods Conference to the problem of market instability in international trade indicate an incorrect interpretation of Ricardo and Mill's theory of comparative advantages.

Balancing the balance of payments of the countries only serves to maintain the most fixed as possible the exchange rates. But maintaining fixed the exchange rates is an objective contrary to the objective of balancing the trade balances of countries. It is the imbalances in the trade balances of the countries that produce instability in the international trade and that is precisely caused by preventing exchange rates from changing.

Consequently, in the case of the Common Market of the American Hemisphere that concerns us, we only have the option of maintaining the equilibrium in the trade balance of the 22 countries instead of the equilibrium in the balance of payments. As we have pointed out, in order to maintain the equilibrium in the trade balance, the flotation of all 21 currencies against the dollar is required, maintaining the restrictions on the mentioned components of the balance of payments.

### Disconnection of the Monetary Policy and the Interest Rate of the Mobility of International Capitals

As we have already pointed out, the regulation of capital movements to maintain equilibrium in the financial, capital and income balances of each country is necessary to ensure that the floating primary exchange rate regime produce the set of universal equilibrium primary exchange rates that in turn match the equivalent exchange rates. However, there are other reasons why it would be desirable to maintain the regulation of capital movements. For example, one reason why it is desirable to maintain balance in the financial, capital and income balances of each country is to disconnect the monetary policy and interest rate of each country from the fluctuations in international capital markets. This disconnection allows countries to have a very important tool to combat inflation and low levels of production and employment through the monetary policy and interest rate of each country.

## A MATHEMATICAL MODEL ON THE THEORY OF COMPARATIVE ADVANTAGES IN THE CASE OF MORE THAN TWO COUNTRIES AND A PERFECT COMPETITION MARKET

How to demonstrate that there is a set of equivalent universal equilibrium primary exchange rates that balances the trade balances of all countries that trade with each other within a common market? To make this demonstration we need to formulate a mathematical model that describes the behavior of international trade that is generated between countries based on the set of exchange rates of those countries.

In order to simplify the analysis, we start from the assumption of a common market of three countries A, B and C and four goods 1, 2, 3 and 4. We assume that what is true for those three countries and those four goods must be true for more than three countries and for more than four goods. The model should describe the behavior of the trade balance of each of the three countries with the other two BCa.bc, BCb.ac and BCc.ab based on the exchange rates of the countries TCab, TCac, TCba, TCca, TCbc and TCcb. Using this mathematical model, we will have to demonstrate that there is a set of primary exchange rates TCab and TCac of equivalent universal equilibrium that equal all three trade balances to zero. That is to say:

$$BCa.bc = F1(TCab, TCac) = 0$$

$$BCb.ac = F2(TCba, TCbc) = 0$$

$$BCc.ab = F3(TCca, TCcb) = 0$$

Since $TCba = \frac{1}{TCab}$ , $TCca = \frac{1}{TCac}$ , $TCbc = \frac{TCac}{TCab}$ and $TCcb = \frac{TCab}{TCac}$ , then:

$$BCa.bc = F1(TCab, TCac) = 0$$

$$BCb.ac = F2\left(\frac{1}{TCab} , \frac{TCac}{TCab}\right) = 0$$

$$BCc.ab = F3\left(\frac{1}{TCac} , \frac{TCab}{TCac}\right) = 0$$

We need to specify the three functions F1(TCab, TCac), F2(TCba, TCbc) and F3(TCca, TCcb) that describe the behavior of the three trade balances of the three countries BCa.bc, BCb.ac and BCc.ab with relation to their respective exchange rates. We begin by defining the trade balance of country A with the other countries B and C, as being equal to the trade balance of country A with country B, plus the trade balance of country A with country C. That is:

1.1 BCa.bc = BCab + BCac

Where:

BCa.bc = Trade balance of country A with the other countries B and C
BCab = Trade balance of country A with country B
BCac = Trade balance of country A with country C

We represent the trade balance of country B with the other countries A and C as being equal to the trade balance of country B with country A, plus the trade balance of country B with country C. That is:

1.2 BCb.ac = BCba + BCbc

Wher:

BCb.ac = Trade balance of country B with other countries A and C
BCba = Trade balance of country B with country A
BCbc = Trade balance of country B with country C

Finally, we represent the trade balance of country C with the other countries A and B as being equal to the trade balance of country C with country A, plus the trade balance of country C with country B. That is:

1.3 BCc.ab = BCca + BCcb

Where:

BCc.ab = Trade balance of country C with other countries A and B
BCca = Trade balance of country C with country A
BCcb = Trade balance of country C with country B

With respect to equation 1.1, we postulate that the trade balances of country A with the other countries B and C, that is, BCab and BCac, are in turn composed of the trade balances of each of the four goods 1, 2, 3 and 4 of country A with country B and C respectively, as expressed in equations 1.4 and 1.5.

1.4 BCab = BCab1 + BCab2 + BCab3 + BCab4

1.5 BCac = BCac1 + BCac 2 + BCac3 + BCac4

Where:

BCabj = Trade balance of country A with country B corresponding to good j

BCacj = Trade balance of country A with country C corresponding to good j

A positive BCab1 trade balance of good 1 means that country A export good 1 to country B and if it is negative it import good 1. The same reasoning applies to BCab2, BCab3 and BCab4. In other words, the trade balance of a country A with country B, that is BCab, is equal to the sum of the trade balances of the four goods 1, 2, 3 and 4 of country A with country B. The same reasoning applies to the trade balance of country A with country C.

We postulate the following functional relationship for the trade balance of country A with country B corresponding to good 1:

$$2.1 \quad \mathrm{BCab1} = \left[\frac{(\mathrm{Pa1b}-\mathrm{Pa1})}{\left(\frac{\mathrm{Pa1b}+\mathrm{Pa1}}{2}\right)}\right](\mathrm{Dab1})(\mathrm{TMab1}) = \left[\frac{(\mathrm{Pb1}(\mathrm{TCab})-\mathrm{Pa1})}{\left(\frac{(\mathrm{Pb1}(\mathrm{TCab})+\mathrm{Pa1})}{2}\right)}\right](\mathrm{Dab1})(\mathrm{TMab1})$$

Where:

Pa1 = Price in Country A of good 1
Pa1b = Pb1(TCab) = Price in country A of good 1 from country B in currency of country A
Pb1 = Price in country B of good 1 in currency of country B
TCab = Exchange rate of country A with country B
Dab1 = Demand coefficient that represents the sensitivity of consumers to react to differences between local and foreign prices.
TMab1 = Market size of country A with country B with respect to good 1 expressed in currency of Country A As can be seen, function 2.1 tells us that a country A will import good 1 from country B, that is, the trade balance BCab1 of good 1 will be negative, if the price of that good in currency of country A, that is Pa1b = Pb1(TCab), is lower than the price of that good Pa1 produced in country A. The level of imports from country A corresponding to good 1 (negative value of BCab1) will depend, on the one hand, on the proportional difference between the price Pb1(TCab) of country B and the price Pa1 of country A expressed by the factor $\left[\frac{(Pb1*TCab-Pa1)}{\left(\frac{Pb1*TCab+Pa1}{2}\right)}\right]$ and, for on the other hand, of the demand coefficient expressed by the factor Dab1 multiplied by the size of its market expressed by the factor TMab1.

On the contrary, it can be seen that function 2.1 tells us that a country A will export good 1 to country B, if the price of that good Pa1 produced in country A is lower than the price of that good Pb1 produced in country B in currency of country A, that is Pb1(TCab). The level of exports from country A corresponding to good 1 (positive value of BCab1) will depend, on the one hand, on the proportional difference between the price Pb1(TCab) of country B and the price Pa1 of country A expressed by the factor $\left[\frac{(Pb1*TCab-Pa1)}{\left(\frac{Pb1*TCab+Pa1}{2}\right)}\right]$ and, on the other hand, the demand coefficient expressed by the factor Dab1 multiplied by the size of its market expressed by the factor TMab1.

Note that we can eliminate the demand coefficient Dab1 (or make it equal to 1) in equation 2.1, without altering the logical reasoning behind the formulas. That elimination simplifies the exposure of the analysis.

The above reasoning, applied to good 1 in the case of trade of country A with country B, would apply to all other goods and all other countries with which country A trades. Consequently, eliminating the coefficient Dab1 and generalizing the function 2.1 in the case of the other goods 2, 3 and 4 and the other countries (country C) we have that a trade balance in equilibrium of country A with all other countries B and C, that is, BCa.bc = 0, will imply that:

3.1 $$BCa.bc = BCab + BCac = (BCab1 + BCab2 + BCab3 + BCab4) + (BCac1 + BCac2 + BCac3 + BCac4) = 0$$

3.2 $$BCa.bc = \left\{\left[\frac{(Pb1*TCab-Pa1)}{\left(\frac{(Pb1*TCab+Pa1)}{2}\right)}\right](TMab1) + \left[\frac{(Pb2*TCab-Pa2)}{\left(\frac{(1Pb2*TCab+Pa2)}{2}\right)}\right](TMab2) + \left[\frac{(Pb3*TCab-Pa3)}{\left(\frac{(Pb3*TCab+Pa3)}{2}\right)}\right](TMab3) + \left[\frac{(Pb4*TCab-Pa4)}{\left(\frac{(Pb4*TCab+Pa4)}{2}\right)}\right](TMab4)\right\} + \left\{\left[\frac{(Pc1*TCac-Pa1)}{\left(\frac{(Pc1*TCac+Pa1)}{2}\right)}\right](TMac1) + \left[\frac{(Pc2*TCac-Pa2)}{\left(\frac{(Pc2*TCac+Pa2)}{2}\right)}\right](TMac2) + \left[\frac{(Pb3*TCac-Pa3)}{\left(\frac{(Pb3*TCac+Pa3)}{2}\right)}\right](TMac3) + \left[\frac{(Pc4*TCac-Pa4)}{\left(\frac{(Pc4*TCac+Pa4)}{2}\right)}\right](TMac4)\right\} = 0$$

The same reasoning would apply to the trade balance of country B with all other countries and country C with all other countries. That is, BCb.ac = 0 and BCc.ab = 0, will imply that:

3.3. $$BCb.ac = \left\{\left[\frac{(Pa1*TCba-Pb1)}{\left(\frac{Pa1*TCba+Pb1}{2}\right)}\right](TMba1) + \left[\frac{(Pa2*TCba-Pb2)}{\left(\frac{Pa2*TCba+Pb2}{2}\right)}\right](TMba2) + \left[\frac{(Pa3*TCba-Pb3)}{\left(\frac{Pa3*TCba+Pb3}{2}\right)}\right](TMba3) + \left[\frac{(Pa4*TCba-Pb4)}{\left(\frac{Pa4*TCba+Pb4}{2}\right)}\right](TMba4)\right\} + \left\{\left[\frac{(Pc1*TCbc-Pb1)}{\left(\frac{Pc1*TCbc+Pb1}{2}\right)}\right](TMbc1) + \left[\frac{(Pc2*TCbc-Pb2)}{\left(\frac{Pc2*TCbc+Pb2}{2}\right)}\right](TMbc2) + \left[\frac{(Pc3*TCbc-Pb3)}{\left(\frac{Pc3TCbc+Pb3}{2}\right)}\right](TMbc3) + \left[\frac{(Pc4*TCbc-Pb4)}{\left(\frac{Pc4*TCbc+Pb4}{2}\right)}\right](TMbc4)\right\} = 0$$

$$3.4.\quad BCc.ab = \left\{\left[\frac{(Pa1*TCca-Pc1)}{\left(\frac{Pa1*TCca+Pc1}{2}\right)}\right](TMca1) + \left[\frac{(Pa2*TCca-Pc2)}{\left(\frac{Pa2*TCca+Pc2}{2}\right)}\right](TMca2) + \left[\frac{(Pa3*TCca-Pc3)}{\left(\frac{Pa3*TCca+Pc3}{2}\right)}\right](TMca3) + \left[\frac{(Pa4*TCca-Pc4)}{\left(\frac{Pa4*TCca+Pc4}{2}\right)}\right](TMca4)\right\}$$
$$+ \left\{\left[\frac{(Pb1*TCcb-Pc1)}{\left(\frac{Pb1*TCcb+Pc1}{2}\right)}\right](TMcb1) + \left[\frac{(Pb2*TCcb-Pc2)}{\left(\frac{Pb2*TCcb+Pc2}{2}\right)}\right](TMcb2) + \left[\frac{(Pb3*TCcb-Pc3)}{\left(\frac{Pb3*TCcb+Pc3}{2}\right)}\right](TMcb3) + \left[\frac{(Pb4*TCcb-Pc4)}{\left(\frac{Pb4*TCcb+Pc4}{2}\right)}\right](TMcb4)\right\} = 0$$

What does the previous model tell us in equation 3.2? It tells us that, if we start from the assumption of a perfect competition market between foreign and domestic products, where goods of the same class are homogeneous and, therefore, are not differentiated by consumers, then country A will import good 1 from all other countries B and C, if their prices Pb1(TCab) and Pc1(TCac) are lower than the price of that same good produced locally Pa1. The level of importation of country A corresponding to good 1 from all other countries B and C, will depend on the one hand, on the proportional size of the difference between the prices of foreign countries and the local price $\left[\frac{(Pb1*TCab-Pa1)}{\left(\frac{(Pb1*TCab+Pa1)}{2}\right)}\right]$ and $\left[\frac{(Pc1*TCac-Pa1)}{\left(\frac{(Pc1*TCac+Pa1)}{2}\right)}\right]$, and , on the other hand, of the size of the market of good 1 between the countries TMab1 and TMac1.

The same reasoning applies to the other goods 2, 3 and 4.

How is the market size between country A and B corresponding to good 1 determined, that is, how is determined TMab1?

John Stuart Mill noted the role of the size of the smallest country in determining the terms of the exchange. In the model we have developed, the determination of the exchange rate between two countries TCab and, consequently, the terms of trade, are determined, not only by the proportional difference between the price of the foreign good Pb1*TCab and the national price Pa1, that is $\left[\frac{(Pb1*TCab-Pa1)}{\left(\frac{Pb1*TCab+Pa1}{2}\right)}\right]$, but also for the size of the market between the two countries TMab1. This market size in our model is determined in turn by the size of the country with the smallest market. We can assume that the size of the market in country A of good 1 is given by the value of production in country A of good 1, that is VPa1 = Pa1(Qa1). The same would apply to the size of the market in country B of good 1 which is given by the value of production in country B of good 1, that is VPb1 = Pb1*TCab (Qb1). Consequently, the size of the good 1 market between country A and B would be given by:

3.5 TMab1 = Min(VPa1, VPb1) = Min(Pa1(Qa1), Pb1*TCab(Qb1))

Note that the size of the market between country A and B corresponding to good 1 is expressed in currency of country A. Note also, that the size of the market between country A and B corresponding to good 1 must be equal to the size of the market between country B and A corresponding to good 1. That is to say:

3.6 TMab1 = TMba1= Min(Pa1(Qa1), Pb1*TCab(Qb1)

The same applies to the size of the market between country A and C corresponding to good 1 which has to be equal to the size of the market between country C and A corresponding to good 1. That is to say:

3.7 TMac1 = TMca1 = Min(Pa1(Qa1), Pc1*TCac(Qc1)

The same applies to the size of the market between country B and C corresponding to good 1 which has to be equal to the size of the market between country C and B corresponding to good 1. That is to say:

3.8 TMbc1 = TMcb1 = Min(Pb1*TCab(Qb1), Pc1*TCac(Qc1)

Note that all sizes of markets TMihj are expressed in currency of country A, that is US $. Note also, that the exchange rates TCba in equation 3.3 and TCca in equation 3.4 are counter-primary and the exchange rates TCbc in equation 3.3 and TCcb in equation 3.4 are non-primary. Consequently, as discussed in a previous section, in order to maintain the equivalence between the values of the currencies with respect to the primary exchange rates, they have to be defined in terms of the primary exchange rates as:

4.1. $\mathrm{TCba} = \frac{1}{\mathrm{TCab}}$

4.2. $\mathrm{TCca} = \frac{1}{\mathrm{TCac}}$

4.3. $\mathrm{TCbc} = \frac{\mathrm{TCac}}{\mathrm{TCab}}$

4.4. $\mathrm{TCcb} = \frac{\mathrm{TCab}}{\mathrm{TCac}}$

Therefore, applying 4.1, 4.2, 4.3 and 4.4 in equations 3.2, 3.3 and 3.4 would be expressed as:

5.1.
$$\mathrm{BCa.bc} = \left\{\left[\frac{(\mathrm{Pb1{*}TCab} - \mathrm{Pa1})}{\left(\frac{(\mathrm{Pb1{*}TCab} + \mathrm{Pa1})}{2}\right)}\right](\mathrm{TMab1}) + \left[\frac{(\mathrm{Pb2{*}TCab} - \mathrm{Pa2})}{\left(\frac{(\mathrm{Pb2{*}TCab} + \mathrm{Pa2})}{2}\right)}\right](\mathrm{TMab2}) + \left[\frac{(\mathrm{Pb3{*}TCab} - \mathrm{Pa3})}{\left(\frac{(\mathrm{Pb3{*}TCab} + \mathrm{Pa3})}{2}\right)}\right](\mathrm{TMab3}) + \left[\frac{(\mathrm{Pb4{*}TCab} - \mathrm{Pa4})}{\left(\frac{(\mathrm{Pb4{*}TCab} + \mathrm{Pa4})}{2}\right)}\right](\mathrm{TMab4})\right\} +$$
$$\left\{\left[\frac{(\mathrm{Pc1{*}TCac} - \mathrm{Pa1})}{\left(\frac{(\mathrm{Pc1{*}TCac} + \mathrm{Pa1})}{2}\right)}\right](\mathrm{TMac1}) + \left[\frac{(\mathrm{Pc2{*}TCac} - \mathrm{Pa2})}{\left(\frac{(\mathrm{Pc2{*}TCac} + \mathrm{Pa2})}{2}\right)}\right](\mathrm{TMac2}) + \left[\frac{(\mathrm{Pb3{*}TCac} - \mathrm{Pa3})}{\left(\frac{(\mathrm{Pb3{*}TCac} + \mathrm{Pa3})}{2}\right)}\right](\mathrm{TMac3}) + \left[\frac{(\mathrm{Pc4{*}TCac} - \mathrm{Pa4})}{\left(\frac{(\mathrm{Pc4{*}TCac} + \mathrm{Pa4})}{2}\right)}\right](\mathrm{TMac4})\right\} = 0$$

5.2.
$$\mathrm{BCb.ac} = \left\{\left[\frac{\left((\mathrm{Pa1}{*}\frac{1}{\mathrm{TCab}} - \mathrm{Pb1}\right)}{\left(\frac{(\mathrm{Pa1}{*}\frac{1}{\mathrm{TCab}} + \mathrm{Pb1})}{2}\right)}\right](\mathrm{TMab1}) + \left[\frac{\left(\mathrm{Pa2}{*}\frac{1}{\mathrm{TCab}} - \mathrm{Pb2}\right)}{\left(\frac{(\mathrm{Pa2}{*}\frac{1}{\mathrm{TCab}} + \mathrm{Pb2})}{2}\right)}\right](\mathrm{TMab2}) + \left[\frac{\left(\mathrm{Pa3}{*}\frac{1}{\mathrm{TCab}} - \mathrm{Pb3}\right)}{\left(\frac{(\mathrm{Pa3}{*}\frac{1}{\mathrm{TCab}} + \mathrm{Pb3})}{2}\right)}\right](\mathrm{TMab3}) + \left[\frac{\left((\mathrm{Pa4}{*}\frac{1}{\mathrm{TCab}} - \mathrm{Pb4}\right)}{\left(\frac{(\mathrm{Pa4}{*}\frac{1}{\mathrm{TCab}}\mathrm{Pb4}}{2}\right)}\right](\mathrm{TMab4})\right\} +$$
$$\left\{\left[\frac{\left(\mathrm{Pc1}{*}\frac{\mathrm{TCac}}{\mathrm{TCab}} - \mathrm{Pb1}\right)}{\left(\frac{(\mathrm{Pc1}{*}\frac{\mathrm{TCac}}{\mathrm{TCab}} + \mathrm{Pb1})}{2}\right)}\right](\mathrm{TMbc1}) + \left[\frac{\left(\mathrm{Pc2}{*}\frac{\mathrm{TCac}}{\mathrm{TCab}} - \mathrm{Pb2}\right)}{\left(\frac{(\mathrm{Pc2}{*}\frac{\mathrm{TCac}}{\mathrm{TCab}} + \mathrm{Pb2})}{2}\right)}\right](\mathrm{TMbc2}) + \left[\frac{\left((\mathrm{Pc3}{*}\frac{\mathrm{TCac}}{\mathrm{TCab}} - \mathrm{Pb3}\right)}{\left(\frac{(\mathrm{Pc3}\frac{\mathrm{TCac}}{\mathrm{TCab}} + \mathrm{Pb3})}{2}\right)}\right](\mathrm{TMbc3}) + \left[\frac{\left(\mathrm{Pc4}{*}\frac{\mathrm{TCac}}{\mathrm{TCab}} - \mathrm{Pb4}\right)}{\left(\frac{(\mathrm{Pc4}{*}\frac{\mathrm{TCac}}{\mathrm{TCab}} + \mathrm{Pb4})}{2}\right)}\right](\mathrm{TMbc4})\right\} = 0$$

5.3.
$$\mathrm{BCc.ab} = \left\{\left[\frac{\left(\mathrm{Pa1}{*}\frac{1}{\mathrm{TCac}} - \mathrm{Pc1}\right)}{\left(\frac{(\mathrm{Pa1}{*}\frac{1}{\mathrm{TCac}} + \mathrm{Pc1})}{2}\right)}\right](\mathrm{TMac1}) + \left[\frac{\left(\mathrm{Pa2}{*}\frac{1}{\mathrm{TCac}} - \mathrm{Pc2}\right)}{\left(\frac{(\mathrm{Pa2}{*}\frac{1}{\mathrm{TCac}} + \mathrm{Pc2})}{2}\right)}\right](\mathrm{TMac2}) + \left[\frac{\left(\mathrm{Pa3}{*}\frac{1}{\mathrm{TCac}} - \mathrm{Pc3}\right)}{\left(\frac{(\mathrm{Pa3}{*}\frac{1}{\mathrm{TCac}} + \mathrm{Pc3})}{2}\right)}\right](\mathrm{TMac3}) + \left[\frac{\left((\mathrm{Pa4}{*}\frac{1}{\mathrm{TCac}} - \mathrm{Pc4}\right)}{\left(\frac{(\mathrm{Pa4}{*}\frac{1}{\mathrm{TCac}} + \mathrm{Pc4})}{2}\right)}\right](\mathrm{TMac4})\right\} +$$
$$\left\{\left[\frac{\left(\mathrm{Pb1}{*}\frac{\mathrm{TCab}}{\mathrm{TCac}} - \mathrm{Pc1}\right)}{\left(\frac{(\mathrm{Pb1}{*}\frac{\mathrm{TCab}}{\mathrm{TCac}} + \mathrm{Pc1})}{2}\right)}\right](\mathrm{TMbc1}) + \left[\frac{\left(\mathrm{Pb2}{*}\frac{\mathrm{TCab}}{\mathrm{TCac}} - \mathrm{Pc2}\right)}{\left(\frac{(\mathrm{Pb2}{*}\frac{\mathrm{TCab}}{\mathrm{TCac}} + \mathrm{Pc2}}{2}\right)}\right](\mathrm{TMbc2}) + \left[\frac{\left(\mathrm{Pb3}{*}\frac{\mathrm{TCab}}{\mathrm{TCac}} - \mathrm{Pc3}\right)}{\left(\frac{(\mathrm{Pb3}{*}\frac{\mathrm{TCab}}{\mathrm{TCac}} + \mathrm{Pc3})}{2}\right)}\right](\mathrm{TMbc3}) + \left[\frac{\left(\mathrm{Pb4}{*}\frac{\mathrm{TCab}}{\mathrm{TCac}} - \mathrm{Pc4}\right)}{\left(\frac{(\mathrm{Pb4}{*}\frac{\mathrm{TCab}}{\mathrm{TCac}} + \mathrm{Pc4}}{2}\right)}\right](\mathrm{TMbc4})\right\} = 0$$

As will be recalled, all sizes of the markets TMahj and, consequently, all trade balances Ba.ba, BCb.ab and BCc.ab are expressed in US $.

As you can see, we have a model of three equations and two unknowns that are the two primary exchange rates TCab and TCac. We want to determine the equivalent universal primary exchange rates $\overline{\text{TC}}$ab and $\overline{\text{TC}}$ac that balance the trade balances of the three countries simultaneously. That is, determine the values of TCab and TCac that equal to zero equations 5.1, 5.2 and 5.3.

To find the values of TCab and TCac that balance the three trade balances simultaneously, we can use two procedures.

1. Solution of the model by trial and error. The first procedure consists in introducing different values for the set of the two primary exchange rates TCab and TCac in equations 5.1, 5.2 and 5.3, simulating what happens when we float both currencies B and C with respect to currency A. This trial and error would continue until successive approximations obtain the primary exchange rates TCab and TCac that in Equations 5.1, 5.2 and 5.3 balance the trade balances of the three countries, BCa.bc = 0, BCb.ab = 0 and BCc.ab = 0.

2. Solution of the model by algebraic arrangement. The second procedure consists in solving the system of equations by clearing with respect to the primary exchange rates TCab and TCac to obtain the values that balance the trade balances of the three countries BCa.bc = 0, BCb.ab = 0 and BCc.ab = 0

### Model Solution by Trial and Error

Let's start by using the first procedure to determine the equivalent universal equilibrium primary exchange rates. Assume the following values for the prices of the four goods 1, 2, 3 and 4 in the three countries A, B and C presented in Table 1 and assume the following market sizes for the four goods 1, 2, 3 and 4 in the three countries A, B and C presented in Table 2:

Tabla 1

| | | |
|---|---|---|
| Pa1 = 23.08 | Pb1 = 729.93 | Pc1 = 48.61 |
| Pa2 = 15.38 | Pb2 = 547.45 | Pc2 = 34.72 |
| Pa3 = 7.69 | Pb3 = 364.96 | Pc3 = 20.83 |
| Pa4 = 3.85 | Pb4 = 182.48 | Pc4 = 10.42 |

Tabla 2

| | | |
|---|---|---|
| TMab1 = TMba1 = 137.00 | TMac1 = TMca1 = 43.64 | TMbc1 = TMcb1 = 43.64 |
| TMab2 = TMba2 = 205.50 | TMac2 = TMca2 = 65.45 | TMbc2 = TMcb2 = 65.45 |
| TMab3 = TMba3 = 205.50 | TMac3 = TMca3 = 65.45 | TMbc3 = TMcb3 = 65.45 |
| TMab4 = TMba4 = 137.00 | TMac4 = TMca4 = 43.64 | TMbc4 = TMcb4 = 43.64 |

Assume the following initial values for the two primary exchange rates TCab and TCac between countries A, B and C based on the quotient of the sum of the prices of goods in each country that was presented in Table 1:

$$6.1. \quad TCab = \frac{(Pa1 + Pa2 + Pa3 + Pa4)}{(Pb1 + Pb2 + Pb3 + Pb4)} = \frac{(23.08 + 15.38 + 7.69 + 3.85)}{(729.93 + 547.45 + 364.96 + 182.48)} = 0.0274$$

$$6.2. \quad TCac = \frac{(Pa1 + Pa2 + Pa3 + Pa4)}{(Pc1 + Pc2 + Pc3 + Pc4)} = \frac{(23.08 + 15.38 + 7.69 + 3.85)}{(48.61 + 34.72 + 20.83 + 10.42)} = 0.436364$$

If we introduce the 12 prices of Table 1, the 12 market sizes of Table 2 and the two initial primary exchange rates of formulas 6.1 and 6.2 in the model described in the three equations 5.1, 5.2 and 5.3 we obtain the result that It is presented in the Appendix Spreadsheet 1, Image Section 1.1.

As can be seen from Image Section 1.1, the model shows a surplus of 78.08 in the trade balance of country A, a deficit of -71.74 in the trade balance of country B and a deficit of -6.34 in the trade balance of country C .

The Appendix Spreadsheet 1, Image Section 1.1 was obtained from the Spreadsheet 1 shown below and where the international trade model described in equations 5.1, 5.2 and 5.3 was introduced, in conjunction with the initial primary exchange rates shown in equations 6.1 and 6.2.

Spreadsheet 1

Microsoft
Excel Worksheet

In the Appendix Spreadsheet 1, in addition to Section 1.1, six other sections are presented.

In the Appendix Spreadsheet 1, Image Section 1.1, the trade balances of the three countries are presented. The first trade balance (equation 5.1), corresponding to country A with all other countries, consists of 8 terms BCa.bc = BCab1 + BCab2 + BCab3 + BCab4 + BCac1 + BCac2 + BCac3 + BCac4. These eight terms are the sum of the 4 trade balances of country A with country B for each of the four goods (first four terms of equation 5.1: BCab1 + BCab2 + BCab3 + BCab4) plus the sum of the 4 trade balances from country A with country C for each of the four goods (last four terms of equation 5.1: BCac1 + BCac2 + BCac3 + BCac4). In the Appendix Spreadsheet 1, Image Section 1.2 are presented the 8 formulas that describe these 8 trade balances that we indicated in Section 1.1 corresponding to equation 5.1. The 8 values that appear in Section 1.1 and that define Equation 5.1 are obtained from the calculation made from the 8 formulas in Section 1.2.

The prices Paj, the sizes of markets TMabj, TMacj and TMbcj and all other values used to estimate the 8 trade balances that appear in Section 1.2 are presented just below the formulas that define trade balances. These values constitute the model parameters that are extracted in turn from Section 1.5: Model Parameters and which are shown in the Appendix Spreadsheet 1 Image:

Section 1.5. Finally, the initial values of the variables of the model, that is, the primary exchange rate TCab and TCac in equation 5.1 of Section 1.1 are given by the formulas presented in Section 1.7. The initial values of Section 1.7 are entered in Section 1.6 because it is from Section 1.6 that the values assumed by the primary exchange rates TCab and TCac that are entered in equation 5.1 are read.

The above reasoning about the relationship between equation 5.1 in Section 1.1 and Section 1.2, Section 1.5 and Section 1.7 applies equally to the case of the other two equations 5.2 and 5.3.

With respect to Section 1.2 we make the following observation. The determination of the price in country A of good 1 is given by the formula:

$$Pa1 = Wa*Za1 + Ga1$$

According to the model of the comparative advantages of Ricardo and Mill, the prices of goods are determined by labor productivity and the salary Wa1*Za1. This assumption ignores the income of the entrepreneur factor Ga1 and is a simplification made with the purpose of facilitating the analysis. We make the price of the goods functions, both of the costs of the labor factor and the entrepreneur factor. As can be seen in Section 1.5, to determine the income of the labor factor or salary Wa and the income or profit or gain of the entrepreneur factor Ga1 in the determination of the price, we start from the assumption of an average margin profit or gain MGa in the economy of country A. That average margin profit in the economy of country A is used to determine both the income of the labor or wage Wa production factor and the income of the entrepreneur or profit Gai factor. In the Image: Section 1.5 Model Parameters, the productivity of work Zij is presented for each country i and for each good j. The productivity of work Zij indicates the work units that are used in the country i in the production of one unit of good j. Therefore, the price in country i of good j, that is to say Pij is equal to the cost of the production factor work in the production of one unit of good j and that is equal to Wi(Zij), plus the cost of the entrepreneur factor for each unit of the good produced and that is equal to Gij. As can be seen in Section 1.5, the wage Wi level and the profit Gij in each country i and for each good j is a function of the money supply of each country OMi, the average margin of profit in the economy of each country MGi , the size of the population of each country Ni, the proportion of the labor force employed in the production of each good in each country FTji and the level of production of each country for each good Qij. That is, Wa = (OMa(1-MGa))/Na and Ga1 = (OMa(MGa)(FTa1))/Qa1. These are the model parameters used to determine the prices of the four goods in the three countries shown in Table 1.

The Image Section 1.6 shows the exchange rates used in the model. Section 1.7 presents the initial exchange rates used in Section 1.6. These initial exchange rates are obtained from the calculation of the quotients between the sum of the prices of the goods of each country. The first and second rows of these initial exchange rates of Section 1.7 are those presented in equations 6.1 and 6.2 some paragraphs back. These ratios are used as initial values of the exchange rates in Section 1.6. From these initial values, the two primary exchange rates TCab and TCac (first and second row in Section 1.6) are changed, looking by trial and error the two primary exchange rates that generate the equivalent universal equilibrium in the trade balances of the three countries presented in Section 1.1 (equations 5.1, 5.2 and 5.3).

Suppose we put to float the currencies of countries B and C with respect to the currency of country A. It is expected that the successive adjustments of the primary exchange rates TCab and TCac produced by the floating of the currencies, will generate the equivalent universal equilibrium primary exchange rates. We can simulate this flotation process of currencies by gradually modifying primary exchange rates, trying to reduce deficits and surpluses in the three trade balances BCa.bc, BCb.ac and BCc.ab that appear in Image Section 1.1 of the Appendix Spreadsheet 1 (equations 5.1, 5.2 and 5.3). In this way, by trial and error and successive approximations, we obtain the equivalent universal equilibrium primary exchange rates shown below:

Equivalent Universal Equilibrium Primary Exchange Rates

7.1. TCab = 0.0249159
7.2. TCac = 0.410078

When we introduce these two primary exchange rates in Section 1.6 of Worksheet 1 and, as a consequence, in the three equations of model 5.1, 5.2 and 5.3 corresponding to Section 1.1 and sections 1.2, 1.3 and 1.4, it is generated the result presented in the Appendix Spreadsheet 2 Image Section 2.1.

As can be seen, the trade balances of the three countries BCa.bc, BCb.ac and BCc.ab are simultaneously balanced at zero.

This result shows that, for a common market in more than two countries, there is a set of primary exchange rates that balance the trade balances of all countries that trade with each other.

The Image Section 2.1 of the Appendix Spreadsheet 2 was obtained from the Spreadsheet 2 shown below. The Spreadsheet 2 is the result of entering in the Spreadsheet 1 the equivalent universal equilibrium primary exchange rates shown in equations 7.1 and 7.2.

Spreadsheet 2

Microsoft
Excel Worksheet

The previous result allows us to suggest that, in the case of the common market of the American hemisphere, it is feasible to achieve the equilibrium of all trade balances by letting the 21 primary exchange rates be obtained by freely floating the 21 currencies with the reference currency (the dollar). All other 441 exchange rates between countries, which we have called counter-primary and non-primary, would be determined by the algebraic formulas explained above. As we indicated, these would be the 462 universal equilibrium exchange rates that maintain the equivalence between the values of the 21 currencies with the dollar.

This free flotation of the 21 currencies must be carried out without capital flows in the balance of payments being able to affect the determination of the equivalent universal equilibrium primary exchange rates. For this, it would be necessary to end the policy of maintaining free capital mobility. In other words, it would be necessary to let the free flotation of the 21 currencies with the dollar determine the equivalent universal equilibrium primary exchange rates, while maintaining strict regulation of capital movements so that they do not affect the determination of the primary exchange rate with the dollar. As you can see these are measures completely contrary to those established in Bretton Woods.

### MODEL SOLUTION BY ALGEBRAIC ARRANGEMENT

In the previous topic we determined by trial and error process the equivalent universal equilibrium primary exchange rates that were presented in 7.1 and 7.2. The second way in which we can determine the equivalent universal equilibrium primary exchange rates is by algebraic rearrangement, clearing the model presented in equations 5.1, 5.2 and 5.3 with respect to the primary exchange rate variables TCab and TCac. We have tried to do that but without success. If in equation 5.2 we multiply the first four terms by $\frac{\text{TCab}}{\text{TCab}} = 1$, we find that:

$$\frac{\text{TCab}}{\text{TCab}}\left\{\left[\frac{\left((\text{Pa1}*\frac{1}{\text{TCab}} - \text{Pb1}\right)}{\left(\frac{(\text{Pa1}*\frac{1}{\text{TCab}} + \text{Pb1})}{2}\right)}\right](\text{TMab1}) + \left[\frac{\left(\text{Pa2}*\frac{1}{\text{TCab}} - \text{Pb2}\right)}{\left(\frac{(\text{Pa2}*\frac{1}{\text{TCab}} + \text{Pb2})}{2}\right)}\right](\text{TMab2}) + \left[\frac{\left(\text{Pa3}*\frac{1}{\text{TCab}} - \text{Pb3}\right)}{\left(\frac{(\text{Pa3}*\frac{1}{\text{TCab}} + \text{Pb3})}{2}\right)}\right](\text{TMab3}) + \left[\frac{\left((\text{Pa4}*\frac{1}{\text{TCab}} - \text{Pb4}\right)}{\left(\frac{(\text{Pa4}*\frac{1}{\text{TCab}}\text{Pb4}}{2}\right)}\right](\text{TMab4})\right\} =$$
$$-\left\{\left[\frac{(\text{Pb1}*\text{TCab} - \text{Pa1})}{\left(\frac{(\text{Pb1}*\text{TCab} + \text{Pa1})}{2}\right)}\right](\text{TMab1}) + \left[\frac{(\text{Pb2}*\text{TCab} - \text{Pa2})}{\left(\frac{(\text{Pb2}*\text{TCab} + \text{Pa2})}{2}\right)}\right](\text{TMab2}) + \left[\frac{(\text{Pb3}*\text{TCab} - \text{Pa3})}{\left(\frac{(\text{Pb3}*\text{TCab} + \text{Pa3})}{2}\right)}\right](\text{TMab3}) + \left[\frac{(\text{Pb4}*\text{TCab} - \text{Pa4})}{\left(\frac{(\text{Pb4}*\text{TCab} + \text{Pa4})}{2}\right)}\right](\text{TMab4})\right\}$$

That is, the sum of the first four terms on the left in equation 5.2 is equal to the negative of the sum of the first four terms on the left in equation 5.1. This is logical if we visualize that imports of country A from Country B are equal to exports of country B to country A and vice versa. Therefore, we can add equation 5.2 to equation 5.1 and that sum of terms would be eliminated. We can do the same by multiplying by $\frac{\text{TCac}}{\text{TCac}} = 1$ the sum of the first four terms on the left in equation 5.3, which is equal to the negative of the sum of the last four terms on the right in the equation 5.1. Therefore, we can add equation 5.3 to equation 5.1 and that sum of terms would be eliminated. Despite these simplifications we could not clear the model with respect to the variables TCab and TCac. We understand that we do not have the mathematical skills to carry out this work, which is why we leave it raised, waiting for some colleague to generate the solution.

## A MATHEMATICAL MODEL ON THE THEORY OF COMPARATIVE ADVANTAGES IN THE CASE OF MORE THAN TWO COUNTRIES AND A MONOPOLISTIC COMPETITION MARKET

In order to introduce the assumption of a monopolistic competition market into the model, we need to modify equations 5.1, 5.2 and 5.3 that define the model. As we saw these functions are of a continuous nature.

In order for the trade balances of the four goods between two countries BCabj, BCacj and BCbcj to be continuous functions, we have had to start from the assumption of a perfect competition market in which foreign goods are homogeneous in relation to national goods, although differentiable of the other foreign goods of the same class that compete in the market of each country. That is, consumers do not have preferences between the foreign and national goods, although they do have preferences between two foreign goods. That is what is implied in function 3.2 that describes the trade balance of country A with countries B and C corresponding to good 1. The same reasoning applies to the other two functions 3.3 and 3.4.

In function 7.3 below we reproduce the terms of function 3.2 that we are interested in analyzing:

$$7.3 \qquad BCa.bc1 = BCab1 + BCac1 = \left[\frac{(Pb1(TCab) - Pa1)}{\left(\frac{(Pb1(TCab)+Pa1)}{2}\right)}\right](TMab1) + \left[\frac{(Pc1(TCac) - Pa1)}{\left(\frac{(Pc1(TCac)+Pa1)}{2}\right)}\right](TMac1)$$

The second member of the function 7.3 assumes that the consumer has preferences for the good 1 of country C independent of his preference for the good 1 of country B. This is an unrealistic assumption. Assume that the consumer does not distinguish between the foreign good and the national good, but that it does between the two foreign goods of the same type, it is not a very realistic assumption, but, nevertheless, it allows to express the model in the form of continuous functions .

In addition to the previous approach, the assumption of perfect competition is unrealistic, since most of the goods markets are of monopolistic competition, where consumers have preferences among goods of the same class.

So that we can introduce in the model on international trade that we have developed the Paul Krugman's ideas on the effect of consumer tastes and preferences as determinants of international trade and also introduce into the model the effect of tariffs as determinants of international trade, it is necessary to modify the equations of continuous type to adopt mathematical functions of conditional type that are not continuous.

In the field of social sciences, unlike in the field of physical sciences, the behavior of phenomena does not seem to be better described by continuous mathematical functions. On the contrary, mathematics based on continuous functions so perfectly describe physical phenomena that it has been suggested that more than an invention to describe physical relationships is a discovery resulting from man's sensory experience of the physical world. Unlike the field of physical sciences, in the field of social sciences phenomena do not seem to be described by continuous mathematical functions. It is a more chaotic world, where the sequence of behaviors is not continuous, but conditional. That is, if X occurs, then Y will occur, or otherwise Z.

In keeping with the above, we will have to modify the continuous equation 7.3. This equation expresses the trade balance of country A with countries B and C with respect to good 1, that is BCa.bc1 = BCab1 + BCac1, as continuous functions of the exchange rates TCab and TCac. We need to modify that equation to introduce conditional type equations that treat the trade balance of a good BCabj as a function of the difference between exports EXajb of a good j and its imports IMajb. That is: BCabj = (EXajb - IMajb). Consequently, the first term of the right member of the continuing function 7.3 in the previous model is replaced by conditional function 8.1 below:

8.1 $\mathrm{BCab1} = \mathrm{EXa1b} - \mathrm{IMa1b}$

Where:

8.2 EXa1b = IMb1a = Exports of country A of good 1 to country B is equal to import of country B of good 1 from country A.

8.3 $\mathrm{IMa1b} = \mathrm{IF}\left\{\left[\frac{(\mathrm{Pa1}-(1-\mathrm{PR1ab})(1+\mathrm{AR1a})\mathrm{Pb1}*\mathrm{TCab})}{\left(\frac{(\mathrm{Pa1}+(1-\mathrm{PR1ab})(1+\mathrm{AR1a})\mathrm{Pb1}*\mathrm{TCab})}{2}\right)}\right] > 0\,,\left[\frac{(\mathrm{Pa1}-(1-\mathrm{PR1ab})(1+\mathrm{AR1a})\mathrm{Pb1}*\mathrm{TCab})}{\left(\frac{(\mathrm{Pa1}+(1-\mathrm{PR1ab})(1+\mathrm{AR1a})\mathrm{Pb1}*\mathrm{TCab})}{2}\right)}\right]\mathrm{TMab1}\,,0\right\} =$ Imports of country A of good 1 from country B

Therefore, function 8.1 would remain as conditional function 8.4 below:

8.4 $\mathrm{BCab1} = \mathrm{EXa1b} - \mathrm{IMa1b} = \mathrm{IMb1a} - \mathrm{IF}\left\{\left[\frac{(\mathrm{Pa1}-(1-\mathrm{PR1ab})(1+\mathrm{AR1a})\mathrm{Pb1}*\mathrm{TCab})}{\left(\frac{(\mathrm{Pa1}+(1-\mathrm{PR1ab})(1+\mathrm{AR1a})\mathrm{Pb1}*\mathrm{TCab})}{2}\right)}\right] > 0\,,\left[\frac{(\mathrm{Pa1}-(1-\mathrm{PR1ab})(1+\mathrm{AR1a})\mathrm{Pb1}*\mathrm{TCab})}{\left(\frac{(\mathrm{Pa1}+(1-\mathrm{PR1ab})(1+\mathrm{AR1a})\mathrm{Pb1}*\mathrm{TCab})}{2}\right)}\right]\mathrm{TMab1}\,,0\right\}$

Function 8.2 tells us that exports EXa1b of country A of good 1 to country B must be equal to the imports IMb1a of country B of good 1 from country A.

Conditional function 8.3 tells us that, if the local price of good 1, that is Pa1, is greater than the foreign price (1-PR1ab)(1 + AR1a)Pb1*TCab, then country A will import good 1 of the country B for the amount of $\left[\frac{(\mathrm{Pa1}-(1-\mathrm{PR1ab})(1+\mathrm{AR1a})\mathrm{Pb1}*\mathrm{TCab})}{\left(\frac{(\mathrm{Pa1}+(1-\mathrm{PR1ab})(1+\mathrm{AR1a})\mathrm{Pb1}*\mathrm{TCab})}{2}\right)}\right]\mathrm{TMab1}$. Otherwise its import will be zero. As can be seen, the level of importation in country A of good 1 from country B will depend on the proportional difference between the price of good 1 produced nationally Pa1 and the price in local currency of the good 1 produced abroad (1-PR1ab)(1 + AR1a)Pb1*TCab, multiplied by the market size TMab1. But this equation is conditional. That is, it applies in case the difference shows a positive value. When this is not the case and the difference show a negative value, the import level will be zero.

By generalizing these results, we have that the continuous equations 5.1, 5.2 and 5.3 in the previous model would be replaced by the conditional equations 8.7, 8.8 and 8.9 below:

$$8.7.\ \mathrm{BCa.bc} = \left(\mathrm{IMb1a} - \mathrm{IF}\left\{\left[\frac{(\mathrm{Pa1} - (1 - \mathrm{PR1ab})(1 + \mathrm{AR1a})\mathrm{Pb1} * \mathrm{TCab})}{\left(\frac{(\mathrm{Pa1} + (1 - \mathrm{PR1ab})(1 + \mathrm{AR1a})\mathrm{Pb1} * \mathrm{TCab})}{2}\right)}\right] > 0, \quad \left[\frac{(\mathrm{Pa1} - (1 - \mathrm{PR1ab})(1 + \mathrm{AR1a})\mathrm{Pb1} * \mathrm{TCab})}{\left(\frac{(\mathrm{Pa1} + (1 - \mathrm{PR1ab})(1 + \mathrm{AR1a})\mathrm{Pb1} * \mathrm{TCab})}{2}\right)}\right] TMab1, 0\right\}\right)$$

$$+ \left(\mathrm{IMb2a} - \mathrm{IF}\left\{\left[\frac{(\mathrm{Pa2} - (1 - \mathrm{PR2ab})(1 + \mathrm{AR2a})\mathrm{Pb2} * \mathrm{TCab})}{\left(\frac{(\mathrm{Pa2} + (1 - \mathrm{PR2ab})(1 + \mathrm{AR2a})\mathrm{Pb2} * \mathrm{TCab})}{2}\right)}\right] > 0, \quad \left[\frac{(\mathrm{Pa2} - (1 - \mathrm{PR2ab})(1 + \mathrm{AR2a})\mathrm{Pb2} * \mathrm{TCab})}{\left(\frac{(\mathrm{Pa2} + (1 - \mathrm{PR2ab})(1 + \mathrm{AR2a})\mathrm{Pb2} * \mathrm{TCab})}{2}\right)}\right] TMab2, 0\right\}\right)$$

$$+ \left(\mathrm{IMb3a} - \mathrm{IF}\left\{\left[\frac{(\mathrm{Pa3} - (1 - \mathrm{PR3ab})(1 + \mathrm{AR3a})\mathrm{Pb3} * \mathrm{TCab})}{\left(\frac{(\mathrm{Pa3} + (1 - \mathrm{PR3ab})(1 + \mathrm{AR3a})\mathrm{Pb3} * \mathrm{TCab})}{2}\right)}\right] > 0, \quad \left[\frac{(\mathrm{Pa3} - (1 - \mathrm{PR3ab})(1 + \mathrm{AR3a})\mathrm{Pb3} * \mathrm{TCab})}{\left(\frac{(\mathrm{Pa3} + (1 - \mathrm{PR3ab})(1 + \mathrm{AR3a})\mathrm{Pb3} * \mathrm{TCab})}{2}\right)}\right] TMab3, 0\right\}\right)$$

$$+ \left(\mathrm{IMb4a} - \mathrm{IF}\left\{\left[\frac{(\mathrm{Pa4} - (1 - \mathrm{PR4ab})(1 + \mathrm{AR4a})\mathrm{Pb4} * \mathrm{TCab})}{\left(\frac{(\mathrm{Pa4} + (1 - \mathrm{PR4ab})(1 + \mathrm{AR4a})\mathrm{Pb4} * \mathrm{TCab})}{2}\right)}\right] > 0, \quad \left[\frac{(\mathrm{Pa4} - (1 - \mathrm{PR4ab})(1 + \mathrm{AR4a})\mathrm{Pb4} * \mathrm{TCab})}{\left(\frac{(\mathrm{Pa4} + (1 - \mathrm{PR4ab})(1 + \mathrm{AR4a})\mathrm{Pb4} * \mathrm{TCab})}{2}\right)}\right] TMab4, 0\right\}\right)$$

$$+ \left(\mathrm{IMc1a} - \mathrm{IF}\left\{\left[\frac{(\mathrm{Pa1} - (1 - \mathrm{PR1ac})(1 + \mathrm{AR1a})\mathrm{Pc1} * \mathrm{TCac})}{\left(\frac{(\mathrm{Pa1} + (1 - \mathrm{PR1ac})(1 + \mathrm{AR1a})\mathrm{Pc1} * \mathrm{TCac})}{2}\right)}\right] > 0, \quad \left[\frac{(\mathrm{Pa1} - (1 - \mathrm{PR1ac})(1 + \mathrm{AR1a})\mathrm{Pc1} * \mathrm{TCac})}{\left(\frac{(\mathrm{Pa1} + (1 - \mathrm{PR1ac})(1 + \mathrm{AR1a})\mathrm{Pc1} * \mathrm{TCac})}{2}\right)}\right] TMac1, 0\right\}\right)$$

$$+ \left(\mathrm{IMc2a} - \mathrm{IF}\left\{\left[\frac{(\mathrm{Pa2} - (1 - \mathrm{PR2ac})(1 + \mathrm{AR2a})\mathrm{Pc2} * \mathrm{TCac})}{\left(\frac{(\mathrm{Pa2} + (1 - \mathrm{PR2ac})(1 + \mathrm{AR2a})\mathrm{Pc2} * \mathrm{TCac})}{2}\right)}\right] > 0, \quad \left[\frac{(\mathrm{Pa2} - (1 - \mathrm{PR2ac})(1 + \mathrm{AR2a})\mathrm{Pc2} * \mathrm{TCac})}{\left(\frac{(\mathrm{Pa2} + (1 - \mathrm{PR2ac})(1 + \mathrm{AR2a})\mathrm{Pc2} * \mathrm{TCac})}{2}\right)}\right] TMac2, 0\right\}\right)$$

$$+ \left(\mathrm{IMc3a} - \mathrm{IF}\left\{\left[\frac{(\mathrm{Pa3} - (1 - \mathrm{PR3ac})(1 + \mathrm{AR3a})\mathrm{Pc3} * \mathrm{TCac})}{\left(\frac{(\mathrm{Pa3} + (1 - \mathrm{PR3ac})(1 + \mathrm{AR3a})\mathrm{Pc3} * \mathrm{TCac})}{2}\right)}\right] > 0, \quad \left[\frac{(\mathrm{Pa3} - (1 - \mathrm{PR3ac})(1 + \mathrm{AR3a})\mathrm{Pc3} * \mathrm{TCac})}{\left(\frac{(\mathrm{Pa3} + (1 - \mathrm{PR3ac})(1 + \mathrm{AR3a})\mathrm{Pc3} * \mathrm{TCac})}{2}\right)}\right] TMac3, 0\right\}\right)$$

$$+ \left(\mathrm{IMc4a} - \mathrm{IF}\left\{\left[\frac{(\mathrm{Pa4} - (1 - \mathrm{PR4ac})(1 + \mathrm{AR4a})\mathrm{Pc4} * \mathrm{TCac})}{\left(\frac{(\mathrm{Pa4} + (1 - \mathrm{PR4ac})(1 + \mathrm{AR4a})\mathrm{Pc4} * \mathrm{TCac})}{2}\right)}\right] > 0, \quad \left[\frac{(\mathrm{Pa4} - (1 - \mathrm{PR4ac})(1 + \mathrm{AR4a})\mathrm{Pc4} * \mathrm{TCac})}{\left(\frac{(\mathrm{Pa4} + (1 - \mathrm{PR4ac})(1 + \mathrm{AR4a})\mathrm{Pc4} * \mathrm{TCac})}{2}\right)}\right] TMac4, 0\right\}\right)$$

$$8..8.\ \mathrm{BCb.ac} = \left(\mathrm{IMa1b} - \mathrm{IF}\left\{\left[\frac{(\mathrm{Pb1} - (1 - \mathrm{PR1ba})(1 + \mathrm{AR1b})\mathrm{Pa1} * \mathrm{TCba})}{\left(\frac{(\mathrm{Pb1} + (1 - \mathrm{PR1ba})(1 + \mathrm{AR1b})\mathrm{Pa1} * \mathrm{TCba})}{2}\right)}\right] > 0, \quad \left[\frac{(\mathrm{Pb1} - (1 - \mathrm{PR1ba})(1 + \mathrm{AR1b})\mathrm{Pa1} * \mathrm{TCba})}{\left(\frac{(\mathrm{Pb1} + (1 - \mathrm{PR1ba})(1 + \mathrm{AR1b})\mathrm{Pa1} * \mathrm{TCba})}{2}\right)}\right] TMba1, 0\right\}\right)$$

$$+ \left(\mathrm{IMa2b} - \mathrm{IF}\left\{\left[\frac{(\mathrm{Pb2} - (1 - \mathrm{PR2ba})(1 + \mathrm{AR2b})\mathrm{Pa2} * \mathrm{TCba})}{\left(\frac{(\mathrm{Pb2} + (1 - \mathrm{PR2ba})(1 + \mathrm{AR2b})\mathrm{Pa2} * \mathrm{TCba})}{2}\right)}\right] > 0, \quad \left[\frac{(\mathrm{Pb2} - (1 - \mathrm{PR2ba})(1 + \mathrm{AR2b})\mathrm{Pa2} * \mathrm{TCba})}{\left(\frac{(\mathrm{Pb2} + (1 - \mathrm{PR2ba})(1 + \mathrm{AR2b})\mathrm{Pa2} * \mathrm{TCba})}{2}\right)}\right] TMba2, 0\right\}\right)$$

$$+ \left(\mathrm{IMa3b} - \mathrm{IF}\left\{\left[\frac{(\mathrm{Pb3} - (1 - \mathrm{PR3ba})(1 + \mathrm{AR3b})\mathrm{Pa3} * \mathrm{TCba})}{\left(\frac{(\mathrm{Pb3} + (1 - \mathrm{PR3ba})(1 + \mathrm{AR3b})\mathrm{Pa3} * \mathrm{TCba})}{2}\right)}\right] > 0, \quad \left[\frac{(\mathrm{Pb3} - (1 - \mathrm{PR3ba})(1 + \mathrm{AR3b})\mathrm{Pa3} * \mathrm{TCba})}{\left(\frac{(\mathrm{Pb3} + (1 - \mathrm{PR3ba})(1 + \mathrm{AR3b})\mathrm{Pa3} * \mathrm{TCba})}{2}\right)}\right] TMba3, 0\right\}\right)$$

$$+ \left(\mathrm{IMa4b} - \mathrm{IF}\left\{\left[\frac{(\mathrm{Pb4} - (1 - \mathrm{PR4ba})(1 + \mathrm{AR4b})\mathrm{Pa4} * \mathrm{TCba})}{\left(\frac{(\mathrm{Pb4} + (1 - \mathrm{PR4ba})(1 + \mathrm{AR4b})\mathrm{Pa4} * \mathrm{TCba})}{2}\right)}\right] > 0, \quad \left[\frac{(\mathrm{Pb4} - (1 - \mathrm{PR4ba})(1 + \mathrm{AR4b})\mathrm{Pa4} * \mathrm{TCba})}{\left(\frac{(\mathrm{Pb4} + (1 - \mathrm{PR4ba})(1 + \mathrm{AR4b})\mathrm{Pa4} * \mathrm{TCba})}{2}\right)}\right] TMba4, 0\right\}\right)$$

$$+ \left(\mathrm{IMc1b} - \mathrm{IF}\left\{\left[\frac{(\mathrm{Pb1} - (1 - \mathrm{PR1bc})(1 + \mathrm{AR1b})\mathrm{Pc1} * \mathrm{TCbc})}{\left(\frac{(\mathrm{Pb1} + (1 - \mathrm{PR1bc})(1 + \mathrm{AR1b})\mathrm{Pc1} * \mathrm{TCbc})}{2}\right)}\right] > 0, \quad \left[\frac{(\mathrm{Pb1} - (1 - \mathrm{PR1bc})(1 + \mathrm{AR1b})\mathrm{Pc1} * \mathrm{TCbc})}{\left(\frac{(\mathrm{Pb1} + (1 - \mathrm{PR1bc})(1 + \mathrm{AR1b})\mathrm{Pc1} * \mathrm{TCbc})}{2}\right)}\right] TMbc1, 0\right\}\right)$$

$$+ \left(\mathrm{IMc2b} - \mathrm{IF}\left\{\left[\frac{(\mathrm{Pb2} - (1 - \mathrm{PR2bc})(1 + \mathrm{AR2b})\mathrm{Pc2} * \mathrm{TCbc})}{\left(\frac{(\mathrm{Pb2} + (1 - \mathrm{PR2bc})(1 + \mathrm{AR2b})\mathrm{Pc2} * \mathrm{TCbc})}{2}\right)}\right] > 0, \quad \left[\frac{(\mathrm{Pb2} - (1 - \mathrm{PR2bc})(1 + \mathrm{AR2b})\mathrm{Pc2} * \mathrm{TCbc})}{\left(\frac{(\mathrm{Pb2} + (1 - \mathrm{PR2bc})(1 + \mathrm{AR2b})\mathrm{Pc2} * \mathrm{TCbc})}{2}\right)}\right] TMbc2, 0\right\}\right)$$

$$+ \left(\mathrm{IMc3b} - \mathrm{IF}\left\{\left[\frac{(\mathrm{Pb3} - (1 - \mathrm{PR3bc})(1 + \mathrm{AR3b})\mathrm{Pc3} * \mathrm{TCbc})}{\left(\frac{(\mathrm{Pb3} + (1 - \mathrm{PR3bc})(1 + \mathrm{AR3b})\mathrm{Pc3} * \mathrm{TCbc})}{2}\right)}\right] > 0, \quad \left[\frac{(\mathrm{Pb3} - (1 - \mathrm{PR3bc})(1 + \mathrm{AR3b})\mathrm{Pc3} * \mathrm{TCbc})}{\left(\frac{(\mathrm{Pb3} + (1 - \mathrm{PR3bc})(1 + \mathrm{AR3b})\mathrm{Pc3} * \mathrm{TCbc})}{2}\right)}\right] TMbc3, 0\right\}\right)$$

$$+ \left(\mathrm{IMc4b} - \mathrm{IF}\left\{\left[\frac{(\mathrm{Pb4} - (1 - \mathrm{PR4bc})(1 + \mathrm{AR4b})\mathrm{Pc4} * \mathrm{TCbc})}{\left(\frac{(\mathrm{Pb4} + (1 - \mathrm{PR4bc})(1 + \mathrm{AR4b})\mathrm{Pc4} * \mathrm{TCbc})}{2}\right)}\right] > 0, \quad \left[\frac{(\mathrm{Pb4} - (1 - \mathrm{PR4bc})(1 + \mathrm{AR4b})\mathrm{Pc4} * \mathrm{TCbc})}{\left(\frac{(\mathrm{Pb4} + (1 - \mathrm{PR4bc})(1 + \mathrm{AR4b})\mathrm{Pc4} * \mathrm{TCbc})}{2}\right)}\right] TMbc4, 0\right\}\right)$$

$$
\begin{aligned}
8..9.\ \mathrm{BCc.ab} = &\left(\mathrm{IMa1c} - \mathrm{IF}\left\{\left[\frac{(\mathrm{Pc1}-(1-\mathrm{PR1ca})(1+\mathrm{AR1c})\mathrm{Pa1}*\mathrm{TCca})}{\left(\frac{(\mathrm{Pc1}+(1-\mathrm{PR1ca})(1+\mathrm{AR1c})\mathrm{Pa1}*\mathrm{TCca})}{2}\right)}\right]>0,\ \left[\frac{(\mathrm{Pc1}-(1-\mathrm{PR1ca})(1+\mathrm{AR1c})\mathrm{Pa1}*\mathrm{TCca})}{\left(\frac{(\mathrm{Pc1}+(1-\mathrm{PR1ca})(1+\mathrm{AR1c})\mathrm{Pa1}*\mathrm{TCca})}{2}\right)}\right]TMca1, 0\right\}\right)\\
&+\left(\mathrm{IMa2c} - \mathrm{IF}\left\{\left[\frac{(\mathrm{Pc2}-(1-\mathrm{PR2ca})(1+\mathrm{AR2c})\mathrm{Pa2}*\mathrm{TCca})}{\left(\frac{(\mathrm{Pc2}+(1-\mathrm{PR2ca})(1+\mathrm{AR2c})\mathrm{Pa2}*\mathrm{TCca})}{2}\right)}\right]>0,\ \left[\frac{(\mathrm{Pc2}-(1-\mathrm{PR2ca})(1+\mathrm{AR2c})\mathrm{Pa2}*\mathrm{TCca})}{\left(\frac{(\mathrm{Pc2}+(1-\mathrm{PR2ca})(1+\mathrm{AR2c})\mathrm{Pa2}*\mathrm{TCca})}{2}\right)}\right]TMca2, 0\right\}\right)\\
&+\left(\mathrm{IMa3c} - \mathrm{IF}\left\{\left[\frac{(\mathrm{Pc3}-(1-\mathrm{PR3ca})(1+\mathrm{AR3c})\mathrm{Pa3}*\mathrm{TCca})}{\left(\frac{(\mathrm{Pc3}+(1-\mathrm{PR3ca})(1+\mathrm{AR3c})\mathrm{Pa3}*\mathrm{TCca})}{2}\right)}\right]>0,\ \left[\frac{(\mathrm{Pc3}-(1-\mathrm{PR3ca})(1+\mathrm{AR3c})\mathrm{Pa3}*\mathrm{TCca})}{\left(\frac{(\mathrm{Pc3}+(1-\mathrm{PR3ca})(1+\mathrm{AR3c})\mathrm{Pa3}*\mathrm{TCca})}{2}\right)}\right]TMca3, 0\right\}\right)\\
&+\left(\mathrm{IMa4c} - \mathrm{IF}\left\{\left[\frac{(\mathrm{Pc4}-(1-\mathrm{PR4ca})(1+\mathrm{AR4c})\mathrm{Pa4}*\mathrm{TCca})}{\left(\frac{(\mathrm{Pc4}+(1-\mathrm{PR4ca})(1+\mathrm{AR4c})\mathrm{Pa4}*\mathrm{TCca})}{2}\right)}\right]>0,\ \left[\frac{(\mathrm{Pc4}-(1-\mathrm{PR4ca})(1+\mathrm{AR4c})\mathrm{Pa4}*\mathrm{TCca})}{\left(\frac{(\mathrm{Pc4}+(1-\mathrm{PR4ca})(1+\mathrm{AR4c})\mathrm{Pa4}*\mathrm{TCca})}{2}\right)}\right]TMca4, 0\right\}\right)\\
&+\left(\mathrm{IMb1c} - \mathrm{IF}\left\{\left[\frac{(\mathrm{Pc1}-(1-\mathrm{PR1cb})(1+\mathrm{AR1c})\mathrm{Pb1}*\mathrm{TCcb})}{\left(\frac{(\mathrm{Pc1}+(1-\mathrm{PR1cb})(1+\mathrm{AR1c})\mathrm{Pb1}*\mathrm{TCcb})}{2}\right)}\right]>0,\ \left[\frac{(\mathrm{Pc1}-(1-\mathrm{PR1cb})(1+\mathrm{AR1c})\mathrm{Pb1}*\mathrm{TCcb})}{\left(\frac{(\mathrm{Pc1}+(1-\mathrm{PR1cb})(1+\mathrm{AR1c})\mathrm{Pb1}*\mathrm{TCcb})}{2}\right)}\right]TMcb1, 0\right\}\right)\\
&+\left(\mathrm{IMb2c} - \mathrm{IF}\left\{\left[\frac{(\mathrm{Pc2}-(1-\mathrm{PR2cb})(1+\mathrm{AR2c})\mathrm{Pb2}*\mathrm{TCcb})}{\left(\frac{(\mathrm{Pc2}+(1-\mathrm{PR2cb})(1+\mathrm{AR2c})\mathrm{Pb2}*\mathrm{TCcb})}{2}\right)}\right]>0,\ \left[\frac{(\mathrm{Pc2}-(1-\mathrm{PR2cb})(1+\mathrm{AR2c})\mathrm{Pb2}*\mathrm{TCcb})}{\left(\frac{(\mathrm{Pc2}+(1-\mathrm{PR2cb})(1+\mathrm{AR2c})\mathrm{Pb2}*\mathrm{TCcb})}{2}\right)}\right]TMcb2, 0\right\}\right)\\
&+\left(\mathrm{IMb3c} - \mathrm{IF}\left\{\left[\frac{(\mathrm{Pc3}-(1-\mathrm{PR3cb})(1+\mathrm{AR3c})\mathrm{Pb3}*\mathrm{TCcb})}{\left(\frac{(\mathrm{Pc3}+(1-\mathrm{PR3cb})(1+\mathrm{AR3c})\mathrm{Pb3}*\mathrm{TCcb})}{2}\right)}\right]>0,\ \left[\frac{(\mathrm{Pc3}-(1-\mathrm{PR3cb})(1+\mathrm{AR3c})\mathrm{Pb3}*\mathrm{TCcb})}{\left(\frac{(\mathrm{Pc3}+(1-\mathrm{PR3cb})(1+\mathrm{AR3c})\mathrm{Pb3}*\mathrm{TCcb})}{2}\right)}\right]TMcb3, 0\right\}\right)\\
&+\left(\mathrm{IMb4c} - \mathrm{IF}\left\{\left[\frac{(\mathrm{Pc4}-(1-\mathrm{PR4cb})(1+\mathrm{AR4c})\mathrm{Pb4}*\mathrm{TCcb})}{\left(\frac{(\mathrm{Pc4}+(1-\mathrm{PR4cb})(1+\mathrm{AR4c})\mathrm{Pb4}*\mathrm{TCcb})}{2}\right)}\right]>0,\ \left[\frac{(\mathrm{Pc4}-(1-\mathrm{PR4cb})(1+\mathrm{AR4c})\mathrm{Pb4}*\mathrm{TCcb})}{\left(\frac{(\mathrm{Pc4}+(1-\mathrm{PR4cb})(1+\mathrm{AR4c})\mathrm{Pb4}*\mathrm{TCcb})}{2}\right)}\right]TMcb4, 0\right\}\right)
\end{aligned}
$$

If in the Spreadsheet 1, in sections 1.1, 1.2, 1.3 and 1.4, where we have expressed the model, we substitute the continuous functions expressed by the trade balances, equations 5.1, 5.2 and 5.3 by the new conditional functions described in the equations 8.7, 8.8 and 8.9, we would result in a new version of the model shown in Appendix Spreadsheet 3 Images Sections 3.1, 3.2, 3.3 and 3.4.

The Images Section 3.1, 3.2, 3.3, 3.4, 3.5, 3.6 and 3.7 in the Appendix Spreadsheet 3 were obtained from the Spreadsheet 3 shown below and where the international trade model described in equations 8.7, 8.8 and 8.9 was introduced.

Spreadsheet 3

Microsoft Excel Worksheet

In the Image Section 3.1 of the Appendix Spreadsheet 3 it can be seen that the model shows a surplus of 78.08 in the trade balance of country A, a deficit of -71.74 in the trade balance of country B and a deficit of -6.34 in the trade balance of country C. As can be seen, these values that arise from the new conditional equations 8.7, 8.8 and 8.9 are the same as those obtained with continuous equations 5.1, 5.2 and 5.3 of the previous model. In the lower part of Section 3.1, the trade balances of each country with each country are broken down individually. For example, it can be seen that the US surplus of 78.08 is made up of the surplus with Mexico of 64.57 plus the surplus with Canada of 13.51.

As you can see from the images, two things have changed with respect to the previous model.

First, it can be seen in the Spreadsheet 3 Image Section 3.5 MODEL PARAMETERS, that is added the breakdown of consumer preferences PRjab, PRjac…..PRjcb from each of the three countries for each of the four goods and, in addition, tariff rates have been added ARja, ARjb and ARjc of each of the three countries for each of the four goods.

Second, it can be seen in the Appendix Spreadsheet 3 Images Sections 3.2, 3.3 and 3.4, that the definitions of trade balances given by equations 5.1, 5.2 and 5.3 have been changed and replaced by the new definitions given by equations 8.7, 8.8 and 8.9.

As we noted, the Spreadsheet 3 Image Section 3.5 shows the coefficients used to introduce the effect of consumer tastes and preferences for foreign goods on the demand for imported goods into the model. These coefficients PR1ab, PR2ab ... PR4ab and PR1ac, PR2ac ... PR4ac represent the tastes and preferences in percentage terms of consumers in country A (United States) for goods from foreign countries B (Mexico) and C (Canada) respectively.

Assigning a value of 0 to these coefficients means neutrality, that is, zero preferences. For example, a PR1ab = 0.25 coefficient implies that consumers in country A = USA express a preference for good 1 from country B = Mexico that encourages them to prefer that good, even if the price is 25% higher than that whose coefficient is 0 (neutral). When the coefficients on tastes and preferences are at zero, that implies that, since consumers do not have preferences for foreign goods, they are indifferent to buying the national or foreign good. Consequently, they will buy the foreign good as long as its price is lower than the national one. On the contrary, if in country A (United States) they have preferences for the design of a foreign product or for the image or for the quality, then they will be willing to pay more for that good, even if it has a higher price than the national product . Such is the case with imports of Mercedes Benz or Jaguar despite the fact that the United States produces Cadillac. Consequently, when these coefficients are assigned values above zero, it may happen that certain goods are imported despite having higher prices than national goods.

In the same Section 3.5 PARAMETERS OF THE MODEL of the Spreadsheet 3, the other coefficients AR1a, AR2a ... ARb1 ... AR4c are presented, which represent the proportion of import duties that is applied to the goods 1, 2, ... 4 from any foreign country. A value greater than zero in the tariff coefficients will imply an increase in the price that the consumer will have to pay for the foreign good. If the tariff is low, the increase in the price will cause imports to be reduced. If the tariff is high enough, so as to make the price of the foreign good higher in relation to the price of the national good, then no amount of the good will be imported unless consumers have tastes and preferences for those goods produced abroad. In this case the corresponding PR1ab, PR2ab ... PR4ab and PR1ac, PR2ac ... PR4ac coefficient will be greater than zero. In the event that consumers have a preference for the imported good, the high price due to the tariff could be offset by the consumer's preference, depending on the value of the coefficients PR1ab, PR2ab ... PR4ab and PR1ac, PR2ac ... PR4ac. If the tariff were too high, the price of the foreign good would be prohibitive even for consumers who have a preference for that product.

It can be seen that the conditional equations 8.7, 8.8 and 8.9 corresponding to the trade balances of the new model yield the same result as the continuous equations 5.1, 5.2 and 5.3 corresponding to the previous model. That is, both systems of equations result in a surplus of 78.08 in the trade balance of country A, a deficit of -71.74 in the trade balance of country B and a deficit of -6.34 in the trade balance of country C. However, unlike the previous model, the new model allows us to ask questions about the effect of consumer preferences and/or tariff rates on international trade that cannot be done to the previous model. Consequently, the new model based on conditional equations is superior as it best describes reality.

### THE NEW MODEL STARTING FROM THE ASSUMPTION OF A PERFECT COMPETITION MARKET

Assume that the PRjih coefficients that represent the preferences of the consumers of the different countries for the different goods are zero and that the ARji tariffs are zero, as illustrated in Section 3.5 of the Spreadsheet 3. Under such circumstances the new model would be based on the assumption of a perfect competition market. When, on the contrary, the PRjih coefficients are not zero, the new model would be based on the assumption of a monopolistic competition market.

Suppose that we put the currencies of countries B and C to float with respect to the currency of country A. It is expected that the successive adjustments of the primary exchange rates TCab and TCac produced by the floating of the currencies will generate the equivalent universal equilibrium primary exchange rates. We can simulate this flotation process of currencies by gradually modifying primary exchange rates in Section 3.6, trying to reduce deficits and surpluses in the three trade balances BCa.bc, BCb.ac and BCc.ab that appear in Section 3.1 of the Spreadsheet 3. In this way, by trial and error and successive approximations, are obtained the equivalent universal equilibrium primary exchange rates shown below

9.1 TCab = 0.0249159
9.2 TCac = 0.410078

When we introduce these two primary exchange rates in the three equations of model 8.7, 8.8 and 8.9 of the Spreadsheet 3, by introducing them in Section 3.6, the result presented in the Appendix Spreadsheet 4, Image Section 4.1 is generated.

The Image Section 4.1 in the Appendix Spreadsheet 4 was obtained from the Spreadsheet 4 shown below.

Spreadsheet 4

Microsoft
Excel Worksheet

As can be seen in the Appendix Spreadsheet 4 Image Section 4.1, the trade balances of the three countries BCa.bc, BCb.ac and BCc.ab are simultaneously balanced at zero.

With this result it is demonstrated that, for a common market of more than two countries, under the assumption of a perfect competition market, there is a set of primary exchange rates that balance the trade balances of all countries that trade with each other.

### THE NEW MODEL STARTING FROM THE ASSUMPTION OF MONOPOLISTIC COMPETITION MARKET

The model we have developed in equations 8.7, 8.8 and 8.9 allows us to analyze the balance of international trade, both under conditions of perfect competition, and under conditions of monopolistic competition. In the previous topic, all the coefficients that represent the preferences of the consumers were assigned a value of zero, which supposes a situation of perfect competition in which the goods that are traded are homogeneous or undifferentiated. Consequently, countries will import only those foreign goods whose prices are lower than the prices of national goods. In this topic we will run the model under the assumption of Paul Krugman regarding the existence of an international trade where monopolistic competition exists and, therefore, in which the goods traded are not homogeneous but differentiated. Suppose that we introduce in the Spreadsheet 3 the preferences of consumers for foreign goods presented in Table 4.

Table 4
Consumer Preferences

| | |
|---|---|
| PR1ab = .25 | PR1ac = .15 |
| PR2ab = .50 | PRa2c = .05 |
| PR3ab = .75 | PRa3c = .15 |
| PR4ab = .38 | PRa4c = .12 |
| | |
| PR1ba = .25 | PR1bc = .05 |
| PR2ba = .18 | PR2bc = .16 |
| PR3ba = .14 | PR3bc = .21 |
| PR4ba = .22 | PR1bc = .17 |
| | |
| PR1ca = .15 | PR1cb = .19 |
| PR2ca = .25 | PR2cb = .15 |
| PR3ca = .15 | PR3cb = .14 |
| PR4ca = .18 | PR4cb = .14 |

If we introduce in Section 3.5 of the Spreadsheet 3, the preferences of consumers PRijh shown in Table 4, we obtain the result presented in the Appendix Spreadsheet 5 Image Section 5.1.

As can be seen, when starting from the assumption of the existence of consumer preferences offered in Image Section 5.5 of the Appendix Spreadsheet 5 in the case of countries A, B and C, the model throws in the Image Section 5.1 a deficit of -187.83 in the trade balance of country A, a surplus of 215.65 in the trade balance of country B and a deficit of -27.82 in the trade balance of country C.

Images 5.1 and 5.5 were obtained from Section 5.1 and 5.5 of the Spreadsheet 5 shown below and where the international trade model was introduced.

Spreadsheet 5

Microsoft
Excel Worksheet

Suppose that we set the currencies of countries B and C to float with respect to the currency of country A. It is expected that the successive adjustments of the primary exchange rates TCab and TCac produced by the floating of the currencies will generate the equivalence universal equilibrium primary exchange rates. We can simulate this flotation process of currencies by gradually modifying primary exchange rates in Section 5.6, trying to reduce deficits and surpluses in the three trade balances BCa.bc, BCb.ac and BCc.ab that appear in Section 5.1 of the Spreadsheet 5. In this way, by trial and error and successive approximations, are obtained the equivalence universal equilibrium primary exchange rates shown below:

10.1 TCab = 0.03160945
10.2 TCac = 0.452894

When we introduce in these Spreadsheet 5 these two primary exchange rates given by 10.1 and 10.2 in the three equations of model 8.7, 8.8 and 8.9 presented above, the result presented in the Appendix Spreadsheet 6 Image Section 6.1 is generated.

The spreadsheet where the previous model was introduced and the results presented in Image Sections 6.1, 6.2, 6.3, 6.4 and 6.5 were obtained is shown in the Spreadsheet 6

Spreadsheet 6

Microsoft
Excel Worksheet

As can be seen, the trade balances of the three countries BCa.bc, BCb.ac and BCc.ab are simultaneously balanced at zero. This result shows that, for a common market of more than two countries, under the assumption of a monopolistic competition market, there is a set of primary exchange rates that balance the trade balances of all the countries that trade with each other.

## MR. KRUGMAN AND THE CLASSICS RICARDO AND MILL REGARDING THE EFFECT OF MONOPOLISTIC COMPETITION ON THE PATTERN OF INTERNATIONAL TRADE

What is the synthesis between Ricardo and Mill's theory of comparative advantages and Krugman's theory?

The benefit obtained by the countries that trade based on the relative productivity of labor is that obtained from the increase in the joint production of all countries as a result of each country specializing in the production of those goods where it has comparative advantages. The benefit obtained by the countries that trade based on consumer preferences is that obtained from the increase in the quality of the goods consumed, that is, the value obtained from the goods for the money paid ("Value pricing").

You can see in the images sections 4.2, 4.3 and 4.4 of the Appendix Spreadsheet 4 that, due to the assumption of perfect competition where there are no consumer preferences between domestic and foreign goods, assuming homogeneous goods, trade balances BCihj from the three i countries for each of the four j goods can only show one of two possible outcomes. Or the country imports good j when the local price is higher than the price of the foreign good or the country exports good j when the local price is lower than the price of the foreign good. This is to be expected when starting from the premise that consumers have no preference and, therefore, do not distinguish between local and foreign goods. In such a case, the only consideration of the consumer at the time of acquiring the good is to choose the one with the lowest price. On the contrary, it can be seen in the images section 6.2, 6.3 and 6.4 that, when assuming the existence of consumer preferences for goods, under the assumption of a monopolistic competition market, a country can import the good j from abroad while exporting the same local good j to foreign countries. In this way it is proved how Krugman's claims about the existence of an international trade based on the existence of monopolistic competition are perfectly compatible with the theory of comparative advantages of Ricardo and Mill.

We can infer that when there are no consumer preferences, that is, when goods are homogeneous and the market is perfectly competitive, trade between countries will be determined by the comparative advantages generated by differences in labor productivity in different countries, as established by Ricardo and Mill's theory of comparative advantages. This labor productivity is what will determine the prices of goods and, consequently, the differences between the prices of foreign goods and their national goods of the same class. Therefore, since foreign and domestic goods are presumed undifferentiated, the price, which is determined by the relative productivity in each country, would be the only factor that would determine that the consumers in a country import a good. Consequently, since the price will be determined by the relative productivity in each country, it will be the relative productivity in each country, together with the determination of the exchange rate, which determines what goods a country will import and the benefit it will obtain in terms of obtaining higher levels of production and consumption. On the contrary, we can infer that under monopolistic competition where there are similar but differentiated goods and, consequently, there are consumer preferences, trade between countries will be determined, both by consumer preferences for similar but differentiated goods, such and as Krugman's theory establishes, as by labor productivity, as Ricardo and Mill establish.

We will demonstrate that Krugman's theory is correct in affirming that, under a monopolistic competition framework where countries develop economies of scale in different industrial sectors and achieve production costs in different goods that are similar to those of other countries, that is to say equal labor productivity, trade between countries will be determined for consumer preferences in a monopolistic competition market. However, we will demonstrate that the theory of comparative advantages is equally correct and that, in the absence of equal work productivity, which is most plausible, trade between countries will be determined, both by the relative productivity of labor, and by Consumer preferences.

The theory of comparative advantages suggests that, in the absence of different relative labor productivity, trade between countries would be zero. In Spreadsheet 4, the assumption of absence of preferences in consumers is based. Therefore, if we introduce equal work productivity in the three countries in Section 4.5 of Worksheet 4, we should expect that trade between the three countries will be zero since there will be no comparative advantages nor will there be consumer preferences. Suppose we match the labor productivity of countries B and C to that of Country A, as shown in Table 5.

Tabla 5

| Productividad del Trabajo País A | Productividad del Trabajo País B | Productividad del Trabajo País C |
|---|---|---|
| Za1 = 3.00 | Zb1 = 3.00 | Zc1 = 3.00 |
| Za1 = 2.00 | Zb1 = 2.00 | Zc1 = 2.00 |
| Za1 = 1.00 | Zb1 = 1.00 | Zc1 = 1.00 |
| Za1 = 0.50 | Zb1 = 0.50 | Zc1 = 0.50 |

When we enter in Section 4.5 of the Spreadsheet 4 the same work productivity as shown in Table 5, the result presented in the Appendix Spreadsheet 7 Image Section 7.1 is generated.

As can be seen, when starting from the assumption of the non-existence of consumer preferences and the assumption of equal labor productivity in the production of all products among all countries that trade with each other, as shown in the Appendix Spreadsheet 7 Image Section 7.5, the model shows in the Appendix Spreadsheet 7 Image Section 7.1 a deficit of -381.35 in the trade balance of country A, a surplus of 366.19 in the trade balance of country B and a surplus of 15.17 in the trade balance of country C .

The Images Section 7.1 and 7.5 were obtained from Section 7.1 and 7.5 of the Spreadsheet 7 shown below and where the international trade model was introduced.

Spreadsheet 7

Microsoft
Excel Worksheet

Suppose we set the currencies of countries B and C to float with respect to the currency of country A. We can simulate this process of flotation of currencies by gradually modifying in section 7.6 the primary exchange rates, trying to reduce deficits and surpluses in the three trade balances BCa.bc, BCb.ac and BCc.ab that appear in Section 7.1 of the Spreadsheet 7, model equations 8.7, 8.8 and 8.9. In this way, by means of trial and error and successive approximations, the equivalent universal equilibrium primary exchange rates shown below are obtained:

11.1 TCab = 0.04215376
11.2 TCac = 0.5538495

When we enter in the Spreadsheet 7 Section 7.6 these two primary exchange rates given by 11.1 and 11.2 in the three equations of model 8.7, 8.8 and 8.9, the result presented in the Appendix Spreadsheet 8 Image Section 8.1 is generated.

The spreadsheet where the previous model was introduced and the results presented in Image 8.1 were obtained, is shown in the Spreadsheet 8 below.

Spreadsheet 8

Microsoft
Excel Worksheet

As can be seen, the trade balances of the three countries BCa.bc, BCb.ac and BCc.ab are simultaneously balanced at zero. In the lower part of Section 8.1 of the Spreadsheet 8 it can be seen that, in effect, both exports and imports from all countries are reduced to zero and trade between countries disappears as suggested by the theory of the comparative advantages of Ricardo and Mill.

With this result it is demonstrated that, under the assumption that the economies of all countries that trade with each other generate equal labor productivity in the production of all products, then there would be no international trade based on comparative advantages. But this assumption as we know is not real.

Now consider the different situation in which the labor productivity in the three countries for the different goods is equal, but there are consumer preferences for similar but differentiated goods.

When you enter in Section 8.5 of the Spreadsheet 8 equal work productivity in all the countries shown in Table 5 and the consumer preferences shown in Table 4, is generated the result presented in the Appendix Spreadsheet 9 Image Section 9.1 and Section 9.5.

As can be seen, when starting from the assumption of the existence of consumer preferences and the assumption of equal labor productivity among all countries that trade with each other, as shown in Image Section 9.5 of the Appendix Spreadsheet 9, the model shows in the Image Section 9.1 a deficit of -503.44 in the trade balance of country A, a surplus of 525.68 in the trade balance of country B and a deficit of -22.24 in the trade balance of country C.

The images section 9.1 and 9.5 were obtained from Section 9.1 and 9.5 of the Spreadsheet 9 shown below and where the international trade model was introduced.

Spreadsheet 9

Microsoft
Excel Worksheet

Suppose we set the currencies of countries B and C to float with respect to the currency of country A. We can simulate this process of flotation of currencies by gradually modifying in Section 9.6 the primary exchange rates, trying to reduce deficits and surpluses in the three trade balances BCa.bc, BCB.ac and BCc.ab that appear in Section 9.1 of the Spreadsheet 9. In this way, through trial and error and successive approximations, equivalent universal equilibrium primary exchange rates shown below are obtained:

12.1 TCab = 0.0534337
12.2 TCac = 0.6134285

When we enter in the Spreadsheet 9 Section 9.5 these two primary exchange rates given by 12.1 and 12.2 in the three equations of model 8.7, 8.8 and 8.9 presented above, the result presented in the Appendix Spreadsheet 10 Image Section 10.1 is generated.

The spreadsheet where the previous model was introduced and the results presented in Image 10.1 were obtained, is shown in the Spreadsheet 10.

Spreadsheet 10

Microsoft
Excel Worksheet

As can be seen, the trade balances of the three countries BCa.bc, BCb.ac and BCc.ab are simultaneously balanced at zero. In the lower part of Section 10.1 of the Spreadsheet 10 it can be seen that exports and imports from all countries are not zero, but on the contrary, a trade between countries greater than zero is generated, as Krugman's theory suggests.

With this result it is demonstrated that, under the assumption that economies of scale of all countries that trade with each other generate equal work productivity in all countries, then international trade would be determined solely by consumer preferences and not determined at all by comparative advantages, that is, not determined by labor productivity in each country.

The situation described in the previous paragraphs is not realistic. The reality is that, except for the production of some goods, a large part of the goods produced in the different countries do not have equal labor productivity and, consequently, the same real production costs. Consequently, since the most plausible hypothesis is that there are differences in the relative labor productivity of the countries with respect to many of the goods that are produced and that there are also consumer preferences for similar but differentiated goods, we must conclude that the Krugman's theory complements and does not contradict Ricardo and Mill's theory of comparative advantages.

## MR. KRUGMAN AND THE CLASSICS RICARDO AND MILL REGARDING THE EFFECT OF SCALE ECONOMIES ON INTERNATIONAL TRADE

According to Krugman, economies of scale in the production of different goods suggest that the governments of the countries should help the economic development of certain strategic sectors with the porpoise of achieving economies of scale that allow that country to compete with other countries in those strategic Item. Using the model that we have developed we can show that Krugman's theory is correct by stating that a country can benefit from an increase in international trade as a result of a policy aimed at the development of a particular industrial sector. However, this is achieved at the cost of increasing its dependence on international trade. To demonstrate, we must introduce the existence of economies of scale into the model. This is achieved by making the productivity of the Zij labor of each country i in each good or industrial sector j functions of the scale of production in those industrial sectors.

In Table 6 we have expressed the productivity of work Zij as a function of the workforce FTji and, consequently, of the number of workers Nij employed in the production of each country i for each good j. The productivity Zij values shown in Table 6 were calculated from the data provided in section 5.5 Parameters of the Model of Spreadsheet 5, for the kij and Nij values.

Table 6

| Labor Productivity Country A | Labor Productivity Country B | Labor Productivity Country C |
|---|---|---|
| $Za1 = ka1\left(1-\frac{Na1}{Na1+Nb1+Nc1}\right)$ = 1.0422 | $Zb1 = kb1\left(1-\frac{Nb1}{Na1+Nb1+Nc1}\right)$ = 2.8996 | $Zc1 = kc1\left(1-\frac{Nc1}{Na1+Nb1+Nc1}\right)$ = 3.2470 |
| $Za2 = ka2\left(1-\frac{Na2}{Na2+Nb2+Nc2}\right)$ = 0.6948 | $Zb2 = kb2\left(1-\frac{Nb2}{Na2+Nb2+Nc2}\right)$ = 2.1747 | $Zc2 = kc2\left(1-\frac{Nc2}{Na2+Nb2+Nc2}\right)$ = 2.3193 |
| $Za3 = ka3\left(1-\frac{Na3}{Na3+Nb3+Nc3}\right)$ = 0.3474 | $Zb3 = kb3\left(1-\frac{Nb3}{Na3+Nb3+Nc3}\right)$ = 1.4498 | $Zc3 = kc3\left(1-\frac{Nc3}{Na3+Nb3+Nc3}\right)$ = 1.3916 |
| $Za4 = ka4\left(1-\frac{Na4}{Na4+Nb4+Nc4}\right)$ = 0.1737 | $Zb4 = kb4\left(1-\frac{Nb4}{Na1+Nb1+Nc1}\right)$ = 0.7149 | $Zc1 = kc1\left(1-\frac{Nc1}{Na1+Nb1+Nc1}\right)$ = 0.6958 |

In Spreadsheet 5, we start from the premise of an international trade based or determined, so much for the existence of preferences in consumers, as for labor productivity. When we introduce in Section 5.5 of the Spreadsheet 5 the work productivity shown in Table 6, the result presented in the Appendix Spreadsheet 11 Image Section 11.1 is generated.

As can be seen, in Spreadsheet 11 in Section 11.5 Model Parameters, we have made labor productivity Zij in each industrial sector j of each country i a function of the scale of production. The scale of production is determined by the proportion of the labor force FTji employed in the production of each good j in each country i and its corresponding number of workers Nij dedicated to the production of each good j in each country i. That is: FTji(Ni) = Nij where Ni is the number of workers in the country i.

As shown in the Appendix Spreadsheet 11 Image Section 11.1, the model shows a deficit of -94.54 in the trade balance of country A, a surplus of 104.96 in the trade balance of country B and a deficit of -10.42 in the trade balance from country C.

The images section 11.1 and 11.5 were obtained from Section 11.1 and 11.5 of the Spreadsheet 11 shown below and where the international trade model was introduced.

Spreadsheet 11

Microsoft
Excel Worksheet

Suppose that we put the currencies of countries B and C to float with respect to the currency of country A. We can simulate this process of flotation of currencies by gradually modifying in Section 11.6 the primary exchange rates, trying to reduce deficits and surpluses in the three trade balances BCa.bc, BCb.ac and BCc.ab that appear in Section 11.1 of the Spreadsheet 11, equations of model 8.7, 8.8 and 8.9. In this way, by trial and error and successive approximations, the equivalent universal equilibrium primary exchange rates shown below are obtained:

13.1 TCab = 0.0152689
13.2 TCac = 0.170267

When we enter in the Spreadsheet 11 Section 11.6 these two primary exchange rates given by 13.1 and 13.2 in the three equations of model 8.7, 8.8 and 8.9, the result presented in the Appendix Spreadsheet 12 Image Section 12.1 is generated. As can be seen, the trade balances of the three countries BCa.bc, BCb.ac and BCc.ab are simultaneously balanced at zero.

The spreadsheet where the previous model was introduced and the results presented in Image 12.1 were obtained, is shown in the Spreadsheet 12 below.

Spreadsheet 12

Microsoft
Excel Worksheet

Now suppose that the government of country C decides to help the development of industrial sector 1 with the pourpose to developing economies of scale. Assume that after several years the proportion of labor force FT1c employed in that sector increases from FT1c = 20% in Section 11.5 of the Spreadsheet 11 to FT1c = 80% as shown in Image Section 13.5 in the Appendix Spreadsheet 13. This generates in the Appendix Spreadsheet 13 Image Section 13.1 a deficit in the trade balance of country A of -127.80 a surplus in the trade balance of country B of 73.05 and a surplus in the trade balance of country C of 54.75.

The spreadsheet where the previous model was introduced and the results presented in Image 13.1 were obtained, is shown in the Spreadsheet 13 below.

Spreadsheet 13

Microsoft
Excel Worksheet

Suppose we set the currencies of countries B and C to float with respect to the currency of country A. We can simulate this process of flotation of currencies by gradually modifying in Section 13.6 the primary exchange rates, trying to reduce deficits and surpluses in the three trade balances BCa.bc, BCb.ac and BCc.ab that appear in Section 13.1 of the Spreadsheet 13, equations of model 8.7, 8.8 and 8.9. In this way, by means of trial and error and successive approximations, the equivalent universal equilibrium primary exchange rates shown below are obtained:

14.1 TCab = 0.0157254
14.2 TCac = 0.23435

When we enter in the Spreadsheet 13 Section 13.6 these two primary exchange rates given by 14.1 and 14.2 in the three equations of model 8.7, 8.8 and 8.9, the result presented in the Appendix Spreadsheet 14 Image Section 14.1 is generated. As can be seen, the trade balances of the three countries BCa.bc, BCb.ac and BCc.ab are simultaneously balanced at zero.

The spreadsheet where the previous model was introduced and the results presented in Image 14.1 were obtained, is shown in the Spreadsheet 14 below.

Spreadsheet 14

Microsoft Excel Worksheet

In the lower part of Section 14.1 of Worksheet 14 it can be seen that, in effect, the trade balances of all countries are reduced to zero and trade between countries increases. Trade in country A increased from EXa = 197.02 on Spreadsheet 12, Section 12.1 to EXa = 215.05 on Spreadsheet 14, Section 14.1. Trade in country B increased from EXb = 197.21 in Spreadsheet 12, Section 12.1 to EXb = 203.14 in Spreadsheet 14, Section 14.1. Trade in country C increased from EXc = 30.49 in Spreadsheet 12, Section 12.1 to EXc = 41.17 in Spreadsheet 14, Section 14.1.

What happened? It can be seen that the labor force in country C has shifted from sectors 2, 3 and 4 to sector 1, increasing the productivity of sector 1 and reducing the productivity of sectors 2, 3 and 4. Consequently, it has been reduced the work force and, therefore, the production in the less productive sectors and it has been increase the work force and, therefore, the production in the most productive sector. The result is an increase in labor productivity in 80% of the workforce and a reduction in labor productivity in 20% of the workforce. The net result is an increase in the overall or total productivity of the workforce and an increase in the country's production. However, this is achieved at the cost of increasing the dependence of country C on international trade.

With this result it is demonstrated that, under the assumption of the existence of economies at scale of all countries that trade among themselves, the country that develops the growth of a strategic sector will increase its foreign trade, as well as that of the other countries.

## THE SIMULTANEOUS EQUILIBRIUM OF COMMERCIAL BALANCES OF THE COUNTRIES THROUGH TARIFFS INSTEAD OF EXCHANGE RATES

Recently we have seen that the United States on the initiative of President Donald Trump has begun a tariff policy to reduce the US trade deficit with Mexico, Canada and China. The question that arises is: Can the equilibrium in the US trade balance be achieved through the application of tariffs? The model we have developed presents evidence that it is so. That is, you can replace the change in exchange rates for the imposition of tariffs, as an instrument to balance the trade balance of the countries. However, although the trade balances of the countries can be balanced using the tariff rates, it is not convenient to do so because trade is reduced and with it the increases in joint productivity that arise from the exploitation of comparative advantages. This reduction in trade is due to the fact that when the deficit in the trade balance of a country is reduced by using increases in tariffs, imports are reduced, but exports do not increase. On the contrary, when the deficit in the trade balance of a country is reduced by increasing the exchange rate, the imports of the country in question are reduced, but exports increase in proportion.

Unlike the set of equivalent universal equilibrium primary exchange rates, which is unique, in the case of tariff rates the set that balances the trade balance of all countries is not unique, but there are multiple sets that balance the trade balances of all countries. Therefore, although the trade balances of all countries can be balanced using tariffs, however, changes in tariffs between one country and another would have to be coordinated, either through negotiations or by a supranational institution. Otherwise, if each country is dedicated to respond in retaliation to the tariffs imposed by the other country, establishing a tariff trade war, the result is uncertain.

We can demonstrate the above, as follows. Assume that we are in an initial situation equal to that described in Spreadsheet 5 where the United States is in deficit with Mexico and Canada for the magnitude of -187.83. Suppose that, instead of modifying the exchange rate of the United States with Mexico, we modify the tariffs with the objective of reducing and eventually eliminating the deficit in the United States trade balance.

As we have already pointed out, in the model presented in Spreadsheet 5, tariffs are represented by the coefficients AR1a, AR2a, ... AR 4a that appear in Section 5.5.

In the Appendix Spreadsheet 5, Image Section 5.1 we can see that the United States has a deficit in its trade balance of -187.83. In the lower part of Section 5.1 we can see that this deficit is divided into a deficit with Mexico of -231.05 and a surplus with Canada of 43.22. If you study Spreadsheet 5, Section 5.2, you can see that the main source of this deficit in the United States originates from imports of goods 1, 2 and 3 from Mexico. Assume that the United States applies a tariff to good 1 of 0.4801, a tariff to good 2 of 0.4712 and a tariff to good 3 of 0.53 so as to reduce the demand of consumers in the United States for those goods and thus reduce the deficit of U.S. Now suppose that Mexico does not retaliate and that it accepts tariffs in a coordinated manner with the US. However, suppose that Canada applies a tariff to good 1 of 0.13, a tariff to good 2 of 0.10, a tariff to good 3 of 0.1015 and a tariff to good 4 of 0.25.
Table 7 summarizes these increases in country tariffs.

Table 7

| |
|---|
| Tariffs Country A |
| ARa1 = 0.4801 |
| ARa2 = 0.4712 |
| ARa3 = 0.5300 |
| ARa4 = 0.0000 |
| Tariffs Country B |
| ARb1 = 0.0000 |
| ARb2 = 0.0000 |
| ARb3 = 0.0000 |
| ARb4 = 0.0000 |
| Tariffs Country C |
| ARc1 = 0.1300 |
| ARc2 = 0.1000 |
| ARc3 = 0.1015 |
| ARc4 = 0.2500 |

When we enter in the Calculation Sheet 5 Section 5.5 the tariffs of Table 7, the result presented in the Appendix Spreadsheet 15 Image Section 15.1 and image Section 15.5 are generated.

The spreadsheet where the previous model was introduced, and the results presented in the Images Section 15.1 and 15.5 above are obtained is shown in the Spreadsheet 15.

Spreadsheet 15

Microsoft
Excel Worksheet

As can be seen in the Appendix Spreadsheet 15, Image Section 15.1, the trade balances of all countries would be perfectly balanced, in the same way as was done with the changes in the exchange rates in the Spreadsheet 6. However, when in the Appendix Spreadsheet 6, Image Section 6.1 (lower part) compares the volume of foreign trade that is generated when the equivalent universal equilibrium primary exchange rates are used instead of the tariff rates, we see that the volume of trade between the three countries has been reduced significantly. For example, in the case of the United States of $ 416.49 in the Appendix Spreadsheet 6, Image Section 6.1 to $240.42 in the Appendix Spreadsheet 15, Image Section 15.1, in the case of Mexico of $416.98 in the Appendix Spreadsheet 6, Image Section 6.1 to $251.89 in the Appendix Spreadsheet 15, Image Section 15.1 and in the case of Canada from $80.67 in the Appendix Spreadsheet 6, Image Section 6.1 to $44.41 in the Appendix Spreadsheet 15, Image Section 15.1.

We see then that, although the trade balances of the countries can be balanced using the tariff rates, it is not convenient to do so as trade is reduced and with it the increases in joint productivity that arise from the exploitation of comparative advantages.

In conclusion, Ricardo and Mill's theory of comparative advantages, once have been added the cost or price of the business production factor Gij, as well as how to achieve the equivalent universal equilibrium primary exchange rates and Krugman's contributions on the existence of monopolistic competition markets in international trade, it constitutes the strongest foundation today to explain why international trade occurs and how can it develop an international trade policy that generates stability. Krugman's theory of international trade complements the theory of comparative advantages by making it more complete and more solid.

## A MODEL ON INTERNATIONAL TRADE OF FOUR COUNTRIES

The model we developed simulating trade between 3 countries, can be generalized in the case of more than 3 countries. We have developed the previous model for the case of 4 countries. In the Appendix Spreadsheet 16, Image Section 16.1 and 16.5 you can see the result. These images are obtained from the Spreadsheet 16 where the model is presented with the results of the trade balances of the 4 countries BCa.bcd, BCb.acd, BCc.abd, BCd.abc in the case that the exchange rates TCab, TCac and TCad introduced in Section 16.6 are those corresponding to the ratios of the 4 prices of the countries shown in Section 16.7 and assuming the existence of a monopolistic competition market where goods are not homogeneous and, therefore, the coefficients that indicate the preferences of consumers PRjih in Section 16.5 are different from zeros. In addition, it is based on the assumption of the existence of economies of scale as can be seen in the definitions of labor productivity Zij in Section 16.5.

As can be seen in Appendix Spreadsheet 16, Image Section 16.1, the model shows a deficit of -172.22 in the trade balance of country A, a surplus of 214.36 in the trade balance of country B, a deficit of -22.38 in the trade balance of country C and a deficit of -19.76 in the trade balance of country D.

The Appendix Spreadsheet 16, Image Section 16.1 was obtained from the Spreadsheet 16 shown below.

Spreadsheet 16
Trade Balance of 4 Countries

Microsoft
Excel Worksheet

Suppose that we put the currencies of countries B, C and D to float with respect to the currency of country A. We can simulate this process of flotation of currencies by gradually modifying in Section 16.6 the three primary exchange rates, trying to reduce deficits and surpluses in the 4 trade balances BCa.bcd, BCb.acd, BCc.abd and BCd.abc that appear in the lower part of Section 16.1 of the Spreadsheet 16. In this way, through trial and error and successive approximations, the three equivalent universal equilibrium primary exchange rates shown below are obtained:

15.1 TCab = 0.031438
15.2 TCac = 0.452446
15.3 TCad = 0.009936

In the Appendix Spreadsheet 17 Image Section 17.1 the model with the results of the trade balances of the 4 countries is presented in case the exchange rates introduced in Section 16.6 of the Spreadsheet 16 are those corresponding to the set of the three equivalent universal equilibrium primary exchange rates obtained by trial and error and shown in 15.1, 15.2 and 15.3. As can be seen, the trade balances of the 4 countries BCa.bcd, BCb.acd, BCc.abd and BCd.abc are simultaneously balanced at zero.

The Appendix Spreadsheet 17, Image Section 17.1 was obtained from the Spreadsheet 17 shown below and where the international trade model was introduced.

Spreadsheet 17
Trade Balance of 4 Countries

Microsoft
Excel Worksheet

In the Appendix Spreadsheet 18 Image Section 18.1 the model with the results of the trade balances of the 4 countries is presented in the event that the existence of a market of perfect competition is assumed instead of monopolistic competition and, consequently, the goods are homogeneous. Under such circumstance, the coefficients that indicate consumer preferences PRjih in Section 18.5 of Spreadsheet 18 are all zeros. There for, the exchange rates introduced in Section 18.6 are those corresponding to the set of three equivalent universal equilibrium primary exchange rates obtained by trial and error.

The Appendix Spreadsheet 18, Image Section 18.1 was obtained from the Spreadsheet 18 shown below and where the international trade model was introduced.

Spreadsheet 18
Trade Balance of 4 Countries

Microsoft
Excel Worksheet

## PROPOSAL FOR A SECOND BRETTON WOODS CONFERENCE

In the previous sections, we have demonstrated two things:

1. It has been demonstrated that, if we start from the premise of more than two countries and international trade based on a perfectly competitive market where purchases of foreign goods or imports depend inversely on foreign prices, or if we start from the premise of more than two countries and international trade based on a monopolistic competition market where the purchases of goods or imports depend both on the tastes and preferences of consumers and on foreign prices, then there will be a set of exchange rates that balance the trade balances of all the countries that trade with each other and that we have called the set of equivalent universal equilibrium exchange rates.

2. It has been demonstrated that, both when starting from the premise of a perfect competition market, and when starting from the premise of a monopolistic competition market, the equilibrium exchange rate with the dollar of each country is not independent of the equilibrium exchange rates with the dollar of the other countries.

From the above, the question that arises is how can the foreign exchange markets be organized among the countries that trade among themselves, to generate the set of equivalent universal equilibrium exchange rates? We have noted that a necessary condition for the floating of currencies to generate equivalent universal equilibrium exchange rates is that there be equilibrium in the financial and capital balances. Consequently, the institution or institutions that are created to organize and regulate international trade must be structured so that they can control the flow of international capital and maintain the equilibrium in these balances. Nonetheless, even assuming that the objective of countries to maintain their financial and capital balances individually in equilibrium is achieved, the floating of the currencies against the dollar will not generate the equivalent universal equilibrium primary exchange rates if the governments of the countries decide to accumulate wealth in dollars and keep their currencies undervalued with exchange rates TCab, TCac below their equilibrium exchange rate. Consequently, the institution or institutions that are created to organize and regulate international trade must be structured so that they can control that countries accumulate wealth in the most widely accepted currency, the dollar.

Therefore, given that the tendency of governments in undeveloped countries is to borrow until they are in debt and the tendency of governments in all countries, developed or undeveloped, is to keep their currencies undervalued in order to generate a surplus in their trade balance, it is obvious that in order to achieve the objective of maintaining the equilibrium of the trade, financial and capital balances of all the countries that trade with each other, the administration of the foreign exchange markets of each country cannot be left to the governments of each country.

Taking into consideration the above, the most sensible it would be to centralize those markets through one or several institutions that respond to the collective objective of balancing the trade balances of all countries. This would be an idea similar to the one proposed by John Maynard Keynes in Bretton Wood regarding the creation of an International Clearing Union. However, in this paper we have shown that balancing the balance of payments as proposed by Keynes and Dexter White in Bretton Wood is a mistake since the goal must be to balance the trade balance.

Therefore, it is necessary to create a supranational institution that has the function of regulating the currency markets of all the countries that trade with each other. That institution could be called the Bank of International Commerce (BIC).

The International Monetary Fund (IMF) and the World Bank (WB) could be transformed to become the BIC. Its main task would be to determine the set of equivalent universal equilibrium primary exchange rates that balance the trade balances of all the countries that trade with each other. The question that arises is how the BIC or the IMF and the WB could be organized to fulfill the purpose described above. Let's see.

The BIC or the IMF would function as a clearinghouse. The main task of the BIC or the IMF would be to determine the equilibrium exchange rates of any country with the most widely accepted currency, that is, the dollar. These are the exchange rates that we have called equivalent universal equilibrium primary exchange rates. To determine the equivalent universal equilibrium primary exchange rate of each country with the dollar, the BIC will indicate to each country the balances or reserves of dollars and national currencies that each country must have in the BIC to attend to its foreign trade. To determine the equivalent universal equilibrium primary exchange rate of each country with the dollar, it is necessary that the BIC be the only institution through which dollars can be exchanged for national currencies with the purpose of buying or selling goods and services between countries

## FORMULA FOR THE BIC TO DETERMINE THE SET OF THE EQUIVALENT UNIVERSAL EQUILIBRIUM PRIMARY EXCHANGE RATES

There are two formulas or procedures that the BIC can use to determine equivalent universal equilibrium exchange rates.

### First Formula Or Procedure For The BIC To Determine The Equivalent Universal Equilibrium Primary Exchange Rates

The BIC could float currencies against the dollar by keeping the dollar reserves used to transact goods and services in the trade balance, separate from the dollar reserves used to make financial or direct investment transactions or payments in the income balance within the current account balance. The first step of the BIC would be to determine what is the size of the international trade of each country, so that it can be determined what is the optimal size of the foreign exchange reserves that each country must have in the BIC so that the BIC can manage said trade. For example, the BIC could require from each country a reserve of its national currency equivalent to the average of the sum of its exports plus its imports of the last three years. Furthermore, the BIC may require each country to annually increase these reserves by the same proportion of its annual inflation rate. The BIC will convert half of each country's reserves into dollars and use them to process its imports. The other half of each country's reserves will be held in national currency to process its exports. This requirement would apply to all countries belonging to the BIC, except the country with the most widely accepted currency. The country with the most widely accepted currency, in our example the United States, will be the one that provides the base currency that will be used to express the value of all other currencies (primary exchange rates) and also generate liquidity. Consequently, the reserves that this country would have to supply the BIC in its national currency (the dollar) would be equal to the average of the sum of its exports plus its imports of the last three years, plus the average exports of the last three years of all other countries within BIC minus the last three years' average imports of all other countries from the United States. That is, apart from the reserves that the United States would have to provide to the BIC to attend to its foreign trade with other countries, the United States would have to provide the BIC with the reserves to attend to the foreign trade of all other countries within the BIC that are not the United States. These additional reserves would be equal to the average exports of the last three years of the other countries less imports from the United States or, what is the same, the average exports of the last three years of the other countries less the exports of the other countries to the United States. In addition, the BIC may require the United States to increase those reserves by the same proportion of its inflation rate.

Those additional dollars that the United States would have to provide so that the other countries can process their exports with the other countries within the BIC would not affect their economy since they would not be used to acquire merchandise in the United States (exports from the United States to the BIC countries) but to acquire merchandise in the other countries within the BIC (exports of the other countries belonging to the BIC). Consequently, those dollars would be used to provide liquidity to the common market of the countries belonging to the BIC and would not be returning to the United States economy as imports from the other countries. In exchange for the extra dollars that the United States would be providing to give liquidity to the common market of the countries belonging to the BIC, the United States would have an increase in its international currency reserves of the other countries belonging to the BIC equivalent to the additional dollars that it would be providing to give liquidity to the common market. On the day that those dollars were used to import merchandise from the United States instead of serving as an international means of payment among the BIC countries, the United States would have the option of using those international currencies to import goods from the other BIC countries in the same magnitude so that it could keep its trade balance in equilibrium.

All countries would be obliged to increase the share of their national currency in the BIC in the same proportion as their inflation. The BIC will distribute these increments of the national currencies of each country proportionally between the reserves that those countries maintain in the BIC and the reserves of these currencies the United States maintains in the BIC. Consequently, the reserves that the United States has of foreign currency would not lose purchasing power over time. The United States would be obliged to increase the share of its national currency (the dollar) in the BIC in the same proportion as its inflation. The BIC will distribute those increases in the national currency of the United States proportionally between the dollar reserves that the United States maintains in the BIC and the dollar reserves that the other countries maintain in the BIC. Consequently, the reserves that other countries have of dollars would not lose purchasing power over time.

The fact that each country has to increase its reserves within the BIC in the same proportion as its inflation rate allows each country to maintain the required reserves in the currencies of other countries without fear of losing purchasing value. Furthermore, the BIC may require from each country a limit to its inflation rate so that the country that does not comply is expelled from the BIC.

To achieve the objective of balancing the trade balances of all countries within the BIC, the bank would have the power to adjust the exchange rates of the different currencies with the dollar.

Said power of the BIC to adjust and determine the exchange rates of each country would be granted by the member countries of the BIC. The country that does not subscribe to the BIC would be outside the Common Market made up of these countries. If the size of a country's B dollar reserves is reduced below its optimum level, that is, its imports are greater than its exports, the BIC will decrease that country's exchange rate with the dollar TCab. The decrease in the exchange rate will decrease imports and increase exports of that country B, not only with the United States (country A) but with the other countries. The decline in the optimal level of a country's reserves occurs when the value of dollar reserves is less than the value of reserves in national currency. If, on the other hand, the size of a country's B dollar reserves increases above its optimum level, that is, its imports are less than its exports, the BIC will increase that country's exchange rate with the dollar TCab, which will increase imports and decrease exports of that country B, not only with the United States (country A) but with the other countries.

In carrying out the above work with all the countries that trade with each other, the BIC would be determining the set of equivalent universal equilibrium primary exchange rates. That is, in this way the BIC would be determining, by trial and error, the set of equivalent universal equilibrium primary exchange rates.

The primary exchange rates of each currency with the base currency (the dollar) must be modified by the BIC in the corresponding proportion to offset the effect of inflation. If the dollar (Country A) has an inflation of 2% and the currency of Country B has an inflation rate of 5%, then the exchange rate of that currency with the dollar TCab should be modified in the corresponding proportion ($TCab_{t2}=TCab_{t1}*\left(\frac{1+0.02}{1+0.05}\right)=TCab_{t1}*0.971429$).

Finally, the base currency that is used to define the value of the other currencies should not have a fixed exchange rate with gold. This was a mistake of the Bretton Woods conference. Maintaining a fixed exchange rate with the gold of the base currency is equivalent to generating an international market based on a single currency. This has the effect of allowing trade based on absolute advantages, but it hinders trade based on comparative advantages. That was what happened between 1947 and 1971, where the support of the dollar in gold limited the growth of international trade based on goods where countries had comparative advantages. As it has already been pointed out at the beginning of this writing, the trade based on the absolute advantages does not tend towards the balance of the trade balances. It is trade based on comparative advantages that tends towards balance in trade balances and also that which generates the maximum productivity possible with stability in international trade.

## Second Formula Or Procedure For The BIC To Determine The Equivalent Universal Equilibrium Primary Exchange Rates

As we have pointed out, the function of the BIC would be to determine the set of equivalent universal equilibrium primary exchange rates. To achieve this objective, the bank would have the power to adjust the exchange rates of the different currencies with the dollar. In the previous section we have seen how the BIC can adjust the exchange rates with the dollar simulating a float process. The question that arises is: is there a formula, that is to say a procedure, that allows the BIC to fulfill as quickly and efficiently as possible its goal of determining the set of equivalent universal equilibrium primary exchange rates? Let's see.

## THE SET OF EXCHANGE RATE CORRECTION UNITS

Suppose the BIC establishes the goal of balancing the trade balances of all countries within the BIC over a 10-year period. The task that the BIC would have to perform is to determine the set of equivalent universal equilibrium primary exchange rates and apply them. To achieve this objective, the BIC must determine the number of units and the sign in which each exchange rate with the dollar of each country will have to change (increase or decrease) so that the imbalance in its trade balance is reduced to zero. Let us call that set of correction units, ***Set of Exchange Rate Correction Units*** (SERCU). Estimating that SERCU set is the task of the BIC and may require multiple attempts before a satisfactory estimate is achieved that is capable of reducing the imbalance in the trade balances of all countries to zero or at least to imbalances close to zero.

As we have pointed out, the BIC's task is to determine for all countries the number of units in which their primary exchange rates with the dollar must decrease or increase so that these countries reduce their imbalances in their trade balances to zero or to values as close to zero as possible. For this, the BIC must use the statistical information available on the changes that have occurred in the exchange rates with the dollar of each country and the result of those changes on their respective increases or decreases in the trade balances of each country in the long term (2 years). With this information, the BIC must estimate the SERCU set that contains the correction units that must be added to the exchange rates of all the countries to obtain the estimate of the set of equivalent universal equilibrium primary exchange rates. The application of this estimation of the equivalent universal equilibrium primary exchange rates is supposed to correct the imbalances in all the trade balances of all the countries reducing them to zero or to some value close to zero.

Once the equivalent universal equilibrium exchange rate estimate is applied, the BIC will have to wait 2 years to see the results. Based on the differences that the BIC finds after 2 years between the new values assumed by the imbalances in the trade balances and the zero values that the trade balances are supposed to have when they are in equilibrium, a new estimate or correction of the SERCU set would be made again. The procedure would be repeated every 2 years until it gets closer to an increasingly precise estimate of the SERCU set and, therefore, of the equivalent universal equilibrium primary exchange rates.

## HOW TO SIMPLIFY THE DETERMINATION OF THE SERCU SET?

One procedure that could decrease the number of adjustment periods that the BIC would have to carry out to reach the objective of estimating the SERCU set as precisely as possible and thereby achieve the most accurate determination of equivalent universal equilibrium exchange rates is the following. Suppose we divide the imbalance in the trade balances of all countries into 100 parts. Then the BIC must determine the set of $SERCU_{(1/100)}$ exchange rate correction units to reduce by one hundredth the imbalances in the trade balances of all countries. Once this $SERCU_{(1/100)}$ set is obtained, which reduces the imbalances in the countries' trade balances by one hundredth, then it is multiplied by 100 to obtain an estimate of the SERCU set, which theoretically should reduce the imbalances in the balances of all countries at zero or near zero. This estimate is the one that would be applied to all the current primary exchange rates with the dollar and it would take 2 years to verify the result. Let's see an example.

Suppose a Common Market of 4 countries A, B, C and D, under a condition of monopolistic competition, where country A is the one with the most widely accepted currency, in the case of our example United States. Suppose that the primary exchange rates are those that correspond to the quotient of the sum of the prices of country A divided by the sum of the prices of the country under consideration. In Spreadsheet 16 we present this situation.

In Table 8 Column 2 we have detailed the imbalances in the trade balances of the 4 countries that appear in Spreadsheet 16 Lower Part of Section 16.1. Column 3 of this table presents Column 2 divided by 100. Column 4 presents the values that the imbalances in the trade balances of the countries must assume, in order to reduce the imbalances in the trade balances by one hundredth. Column 5 presents the estimate of the $SERCU_{(1/100)}$ set that is supposed to reduce the imbalances in the trade balances of the three countries by one hundredth or as close as possible to one hundredth. This estimate must be made based on the statistical information that exists on the exchange rates of each country and their corresponding trade balances 2 years after the changes in the exchange rates have occurred. Column 6 presents the current exchange rates. Column 7 presents the exchange rates resulting from adding to column 6 the $SERCU_{(1/100)}$ set in column 5. The exchange rates in Column 7 are those that the BIC must apply to reduce the imbalances in the countries' trade balances by one hundredth or as close as possible to one hundredth.

Table 8

| Column 1 | Column 2 | Column 3 | Column 4 | Column 5 | Column 6 | Column 7 |
|---|---|---|---|---|---|---|
| | TB | TB/100 | TB-(TB/100) | $SERCU_{(1/100)}$ | Current Exchange Rates (CER) | Resulting Exchange Rates CER+ $SERCU_{(1/100)}$ |
| Country B | 214.33 | 2.1433 | 212.18 | 0.00001000 | 0.027541 | 0.027551 |
| Country C | -22.37 | (0.2237) | (22.15) | 0.00010000 | 0.437717 | 0.437817 |
| Country D | -19.75 | (0.1975) | (19.55) | -0.00001000 | 0.010079 | 0.010069 |

In the Appendix Spreadsheet 19 Page 1 Section 19.1 and in the Appendix Spreadsheet 19 Page 1 Section 19.7 and 19.8 we have presented the Spreadsheet 16. The calculation of columns 2, 3 and 4 of Table 8 is what it is presented in between sections 19.7 and 19.8. In the Appendix Spreadsheet 19 Page 2 Section 19.1 and in the Appendix Spreadsheet 19 Page 2 Section 19.7 and 19.8 the model of the Spreadsheet 19 Page 1 is presented but with the new rates of change of the Spreadsheet 19 Page 2 Section 19.7. These new exchange rates on Page 2 are the result of adding to the current exchange rates on Page 1 the estimate of the $SERCU_{(1/100)}$ set that appears on Page 2, between sections 19.7 and 19.8. In Table 8 Column 5 this estimate of the $SERCU_{(1/100)}$ set appears, in Column 6 these current exchange rates appear, and in Column 7 said resulting exchange rates. As can be seen in Appendix Spreadsheet 19 Page 2 Lower Part Section 19.1, the imbalances in the trade balances of the countries A, B, C and D (213.75, -22.53 and -19.40) are not exactly the same as the projected imbalances in Column 4 of Table 8 (212.18, -22.15 and -19.55). Therefore, we must correct again the estimate of the $SERCU_{(1/100)}$ set. This correction procedure in the estimation of the SERCU$_{(1/100)}$ set is continued until the SERCU$_{(1/100)}$ set is obtained by successive approximations, which reduces the imbalances in the trade balances of all countries to one hundredth part or closest possible to one hundredth (Table 8 Column 4).

The Appendix Spreadsheet 19 was obtained from the Spreadsheet 19 shown below and where the international trade model was introduced.

Spreadsheet 19

Microsoft
Excel Worksheet

Every time a new estimate of SERCU$_{(1/100)}$ is obtained, it is necessary to wait 2 years to verify the result. Therefore, the adjustment procedure would be repeated every 2 years until the most possible accurate estimate of the set of correction units SERCU$_{(1/100)}$ is obtained to reduce trade balance imbalances by one hundredth. As can be seen on Page 3, Lower Part of Section 19.1, the values of the trade balances of countries B, C and D (212.18, -22.15 and -19.55) are equal to the corresponding values in Table 8 Column 4 (212.18, -22.15 and -19.55). Consequently, that SERCU$_{(1/100)}$ set of Page 3 is the best estimate because it achieves the objective of reducing the trade balances of all countries by one hundredth or as close as possible to one hundredth.

Once the best estimate of that SERCU$_{(1/100)}$ set has been obtained, which reduces the imbalances in the trade balances of all the countries by one hundredth, it is multiplied by 100 to obtain an estimate of the SERCU set that reduces to zero or as close as possible to zero the imbalance in the trade balances of all countries. This estimate of SERCU set which reduces the imbalances in the trade balances of all countries to values close to zero, is presented in Appendix Spreadsheet 19 Page 4. As can be seen on Appendix Spreadsheet 19 Page 4, Lower Part of Section 19.1, the values of the trade balances of countries B, C and D are close to zero. Country A trade balance reduced its deficit from -170.48 in Appendix 19 Page 3 Lower Part Section 19.1 to a relatively small surplus of 14.16 in Appendix 19 Page 4 Lower Part Section 19.1, Country B trade balance reduced its surplus from 212.18 to a relatively small deficit of -9.61, Country C trade balance reduced its deficit from -22.15 to a relatively small deficit of -2.32 and Country D trade balance reduced its deficit from -19.55 to a relatively small deficit of -2.23.

There is no doubt that, with this result, the countries belonging to the BIC would be in a much better situation than under the original situation. Consequently, even if the BIC did not make corrections again on the SERCU set, it would be achieving a great improvement in reducing the imbalances in the trade balances of all countries. However, the BIC's task must be a continuous one over time, so that as it makes new corrections every 2 years in the SERCU set, the countries trade balances will come closer to the perfect balance.

Consequently, on these results the BIC can readjust the exchange rates every 2 years to approximate an increasingly accurate estimate of the SERCU set, which when added to the original exchange rates generates the equivalent universal equilibrium primary exchange rates. In the Appendix Spreadsheet 19 Page 5 Lower Part Section 19.1 the SERCU set is presented, which when added to the original exchange rates generates the equivalent universal equilibrium primary exchange rates that produce equilibrium in the trade balances of all countries.

Note that multiplying the $SERCU_{(1/100)}$ set presented on Page 3 by 100 gives an inaccurate estimate of the SERCU set corresponding to the equivalent universal equilibrium primary exchange rates. This margin of error in the SERCU set estimate is due to changes that occur in the elasticities of trade balances as imbalances are reduced.

If the elasticities of the trade balances with respect to the corresponding exchange rates were constant, then all that would have to be done would be to multiply the $SERCU_{(1/100)}$ set on Page 3 by 100 and apply it to all the exchange rates to obtain the set of equivalent universal equilibrium exchange rates. In such a case, since the elasticities are assumed to be constant, the relationship between $SERCU_{(1/100)}$ and SERCU would be linear. However, since we know that these elasticities are not constant, we should expect that our estimate of SERCU based on the multiplication of $SERCU_{(1/100)}$ by 100 should be corrected. The time it takes to refine these estimates for the SERCU set should be considered an ongoing task of the BIC.

## THE STABILITY IN THE GROWTH OF INTERNATIONAL TRADE AND THE EQUILIBRIUM IN THE FINANCIAL AND CAPITAL BALANCE IN THE SHORT AND MEDIUM TERM

Note that in the short and medium term, it would not be necessary for each country to maintain its financial and capital balance in equilibrium to ensure that the adjustments in the currency exchange rate, administered by the BIC, generate the equivalent universal equilibrium primary exchange rates. However, in the long term, the country that has maintained a continuous surplus in its financial or capital balance would be forced to generate a surplus in its trade balance with which to pay the principal and interest on loans made and/or earnings to be repatriated from foreign investments. That is to say, it would have to pay by depreciating its currency and thus generate a surplus in its trade balance with which to pay its debt. It is to be expected that the surplus in the trade balance of that country will simultaneously generate a deficit in the trade balance, not only of the country that loaned it the money or provided the investment, but also generates a deficit in the trade balance of all other countries with which that country trades, even if they are not countries that have lent it the money or have made the investment.

The demonstration of the statement that we make in the previous paragraph is as follows. Suppose we are in an equilibrium situation in a market of 4 countries A, B, C and D under a monopolistic competition market, as shown in Spreadsheet 17. Suppose that country D, as a consequence of the continuous surpluses in its financial and of capital balances, accumulates in the long term a debt in the payment of the principal and the interests, as well as in the payment of dividends and earnings. Suppose that country D, in order to pay its foreign debt, decides to depreciate its currency to generate a surplus in its trade balance with which to pay its debt. Lastly, suppose that for this purpose it decides to decrease its exchange rate with the dollar from TCad = 0.009936 in Spreadsheet 17 to TCad = 0.008936.

In the Appendix Spreadsheet 20 Image Section 20.1 the model with the results of the trade balances of the 4 countries is presented.

Appendix Spreadsheet 20, Image Section 20.1 was obtained from Spreadsheet 20 shown below and where the model on international trade was introduced.

Spreadsheet 20
Trade Balance of 4 Countries

Microsoft
Excel Worksheet

In the Appendix Spreadsheet 20 Section 20.1 lower part, it can be seen that the surplus in the trade balance generated by country D was BCd.abc = 22.11. Said surplus in its trade balance generated a deficit of -6.82 in the trade balance of country A that loaned the money or made the investment in country D. But not only did it generate a deficit in the trade balance of Country A that lent the money or made the investment, but also generated a deficit in the trade balances of the other countries with which country D trades and that did not lend it the money or made the investment. The deficit it generated in country B was -6.74 and in country C it was -8.55. Note that the level of the surplus in the trade balance of country D generated by the devaluation of its currency against the dollar, is equal to the sum of the deficits in the trade balances of all the other countries that trade with country D and each other and whose equilibrium exchange rates with the dollar will no longer generate the equilibrium of their respective trade balances. In this way, it is verified that the equilibrium exchange rate with the dollar of one country is not independent of the equilibrium exchange rates with the dollar of the other countries with which it trades and that long-term equilibrium in the trade balances of all countries that trade with each other can only be achieved collectively when financial and capital balances are kept in equilibrium. Consequently, the function of the BIC would be to determine and maintain the set of equivalent universal equilibrium primary exchange rates of all the currencies of the countries that trade among themselves with respect to the dollar, while the function of the WB would be to maintain the equilibrium of the financial and capital balances of all countries that trade with each other.

Note that, based on the results presented in the Appendix Spreadsheet 20 Section 20.1 lower part, that we have just discussed, we can conclude that as long as all countries do not establish and achieve the common objective of maintaining the equilibrium of their financial and capital balances, the external debts generated by the deficits in the trade balance of some countries, among which the main one would be the United States, will be condemned to grow indefinitely. And the worst thing is that these external debts of the United States and of the countries with deficit trade balances, will be condemned to grow indefinitely to feed the surpluses in the trade balances of the countries that keep their currencies undervalued with respect to the dollar, in order to generate surpluses in their trade balances with which to pay the interest and the principal of the loans made, as well as the repatriation of the profits corresponding to foreign investments in their territory.

Note also that when a country like China maintains its currency devalued to generate a surplus in its trade balance and, as a consequence, generates deficits in the trade balance of the United

States and other countries, it happens that this leads other countries to devalue their currencies also against the dollar to offset their deficit. Consequently, the United States deficit will increase further.

## THE REGULATION OF LOANS AND FOREIGN INVESTMENTS OF EACH COUNTRY

In addition to the previous task of determining the equivalent universal equilibrium primary exchange rates, the BIC or any other designated institution, such as the WB, would be in charge of regulating the loans and foreign investments of each country, ensuring that the equilibrium of the financial and capital balances is maintained. For this, it must be established that all foreign loans are made through the WB that would convert the foreign currency in which the loan is made into national currency, functioning as a currency clearinghouse of foreign currencies. The same procedure would be used in the case of direct investments. Among the objectives that the WB would have, would be to ensure that for each foreign loan or investment received by country B, a loan or investment from that country B is generated abroad, so as to maintain the equilibrium in the financial balance and the capital balance of that country. The interest to pay on both loans should ideally be the same, so that the equilibrium in the income balance within the current account balance is maintained. Each investment from one country to another must be processed through the WB and the country that receives the investment must use the corresponding increase in its foreign currency to make an equivalent investment in the country from which it received the investment.

In other words, the IMF or BIC would function as a supranational institution (clearinghouse) that would be in charge of determining equilibrium exchange rates with the dollar of each country and the World Bank, would function as a supranational institution (clearinghouse) that would be in charge of supervising the financial and capital balances of each country in order to avoid imbalances in the financial and capital balances. By maintaining the equilibrium of these balances, the World Bank would be achieving a second objective, which would be to avoid the destabilization of national economies by allowing countries with equilibrated financial and capital balances to acquire control over their monetary policy and interest rate, as well as its fiscal policy.

To achieve the above objectives, it would be necessary to convene a second Breton Woods conference that would be in charge of the creation of the BIC or the transformation of the IMF and the WB.

In the event that the United States decides not to participate in the creation of the BIC or in the redefinition of the IMF and the WB, the other countries could choose to create an international currency similar to the ***bancor*** proposed by John Maynard Keynes at the Bretton Woods conference. This currency would be backed by the foreign exchange reserves that the countries have in the BIC of the most widely accepted currencies, including the dollar.

## RULES AND CONDITIONS NEEDED FOR THE AMERICAN HEMISPHERE COMMON MARKET TO WORK

In the previous sections we have explained the way in which a common market can be achieved in which all the countries that trade with each other manage to keep their trade balance in equilibrium. We have shown that if we start from the premise of a perfect competition market or a monopolistic competition market, we can obtain the set of primary exchange rates that balance the trade balances of all the countries that trade with each other.

Note that it starts from the premise that countries within that common market do not trade with third countries outside that common market or with other common markets. That is a necessary condition for determining the set of primary exchange rates that balance the trade balances of all countries that trade with each other.

The reason for maintaining the previous restriction is as follows. If a country within the common market negotiates with third countries outside the common market, it affects the determination of all equivalent universal equilibrium primary exchange rates that balances the trade balances of all other countries within the common market. In such a case, the only way to determine the equivalent universal equilibrium primary exchange rates is if these third countries establish their exchange rates with the dollar and are subject to the same rules that regulate the trade of countries within that common market.

In other words, if we define the set of equivalent universal equilibrium primary exchange rates with the dollar, as one that balances the trade balances of all countries that trade with each other, including the United States, then no country can set its equilibrium exchange rate with the dollar regardless of the equilibrium exchange rates with the dollar of other countries, because the exchange rate of equilibrium with the dollar of any country is not independent of the exchange rate of equilibrium with the dollar of the other countries with which that country trades, apart from its trade with the United States. This is why neither China, nor Mexico, nor Canada, can set their equilibrium exchange rate with the dollar regardless of the equilibrium exchange rate with the dollar of other countries. The only way to do this is that all countries that trade with the United States belong to the same common market and then determine their exchange rate with the dollar through put their currency to float against the dollar, while determining the rest of the exchange rates among the other countries by applying the formulas that define the set of equivalent exchange rates.

Note that the equilibrium exchange rate with the dollar of a country is not the one that balances the trade balance of that country with the United States, but the one that balances the trade balance of that country with all the other countries with which it trades, including the United States. Consequently, the equilibrium exchange rate of that country with the United States can generate surpluses in its trade balance with the United States, as long as it is offset by the deficits of that country with the other countries with which it trades.

## How to Achieve The Countries That Integrate The Common Market Stop Trading With Third Countries

### The Paulatine Establishment of a General Tariff

A rule that the countries of the American Hemisphere Common Market (AHCM) must have is that they do not trade with third countries. That is, the above formula presupposes that the countries that make up the common market do not trade with third countries that are not within the common market.

One way to proceed is for the American Hemisphere Common Market to establish a general tariff that will isolate it from trade with countries that do not belong to the common market.

The tariff should be applied gradually to avoid causing abrupt changes in the economies of the countries. For example, a 2% tariff can be imposed on all merchandise from countries outside the American Hemisphere Common Market. That tariff would increase at a rate of 2% per year for 10 years to reach a maximum of 20%.

### Exceptions to The Application of a General Tariff

Because there are third countries with key goods that cannot be or are difficult or that should not be produced within the countries that make up the common market, the general tariff would not apply to those key goods. For example, in the case of oil, it is a key asset. Therefore, the general tariff would not apply to imports of oil from third countries outside the common market.

What effect would it have not applied the general tariff to countries that export oil or key goods who are not subscribed to the BIC on the equilibrium in the trade balance of the countries that are subscribed to the BIC and what should the BIC do? As we have already discussed in a previous topic, the surpluses in the trade balance of one country are financed with the deficits in the trade balance of the other countries, preventing the determination of the equivalent universal equilibrium primary exchange rates. However, as we will see later in Spreadsheed 21, the foregoing does not prevent the BIC from achieving the determination of the set of equivalent universal equilibrium primary exchange rates that balance the trade balance of all countries within the common market subscribed to the BIC, except for the country's trade balance with the most widely accepted currency. In such a case it would be the United States that would have to assume in a deficit, the surplus of the oil exporting countries or of key goods that do not subscribe to the BIC and that export their product to the BIC countries. However, this deficit that the United States would have to assume in its trade balance would not have great repercussion for the reasons that we will explain in the following paragraphs.

Under the circumstance described in the previous paragraph, the formula or procedure that the BIC would have to follow to determine the equivalent universal equilibrium primary exchange rates would exclude the country or countries with key goods from the adjustments in the exchange rate since they are not subscribed to the BIC. This would result in the BIC determining the set of equivalent universal equilibrium exchange rates of all countries within the common market subscribed to the BIC, except that of the country with the most widely accepted currency, that is, the United States. Countries with key goods that do not belong to the BIC but that export their key product to the BIC countries would have a surplus in their trade balances that would be financed with a deficit in the United States trade balance. The country or countries with key goods, not belonging to the BIC, would end up with a trade balance in surplus, while the other countries within the common market subscribed to the BIC would end up with their trade balances in equilibrium, except for the country with the most widely accepted currency that would absorb said surpluses in a deficit. As we have pointed out, under that circumstance, the country with the most widely accepted currency in which the primary exchange rates are established, in the case that concerns us the United States, would be the only one that would end up absorbing in its deficit the surpluses in the balance of the oil exporting countries that do not subscribe to the BIC. However, as we have pointed out, this would not be a very serious problem since, the common market defined by the countries that subscribe to the BIC can take measures that neutralize the harmful effect of this deficit on the United States trade balance, as we will see later in the next topic.

The demonstration of the statements we make in the previous paragraph is as follows. Suppose that we are in a situation of a market of 4 countries under monopolistic competition, as shown in Spreadsheed 16. Suppose that country B that exhibits a surplus of 214.36 is an oil country like Saudi Arabia that is not subscribed to the BIC. In the Appendix Spreadsheet 21 Page 1 Image Section 21.1 the model with the results of the trade balances of the 4 countries is presented. Finally, suppose that the BIC cannot include country B in determining the set of universal equilibrium exchange rates because it does not belong to the BIC.

Appendix Spreadsheet 21 Page Image Section 21.1 was obtained from Spreadsheet 21 shown below and where the model on international trade was introduced.

Spreadsheet 21
Trade Balance of 4 Countries

Microsoft
Excel Worksheet

In this case, the BIC will only focus on determining the set of equivalent universal equilibrium exchange rates TCac and TCad of all the other countries subscribed to the BIC (countries C and D) minus country B. In such case, the BIC will estimate the SERCU set that when added to the current exchange rates generate an estimate of the set of equivalent general equilibrium exchange rates of all the other countries subscribed to the BIC (countries C and D) minus country B (see Appendix Spreadsheet 21 Page 2 Image Sections 21.1 and 21.7). The BIC would apply these exchange rates and would wait 2 years to verify the results. Based on what has been observed, the BIC may make corrections to the estimate of the SERCU set (see Appendix Spreadsheet 21 Page 3 Image Sections 21.1 and 21.7) and so on until the equivalent universal equilibrium primary exchange rates shown in the Appendix Spreadsheet 21 Page 4 Image Sections 21.1 and 21.7 are determined.

As can be seen in Page 4, countries C and D belonging to the common market subscribed to the BIC end with their trade balances in equilibrium and country A with the most widely accepted currency, in the case of our example, the United States, end with a deficit -192.23 equal to the surplus 192.23 from country B, in the case of our example Saudi Arabia.

In this way, it is demonstrated that a general tariff can be applied to all countries that do not belong to the BIC and leave the exporting countries of key goods exempt from the application of that tariff in the export of their key goods. This measure would not prevent the universal equivalent equilibrium primary exchange rates of all the countries belonging to the BIC from being determined and thereby achieve equilibrium in the trade balances of all countries, except the country with the most widely accepted currency.

Table 9 shows the 20 countries with the largest trade balance imbalances (surpluses or deficits) in order of magnitude measured in dollars. It can be seen that the greatest imbalance in the trade balance is that of the European Union, followed by Russia, India and Saudi Arabia. Russia and Saudi Arabia are oil exporting countries. If we assume that Russia subscribes to the BIC but Saudi Arabia does not, then the BIC should appreciate the currency of the European Union, the Euro, against the dollar as well as the currencies of the other countries with large surpluses, except the currency of Saudi Arabia. Regarding Saudi Arabia, the BIC will not adjust its exchange rate since it does not belong to the BIC. In the case of countries with large deficits in their trade balance, the BIC will depreciate their currencies against the dollar. The procedure would continue until the equivalent universal equilibrium primary exchange rates have been obtained for all countries belonging to the BIC, except for Saudi Arabia that does not belong to the BIC.

Table 9

| | | | Superavit or Deficit Trade Balance |
|---|---|---|---|
| 1 | Euro area | 2018 | 483,094,018.84 |
| 2 | Russian Federation | 2018 | 164,490,090.00 |
| 3 | India | 2018 | -105,917,684.94 |
| 4 | Saudi Arabia | 2018 | 104,559,202.36 |
| 5 | China | 2018 | 102,921,464.81 |
| 6 | Singapore | 2018 | 96,726,920.01 |
| 7 | Korea, Rep. | 2018 | 82,129,500.00 |
| 8 | United Kingdom | 2018 | -50,188,171.11 |
| 9 | Thailand | 2018 | 48,545,384.21 |
| 10 | Philippines | 2018 | -39,349,787.74 |
| 11 | Pakistan | 2018 | -37,738,340.00 |
| 12 | Qatar | 2018 | 36,750,000.00 |
| 13 | Canada | 2018 | -36,362,039.63 |
| 14 | Iraq | 2018 | 35,054,700.00 |
| 16 | Macao SAR, China | 2018 | 27,567,758.71 |
| 17 | Malaysia | 2018 | 25,132,636.09 |
| 18 | Egypt, Arab Rep. | 2018 | -24,708,593.00 |
| 19 | Mexico | 2018 | -22,719,234.60 |
| 20 | Algeria | 2017 | -22,556,190.37 |

### Effect of Trade of Key Goods With Third Countries on Equilibrium in Commercial Balances of Common Market Countries

In the previous paragraphs we have explained how, in a Common Market of the American Hemisphere (CMAH) where the member countries do not trade with third countries outside the common market, it is possible to determine the equivalent universal equilibrium primary exchange rates that balance the trade balances of all countries that trade with each other.

The question that arises at this moment is: how to avoid that the trade of key goods of the countries belonging to the CMAH with third countries that are not members of the CMAH affect the determination of the equivalent universal equilibrium primary exchange rates that balance the trade balances from all CMAH countries? We have already seen in a previous topic that the solution to this problem consists in establishing as an objective that the BIC determines the equivalent universal equilibrium primary exchange rates of all the countries belonging to the BIC except for the countries that export key goods that do not belong to the BIC. In this way, the BIC would be generating that all the countries belonging to the BIC can have their trade balances in balance, while the surplus of the countries exporting key goods that do not belong to the BIC is financed with the country's deficit with the currency of greater acceptance, in the case of our example, the dollar. However, this solution does not consider how to solve the problem that would be generated by determining the equivalent universal equilibrium primary exchange rates if the countries that export key goods want to trade with the countries belonging to the BIC without belonging to the BIC.

As we have pointed out, the goods of third countries to which the CMAH does not impose the general tariff because they are key goods, may be freely purchased by the countries belonging to the CMAH, including the United States itself. In order that the trade of these non-BIC countries with the BIC countries not to affect the determinations of the equilibrium exchange rates with the dollar of the CMAH countries belonging to the BIC, it is necessary to take some measures. First, these imports must be paid in dollars as that would facilitate maintaining balance in the trade balances of all countries within the CMAH. It is expected that those imports that are paid in dollars will generate a deficit in the US trade balance due to the magnitude of those CMAH imports. Those dollars that leave outside the CMAH economies have one of three destinations: 1. they return as loans or as investments of those third countries in to the CMAH countries including the United States, 2. they return as imports from those third countries in to the CMAH countries including the United States, 3. they remain outside the CMAH as treasured money in third countries.

In the first case, the surpluses in dollars of third countries should not be allowed to return as loans or as investments of those third countries in to the CMAH countries. That is, these loans or investments from third countries in the CMAH countries should be prohibited. They should not be allowed, because such loans or investments would prevent keeping the financial balance, the capital balance and the income balance of the CMAH countries in equilibrium. As discussed in previous sections, in order to make possible generate the equivalent universal equilibrium primary exchange rates that balance the trade balances of all the CMAH countries simultaneously, it is necessary maintain equilibrium in those three balances. Consequently, the only two viable alternatives that third countries would have with the dollars they accumulate would be to use them to buy goods in the CMAH countries or keep them out of the CMAH countries (as an accumulator of wealth or as a means of international payment).

In the second case, if all the dollars paid for the imports made by the CMAH countries return as imports from those third countries in the CMAH countries, that is, as exports from the CMAH countries to those third countries, then nothing would have to be done. In such a case, the trade balances of the CMAH countries with the third countries would be in equilibrium and, consequently, we can expect that the deficit of the United States with the exporting countries of key goods is reduced by the same proportion. Consequently, the equivalent universal equilibrium primary exchange rates that balance the trade balances of all countries within the CMAH would not be affected.

In the third case, if the dollars are kept out of the CMAH as treasured money, the monetary mass of dollars should only be increased by that amount. Consequently, dollars that do not return as exports of the countries within the CMAH to third countries must be replaced by the United States government.

### The Incorporation of New Countries to The American Hemisphere Common Market

The incorporation of new countries or common markets in the American Hemisphere Common Market presupposes that the country or the common market incorporated accepts not to trade with third countries without applying the general tariff established by the American Hemisphere Common Market.

The countries that could be incorporated would be those of the European Common Market, Japan, South Korea and Taiwan.

### The Use of Computer Programs so that The BIC Can Speed Up the Procedure to Determine the Equivalent Universal Equilibrium Primary Exchange Rates by Trial and Error Process

The question that remains to be answered: is there a possibility to create a computer program that can speed up the trial and error process that the BIC has to carry out to determine the equivalent universal equilibrium primary exchange rates?

# Appendix Spreadsheet 1

## Image Section 1.1
## COMMERCIAL BALANCE. MODEL EQUATIONS

| | | | | | | | | | | | | | | | | |
|---|---|---|---|---|---|---|---|---|---|---|---|---|---|---|---|---|
| **Spreadsheet 1 - MODEL SIMULATION INTERNATIONAL TRADE OF 3 COUNTRIES:** | | | | | | | | | | | | | | | | |
| | | | | | | | | | | **SECTION 1.1: COMMERCIAL BALANCES. MODEL EQUATIONS** | | | | | | |
| | | **Equation 5.1 Defining the Trade Balance of Country A (United States) with All Other Countries (BCa.bc)** | | | | | | | | | | | | | | |
| | | **Bca.bc = BCab1 + BCab2 + BCab3 + BCab4 + BCac1 + BCac2 + BCac4 + BCac4** | | | | | | | | | | | | | | |
| **BCa.bc** | **=** | **BCab1** | **+** | **BCab2** | **+** | **BCab3** | **+** | **BCab4** | **+** | **BCac1** | **+** | **BCac2** | **+** | **BCac3** | **+** | **BCac4** |
| **78.08** | **=** | **-19.57** | **+** | **(5.20)** | **+** | **53.61** | **+** | **35.74** | **+** | **(3.67)** | **+** | **(1.00)** | **+** | **10.91** | **+** | **7.27** |
| | | **Equation 5.2 Defining the Trade Balance of Country B (Mexico) with All Other Countries (BCb.ac)** | | | | | | | | | | | | | | |
| | | | | **BCb.ac = BCba1 + BCba2 + BCba3 + BCba4 + BCbc1 + BCbc2 + BCbc4 + BCbca4** | | | | | | | | | | | | |
| **BCb.ac** | **=** | **BCba1** | **+** | **BCba2** | **+** | **BCba3** | **+** | **BCba4** | **+** | **BCbc1** | **+** | **BCbc2** | **+** | **BCbc3** | **+** | **BCbc4** |
| **-71.74** | **=** | **19.57** | **+** | **5.20** | **+** | **(53.61)** | **+** | **(35.74)** | **+** | **2.57** | **+** | **0.66** | **+** | **(6.23)** | **+** | **(4.16)** |
| | | **Equation 5.3 Defining the Trade Balance of Country C (Canada) with All Other Countries (BCc.ab)** | | | | | | | | | | | | | | |
| | | | | **BCc.ab = BCca1 + BCca2 + BCca3 + BCca4 + BCcb1 + BCcb2 + BCcb3 + BCcb4** | | | | | | | | | | | | |
| **BCc.ab** | **=** | **BCca1** | **+** | **BCca2** | **+** | **BCca3** | **+** | **BCca4** | **+** | **BCcb1** | **+** | **BCcb2** | **+** | **BCcb3** | **+** | **BCcb4** |
| **-6.34** | **=** | **3.67** | **+** | **1.00** | **+** | **(10.91)** | **+** | **(7.27)** | **+** | **(2.57)** | **+** | **(0.66)** | **+** | **6.23** | **+** | **4.16** |
| | | | | | **BCab =** | **64.57** | | | | | | | **BCac =** | **13.51** | | |
| | | | | | **BCba =** | **(64.57)** | | | | | | | **BCbc =** | **(7.16)** | | |
| | | | | | **BCca =** | **(13.51)** | | | | | | | **BCcb =** | **7.16** | | |

# Appendix Spreadsheet 1

## Image Section 1.2
## Determination of the trade balances BCabi and BCaci, the prices Paij and the sizes of the markets TMaij in Country A: UNITED STATES

| | | | | | | | | | | | | | | | | | | |
|---|---|---|---|---|---|---|---|---|---|---|---|---|---|---|---|---|---|---|
| | | | | **SECTION 1.2: Determination of Trade Balances BCabi and BCaci, of prices, market sizes and exports and imports Country A: UNITED STATES** | | | | | | | | | | | | | | |
| **US$ BCab1=((Pb1*TCab - Pa1)/((Pb1*TCab + Pa1)/2))TMab1 = Trade Balance of country A with country B corresponding to good 1** | | | | | | | | | | | | | | | | | | |
| **-19.57** | | | | | | | | | | | | | | | | | | |
| | | **US$ BCac1=((Pc1*TCac - Pa1)/((Pc1*TCac + Pa1)/2))TMac1 = Trade Balance of country A with country C corresponding to good 1** | | | | | | | | | | | | | | | | |
| | | **-3.67** | | | | | | | | | | | | | | | | |
| | **Pa1=Wa*Za1 + Ga1** | | | **Pa1b=Pb1*TCab** | | **Pa1c=Pc1*TCac** | | | **US$ TMab1=Min(VPa1, VPb1*TCab)** | | | | **US$ TMac1=Min(VPa1, VPc1*TCac)** | | | | **VPa1=Pa1*Qa1** | |
| | **23.08** | | | **20.00** | | **21.21** | | | **137.00** | | | | **43.64** | | | | **500.00** | |
| | | **Wa** | | **Za1** | | **Ga1** | | **Qa1=Na1/Za1** | | **Na1=FT1a*Na** | | | | | | | | |
| | | **4.88** | | **3.00** | | **8.42** | | **21.67** | | **65.00** | | | | | | | | |
| **US$ BCab2=((Pb2*TCab - Pa2)/((Pb2*TCab + Pa2)/2))TMab2** | | | | | | | | | | | | | | | | | | |
| **-5.20** | | | | | | | | | | | | | | | | | | |
| | | **US$ BCac2=((Pc2*TCac - Pa2)/((Pc2*TCac + Pa2)/2))TMac2** | | | | | | | | | | | | | | | | |
| | | **-1.00** | | | | | | | | | | | | | | | | |
| | **Pa2=Wa*Za2 + Ga2** | | | **Pa2b=Pb2*TCab** | | **Pa2c=Pc2*TCac** | | | **US$ TMab2=Min(VPa2, VPb2*TCab)** | | | | **US$ TMac2=Min(VPa2, VPc2*TCac)** | | | | **VPa2=Pa2*Qa2** | |
| | **15.38** | | | **15.00** | | **15.15** | | | **205.50** | | | | **65.45** | | | | **750.00** | |
| | | **Wa** | | **Za2** | | **Ga2** | | **Qa2=Na2/Za2** | | **Na2=FT2a*Na** | | | | | | | | |
| | | **4.88** | | **2.00** | | **5.62** | | **48.75** | | **97.50** | | | | | | | | |
| **US$ BCab3=((Pb3*TCab - Pa3)/((Pb3*TCab + Pa3)/2))TMab3** | | | | | | | | | | | | | | | | | | |
| **53.61** | | | | | | | | | | | | | | | | | | |
| | | **US$ BCac3=((Pc3*TCac - Pa3)/((Pc3*TCac + Pa3)/2))TMac3** | | | | | | | | | | | | | | | | |
| | | **10.91** | | | | | | | | | | | | | | | | |
| | **Pa3=Wa*Za3 + Ga3** | | | **Pa3b=Pb3*TCab** | | **Pa3c=Pc3*TCac** | | | **US$ TMab3=Min(VPa3, VPb3*TCab)** | | | | **US$ TMac3=Min(VPa3, VPc3*TCac)** | | | | **VPa3=Pa3*Qa3** | |
| | **7.69** | | | **10.00** | | **9.09** | | | **205.50** | | | | **65.45** | | | | **750.00** | |
| | | **Wa** | | **Za3** | | **Ga3** | | **Qa3=Na3/Za3** | | **Na3=FT3a*Na** | | | | | | | | |
| | | **4.88** | | **1.00** | | **2.81** | | **97.50** | | **97.50** | | | | | | | | |
| **US$ BCab4=((Pb4*TCab - Pa4)/((Pb4*TCab + Pa4)/2))TMab4** | | | | | | | | | | | | | | | | | | |
| **35.74** | | | | | | | | | | | | | | | | | | |
| | | **US$ BCac4=((Pc4*TCac - Pa4)/((Pc4*TCac + Pa4)/2))TMac4** | | | | | | | | | | | | | | | | |
| | | **7.27** | | | | | | | | | | | | | | | | |
| | **Pa4=Wa*Za4 + Ga4** | | | **Pa4b=Pb4*TCab** | | **Pa4c=Pc4*TCac** | | | **US$ TMab4=Min(VPa4, VPb4*TCab)** | | | | **US$ TMac4=Min(VPa4, VPc4*TCac)** | | | | **VPa4=Pa4*Qa4** | |
| | **3.85** | | | **5.00** | | **4.55** | | | **137.00** | | | | **43.64** | | | | **500.00** | |
| | | **Wa** | | **Za4** | | **Ga4** | | **Qa4=Na4/Za4** | | **Na4=FT4a*Na** | | | | | | | | |
| | | **4.88** | | **0.50** | | **1.40** | | **130.00** | | **65.00** | | | | | | | | |

# Appendix Spreadsheet 1
## Image Section 1.3
## Determination of the trade balances BCbai and BCbci, the prices Pbij and the sizes of the markets TMaij in Country B: MEXICO

| | | | | | | | | | | | | | | | | | | |
|---|---|---|---|---|---|---|---|---|---|---|---|---|---|---|---|---|---|---|
| | | | | SECTION 1.3: Determination of Trade Balances BCbai and BCbci, of prices, of market sizes and exports and imports of Country B: MEXICO | | | | | | | | | | | | | | |
| US$ BCba1=((Pa1*TCba - Pb1)/((Pa1*TCba + Pb1)/2))TMab1 | | | | | | | | | | | | | | | | | | |
| 19.57 | | | | | | | | | | | | | | | | | | |
| | | US$ BCbc1=((Pc1*TCbc - Pb1)/((Pc1*TCbc + Pb1)/2))TMbc1 | | | | | | | | | | | | | | | | |
| | | 2.57 | | | | | | | | | | | | | | | | |
| | Pb1=Wb*Zb1 + Gb1 | | | Pb1a=Pa1*TCba | | Pb1c=Pc1*TCbc | | | US$ TMba1=Min(VPb1*Tcab, VPa1) | | | | US$ TMbc1=Min(VPb1*TCab, VPc1*TCac) | | | | VPb1=Pb1*Qb1 | |
| | 729.93 | | | 842.22 | | 774.17 | | | 137.00 | | | | 43.64 | | | | 5,000.00 | |
| | | Wb | | Zb1 | | Gb1 | | Qb1=Nb1/Zb1 | | Nb1=FT1b*Nb | | | | | | | | |
| | | 83.03 | | 4.00 | | 397.81 | | 6.85 | | 27.40 | | | | | | | | |
| US$ BCba2=((Pa2*TCba - Pb2)/((Pa2*TCba + Pb2)/2))TMab2 | | | | | | | | | | | | | | | | | | |
| 5.20 | | | | | | | | | | | | | | | | | | |
| | | US$ BCbc2=((Pc2*TCbc - Pb2)/((Pc2*TCbc + Pb2)/2))TMbc2 | | | | | | | | | | | | | | | | |
| | | 0.66 | | | | | | | | | | | | | | | | |
| | Pb2=Wb2*Zb2 + Gb2 | | | Pb2a=Pa2*TCba | | Pb2c=Pc2*TCbc | | | US$ TMba2=Min(VPb2*Tcab, VPa2) | | | | US$ TMbc2=Min(VPb2*TCab, VPc2*TCac) | | | | VPb2=Pb2*Qb2 | |
| | 547.45 | | | 561.48 | | 552.98 | | | 205.50 | | | | 65.45 | | | | 7,500.00 | |
| | | Wb2 | | Zb2 | | Gb2 | | Qb2=Nb2/Zb2 | | Nb2=FT2b*Nb | | | | | | | | |
| | | 83.03 | | 3.00 | | 298.36 | | 13.70 | | 41.10 | | | | | | | | |
| US$ BCba3=((Pa3*TCba - Pb3)/((Pa3*TCba + Pb3)/2))TMab3 | | | | | | | | | | | | | | | | | | |
| -53.61 | | | | | | | | | | | | | | | | | | |
| | | US$ BCbc3=((Pc3*TCbc - Pb3)/((Pc3*TCbc + Pb3)/2))TMbc3 | | | | | | | | | | | | | | | | |
| | | -6.23 | | | | | | | | | | | | | | | | |
| | Pb3=Wb3*Zb3 + Gb3 | | | Pb3a=Pa3*TCba | | Pb3c=Pc3*TCbc | | | US$ TMba3=Min(VPb3*Tcab, VPa3) | | | | US$ TMbc3=Min(VPb3*TCab, VPc3*TCac) | | | | VPb3=Pb3*Qb3 | |
| | 364.96 | | | 280.74 | | 331.79 | | | 205.50 | | | | 65.45 | | | | 7,500.00 | |
| | | Wb3 | | Zb3 | | Gb3 | | Qb3=Nb3/Zb3 | | Nb3=FT3b*Nb | | | | | | | | |
| | | 83.03 | | 2.00 | | 198.91 | | 20.55 | | 41.10 | | | | | | | | |
| US$ BCba4=((Pa4*TCba - Pb4)/((Pa4*TCba + Pb4)/2))TMab4 | | | | | | | | | | | | | | | | | | |
| -35.74 | | | | | | | | | | | | | | | | | | |
| | | US$ BCbc4=((Pc4*TCbc - Pb4)/((Pc4*TCbc + Pb4)/2))TMbc4 | | | | | | | | | | | | | | | | |
| | | -4.16 | | | | | | | | | | | | | | | | |
| | Pb4=Wb4*Zb4 + Gb4 | | | Pb4a=Pa4*TCba | | Pb4c=Pc4*TCbc | | | US$ TMba4=Min(VPb4*Tcab, VPa4) | | | | US$ TMbc4=Min(VPb4*TCab, VPc4*TCac) | | | | VPb4=Pb4*Qb4 | |
| | 182.48 | | | 140.37 | | 165.89 | | | 137.00 | | | | 43.64 | | | | 5,000.00 | |
| | | Wb4 | | Zb4 | | Gb4 | | Qb4=Nb4/Zb4 | | Nb4=FT4b*Nb | | | | | | | | |
| | | 83.03 | | 1.00 | | 99.45 | | 27.40 | | 27.40 | | | | | | | | |

# Appendix Spreadsheet 1
## Image Section 1.4
## Determination of the trade balances BCcai and BCcbi, the prices Pcij and the sizes of the markets TMcij in Country C: CANADA

| | | | | | | | | | |
|---|---|---|---|---|---|---|---|---|---|
| | | | SECTION 1.4: Determination of Trade Balances BCcai and BCcbi, prices, market sizes and Exports and Imports of Country C: CANADA | | | | | | |
| US$ BCca1=((Pa1*TCca - Pc1)/((Pa1*TCca + Pc1)/2))TMac1 | | | | | | | | | |
| 3.67 | | | | | | | | | |
| | | US$ BCcb1=((Pb1*TCcb - Pc1)/((Pb1*TCcb + Pc1)/2))TMbc1 | | | | | | | |
| | | -2.57 | | | | | | | |
| | Pc1=Wc1*Zc1 + Gc1 | | Pc1a=Pa1*TCca | Pc1b=Pb1*TCcb | | US$ TMca1=Min(VPc1*Tcac, VPa1) | | US$ TMcb1=Min(VPc1*TCac, VPb1*TCab) | VPc1=Pc1*Qc1 |
| | 48.61 | | 52.88 | 45.83 | | 43.64 | | 43.64 | |
| | | | | | | | | | 100.00 |
| | | Wc | Zc1 | Gc1 | Qc1=Nc1/Zc1 | | Nc1=FT1c*Nc | | |
| | | 8.19 | 3.50 | 19.93 | 2.06 | | 7.20 | | |
| US$ BCca2=((Pa2*TCca - Pc2)/((Pa2*TCca+ Pc2)/2))TMac2 | | | | | | | | | |
| 1.00 | | | | | | | | | |
| | | US$ BCcb2=((Pb2*TCcb - Pc2)/((Pb2*TCcb + Pc2)/2))TMbc2 | | | | | | | |
| | | -0.66 | | | | | | | |
| | Pc2=Wc2*Zc2 + Gc2 | | Pc2a=Pa2*TCca | Pc2b=Pb2*TCcb | | US$ TMca2=Min(VPc2*TCac, VPa2) | | US$ TMcb2=Min(VPc2*Tcac, VPb2*TCab) | VPc2=Pc2*Qc2 |
| | 34.72 | | 35.26 | 34.37 | | 65.45 | | 65.45 | 150.00 |
| | | Wc | Zc2 | Gc2 | Qc2=Nc2/Zc2 | | Nc2=FT2c*Nc | | |
| | | 8.19 | 2.50 | 14.24 | 4.32 | | 10.80 | | |
| US$ BCca3=((Pa3*TCca - Pc3)/((Pa3*TCca + Pc3)/2))TMac3 | | | | | | | | | |
| -10.91 | | | | | | | | | |
| | | US$ BCcb3=((Pb3*TCcb - Pc3)/((Pb3*TCcb + Pc3)/2))TMbc3 | | | | | | | |
| | | 6.23 | | | | | | | |
| | Pc3=Wc3*Zc3 + Gc3 | | Pc3a=Pa3*TCca | Pc3b=Pb3*Tccb | | US$ TMca3=Min(VPc3*Tcac, VPa3) | | US$ TMcb3=Min(VPc3*Tcac, VPb3*TCab) | VPc3=Pc3*Qc3 |
| | 20.83 | | 17.63 | 22.92 | | 65.45 | | 65.45 | 150.00 |
| | | Wc | Zc3 | Gc3 | Qc3=Nc3/Zc3 | | Nc3=FT3c*Nc | | |
| | | 8.19 | 1.50 | 8.54 | 7.20 | | 10.80 | | |
| US$ BCca4=((Pa4*TCca - Pc4)/((Pa4*TCca + Pc4)/2))TMac4 | | | | | | | | | |
| -7.27 | | | | | | | | | |
| | | US$ BCcb4=((Pb4*TCcb - Pc4)/((Pb4*Tcb + Pc4)/2))TMbc4 | | | | | | | |
| | | 4.16 | | | | | | | |
| | Pc4=Wc4*Zc4 + Gc4 | | Pc4a=Pa4*TCca | Pc4b=Pb4*TCcb | | US$ TMca4=Min(VPc4*Tcac, VPa4) | | US$ TMcb4=Min(VPc4*Tcac, VPb4*TCab) | VPc4=Pc4*Qc4 |
| | 10.42 | | 8.81 | 11.46 | | 43.64 | | 43.64 | 100.00 |
| | | Wa | Zc4 | Gc4 | Qc4=Nc4/Zc4 | | Nc4=FT4c*Nc | | |
| | | 8.19 | 0.75 | 4.27 | 9.60 | | 7.20 | | |

# Appendix Spreadsheet 1
## Image Section 1.5
## MODEL PARAMETERS

SECTION 1.5: MODEL PARAMETERS

| | | | | | |
|---|---|---|---|---|---|
| | PRODUCTIVITY OF WORK IN THE COUNTRY A | | PRODUCTIVITY OF WORK IN COUNTRY B | | PRODUCTIVITY OF WORK IN COUNTRY C |
| 3.00 | Za1 | 4.00 | Zb1 | 3.50 | Zc1 |
| 2.00 | Za2 | 3.00 | Zb2 | 2.50 | Zc2 |
| 1.00 | Za3 | 2.00 | Zb3 | 1.50 | Zc3 |
| 0.50 | Za4 | 1.00 | Zb4 | 0.75 | Zc4 |

MONETARY OFFER OMi, MARGIN OF GAIN MGi, POPULATION Ni, SALARY Wi, PROFITS Gij, WORK FORCE FT, CONSUMER PREFERENCES PR AND TARIFFS AR

| | | | | | |
|---|---|---|---|---|---|
| 2,500 | OMa = Money Offer Country A US dollar | 0.3650 | MGa = Average Margin of Gain or profit in country A | 325.00 | Na = Population of country A United States |
| 25,000 | OMb = Money Offer Country B Mexico peso | 0.5450 | MGb = Average Margin of Gain or profit in country B | 137.00 | Nb = Population of country B México |
| 500 | OMc = Money Offer Country C Canda dollar | 0.4100 | MGc = Average Margin of Gain or profit in country C | 36.00 | Nc = Population of country C Canada |

| | | | |
|---|---|---|---|
| 4.88 | Wa = Salary in country A for each unit of work hired = (OMa(1-MGa))/Na | 8.42 | Ga1 = Gain or Profit in country A corresponds to good 1 for each unit produced = (OMa(MGa)(FTa1))/Qa1 |
| 83.03 | Wb = Salary in country B for each unit of work hired = (OMb(1-MGb))/Nb | 5.62 | Ga2 = Gain or Profit in country A corresponds to good 2 for each unit produced = (OMa(MGa)(FTa2))/Qa2 |
| 8.19 | Wc = Salary in country C for each unit of work hired = (OMc(1-MGc))/Nc | 2.81 | Ga3 = Gain or Profit in country A corresponds to good 3 for each unit produced = (OMa(MGa)(FTa3))/Qa3 |
| | | 1.40 | Ga4 = Gain or Profit in country A corresponds to good 4 for each unit produced = (OMa(MGa)(FTa4))/Qa4 |
| | | 397.81 | Gb1 = Gain or Profit in country B corresponds to good 1 for each unit produced = (OMb(MGb)(FTb1))/Qb1 |
| | | 298.36 | Gb2 = Gain or Profit in country B corresponds to good 2 for each unit produced = (OMb(MGb)(FTb2))/Qb2 |
| | | 198.91 | Gb3 = Gain or Profit in country B corresponds to good 3 for each unit produced = (OMb(MGb)(FTb3))/Qb3 |
| | | 99.45 | Gb4 = Gain or Profit in country B corresponds to good 3 for each unit produced = (OMb(MGb)(FTb4))/Qb4 |
| | | 19.93 | Gc1 = Gain or Profit in country C corresponds to good 1 for each unit produced = (OMc(MGc)(FTc1))/Qc1 |
| | | 14.24 | Gc2 = Gain or Profit in country C corresponds to good 2 for each unit produced = (OMc(MGc)(FTc2))/Qc2 |
| | | 8.54 | Gc3 = Gain or Profit in country C corresponds to good 3 for each unit produced = (OMc(MGc)(FTc3))/Qc3 |
| | | 4.27 | Gc4 = Gain or Profit in country C corresponds to good 4 for each unit produced = (OMc(MGc)(FTc4))/Qc4 |

DISTRIBUTION OF THE FORCE OF WORK

| | | | | | |
|---|---|---|---|---|---|
| | DISTRIBUTION OF THE COUNTRY A FORCE OF WORK | | DISTRIBUTION OF THE COUNTRY B FORCE OF WORK | | DISTRIBUTION OF THE COUNTRY C FORCE OF WORK |
| 0.20 | FT1a = Proportion force of work producing good 1 in country A | 0.20 | FT1b = Proportion force of work producing good 1 in country B | 0.20 | FT1c = Proportion force of work producing good 1 in country C |
| 0.30 | FT2a = Proportion force of work producing good 2 in country A | 0.30 | FT2b = Proportion force of work producing good 2 in country B | 0.30 | FT2c = Proportion force of work producing good 2 in country C |
| 0.30 | FT3a = Proportion force of work producing good 3 in country A | 0.30 | FT3b = Proportion force of work producing good 3 in country B | 0.30 | FT3c = Proportion force of work producing good 3 in country C |
| 0.20 | FT4a = Proportion force of work producing good 4 in country A | 0.20 | FT4b = Proportion force of work producing good 4 in country B | 0.20 | FT4c = Proportion force of work producing good 4 in country C |

# Appendix Spreadsheet 1
## Image Section 1.6 and 1.7
## MODEL VARIABLES. EXCHANGE RATES

**SECTION 1.6 AND 1.7: MODEL VARIABLES. EXCHANGE RATES. SECTION 1.6 ARE DETERMINED BY TRIAL AND ERROR OF THE 2 PRIMARY EXCHANGE RATES**

The equilibrium can be reached in the three equations of Section 1.1 corresponding to the trade balance of the three countries A, B and C when in this Section 1.6 the set of the 2 primary exchange rates TCab and TCac that generate the simultaneous general balance in the trade balance of each country is established

| | SECTION 1.6: PRIMARY, CONTRA-PRIMARY AND NON-PRIMARY CHANGE RATES | | SECTION 1.7: INITIAL CHANGE RATES |
|---|---|---|---|
| 0.027400 | TCab = Primary and equivalent exchange rate of country A with country B | 0.027400 | TCab In = (Pa1+Pa2+Pa3+Pa4)/(Pb1+Pb2+Pb3+Pb4) = Initial exchange rate of country A with country B |
| 0.436364 | TCac = Primary and equivalent exchange rate of country A with country C | 0.436364 | TCac In = (Pa1+Pa2+Pa3+Pa4)/(Pc1+Pc2+Pc3+Pc4) = Initial exchange rate of country A with country C |
| 36.49635 | TCba = 1/TCab = Initial exchange rate of country B with country A | 36.496350 | TCba In = (Pb1+Pb2+Pb3+Pb4)/(Pa1+Pa2+Pa3+Pa4) = 1/Tcab = Initial exchange rate of country B with country A |
| 2.29166 | TCca = 1/TCac = Initial exchange rate of country C with country A | 2.291667 | TCca In = (Pc1+Pc2+Pc3+Pc4)/(Pa1+Pa2+Pa3+Pa4) = 1/TCac = Initial exchange rate of country C with country A |
| 15.92569 | TCbc = TCac/TCab = Initial exchange rate of country B with country C | 15.925680 | TCbc In = (Pb1+Pb2+Pb3+Pb4)/(Pc1+Pc2+Pc3+Pc4) = TCba/TCca =Initial exchange rate of country B with country C |
| 0.06279 | TCcb = TCab/TCac = Initial exchange rate of country C with country B | 0.062792 | TCcb In = (Pc1+Pc2+Pc3+Pc4)/(Pb1+Pb2+Pb3+Pb4) = TCab/TCac = Initial exchange rate of country C with country B |

# Appendix Spreadsheet 2

## Image Section 2.1
## COMMERCIAL BALANCES. MODEL EQUATIONS

| Spreadsheet 2 - MODEL SIMULATION INTERNATIONAL TRADE OF 3 COUNTRIES: | | | | | | | | | | | | | | | | |
|---|---|---|---|---|---|---|---|---|---|---|---|---|---|---|---|---|
| | | | | | | | | | SECTION 2.1: COMMERCIAL BALANCES. MODEL EQUATIONS | | | | | | | |
| | | Equation 5.1 Defining the Trade Balance of Country A (United States) with All Other Countries (BCa.bc) | | | | | | | | | | | | | | |
| | | Bca.bc = BCab1 + BCab2 + BCab3 + BCab4 + BCac1 + BCac2 + BCac4 + BCac4 | | | | | | | | | | | | | | |
| BCa.bc | = | BCab1 | + | BCab2 | + | BCab3 | + | BCab4 | + | BCac1 | + | BCac2 | + | BCac3 | + | BCac4 |
| 0.00 | = | -29.53 | + | (22.46) | + | 31.20 | + | 20.80 | + | (5.99) | + | (4.76) | + | 6.45 | + | 4.30 |
| | | | | | | | | | | | | | | | | |
| | | Equation 5.2 Defining the Trade Balance of Country B (Mexico) with All Other Countries (BCb.ac) | | | | | | | | | | | | | | |
| | | | | BCb.ac = BCba1 + BCba2 + BCba3 + BCba4 + BCbc1 + BCbc2 + BCbc4 + BCbca4 | | | | | | | | | | | | |
| BCb.ac | = | BCba1 | + | BCba2 | + | BCba3 | + | BCba4 | + | BCbc1 | + | BCbc2 | + | BCbc3 | + | BCbc4 |
| 0.00 | = | 29.53 | + | 22.46 | + | (31.20) | + | (20.80) | + | 3.76 | + | 2.64 | + | (3.84) | + | (2.56) |
| | | | | | | | | | | | | | | | | |
| | | Equation 5.3 Defining the Trade Balance of Country C (Canada) with All Other Countries (BCc.ab) | | | | | | | | | | | | | | |
| | | | | BCc.ab = BCca1 + BCca2 + BCca3 + BCca4 + BCcb1 + BCcb2 + BCcb3 + BCcb4 | | | | | | | | | | | | |
| BCc.ab | = | BCca1 | + | BCca2 | + | BCca3 | + | BCca4 | + | BCcb1 | + | BCcb2 | + | BCcb3 | + | BCcb4 |
| 0.00 | = | 5.99 | + | 4.76 | + | (6.45) | + | (4.30) | + | (3.76) | + | (2.64) | + | 3.84 | + | 2.56 |
| | | | | | | | | | | | | | | | | |
| | | | | | BCab = | 0.00 | | | | | | | BCac = | (0.00) | | |
| | | | | | BCba = | (0.00) | | | | | | | BCbc = | 0.01 | | |
| | | | | | BCca = | 0.00 | | | | | | | BCcb = | (0.01) | | |

# Appendix Spreadsheet 3

## Image Section 3.1
## COMMERCIAL BALANCES. MODEL EQUATIONS

| Spreadsheet 3 - MODEL SIMULATION INTERNATIONAL TRADE OF 3 COUNTRIES: | | | | | | | | | | | | | | | | |
|---|---|---|---|---|---|---|---|---|---|---|---|---|---|---|---|---|
| | | | | | | | | | SECTION 3.1: COMMERCIAL BALANCES. MODEL EQUATIONS | | | | | | | |
| | | Equation 8.7 Defining the Trade Balance of Country A (United States) with All Other Countries (BCa.bc) | | | | | | | | | | | | | | |
| | | | | BCa.bc = BCab1 + BCab2 + BCab3 + BCab4 + BCac1 + BCac2 + BCac3 + BCac4 | | | | | | | | | | | | |
| BCa.bc | = | BCab1 | + | BCab2 | + | BCab3 | + | BCab4 | + | BCac1 | + | BCac2 | + | BCac3 | + | BCac4 |
| 78.08 | = | -19.57 | + | (5.20) | + | 53.61 | + | 35.74 | + | (3.67) | + | (1.00) | + | 10.91 | + | 7.27 |
| | | | | | | | | | | | | | | | | |
| | | Equation 8.8 Defining the Trade Balance of Country B (Mexico) with All Other Countries (BCb.ac) | | | | | | | | | | | | | | |
| | | | | BCb.ac = BCba1 + BCba2 + BCba3 + BCba4 + BCbc1 + BCbc2 + BCbc3 + BCbca4 | | | | | | | | | | | | |
| BCb.ac | = | BCba1 | + | BCba2 | + | BCba3 | + | BCba4 | + | BCbc1 | + | BCbc2 | + | BCbc3 | + | BCbc4 |
| -71.74 | = | 19.57 | | 5.20 | | (53.61) | | (35.74) | | 2.57 | | 0.66 | | (6.23) | | (4.16) |
| | | | | | | | | | | | | | | | | |
| | | Equation 8.9 Defining the Trade Balance of Country C (Canada) with All Other Countries (BCc.ab) | | | | | | | | | | | | | | |
| | | | | BCc.ab = BCca1 + BCca2 + BCca3 + BCca4 + BCcb1 + BCcb2 + BCcb3 + BCcb4 | | | | | | | | | | | | |
| BCc.ab | = | BCca1 | + | BCca2 | + | BCca3 | + | BCca4 | + | BCcb1 | + | BCcb2 | + | BCcb3 | + | BCcb4 |
| -6.34 | = | 3.67 | + | 1.00 | + | (10.91) | + | (7.27) | + | (2.57) | + | (0.66) | + | 6.23 | + | 4.16 |
| | | | | | | | | | | | | | | | | |
| | | | | | COMMERCIAL BALANCES | | | | | | | | | | | |
| | | Country A - United States | | | | | Country B - Mexico | | | | | Country C - Canada | | | | |
| | | | | Deficit or SURPLUS US | | | | | Déficit o Superávit Méx | | | | | Déficit o Superávit Cand | | |
| | | EXa EU | IMa EU | US$ BCa.bc = (Exa - Ima) | | | EXb Mex | IMb Mex | US$ BCb.ac = (EXb - IMb) | | | EXc Cand | IMc Cand | US$ BCc.ab = (EXc - IMc) | | |
| | | 107.53 | 29.45 | 78.08 | | | 28.00 | 99.74 | (71.74) | | | 15.06 | 21.41 | (6.34) | | |
| | | | | | | | | | | | | | | | | |
| | | EXab | IMab | | | | EXba | IMba | | | | EXca | IMca | | | |
| | | 89.35 | 24.77 | 64.57 | | | 24.77 | 89.35 | (64.57) | | | 4.67 | 18.18 | (13.51) | | |
| | | | | | | | | | | | | | | | | |
| | | EXac | IMac | | | | EXbc | IMbc | | | | EXcb | IMcb | | | |
| | | 18.18 | 4.67 | 13.51 | | | 3.22 | 10.39 | (7.16) | | | 10.39 | 3.22 | 7.16 | | |

# Appendix Spreadsheet 3
## Image Section 3.2
## Determination of the Trade Balance BCabi and BCaci, of prices, market sizes and Exports and Imports of Country A: UNITED STATES

| | | | | | | | | | | |
|---|---|---|---|---|---|---|---|---|---|---|
| | | | SECTION 3.2: Determination of Trade Balances BCabi and BCaci, of prices, market sizes and exports and imports Country A: UNITED STATES | | | | | | | |
| US$ BCab1 = (EXa1b - IMa1b) | | | | US$ EXa1b=IMb1a | | | US$ IMa1b=IF((Pa1-(1-PR1ab)(1+AR1a)Pb1*TCab)/((Pa1+(1-PR1ab)(1+AR1a)Pb1*TCab)/2)>0, ((Pa1-(1-PR1ab)(1+AR1a)Pb1*TCab)/((Pa1+(1-PR1a | | | |
| -19.57 | | | | - | | | 19.57 | | | |
| | | US$ BCac1 = (EXa1c - IMa1c) | | US$ EXa1c=IMc1a | | | US$ IMa1c=IF((Pa1-(1-PR1ac)(1+AR1a)Pc1*TCac)/((Pa1+(1-PR1ac)(1+AR1a)Pc1*TCac)/2)>0, ((Pa1-(1-PR1ac)(1+AR1a)Pc1*TCac)/((Pa1+(1-PR1ac) | | | |
| | | (3.67) | | - | | | 3.67 | | | |
| | Pa1=Wa*Za1 + Ga1 | | Pa1b=Pb1*TCab | | Pa1c=Pc1*TCac | | US$ TMab1=Min(VPa1, VPb1*TCab) | | US$ TMac1=Min(VPa1, VPc1*TCac) | VPa1=Pa1*Qa1 |
| | 23.08 | | 20 | | 21.21 | | 137.00 | | 43.64 | 500.00 |
| | | Wa | Za1 | Ga1 | | Qa1=Na1/Za1 | | Na1=FT1a*Na | | |
| | | 4.88 | 3.00 | 8.42 | | 21.67 | | 65.00 | | |
| US$ BCab2 = (EXa2b - IMa2b) | | | | US$ EXa2b=IMb2a | | | US$ IMa2b=IF((Pa2-(1-PR2ab)(1+AR2a)Pb2*TCab)/((Pa2+(1-PR2ab)(1+AR2a)Pb2*TCab)/2)>0, ((Pa2-(1-PR2ab)(1+AR2a)Pb2*TCab)/((Pa2+(1-PR2a | | | |
| (5.20) | | | | - | | | 5.20 | | | |
| | | US$ BCac2 = (EXa2c - IMa2c) | | US$ EXa2c=IMc2a | | | US$ IMa2c=IF((Pa2-(1-PR2ac)(1+AR2a)Pc2*TCac)/((Pa2+(1-PR2ac)(1+AR2a)Pc2*TCac)/2)>0, ((Pa2-(1-PR2ac)(1+AR2a)Pc2*TCac)/((Pa2+(1-PR2ac) | | | |
| | | (1.00) | | - | | | 1.00 | | | |
| | Pa2=Wa*Za2 + Ga2 | | Pa2b=Pb2*TCab | | Pa2c=Pc2*TCac | | US$ TMab2=Min(VPa2, VPb2*TCab) | | US$ TMac2=Min(VPa2, VPc2*TCac) | VPa2=Pa2*Qa2 |
| | 15.38 | | 15 | | 15.15 | | 205.50 | | 65.45 | 750.00 |
| | | Wa | Za2 | Ga2 | | Qa2=Na2/Za2 | | Na2=FT2a*Na | | |
| | | 4.88 | 2.00 | 5.62 | | 48.75 | | 97.50 | | |
| US$ BCab3 = (EXa3b - IMa3b) | | | | US$ EXa3b=IMb3a | | | US$ IMa3b=IF((Pa3-(1-PR3ab)(1+AR3a)Pb3*TCab)/((Pa3+(1-PR3ab)(1+AR3a)Pb3*TCab)/2)>0, ((Pa3-(1-PR3ab)(1+AR3a)Pb3*TCab)/((Pa3+(1-PR3a | | | |
| 53.61 | | | | 53.61 | | | - | | | |
| | | US$ BCac3 = (EXa3c - IMa3c) | | US$ EXa3c=IMc3a | | | US$ IMa3c=IF((Pa3-(1-PR3ac)(1+AR3a)Pc3*TCac)/((Pa3+(1-PR3ac)(1+AR3a)Pc3*TCac)/2)>0, ((Pa3-(1-PR3ac)(1+AR3a)Pc3*TCac)/((Pa3+(1-PR3ac) | | | |
| | | 10.91 | | 10.91 | | | - | | | |
| | Pa3=Wa*Za3 + Ga3 | | Pa3b=Pb3*TCab | | Pa3c=Pc3*TCac | | US$ TMab3=Min(VPa3, VPb3*TCab) | | US$ TMac3=Min(VPa3, VPc3*TCac) | VPa3=Pa3*Qa3 |
| | 7.69 | | 10 | | 9.09 | | 205.50 | | 65.45 | 750.00 |
| | | Wa | Za3 | Ga3 | | Qa3=Na3/Za3 | | Na3=FT3a*Na | | |
| | | 4.88 | 1.00 | 2.81 | | 97.50 | | 97.50 | | |
| US$ BCab4 = (EXa4b - IMa4b) | | | | US$ EXa4b=IMb4a | | | US$ IMa4b=IF((Pa4-(1-PR4ab)(1+AR4a)Pb4*TCab)/((Pa4+(1-PR4ab)(1+AR4a)Pb4*TCab)/2)>0, ((Pa4-(1-PR4ab)(1+AR4a)Pb4*TCab)/((Pa4+(1-PR4a | | | |
| 35.74 | | | | 35.74 | | | - | | | |
| | | US$ BCac4 = (EXa4c - IMa4c) | | US$ EXa4c=IMc4a | | | US$ IMa4c=IF((Pa4-(1-PR4ac)(1+AR4a)Pc4*TCac)/((Pa4+(1-PR4ac)(1+AR4a)Pc4*TCac)/2)>0, ((Pa4-(1-PR4ac)(1+AR4a)Pc4*TCac)/((Pa4+(1-PR4ac) | | | |
| | | 7.27 | | 7.27 | | | - | | | |
| | Pa4=Wa*Za4 + Ga4 | | Pa4b=Pb4*TCab | | Pa4c=Pc4*TCac | | US$ TMab4=Min(VPa4, VPb4*TCab) | | US$ TMac4=Min(VPa4, VPc4*TCac) | VPa4=Pa4*Qa4 |
| | 3.85 | | 5 | | 4.545458 | | 137.00 | | 43.64 | 500.00 |
| | | Wa | Za4 | Ga 4 | | Qa4=Na4/Za4 | | Na4=FT4a*Na | | |
| | | 4.88 | 0.50 | 1.40 | | 130.00 | | 65.00 | | |

# Appendix Spreadsheet 3
## Image Section 3.3
## Determination of the Trade Balance BCbai and BCbci, prices, market sizes and Exports and Imports of Country B: MEXICO

SECTION 3.3: Determination of Trade Balances BCbai and BCbci, of prices, of market sizes and exports and imports of Country B: MEXICO

| | | | | | | | | | | |
|---|---|---|---|---|---|---|---|---|---|---|
| US$ BCba1 = (EXb1a - IMb1a) | | | | US$ EXb1a=IMa1b | | | US$ IMb1a=IF((Pb1-(1-PR1ba)(1+AR1b)Pa1*TCba)/((Pb1-(1-PR1ba)(1+AR1b)Pa1*TCba)/2)>0, ((Pb1-(1-PR1ba)(1+AR1b)Pa1*TCba)/((Pb1-(1-PR1b | | | |
| 19.57 | | | | 19.57 | | | 0 | | | |
| | | US$ BCbc1 = (EXb1c - IMb1c) | | US$ EXb1c=IMc1b | | | US$ IMb1c=IF((Pb1-(1-PR1bc)(1+AR1b)Pc1*TCbc)/((Pb1+(1-PR1bc)(1+AR1b)Pc1*TCbc)/2)>0, ((Pb1-(1-PR1bc)(1+AR1b)Pc1*TCbc)/((Pb1+(1-PR1 | | | |
| | | 2.57 | | 2.57 | | | - | | | |
| | Pb1=Wb*Zb1 + Gb1 | | Pb1a=Pa1*TCba | | Pb1c=Pc1*TCbc | | US$ TMba1=Min(VPb1*TCab, VPa1) | | US$ TMbc1=Min(VPb1*TCab, VPc1*TCac) | VPb1=Pb1*Qb1 |
| | 729.93 | | 842.22 | | 774.1657 | | 137.00 | | 43.64 | 5,000.00 |
| | | Wb | Zb1 | Gb1 | | Qb1=Nb1/Zb1 | | Nb1=FT1b*Nb | | |
| | | 83.03 | 4.00 | 397.81 | | 6.85 | | 27.40 | | |
| US$ BCba2 = (EXb2a - IMb2a) | | | | US$ EXb2a=IMa2b | | | US$ IMb2a=IF((Pb2-(1-PR2ba)(1+AR2b)Pa2*TCba)/((Pb2-(1-PR2ba)(1+AR2b)Pa2*TCba)/2)>0, ((Pb2-(1-PR2ba)(1+AR2b)Pa2*TCba)/((Pb2-(1-PR2b | | | |
| 5.20 | | | | 5.20 | | | - | | | |
| | | US$ BCbc2 = (EXb2c - IMb2c) | | US$ EXb2c=IMc2b | | | US$ IMb2c=IF((Pb2-(1-PR2bc)(1+AR2b)Pc2*TCbc)/((Pb2+(1-PR2bc)(1+AR2b)Pc2*TCbc)/2)>0, ((Pb2-(1-PR2bc)(1+AR2b)Pc2*TCbc)/((Pb2+(1-PR2 | | | |
| | | 0.66 | | 0.66 | | | - | | | |
| | Pb2=Wb2*Zb2 + Gb2 | | Pb2a=Pa2*TCba | | Pb2c=Pc2*TCbc | | US$ TMba2=Min(VPb2*Tcab, VPa2) | | US$ TMbc2=Min(VPb2*TCab, VPc2*TCac) | VPb2=Pb2*Qb2 |
| | 547.45 | | 561.48 | | 552.9755 | | 205.50 | | 65.45 | 7,500.00 |
| | | Wb2 | Zb2 | Gb2 | | Qb2=Nb2/Zb2 | | Nb2=FT2b*Nb | | |
| | | 83.03 | 3.00 | 298.36 | | 13.70 | | 41.10 | | |
| US$ BCba3 = (EXb3a - IMb3a) | | | | US$ EXb3a=IMa3b | | | US$ IMb3a=IF((Pb3-(1-PR3ba)(1+AR3b)Pa3*TCba)/((Pb3-(1-PR3ba)(1+AR3b)Pa3*TCba)/2)>0, ((Pb3-(1-PR3ba)(1+AR3b)Pa3*TCba)/((Pb3-(1-PR3b | | | |
| (53.61) | | | | - | | | 53.61 | | | |
| | | US$ BCbc3 = (EXb3c - IMb3c) | | US$ EXb3c=IMc3b | | | US$ IMb3c=IF((Pb3-(1-PR3bc)(1+AR3b)Pc3*TCbc)/((Pb3+(1-PR3bc)(1+AR3b)Pc3*TCbc)/2)>0, ((Pb3-(1-PR3bc)(1+AR3b)Pc3*TCbc)/((Pb3+(1-PR3 | | | |
| | | (6.23) | | - | | | 6.23 | | | |
| | Pb3=Wb3*Zb3 + Gb3 | | Pb3a=Pa3*TCba | | Pb3c=Pc3*TCbc | | US$ TMba3=Min(VPb3*Tcab, VPa3) | | US$ TMbc3=Min(VPb3*TCab, VPc3*TCac) | VPb3=Pb3*Qb3 |
| | 364.96 | | 280.74 | | 331.7853 | | 205.50 | | 65.45 | 7,500.00 |
| | | Wb3 | Zb3 | Gb3 | | Qb3=Nb3/Zb3 | | Nb3=FT3b*Nb | | |
| | | 83.03 | 2.00 | 198.91 | | 20.55 | | 41.10 | | |
| US$ BCba4 = (EXb4a - IMb4a) | | | | US$ EXb4a=IMa4b | | | US$ IMb4a=IF((Pb4-(1-PR4ba)(1+AR4b)Pa4*TCba)/((Pb4-(1-PR4ba)(1+AR4b)Pa4*TCba)/2)>0, ((Pb4-(1-PR4ba)(1+AR4b)Pa4*TCba)/((Pb4-(1-PR4b | | | |
| (35.74) | | | | - | | | 35.74 | | | |
| | | US$ BCbc4 = (EXb4c - IMb4c) | | US$ EXb4c=IMc4b | | | US$ IMb4c=IF((Pb4-(1-PR4bc)(1+AR4b)Pc4*TCbc)/((Pb4+(1-PR4bc)(1+AR4b)Pc4*TCbc)/2)>0, ((Pb4-(1-PR4bc)(1+AR4b)Pc4*TCbc)/((Pb4+(1-PR4 | | | |
| | | (4.16) | | - | | | 4.16 | | | |
| | Pb4=Wb4*Zb4 + Gb4 | | Pb4a=Pa4*TCba | | Pb4c=Pc4*TCbc | | US$ TMba4=Min(VPb4*Tcab, VPa4) | | US$ TMbc4=Min(VPb4*TCab, VPc4*TCac) | VPb4=Pb4*Qb4 |
| | 182.48 | | 140.3706 | | 165.8926 | | 137.00 | | 43.64 | 5,000.00 |
| | | Wb4 | Zb4 | Gb4 | | Qb4=Nb4/Zb4 | | Nb4=FT4b*Nb | | |
| | | 83.03 | 1.00 | 99.45 | | 27.40 | | 27.40 | | |

# Appendix Spreadsheet 3
## Image Section 3.4
## Determination of the Trade Balances BCcai and Bccbi, prices, market sizes and Exports and Imports of Country C: CANADA

| | | | | | | | | | | |
|---|---|---|---|---|---|---|---|---|---|---|
| | | | SECTION 3.4: Determination of Trade Balances BCcai and BCcbi, prices, market sizes and Exports and Imports of Country C: CANADA | | | | | | | |
| US$ BCca1 = (EXc1a - IMc1a) | | | | US$ EXc1a=IMa1c | | | US$ IMc1a=IF((Pc1-(1-PR1ca)(1+AR1c)Pa1*TCca)/((Pc1+(1-PR1ca)(1+AR1c)Pa1*TCca)/2)>0, ((Pc1-(1-PR1ca)(1+AR1c)Pa1*TCca)/((Pc1+(1-PR1ca) | | | |
| 3.67 | | | | 3.67 | | | 0 | | | |
| | | US$ BCcb1 = (EXc1b - IMc1b) | | US$ EXc1b=IMb1c | | | US$ IMc1b=IF((Pc1-(1-PR1cb)(1+AR1c)Pb1*TCcb)/((Pc1+(1-PR1cb)(1+AR1c)Pb1*TCcb )/2)>0, ((Pc1-(1-PR1cb)(1+AR1c)Pb1*TCcb)/((Pc1+(1-PR1 | | | |
| | | (2.57) | | - | | | 2.57 | | | |
| | Pc1=Wc1*Zc1 + Gc1 | | Pc1a=Pa1*TCca | | Pc1b=Pb1*TCcb | | US$ TMca1=Min(VPc1*TCac, VPa1) | | US$ TMcb1=Min(VPc1*TCac, VPb1*TCab) | VPc1=Pc1*Qc1 |
| | 48.61 | | 52.88457 | | 45.8333 | | 43.64 | | 43.64 | 100.00 |
| | | Wc | Zc1 | Gc1 | | Qc1=Nc1/Zc1 | | Nc1=FT1c*Nc | | |
| | | 8.19 | 3.50 | 19.93 | | 2.06 | | 7.20 | | |
| US$ BCca2 = (EXc2a - IMc2a) | | | | US$ EXc2a=IMa2c | | | US$ IMc2a=IF((Pc2-(1-PR2ca)(1+AR2c)Pa2*TCca)/((Pc2+(1-PR2ca)(1+AR2c)Pa2*TCca)/2)>0, ((Pc2-(1-PR2ca)(1+AR2c)Pa2*TCca)/((Pc2+(1-PR2ca) | | | |
| 1.00 | | | | 1.00 | | | - | | | |
| | | US$ BCcb2 = (EXc2b - IMc2b) | | US$ EXc2b=IMb2c | | | US$ IMc2b=IF((Pc2-(1-PR2cb)(1+AR2c)Pb2*TCcb)/((Pc2+(1-PR2cb)(1+AR2c)Pb2*TCcb )/2)>0, ((Pc2-(1-PR2cb)(1+AR2c)Pb2*TCcb)/((Pc2+(1-PR2 | | | |
| | | (0.66) | | - | | | 0.66 | | | |
| | Pc2=Wc2*Zc2+ Gc2 | | Pc2a=Pa2*TCca | | Pc2b=Pb2*Tccb | | US$ TMca2=Min(VPc2*TCac, VPa2) | | US$ TMcb2=Min(VPc2*TCac, VPb2*TCab) | VPc2=Pc2*Qc2 |
| | 34.72 | | 35.25638 | | 34.37497 | | 65.45 | | 65.45 | 150.00 |
| | | Wc | Zc2 | Gc2 | | Qc2=Nc2/Zc2 | | Nc2=FT2c*Nc | | |
| | | 8.19 | 2.50 | 14.24 | | 4.32 | | 10.80 | | |
| US$ BCca3 = (EXc3a - IMc3a) | | | | US$ EXc3a=IMa3c | | | US$ IMc3a=IF((Pc3-(1-PR3ca)(1+AR3c)Pa3*TCca)/((Pc3+(1-PR3ca)(1+AR3c)Pa3*TCca)/2)>0, ((Pc3-(1-PR3ca)(1+AR3c)Pa3*TCca)/((Pc3+(1-PR3ca | | | |
| (10.91) | | | | - | | | 10.91 | | | |
| | | US$ BCcb3 = (EXc3b - IMc3b) | | EXc3b=IMb3c | | | US$ IMc3b=IF((Pc3-(1-PR3cb)(1+AR3c)Pb3*TCcb)/((Pc3+(1-PR3cb)(1+AR3c)Pb3*TCcb )/2)>0, ((Pc3-(1-PR3cb)(1+AR3c)Pb3*TCcb)/((Pc3+(1-PR3 | | | |
| | | 6.23 | | 6.23 | | | - | | | |
| | Pc3=Wc3*Zc3 + Gc3 | | Pc3a=Pa3*TCca | | Pc3b=Pb3*Tccb | | US$ TMca3=Min(VPc3*TCac, VPa3) | | US$ TMcb3=Min(VPc3*TCac, VPb3*TCab) | VPc3=Pc3*Qc3 |
| | 20.83 | | 17.62819 | | 22.91665 | | 65.45 | | 65.45 | 150.00 |
| | | Wc | Zc3 | Gc3 | | Qc3=Nc3/Zc3 | | Nc3=FT3c*Nc | | |
| | | 8.19 | 1.50 | 8.54 | | 7.20 | | 10.80 | | |
| US$ BCca4 = (EXc4a - IMc4a) | | | | US$ EXc4a=IMa4c | | | US$ IMc4a=IF((Pc4-(1-PR4ca)(1+AR4c)Pa4*TCca)/((Pc4+(1-PR4ca)(1+AR4c)Pa4*TCca)/2)>0, ((Pc4-(1-PR4ca)(1+AR4c)Pa4*TCca)/((Pc4+(1-PR4ca) | | | |
| (7.27) | | | | - | | | 7.27 | | | |
| | | US$ BCcb4 = (EXc4b - IMc4b) | | US$ EXc4b=IMb4c | | | US$ IMc4b=IF((Pc4-(1-PR4cb)(1+AR4c)Pb4*TCcb)/((Pc4+(1-PR4cb)(1+AR4c)Pb4*TCcb )/2)>0, ((Pc4-(1-PR4cb)(1+AR4c)Pb4*TCcb)/((Pc4+(1-PR4 | | | |
| | | 4.16 | | 4.16 | | | - | | | |
| | Pc4=Wc4*Zc4 + Gc4 | | Pc4a=Pa4*TCca | | Pc4b=Pb4*Tccb | | US$ TMca4=Min(VPc4*TCac, VPa4) | | US$ TMcb4=Min(VPc4*TCac, VPb4*TCab) | VPc4=Pc4*Qc4 |
| | 10.42 | | 8.814095 | | 11.45832 | | 43.64 | | 43.64 | 100.00 |
| | | Wa | Zc4 | Gc4 | | Qc4=Nc4/Zc4 | | Nc4=FT4c*Nc | | |
| | | 8.19 | 0.75 | 4.27 | | 9.60 | | 7.20 | | |

# Appendix Spreadsheet 3
## Image Section 3.5
## MODEL PARAMETERS

**SECTION 3.5: MODEL PARAMETERS**

| | PRODUCTIVITY OF WORK IN THE COUNTRY A | | PRODUCTIVITY OF WORK IN COUNTRY B | | PRODUCTIVITY OF WORK IN COUNTRY C |
|---|---|---|---|---|---|
| 3.00 | Za1 | 4.00 | Zb1 | 3.50 | Zc1 |
| 2.00 | Za2 | 3.00 | Zb2 | 2.50 | Zc2 |
| 1.00 | Za3 | 2.00 | Zb3 | 1.50 | Zc3 |
| 0.50 | Za4 | 1.00 | Zb4 | 0.75 | Zc4 |

**MONETARY OFFER OMi, MARGIN OF GAIN MGi, POPULATION Ni, SALARY Wi, PROFITS Gij, WORK FORCE FT, CONSUMER PREFERENCES PR AND TARIFFS AR**

| | | | | | |
|---|---|---|---|---|---|
| 2,500 | OMa = Money Offer Country A US dollar | 0.3650 | MGa = Average Margin of Gain or Profit in country A | 325.00 | Na = Population of country A United States |
| 25,000 | OMb = Money Offer Country B Mexico peso | 0.5450 | MGb = Average Margin of Gain or Profit in country B | 137.00 | Nb = Population of country B México |
| 500 | OMc = Money Offer Country C Canda dollar | 0.4100 | MGc = Average Margin of Gain or Profit in country C | 36.00 | Nc = Population of country C Canada |

| | | | |
|---|---|---|---|
| 4.88 | Wa = Salary in country A for each unit of work hired = (OMa(1-MGa))/Na | 8.42 | Ga1 = Gain or Profit in country A corresponds to good 1 for each unit produced = (OMa(MGa)(FTa1))/Qa1 |
| 83.03 | Wb = Salary in country B for each unit of work hired = (OMb(1-MGb))/Nb | 5.62 | Ga2 = Gain or Profit in country A corresponds to good 2 for each unit produced = (OMa(MGa)(FTa2))/Qa2 |
| 8.19 | Wc = Salary in country C for each unit of work hired = (OMc(1-MGc))/Nc | 2.81 | Ga3 = Gain or Profit in country A corresponds to good 3 for each unit produced = (OMa(MGa)(FTa3))/Qa3 |
| | | 1.40 | Ga4 = Gain or Profit in country A corresponds to good 4 for each unit produced = (OMa(MGa)(FTa4))/Qa4 |
| | | | |
| | | 397.81 | Gb1 = Gain or Profit in country B corresponds to good 1 for each unit produced = (OMb(MGb)(FTb1))/Qb1 |
| | | 298.36 | Gb2 = Gain or Profit in country B corresponds to good 2 for each unit produced = (OMb(MGb)(FTb2))/Qb2 |
| | | 198.91 | Gb3 = Gain or Profit in country B corresponds to good 3 for each unit produced = (OMb(MGb)(FTb3))/Qb3 |
| | | 99.45 | Gb4 = Gain or Profit in country B corresponds to good 3 for each unit produced = (OMb(MGb)(FTb4))/Qb4 |
| | | | |
| | | 19.93 | Gc1 = Gain or Profit in country C corresponds to good 1 for each unit produced = (OMc(MGc)(FTc1))/Qc1 |
| | | 14.24 | Gc2 = Gain or Profit in country C corresponds to good 2 for each unit produced = (OMc(MGc)(FTc2))/Qc2 |
| | | 8.54 | Gc3 = Gain or Profit in country C corresponds to good 3 for each unit produced = (OMc(MGc)(FTc3))/Qc3 |
| | | 4.27 | Gc4 = Gain or Profit in country C corresponds to good 4 for each unit produced = (OMc(MGc)(FTc4))/Qc4 |

**DISTRIBUTION OF THE FORCE OF WORK**

| | DISTRIBUTION OF THE COUNTRY A FORCE OF WORK | | DISTRIBUTION OF THE COUNTRY B FORCE OF WORK | | DISTRIBUTION OF THE COUNTRY C FORCE OF WORK |
|---|---|---|---|---|---|
| 0.20 | FT1a = Proportion force of work producing good 1 in country A | 0.20 | FT1b = Proportion force of work producing good 1 in country B | 0.20 | FT1c = Proportion force of work producing good 1 in country C |
| 0.30 | FT2a = Proportion force of work producing good 2 in country A | 0.30 | FT2b = Proportion force of work producing good 2 in country B | 0.30 | FT2c = Proportion force of work producing good 2 in country C |
| 0.30 | FT3a = Proportion force of work producing good 3 in country A | 0.30 | FT3b = Proportion force of work producing good 3 in country B | 0.30 | FT3c = Proportion force of work producing good 3 in country C |
| 0.20 | FT4a = Proportion force of work producing good 4 in country A | 0.20 | FT4b = Proportion force of work producing good 4 in country B | 0.20 | FT4c = Proportion force of work producing good 4 in country C |

**CONSUMER PREFERENCES AND TARIFF RATES**

**Country A = UNITED STATES**

| | PREFERENCES OF CONSUMERS OF COUNTRY A WITH REGARD TO GOODS OF COUNTRY B | | PREFERENCES OF CONSUMERS OF COUNTRY A WITH REGARD TO GOODS OF COUNTRY C | | TARIFF (ARANCEL) OF COUNTRY A WITH REGARI |
|---|---|---|---|---|---|
| 0 | PR1ab = Preference of good 1 in country A made in country B | 0 | PR1ac = Preference of good 1 in country A made in country C | 0 | AR1a = Tariff (Arancel) of good 1 in country A |
| 0 | PR2ab = Preference of good 2 in country A made in country B | 0 | PR2ac = Preference of good 2 in country A made in country C | 0 | AR2a = Tariff (Arancel) of good 2 in country A |
| 0 | PR3ab = Preference of good 3 in country A made in country B | 0 | PR3ac = Preference of good 3 in country A made in country C | 0 | AR3a = Tariff (Arancel) of good 3 in country A |
| 0 | PR4ab = Preference of good 4 in country A made in country B | 0 | PR4ac = Preference of good 4 in country A made in country C | 0 | AR4a = Tariff (Arancel) of good 4 in country A |

**Country B = MEXICO**

| | PREFERENCES OF CONSUMERS OF COUNTRY B WITH REGARD TO GOODS OF COUNTRY A | | PREFERENCES OF CONSUMERS OF COUNTRY B WITH REGARD TO GOODS OF COUNTRY C | | TARIFF (ARANCEL) OF COUNTRY B WITH REGARI |
|---|---|---|---|---|---|
| 0 | PR1ba = Preference of good 1 in country B made in country A | 0 | PR1bc = Preference of good 1 in country B made in country C | 0 | AR1b = Tariff (Arancel) of good 1 in country B |
| 0 | PR2ba = Preference of good 2 in country B made in country A | 0 | PR2bc = Preference of good 2 in country B made in country C | 0 | AR2b = Tariff (Arancel) of good 2 in country B |
| 0 | PR3ba = Preference of good 3 in country B made in country A | 0 | PR3bc = Preference of good 3 in country B made in country C | 0 | AR3b = Tariff (Arancel) of good 3 in country B |
| 0 | PR4ba = Preference of good 4 in country B made in country A | 0 | PR4bc = Preference of good 4 in country B made in country C | 0 | AR4b = Tariff (Arancel) of good 4 in country B |

**Country C = CANADA**

| | PREFERENCES OF CONSUMERS OF COUNTRY C WITH REGARD TO GOODS OF COUNTRY A | | PREFERENCES OF CONSUMERS OF COUNTRY C WITH REGARD TO GOODS OF COUNTRY B | | TARIFF (ARANCEL) OF COUNTRY C WITH REGARI |
|---|---|---|---|---|---|
| 0 | PR1ca = Preference of good 1 in country C made in country A | 0 | PR1cb= Preference of good 1 in country C made in country B | 0 | AR1c = Tariff (Arancel) of good 1 in country C |
| 0 | PR2ca = Preference of good 2 in country C made in country A | 0 | PR2cb= Preference of good 2 in country C made in country B | 0 | AR2c = Tariff (Arancel) of good 2 in country C |
| 0 | PR3ca = Preference of good 3 in country C made in country A | 0 | PR3cb= Preference of good 3 in country C made in country B | 0 | AR3c = Tariff (Arancel) of good 3 in country C |
| 0 | PR4ca = Preference of good 4 in country C made in country A | 0 | PR4cb= Preference of good 4 in country C made in country B | 0 | AR4c = Tariff (Arancel) of good 4 in country C |

# Appendix Spreadsheet 3
## Image Section 3.6 and 3.7
## MODEL VARIABLES. EXCHANGE RATES

**SECTION 3.6 AND 3.7: MODEL VARIABLES. EXCHANGE RATES. SECTION 3.6 ARE DETERMINED BY TRIAL AND ERROR OF THE 2 PRIMARY EXCHANGE RATES**

The equilibrium can be reached in the three equations of Section 3.1 corresponding to the trade balance of the three countries A, B and C when in this Section 3.6 the set of the 2 primary exchange rates TCab and TCac that generate the simultaneous general balance in the trade balance of each country is established

| | SECTION 3.6: PRIMARY, CONTRA-PRIMARY AND NON-PRIMARY CHANGE RATES | | SECTION 3.7: INITIAL CHANGE RATES |
|---|---|---|---|
| 0.027400 | TCab = Primary and equivalent exchange rate of country A with country B | 0.027400 | TCab In = (Pa1+Pa2+Pa3+Pa4)/(Pb1+Pb2+Pb3+Pb4) = Initial exchange rate of country A with country B |
| 0.436364 | TCac = Primary and equivalent exchange rate of country A with country C | 0.436364 | TCac In = (Pa1+Pa2+Pa3+Pa4)/(Pc1+Pc2+Pc3+Pc4) = Initial exchange rate of country A with country C |
| 36.496350 | TCba = 1/TCab = Contra-Primary and equivalent exchange rate of country B with country A | 36.496350 | TCba In = (Pb1+Pb2+Pb3+Pb4)/(Pa1+Pa2+Pa3+Pa4) = 1/Tcab = Initial exchange rate of country B with country A |
| 2.291665 | TCca = 1/TCac = Contra-Primary and equivalent exchange rate of country C with country A | 2.291667 | TCca In = (Pc1+Pc2+Pc3+Pc4)/(Pa1+Pa2+Pa3+Pa4) = 1/TCac = Initial exchange rate of country C with country A |
| 15.925693 | TCbc = TCac/TCab = Non-Primary and equivalent exchange rate of country B with country C | 15.925680 | TCbc In = (Pb1+Pb2+Pb3+Pb4)/(Pc1+Pc2+Pc3+Pc4) = TCba/TCca =Initial exchange rate of country B with countr |
| 0.062792 | TCcb = TCab/TCac = Non-Primary and equivalent exchange rate of country C with country B | 0.062792 | TCcb In = (Pc1+Pc2+Pc3+Pc4)/(Pb1+Pb2+Pb3+Pb4) = TCab/TCac = Initial exchange rate of country C with count |

# Appendix Spreadsheet 4

## Image Section 4.1
## COMMERCIAL BALANCES. MODEL EQUATIONS

| | | | | | | | | | | | | | | | | |
|---|---|---|---|---|---|---|---|---|---|---|---|---|---|---|---|---|
| **Spreadsheet 4 - MODEL SIMULATION INTERNATIONAL TRADE OF 3 COUNTRIES:** | | | | | | | | | | | | | | | | |
| | | | | | | | | | **SECTION 4.1: COMMERCIAL BALANCES. MODEL EQUATIONS** | | | | | | | |
| | | **Equation 8.7 Defining the Trade Balance of Country A (United States) with All Other Countries (BCa.bc)** | | | | | | | | | | | | | | |
| | | | | **BCa.bc = BCab1 + BCab2 + BCab3 + BCab4 + BCac1 + BCac2 + BCac3 + BCac4** | | | | | | | | | | | | |
| **BCa.bc** | **=** | **BCab1** | **+** | **BCab2** | **+** | **BCab3** | **+** | **BCab4** | **+** | **BCac1** | **+** | **BCac2** | **+** | **BCac3** | **+** | **BCac4** |
| **0.00** | **=** | **-29.53** | **+** | **(22.46)** | **+** | **31.20** | **+** | **20.80** | **+** | **(5.99)** | **+** | **(4.76)** | **+** | **6.45** | **+** | **4.30** |
| | | **Equation 8.8 Defining the Trade Balance of Country B (Mexico) with All Other Countries (BCb.ac)** | | | | | | | | | | | | | | |
| | | | | **BCb.ac = BCba1 + BCba2 + BCba3 + BCba4 + BCbc1 + BCbc2 + BCbc3 + BCbca4** | | | | | | | | | | | | |
| **BCb.ac** | **=** | **BCba1** | **+** | **BCba2** | **+** | **BCba3** | **+** | **BCba4** | **+** | **BCbc1** | **+** | **BCbc2** | **+** | **BCbc3** | **+** | **BCbc4** |
| **0.00** | **=** | **29.53** | | **22.46** | | **(31.20)** | | **(20.80)** | | **3.76** | | **2.64** | | **(3.84)** | | **(2.56)** |
| | | **Equation 8.9 Defining the Trade Balance of Country C (Canada) with All Other Countries (BCc.ab)** | | | | | | | | | | | | | | |
| | | | | **BCc.ab = BCca1 + BCca2 + BCca3 + BCca4 + BCcb1 + BCcb2 + BCcb3 + BCcb4** | | | | | | | | | | | | |
| **BCc.ab** | **=** | **BCca1** | **+** | **BCca2** | **+** | **BCca3** | **+** | **BCca4** | **+** | **BCcb1** | **+** | **BCcb2** | **+** | **BCcb3** | **+** | **BCcb4** |
| **0.00** | **=** | **5.99** | **+** | **4.76** | **+** | **(6.45)** | **+** | **(4.30)** | **+** | **(3.76)** | **+** | **(2.64)** | **+** | **3.84** | **+** | **2.56** |
| | | | | | **COMMERCIAL BALANCES** | | | | | | | | | | | |
| | | **Country A - United States** | | | | | **Country B - Mexico** | | | | | **Country C - Canada** | | | | |
| | | | | **Deficit or SURPLUS US** | | | | | **Déficit o Superávit Méx** | | | | | **Déficit o Superávit Cand** | | |
| | | **EXa EU** | **IMa EU** | **US$ BCa.bc = (Exa - Ima)** | | | **EXb Mex** | **IMb Mex** | **US$ BCb.ac = (EXb - IMb)** | | | **EXc Cand** | **IMc Cand** | **US$ BCc.ab = (EXc - IMc)** | | |
| | | **62.74** | **62.74** | **(0.00)** | | | **58.39** | **58.39** | **0.00** | | | **17.15** | **17.15** | **(0.00)** | | |
| | | **EXab** | **IMab** | | | | **EXba** | **IMba** | | | | **EXca** | **IMca** | | | |
| | | **51.99** | **51.99** | **0.00** | | | **51.99** | **51.99** | **(0.00)** | | | **10.75** | **10.75** | **0.00** | | |
| | | **EXac** | **IMac** | | | | **EXbc** | **IMbc** | | | | **EXcb** | **IMcb** | | | |
| | | **10.75** | **10.75** | **(0.00)** | | | **6.40** | **6.40** | **0.01** | | | **6.40** | **6.40** | **(0.01)** | | |

# Appendix Spreadsheet 5

## Image Section 5.1
## COMMERCIAL BALANCES. MODEL EQUATIONS

| | | | | | | | | | | | | | | | | |
|---|---|---|---|---|---|---|---|---|---|---|---|---|---|---|---|---|
| **Spreadsheet 5 - MODEL SIMULATION INTERNATIONAL TRADE OF 3 COUNTRIES:** | | | | | | | | | | | | | | | | |
| | | | | | | | | | **SECTION 5.1: COMMERCIAL BALANCES. MODEL EQUATIONS** | | | | | | | |
| | | **Equation 8.7 Defining the Trade Balance of Country A (United States) with All Other Countries (BCa.bc)** | | | | | | | | | | | | | | |
| | | | | **BCa.bc = BCab1 + BCab2 + BCab3 + BCab4 + BCac1 + BCac2 + BCac3 + BCac4** | | | | | | | | | | | | |
| **BCa.bc** | **=** | **BCab1** | **+** | **BCab2** | **+** | **BCab3** | **+** | **BCab4** | **+** | **BCac1** | **+** | **BCac2** | **+** | **BCac3** | **+** | **BCac4** |
| **-187.83** | **=** | **-38.35** | **+** | **(106.11)** | **+** | **(125.66)** | **+** | **39.07** | **+** | **(7.30)** | **+** | **13.37** | **+** | **21.38** | **+** | **15.77** |
| | | **Equation 8.8 Defining the Trade Balance of Country B (Mexico) with All Other Countries (BCb.ac)** | | | | | | | | | | | | | | |
| | | | | **BCb.ac = BCba1 + BCba2 + BCba3 + BCba4 + BCbc1 + BCbc2 + BCbc3 + BCbca4** | | | | | | | | | | | | |
| **BCb.ac** | **=** | **BCba1** | **+** | **BCba2** | **+** | **BCba3** | **+** | **BCba4** | **+** | **BCbc1** | **+** | **BCbc2** | **+** | **BCbc3** | **+** | **BCbc4** |
| **215.65** | **=** | **38.35** | | **106.11** | | **125.66** | | **(39.07)** | | **11.69** | | **0.54** | | **(17.84)** | | **(9.79)** |
| | | **Equation 8.9 Defining the Trade Balance of Country C (Canada) with All Other Countries (BCc.ab)** | | | | | | | | | | | | | | |
| | | | | **BCc.ab = BCca1 + BCca2 + BCca3 + BCca4 + BCcb1 + BCcb2 + BCcb3 + BCcb4** | | | | | | | | | | | | |
| **BCc.ab** | **=** | **BCca1** | **+** | **BCca2** | **+** | **BCca3** | **+** | **BCca4** | **+** | **BCcb1** | **+** | **BCcb2** | **+** | **BCcb3** | **+** | **BCcb4** |
| **-27.82** | **=** | **7.30** | **+** | **(13.37)** | **+** | **(21.38)** | **+** | **(15.77)** | **+** | **(11.69)** | **+** | **(0.54)** | **+** | **17.84** | **+** | **9.79** |
| | | | | | **COMMERCIAL BALANCES** | | | | | | | | | | | |
| | | **Country A - United States** | | | | | **Country B - Mexico** | | | | | **Country C - Canada** | | | | |
| | | | | **Deficit or SURPLUS US** | | | | | **Déficit o Superávit Méx** | | | | | **Déficit o Superávit Cand** | | |
| | | **EXa EU** | **IMa EU** | **US$ BCa.bc = (Exa - Ima)** | | | **EXb Mex** | **IMb Mex** | **US$ BCb.ac = (EXb - IMb)** | | | **EXc Cand** | **IMc Cand** | **US$ BCc.ab = (EXc - IMc)** | | |
| | | **265.77** | **453.61** | **(187.83)** | | | **467.55** | **251.90** | **215.65** | | | **59.48** | **87.30** | **(27.82)** | | |
| | | **EXab** | **IMab** | | | | **EXba** | **IMba** | | | | **EXca** | **IMca** | | | |
| | | **207.49** | **438.54** | **(231.05)** | | | **438.54** | **207.49** | **231.05** | | | **15.07** | **58.29** | **(43.22)** | | |
| | | **EXac** | **IMac** | | | | **EXbc** | **IMbc** | | | | **EXcb** | **IMcb** | | | |
| | | **58.29** | **15.07** | **43.22** | | | **29.01** | **44.41** | **(15.40)** | | | **44.41** | **29.01** | **15.40** | | |

# Appendix Spreadsheet 5
# Image Section 5.5
# MODEL PARAMETERS

**SECTION 5.5: MODEL PARAMETERS**

| | | | | | |
|---|---|---|---|---|---|
| | PRODUCTIVITY OF WORK IN THE COUNTRY A | | PRODUCTIVITY OF WORK IN COUNTRY B | | PRODUCTIVITY OF WORK IN COUNTRY C |
| 3.00 | Za1 | 4.00 | Zb1 | 3.50 | Zc1 |
| 2.00 | Za2 | 3.00 | Zb2 | 2.50 | Zc2 |
| 1.00 | Za3 | 2.00 | Zb3 | 1.50 | Zc3 |
| 0.50 | Za4 | 1.00 | Zb4 | 0.75 | Zc4 |

**MONETARY OFFER OMi, MARGIN OF GAIN MGi, POPULATION Ni, SALARY Wi, PROFITS Gij, WORK FORCE FT, CONSUMER PREFERENCES PR AND TARIFFS AR**

| | | | | | |
|---|---|---|---|---|---|
| 2,500 | OMa = Money Offer Country A US dollar | 0.3650 | MGa = Average Margin of Gain or Profit in country A | 325.00 | Na = Population of country A United States |
| 25,000 | OMb = Money Offer Country B Mexico peso | 0.5450 | MGb = Average Margin of Gain or Profit in country B | 137.00 | Nb = Population of country B México |
| 500 | OMc = Money Offer Country C Canda dollar | 0.4100 | MGc = Average Margin of Gain or Profit in country C | 36.00 | Nc = Population of country C Canada |

| | | | |
|---|---|---|---|
| 4.88 | Wa = Salary in country A for each unit of work hired = (OMa(1-MGa))/Na | 8.42 | Ga1 = Gain or Profit in country A corresponds to good 1 for each unit produced = (OMa(MGa)(FTa1))/Qa1 |
| 83.03 | Wb = Salary in country B for each unit of work hired = (OMb(1-MGb))/Nb | 5.62 | Ga2 = Gain or Profit in country A corresponds to good 2 for each unit produced = (OMa(MGa)(FTa2))/Qa2 |
| 8.19 | Wc = Salary in country C for each unit of work hired = (OMc(1-MGc))/Nc | 2.81 | Ga3 = Gain or Profit in country A corresponds to good 3 for each unit produced = (OMa(MGa)(FTa3))/Qa3 |
| | | 1.40 | Ga4 = Gain or Profit in country A corresponds to good 4 for each unit produced = (OMa(MGa)(FTa4))/Qa4 |
| | | 397.81 | Gb1 = Gain or Profit in country B corresponds to good 1 for each unit produced = (OMb(MGb)(FTb1))/Qb1 |
| | | 298.36 | Gb2 = Gain or Profit in country B corresponds to good 2 for each unit produced = (OMb(MGb)(FTb2))/Qb2 |
| | | 198.91 | Gb3 = Gain or Profit in country B corresponds to good 3 for each unit produced = (OMb(MGb)(FTb3))/Qb3 |
| | | 99.45 | Gb4 = Gain or Profit in country B corresponds to good 3 for each unit produced = (OMb(MGb)(FTb4))/Qb4 |
| | | 19.93 | Gc1 = Gain or Profit in country C corresponds to good 1 for each unit produced = (OMc(MGc)(FTc1))/Qc1 |
| | | 14.24 | Gc2 = Gain or Profit in country C corresponds to good 2 for each unit produced = (OMc(MGc)(FTc2))/Qc2 |
| | | 8.54 | Gc3 = Gain or Profit in country C corresponds to good 3 for each unit produced = (OMc(MGc)(FTc3))/Qc3 |
| | | 4.27 | Gc4 = Gain or Profit in country C corresponds to good 4 for each unit produced = (OMc(MGc)(FTc4))/Qc4 |

**DISTRIBUTION OF THE FORCE OF WORK**

| | | | | | |
|---|---|---|---|---|---|
| | DISTRIBUTION OF THE COUNTRY A FORCE OF WORK | | DISTRIBUTION OF THE COUNTRY B FORCE OF WORK | | DISTRIBUTION OF THE COUNTRY C FORCE OF WORK |
| 0.20 | FT1a = Proportion force of work producing good 1 in country A | 0.20 | FT1b = Proportion force of work producing good 1 in country B | 0.20 | FT1c = Proportion force of work producing good 1 in country C |
| 0.30 | FT2a = Proportion force of work producing good 2 in country A | 0.30 | FT2b = Proportion force of work producing good 2 in country B | 0.30 | FT2c = Proportion force of work producing good 2 in country C |
| 0.30 | FT3a = Proportion force of work producing good 3 in country A | 0.30 | FT3b = Proportion force of work producing good 3 in country B | 0.30 | FT3c = Proportion force of work producing good 3 in country C |
| 0.20 | FT4a = Proportion force of work producing good 4 in country A | 0.20 | FT4b = Proportion force of work producing good 4 in country B | 0.20 | FT4c = Proportion force of work producing good 4 in country C |

**CONSUMER PREFERENCES AND TARIFF RATES**

**Country A = UNITED STATES**

| | | | | | |
|---|---|---|---|---|---|
| | PREFERENCES OF CONSUMERS OF COUNTRY A WITH REGARD TO GOODS OF COUNTRY B | | PREFERENCES OF CONSUMERS OF COUNTRY A WITH REGARD TO GOODS OF COUNTRY C | | TARIFF (ARANCEL) OF COUNTRY A WITH REGARI |
| 0.25 | PR1ab = Preference of good 1 in country A made in country B | 0.15 | PR1ac = Preference of good 1 in country A made in country C | 0 | AR1a = Tariff (Arancel) of good 1 in country A |
| 0.50 | PR2ab = Preference of good 2 in country A made in country B | 0.05 | PR2ac = Preference of good 2 in country A made in country C | 0 | AR2a = Tariff (Arancel) of good 2 in country A |
| 0.75 | PR3ab = Preference of good 3 in country A made in country B | 0.15 | PR3ac = Preference of good 3 in country A made in country C | 0 | AR3a = Tariff (Arancel) of good 3 in country A |
| 0.38 | PR4ab = Preference of good 4 in country A made in country B | 0.12 | PR4ac = Preference of good 4 in country A made in country C | 0 | AR4a = Tariff (Arancel) of good 4 in country A |

**Country B = MEXICO**

| | | | | | |
|---|---|---|---|---|---|
| | PREFERENCES OF CONSUMERS OF COUNTRY B WITH REGARD TO GOODS OF COUNTRY A | | PREFERENCES OF CONSUMERS OF COUNTRY B WITH REGARD TO GOODS OF COUNTRY C | | TARIFF (ARANCEL) OF COUNTRY B WITH REGARI |
| 0.25 | PR1ba = Preference of good 1 in country B made in country A | 0.05 | PR1bc = Preference of good 1 in country B made in country C | 0 | AR1b = Tariff (Arancel) of good 1 in country B |
| 0.18 | PR2ba = Preference of good 2 in country B made in country A | 0.16 | PR2bc = Preference of good 2 in country B made in country C | 0 | AR2b = Tariff (Arancel) of good 2 in country B |
| 0.14 | PR3ba = Preference of good 3 in country B made in country A | 0.21 | PR3bc = Preference of good 3 in country B made in country C | 0 | AR3b = Tariff (Arancel) of good 3 in country B |
| 0.22 | PR4ba = Preference of good 4 in country B made in country A | 0.17 | PR4bc = Preference of good 4 in country B made in country C | 0 | AR4b = Tariff (Arancel) of good 4 in country B |

**Country C = CANADA**

| | | | | | |
|---|---|---|---|---|---|
| | PREFERENCES OF CONSUMERS OF COUNTRY C WITH REGARD TO GOODS OF COUNTRY A | | PREFERENCES OF CONSUMERS OF COUNTRY C WITH REGARD TO GOODS OF COUNTRY B | | TARIFF (ARANCEL) OF COUNTRY C WITH REGARI |
| 0.15 | PR1ca = Preference of good 1 in country C made in country A | 0.19 | PR1cb= Preference of good 1 in country C made in country B | 0 | AR1c = Tariff (Arancel) of good 1 in country C |
| 0.25 | PR2ca = Preference of good 2 in country C made in country A | 0.15 | PR2cb= Preference of good 2 in country C made in country B | 0 | AR2c = Tariff (Arancel) of good 2 in country C |
| 0.15 | PR3ca = Preference of good 3 in country C made in country A | 0.14 | PR3cb= Preference of good 3 in country C made in country B | 0 | AR3c = Tariff (Arancel) of good 3 in country C |
| 0.18 | PR4ca = Preference of good 4 in country C made in country A | 0.14 | PR4cb= Preference of good 4 in country C made in country B | 0 | AR4c = Tariff (Arancel) of good 4 in country C |

# Apendix Spreadsheed 6

## Image Sección 6.1
## TRADE BALANCES. MODEL ECUATION

| | | | | | | | | | | | | | | | | |
|---|---|---|---|---|---|---|---|---|---|---|---|---|---|---|---|---|
| Spreadsheet 6 - MODEL SIMULATION INTERNATIONAL TRADE OF 3 COUNTRIES: | | | | | | | | | | | | | | | | |
| | | | | | | | | | SECTION 6.1: COMMERCIAL BALANCES. MODEL EQUATIONS | | | | | | | |
| | | Equation 8.7 Defining the Trade Balance of Country A (United States) with All Other Countries (BCa.bc) | | | | | | | | | | | | | | |
| | | | | BCa.bc = BCab1 + BCab2 + BCab3 + BCab4 + BCac1 + BCac2 + BCac3 + BCac4 | | | | | | | | | | | | |
| BCa.bc | = | BCab1 | + | BCab2 | + | BCab3 | + | BCab4 | + | BCac1 | + | BCac2 | + | BCac3 | + | BCac4 |
| 0.00 | = | -0.06 | + | (58.49) | + | (87.01) | + | 88.30 | + | (4.24) | + | 18.87 | + | 24.64 | + | 17.99 |
| | | Equation 8.8 Defining the Trade Balance of Country B (Mexico) with All Other Countries (BCb.ac) | | | | | | | | | | | | | | |
| | | | | BCb.ac = BCba1 + BCba2 + BCba3 + BCba4 + BCbc1 + BCbc2 + BCbc3 + BCbca4 | | | | | | | | | | | | |
| BCb.ac | = | BCba1 | + | BCba2 | + | BCba3 | + | BCba4 | + | BCbc1 | + | BCbc2 | + | BCbc3 | + | BCbc4 |
| 0.00 | = | 0.06 | | 58.49 | | 87.01 | | (88.30) | | 2.96 | | (13.69) | | (29.21) | | (17.33) |
| | | Equation 8.9 Defining the Trade Balance of Country C (Canada) with All Other Countries (BCc.ab) | | | | | | | | | | | | | | |
| | | | | BCc.ab = BCca1 + BCca2 + BCca3 + BCca4 + BCcb1 + BCcb2 + BCcb3 + BCcb4 | | | | | | | | | | | | |
| BCc.ab | = | BCca1 | + | BCca2 | + | BCca3 | + | BCca4 | + | BCcb1 | + | BCcb2 | + | BCcb3 | + | BCcb4 |
| 0.00 | = | 4.24 | + | (18.87) | + | (24.64) | + | (17.99) | + | (2.96) | + | 13.69 | + | 29.21 | + | 17.33 |
| | | | | | COMMERCIAL BALANCES | | | | | | | | | | | |
| | | Country A - United States | | | | | Country B - Mexico | | | | | Country C - Canada | | | | |
| | | | | Deficit or SURPLUS US | | | | | Déficit o Superávit Méx | | | | | Déficit o Superávit Cand | | |
| | | EXa EU | IMa EU | US$ BCa.bc = (Exa - Ima) | | | EXb Mex | IMb Mex | US$ BCb.ac = (EXb - IMb) | | | EXc Cand | IMc Cand | US$ BCc.ab = (EXc - IMc) | | |
| | | 416.49 | 416.49 | 0.00 | | | 416.98 | 416.98 | (0.00) | | | 80.67 | 80.66 | 0.00 | | |
| | | EXab | IMab | | | | EXba | IMba | | | | EXca | IMca | | | |
| | | 347.77 | 405.03 | (57.27) | | | 405.03 | 347.77 | 57.27 | | | 11.45 | 68.72 | (57.27) | | |
| | | EXac | IMac | | | | EXbc | IMbc | | | | EXcb | IMcb | | | |
| | | 68.72 | 11.45 | 57.27 | | | 11.94 | 69.21 | (57.27) | | | 69.21 | 11.94 | 57.27 | | |

# Appendix Spreadsheet 6
# Image Section 6.2

## Determination of Prices, Market Size, Exports and Imports used in the Trade Balance BCa.ba Of Country A: UNITED STATES

| | | | | | | | | | | |
|---|---|---|---|---|---|---|---|---|---|---|
| | | | SECTION 6.2: Determination of Trade Balances BCabi and BCaci, of prices, market sizes and exports and imports Country A: UNITED STATES | | | | | | | |
| US$ BCab1 = (EXa1b - IMa1b) | | | | US$ EXa1b=IMb1a | | | US$ IMa1b=IF((Pa1-(1-PR1ab)(1+AR1a)Pb1*TCab)/((Pa1+(1-PR1ab)(1+AR1a)Pb1*TCab)/2)>0, ((Pa1-(1-PR1ab)(1+AR1a)Pb1*TCab)/( | | | |
| -0.06 | | | | 45.13 | | | 45.19 | | | |
| | | US$ BCac1 = (EXa1c - IMa1c) | | US$ EXa1c=IMc1a | | | US$ IMa1c=IF((Pa1-(1-PR1ac)(1+AR1a)Pc1*TCac)/((Pa1+(1-PR1ac)(1+AR1a)Pc1*TCac)/2)>0, ((Pa1-(1-PR1ac)(1+AR1a)Pc1*TCac)/((Pa | | | |
| | | (4.24) | | 5.22 | | | 9.46 | | | |
| | Pa1=Wa*Za1 + Ga1 | | Pa1b=Pb1*TCab | | Pa1c=Pc1*TCac | | US$ TMab1=Min(VPa1, VPb1*TCab) | | US$ TMac1=Min(VPa1, VPc1*TCac) | VPa1=Pa1*Qa1 |
| | 23.08 | | 23.07259 | | 22.02 | | 158.05 | | 45.29 | 500.00 |
| | | Wa | Za1 | Ga1 | | Qa1=Na1/Za1 | | Na1=FT1a*Na | | |
| | | 4.88 | 3.00 | 8.42 | | 21.67 | | 65.00 | | |
| US$ BCab2 = (EXa2b - IMa2b) | | | | US$ EXa2b=IMb2a | | | US$ IMa2b=IF((Pa2-(1-PR2ab)(1+AR2a)Pb2*TCab)/((Pa2+(1-PR2ab)(1+AR2a)Pb2*TCab)/2)>0, ((Pa2-(1-PR2ab)(1+AR2a)Pb2*TCab)/( | | | |
| (58.49) | | | | 74.31 | | | 132.80 | | | |
| | | US$ BCac2 = (EXa2c - IMa2c) | | US$ EXa2c=IMc2a | | | US$ IMa2c=IF((Pa2-(1-PR2ac)(1+AR2a)Pc2*TCac)/((Pa2+(1-PR2ac)(1+AR2a)Pc2*TCac)/2)>0, ((Pa2-(1-PR2ac)(1+AR2a)Pc2*TCac)/((Pa | | | |
| | | 18.87 | | 20.87 | | | 2.00 | | | |
| | Pa2=Wa*Za2 + Ga2 | | Pa2b=Pb2*TCab | | Pa2c=Pc2*TCac | | US$ TMab2=Min(VPa2, VPb2*TCab) | | US$ TMac2=Min(VPa2, VPc2*TCac) | VPa2=Pa2*Qa2 |
| | 15.38 | | 17.30444 | | 15.73 | | 237.07 | | 67.93 | 750.00 |
| | | Wa | Za2 | Ga2 | | Qa2=Na2/Za2 | | Na2=FT2a*Na | | |
| | | 4.88 | 2.00 | 5.62 | | 48.75 | | 97.50 | | |
| US$ BCab3 = (EXa3b - IMa3b) | | | | US$ EXa3b=IMb3a | | | US$ IMa3b=IF((Pa3-(1-PR3ab)(1+AR3a)Pb3*TCab)/((Pa3+(1-PR3ab)(1+AR3a)Pb3*TCab)/2)>0, ((Pa3-(1-PR3ab)(1+AR3a)Pb3*TCab)/( | | | |
| (87.01) | | | | 128.54 | | | 215.55 | | | |
| | | US$ BCac3 = (EXa3c - IMa3c) | | US$ EXa3c=IMc3a | | | US$ IMa3c=IF((Pa3-(1-PR3ac)(1+AR3a)Pc3*TCac)/((Pa3+(1-PR3ac)(1+AR3a)Pc3*TCac)/2)>0, ((Pa3-(1-PR3ac)(1+AR3a)Pc3*TCac)/((Pa | | | |
| | | 24.64 | | 24.64 | | | - | | | |
| | Pa3=Wa*Za3 + Ga3 | | Pa3b=Pb3*TCab | | Pa3c=Pc3*TCac | | US$ TMab3=Min(VPa3, VPb3*TCab) | | US$ TMac3=Min(VPa3, VPc3*TCac) | VPa3=Pa3*Qa3 |
| | 7.69 | | 11.5363 | | 9.44 | | 237.07 | | 67.93 | 750.00 |
| | | Wa | Za3 | Ga3 | | Qa3=Na3/Za3 | | Na3=FT3a*Na | | |
| | | 4.88 | 1.00 | 2.81 | | 97.50 | | 97.50 | | |
| US$ BCab4 = (EXa4b - IMa4b) | | | | US$ EXa4b=IMb4a | | | US$ IMa4b=IF((Pa4-(1-PR4ab)(1+AR4a)Pb4*TCab)/((Pa4+(1-PR4ab)(1+AR4a)Pb4*TCab)/2)>0, ((Pa4-(1-PR4ab)(1+AR4a)Pb4*TCab)/( | | | |
| 88.30 | | | | 99.79 | | | 11.49 | | | |
| | | US$ BCac4 = (EXa4c - IMa4c) | | US$ EXa4c=IMc4a | | | US$ IMa4c=IF((Pa4-(1-PR4ac)(1+AR4a)Pc4*TCac)/((Pa4+(1-PR4ac)(1+AR4a)Pc4*TCac)/2)>0, ((Pa4-(1-PR4ac)(1+AR4a)Pc4*TCac)/((Pa | | | |
| | | 17.99 | | 17.99 | | | - | | | |
| | Pa4=Wa*Za4 + Ga4 | | Pa4b=Pb4*TCab | | Pa4c=Pc4*TCac | | US$ TMab4=Min(VPa4, VPb4*TCab) | | US$ TMac4=Min(VPa4, VPc4*TCac) | VPa4=Pa4*Qa4 |
| | 3.85 | | 5.768148 | | 4.717646 | | 158.05 | | 45.29 | 500.00 |
| | | Wa | Za4 | Ga 4 | | Qa4=Na4/Za4 | | Na4=FT4a*Na | | |
| | | 4.88 | 0.50 | 1.40 | | 130.00 | | 65.00 | | |

# Appendix Spreadsheet 6
# Image Section 6.3

## Determination of Prices, Market Size, Exports and Imports used in the Trade Balance BCb.ac Of Country B: MEXICO

| | | | | | | | | | | |
|---|---|---|---|---|---|---|---|---|---|---|
| | | | SECTION 6.3: Determination of Trade Balances BCbai and BCbci, of prices, of market sizes and exports and imports of Country B: MEXICO | | | | | | | |
| US$ BCba1 = (EXb1a - IMb1a) | | | | US$ EXb1a=IMa1b | | | US$ IMb1a=IF((Pb1-(1-PR1ba)(1+AR1b)Pa1*TCba)/((Pb1-(1-PR1ba)(1+AR1b)Pa1*TCba)/2)>0, ((Pb1-(1-PR1ba)(1+AR1b)Pa1*TCba)/( | | | |
| 0.06 | | | | 45.19 | | | 45.12729198 | | | |
| | | US$ BCbc1 = (EXb1c - IMb1c) | | US$ EXb1c=IMc1b | | | US$ IMb1c=IF((Pb1-(1-PR1bc)(1+AR1b)Pc1*TCbc)/((Pb1+(1-PR1bc)(1+AR1b)Pc1*TCbc)/2)>0, ((Pb1-(1-PR1bc)(1+AR1b)Pc1*TCbc)/( | | | |
| | | 2.96 | | 7.40 | | | 4.44 | | | |
| | Pb1=Wb*Zb1 + Gb1 | | Pb1a=Pa1*TCba | | Pb1c=Pc1*TCbc | | US$ TMba1=Min(VPb1*TCab, VPa1) | | US$ TMbc1=Min(VPb1*TCab, VPc1*TCac) | VPb1=Pb1*Qb1 |
| | 729.93 | | 730.06 | | 696.4905 | | 158.05 | | 45.29 | 5,000.00 |
| | | Wb | Zb1 | Gb1 | | Qb1=Nb1/Zb1 | | Nb1=FT1b*Nb | | |
| | | 83.03 | 4.00 | 397.81 | | 6.85 | | 27.40 | | |
| US$ BCba2 = (EXb2a - IMb2a) | | | | US$ EXb2a=IMa2b | | | US$ IMb2a=IF((Pb2-(1-PR2ba)(1+AR2b)Pa2*TCba)/((Pb2-(1-PR2ba)(1+AR2b)Pa2*TCba)/2)>0, ((Pb2-(1-PR2ba)(1+AR2b)Pa2*TCba)/( | | | |
| 58.49 | | | | 132.80 | | | 74.31 | | | |
| | | US$ BCbc2 = (EXb2c - IMb2c) | | US$ EXb2c=IMc2b | | | US$ IMb2c=IF((Pb2-(1-PR2bc)(1+AR2b)Pc2*TCbc)/((Pb2+(1-PR2bc)(1+AR2b)Pc2*TCbc)/2)>0, ((Pb2-(1-PR2bc)(1+AR2b)Pc2*TCbc)/ | | | |
| | | (13.69) | | 4.54 | | | 18.23 | | | |
| | Pb2=Wb2*Zb2 + Gb2 | | Pb2a=Pa2*TCba | | Pb2c=Pc2*TCbc | | US$ TMba2=Min(VPb2*Tcab, VPa2) | | US$ TMbc2=Min(VPb2*TCab, VPc2*TCac) | VPb2=Pb2*Qb2 |
| | 547.45 | | 486.71 | | 497.4932 | | 237.07 | | 67.93 | 7,500.00 |
| | | Wb2 | Zb2 | Gb2 | | Qb2=Nb2/Zb2 | | Nb2=FT2b*Nb | | |
| | | 83.03 | 3.00 | 298.36 | | 13.70 | | 41.10 | | |
| US$ BCba3 = (EXb3a - IMb3a) | | | | US$ EXb3a=IMa3b | | | US$ IMb3a=IF((Pb3-(1-PR3ba)(1+AR3b)Pa3*TCba)/((Pb3-(1-PR3ba)(1+AR3b)Pa3*TCba)/2)>0, ((Pb3-(1-PR3ba)(1+AR3b)Pa3*TCba)/( | | | |
| 87.01 | | | | 215.55 | | | 128.54 | | | |
| | | US$ BCbc3 = (EXb3c - IMb3c) | | US$ EXb3c=IMc3b | | | US$ IMb3c=IF((Pb3-(1-PR3bc)(1+AR3b)Pc3*TCbc)/((Pb3+(1-PR3bc)(1+AR3b)Pc3*TCbc)/2)>0, ((Pb3-(1-PR3bc)(1+AR3b)Pc3*TCbc)/( | | | |
| | | (29.21) | | - | | | 29.21 | | | |
| | Pb3=Wb3*Zb3 + Gb3 | | Pb3a=Pa3*TCba | | Pb3c=Pc3*TCbc | | US$ TMba3=Min(VPb3*Tcab, VPa3) | | US$ TMbc3=Min(VPb3*TCab, VPc3*TCac) | VPb3=Pb3*Qb3 |
| | 364.96 | | 243.35 | | 298.4959 | | 237.07 | | 67.93 | 7,500.00 |
| | | Wb3 | Zb3 | Gb3 | | Qb3=Nb3/Zb3 | | Nb3=FT3b*Nb | | |
| | | 83.03 | 2.00 | 198.91 | | 20.55 | | 41.10 | | |
| US$ BCba4 = (EXb4a - IMb4a) | | | | US$ EXb4a=IMa4b | | | US$ IMb4a=IF((Pb4-(1-PR4ba)(1+AR4b)Pa4*TCba)/((Pb4-(1-PR4ba)(1+AR4b)Pa4*TCba)/2)>0, ((Pb4-(1-PR4ba)(1+AR4b)Pa4*TCba)/( | | | |
| (88.30) | | | | 11.49 | | | 99.79 | | | |
| | | US$ BCbc4 = (EXb4c - IMb4c) | | US$ EXb4c=IMc4b | | | US$ IMb4c=IF((Pb4-(1-PR4bc)(1+AR4b)Pc4*TCbc)/((Pb4+(1-PR4bc)(1+AR4b)Pc4*TCbc)/2)>0, ((Pb4-(1-PR4bc)(1+AR4b)Pc4*TCbc)/( | | | |
| | | (17.33) | | - | | | 17.33 | | | |
| | Pb4=Wb4*Zb4 + Gb4 | | Pb4a=Pa4*TCba | | Pb4c=Pc4*TCbc | | US$ TMba4=Min(VPb4*Tcab, VPa4) | | US$ TMbc4=Min(VPb4*TCab, VPc4*TCac) | VPb4=Pb4*Qb4 |
| | 182.48 | | 121.6773 | | 149.248 | | 158.05 | | 45.29 | 5,000.00 |
| | | Wb4 | Zb4 | Gb4 | | Qb4=Nb4/Zb4 | | Nb4=FT4b*Nb | | |
| | | 83.03 | 1.00 | 99.45 | | 27.40 | | 27.40 | | |

# Appendix Spreadsheet 6
# Image Section 6.4

## Determination of Prices, Market Size, Exports and Imports used in the Trade Balance BCc.ab Of Country C: CANADA

| | | | | | | | | | | |
|---|---|---|---|---|---|---|---|---|---|---|
| | | | SECTION 6.4: Determination of Trade Balances BCcai and BCcbi, prices, market sizes and Exports and Imports of Country C: CANADA | | | | | | | |
| US$ BCca1 = (EXc1a - IMc1a) | | | | US$ EXc1a=IMa1c | | | US$ IMc1a=IF((Pc1-(1-PR1ca)(1+AR1c)Pa1*TCca)/((Pc1+(1-PR1ca)(1+AR1c)Pa1*TCca)/2)>0, ((Pc1-(1-PR1ca)(1+AR1c)Pa1*TCca)/((P | | | |
| 4.24 | | | | 9.46 | | | 5.222444466 | | | |
| | | US$ BCcb1 = (EXc1b - IMc1b) | | US$ EXc1b=IMb1c | | | US$ IMc1b=IF((Pc1-(1-PR1cb)(1+AR1c)Pb1*TCcb)/((Pc1+(1-PR1cb)(1+AR1c)Pb1*TCcb )/2)>0, ((Pc1-(1-PR1cb)(1+AR1c)Pb1*TCcb)/( | | | |
| | | (2.96) | | 4.44 | | | 7.40 | | | |
| | Pc1=Wc1*Zc1 + Gc1 | | Pc1a=Pa1*TCca | | Pc1b=Pb1*TCcb | | US$ TMca1=Min(VPc1*TCac, VPa1) | | US$ TMcb1=Min(VPc1*TCac, VPb1*TCab) | VPc1=Pc1*Qc1 |
| | 48.61 | | 50.95436 | | 50.94479 | | 45.29 | | 45.29 | 100.00 |
| | | Wc | Zc1 | Gc1 | | Qc1=Nc1/Zc1 | | Nc1=FT1c*Nc | | |
| | | 8.19 | 3.50 | 19.93 | | 2.06 | | 7.20 | | |
| US$ BCca2 = (EXc2a - IMc2a) | | | | US$ EXc2a=IMa2c | | | US$ IMc2a=IF((Pc2-(1-PR2ca)(1+AR2c)Pa2*TCca)/((Pc2+(1-PR2ca)(1+AR2c)Pa2*TCca)/2)>0, ((Pc2-(1-PR2ca)(1+AR2c)Pa2*TCca)/((P | | | |
| (18.87) | | | | 2.00 | | | 20.87 | | | |
| | | US$ BCcb2 = (EXc2b - IMc2b) | | US$ EXc2b=IMb2c | | | US$ IMc2b=IF((Pc2-(1-PR2cb)(1+AR2c)Pb2*TCcb)/((Pc2+(1-PR2cb)(1+AR2c)Pb2*TCcb )/2)>0, ((Pc2-(1-PR2cb)(1+AR2c)Pb2*TCcb)/( | | | |
| | | 13.69 | | 18.23 | | | 4.54 | | | |
| | Pc2=Wc2*Zc2+ Gc2 | | Pc2a=Pa2*TCca | | Pc2b=Pb2*Tccb | | US$ TMca2=Min(VPc2*TCac, VPa2) | | US$ TMcb2=Min(VPc2*TCac, VPb2*TCab) | VPc2=Pc2*Qc2 |
| | 34.72 | | 33.96957 | | 38.2086 | | 67.93 | | 67.93 | 150.00 |
| | | Wc | Zc2 | Gc2 | | Qc2=Nc2/Zc2 | | Nc2=FT2c*Nc | | |
| | | 8.19 | 2.50 | 14.24 | | 4.32 | | 10.80 | | |
| US$ BCca3 = (EXc3a - IMc3a) | | | | US$ EXc3a=IMa3c | | | US$ IMc3a=IF((Pc3-(1-PR3ca)(1+AR3c)Pa3*TCca)/((Pc3+(1-PR3ca)(1+AR3c)Pa3*TCca)/2)>0, ((Pc3-(1-PR3ca)(1+AR3c)Pa3*TCca)/((P | | | |
| (24.64) | | | | - | | | 24.64 | | | |
| | | US$ BCcb3 = (EXc3b - IMc3b) | | EXc3b=IMb3c | | | US$ IMc3b=IF((Pc3-(1-PR3cb)(1+AR3c)Pb3*TCcb)/((Pc3+(1-PR3cb)(1+AR3c)Pb3*TCcb )/2)>0, ((Pc3-(1-PR3cb)(1+AR3c)Pb3*TCcb)/( | | | |
| | | 29.21 | | 29.21 | | | - | | | |
| | Pc3=Wc3*Zc3 + Gc3 | | Pc3a=Pa3*TCca | | Pc3b=Pb3*Tccb | | US$ TMca3=Min(VPc3*TCac, VPa3) | | US$ TMcb3=Min(VPc3*TCac, VPb3*TCab) | VPc3=Pc3*Qc3 |
| | 20.83 | | 16.98479 | | 25.4724 | | 67.93 | | 67.93 | 150.00 |
| | | Wc | Zc3 | Gc3 | | Qc3=Nc3/Zc3 | | Nc3=FT3c*Nc | | |
| | | 8.19 | 1.50 | 8.54 | | 7.20 | | 10.80 | | |
| US$ BCca4 = (EXc4a - IMc4a) | | | | US$ EXc4a=IMa4c | | | US$ IMc4a=IF((Pc4-(1-PR4ca)(1+AR4c)Pa4*TCca)/((Pc4+(1-PR4ca)(1+AR4c)Pa4*TCca)/2)>0, ((Pc4-(1-PR4ca)(1+AR4c)Pa4*TCca)/((P | | | |
| (17.99) | | | | - | | | 17.99 | | | |
| | | US$ BCcb4 = (EXc4b - IMc4b) | | US$ EXc4b=IMb4c | | | US$ IMc4b=IF((Pc4-(1-PR4cb)(1+AR4c)Pb4*TCcb)/((Pc4+(1-PR4cb)(1+AR4c)Pb4*TCcb )/2)>0, ((Pc4-(1-PR4cb)(1+AR4c)Pb4*TCcb)/( | | | |
| | | 17.33 | | 17.33 | | | - | | | |
| | Pc4=Wc4*Zc4 + Gc4 | | Pc4a=Pa4*TCca | | Pc4b=Pb4*Tccb | | US$ TMca4=Min(VPc4*TCac, VPa4) | | US$ TMcb4=Min(VPc4*TCac, VPb4*TCab) | VPc4=Pc4*Qc4 |
| | 10.42 | | 8.492393 | | 12.7362 | | 45.29 | | 45.29 | 100.00 |
| | | Wa | Zc4 | Gc4 | | Qc4=Nc4/Zc4 | | Nc4=FT4c*Nc | | |
| | | 8.19 | 0.75 | 4.27 | | 9.60 | | 7.20 | | |

# Appendix Spreadsheet 6
# Image Section 6.5

## MODEL PARAMETERS

**SECTION 6.5: MODEL PARAMETERS**

| | PRODUCTIVITY OF WORK IN THE COUNTRY A | | PRODUCTIVITY OF WORK IN COUNTRY B | | PRODUCTIVITY OF WORK IN COUNTRY C |
|---|---|---|---|---|---|
| 3.00 | Za1 | 4.00 | Zb1 | 3.50 | Zc1 |
| 2.00 | Za2 | 3.00 | Zb2 | 2.50 | Zc2 |
| 1.00 | Za3 | 2.00 | Zb3 | 1.50 | Zc3 |
| 0.50 | Za4 | 1.00 | Zb4 | 0.75 | Zc4 |

**MONETARY OFFER OMi, MARGIN OF GAIN MGi, POPULATION Ni, SALARY Wi, PROFITS Gij, WORK FORCE FT, CONSUMER PREFERENCES PR AND TARIFFS AR**

| | | | | | |
|---|---|---|---|---|---|
| 2,500 | OMa = Money Offer Country A US dollar | 0.3650 | MGa = Average Margin of Gain or Profit in country A | 325.00 | Na = Population of country A United States |
| 25,000 | OMb = Money Offer Country B Mexico peso | 0.5450 | MGb = Average Margin of Gain or Profit in country B | 137.00 | Nb = Population of country B México |
| 500 | OMc = Money Offer Country C Canda dollar | 0.4100 | MGc = Average Margin of Gain or Profit in country C | 36.00 | Nc = Population of country C Canada |

| | |
|---|---|
| 4.88 | Wa = Salary in country A for each unit of work hired = (OMa(1-MGa))/Na |
| 83.03 | Wb = Salary in country B for each unit of work hired = (OMb(1-MGb))/Nb |
| 8.19 | Wc = Salary in country C for each unit of work hired = (OMc(1-MGc))/Nc |
| 8.42 | Ga1 = Gain or Profit in country A corresponds to good 1 for each unit produced = (OMa(MGa)(FTa1))/Qa1 |
| 5.62 | Ga2 = Gain or Profit in country A corresponds to good 2 for each unit produced = (OMa(MGa)(FTa2))/Qa2 |
| 2.81 | Ga3 = Gain or Profit in country A corresponds to good 3 for each unit produced = (OMa(MGa)(FTa3))/Qa3 |
| 1.40 | Ga4 = Gain or Profit in country A corresponds to good 4 for each unit produced = (OMa(MGa)(FTa4))/Qa4 |
| | |
| 397.81 | Gb1 = Gain or Profit in country B corresponds to good 1 for each unit produced = (OMb(MGb)(FTb1))/Qb1 |
| 298.36 | Gb2 = Gain or Profit in country B corresponds to good 2 for each unit produced = (OMb(MGb)(FTb2))/Qb2 |
| 198.91 | Gb3 = Gain or Profit in country B corresponds to good 3 for each unit produced = (OMb(MGb)(FTb3))/Qb3 |
| 99.45 | Gb4 = Gain or Profit in country B corresponds to good 3 for each unit produced = (OMb(MGb)(FTb4))/Qb4 |
| | |
| 19.93 | Gc1 = Gain or Profit in country C corresponds to good 1 for each unit produced = (OMc(MGc)(FTc1))/Qc1 |
| 14.24 | Gc2 = Gain or Profit in country C corresponds to good 2 for each unit produced = (OMc(MGc)(FTc2))/Qc2 |
| 8.54 | Gc3 = Gain or Profit in country C corresponds to good 3 for each unit produced = (OMc(MGc)(FTc3))/Qc3 |
| 4.27 | Gc4 = Gain or Profit in country C corresponds to good 4 for each unit produced = (OMc(MGc)(FTc4))/Qc4 |

**DISTRIBUTION OF THE FORCE OF WORK**

| | DISTRIBUTION OF THE COUNTRY A FORCE OF WORK | | DISTRIBUTION OF THE COUNTRY B FORCE OF WORK | | DISTRIBUTION OF THE COUNTRY C FORCE OF WORK |
|---|---|---|---|---|---|
| 0.20 | FT1a = Proportion force of work producing good 1 in country A | 0.20 | FT1b = Proportion force of work producing good 1 in country B | 0.20 | FT1c = Proportion force of work producing good 1 in country C |
| 0.30 | FT2a = Proportion force of work producing good 2 in country A | 0.30 | FT2b = Proportion force of work producing good 2 in country B | 0.30 | FT2c = Proportion force of work producing good 2 in country C |
| 0.30 | FT3a = Proportion force of work producing good 3 in country A | 0.30 | FT3b = Proportion force of work producing good 3 in country B | 0.30 | FT3c = Proportion force of work producing good 3 in country C |
| 0.20 | FT4a = Proportion force of work producing good 4 in country A | 0.20 | FT4b = Proportion force of work producing good 4 in country B | 0.20 | FT4c = Proportion force of work producing good 4 in country C |

**CONSUMER PREFERENCES AND TARIFF RATES**

**Country A = UNITED STATES**

| | PREFERENCES OF CONSUMERS OF COUNTRY A WITH REGARD TO GOODS OF COUNTRY B | | PREFERENCES OF CONSUMERS OF COUNTRY A WITH REGARD TO GOODS OF COUNTRY C | | TARIFF (ARANCEL) OF COUNTRY A WITH REGARI |
|---|---|---|---|---|---|
| 0.25 | PR1ab = Preference of good 1 in country A made in country B | 0.15 | PR1ac = Preference of good 1 in country A made in country C | 0 | AR1a = Tariff (Arancel) of good 1 in country A |
| 0.50 | PR2ab = Preference of good 2 in country A made in country B | 0.05 | PR2ac = Preference of good 2 in country A made in country C | 0 | AR2a = Tariff (Arancel) of good 2 in country A |
| 0.75 | PR3ab = Preference of good 3 in country A made in country B | 0.15 | PR3ac = Preference of good 3 in country A made in country C | 0 | AR3a = Tariff (Arancel) of good 3 in country A |
| 0.38 | PR4ab = Preference of good 4 in country A made in country B | 0.12 | PR4ac = Preference of good 4 in country A made in country C | 0 | AR4a = Tariff (Arancel) of good 4 in country A |

**Country B = MEXICO**

| | PREFERENCES OF CONSUMERS OF COUNTRY B WITH REGARD TO GOODS OF COUNTRY A | | PREFERENCES OF CONSUMERS OF COUNTRY B WITH REGARD TO GOODS OF COUNTRY C | | TARIFF (ARANCEL) OF COUNTRY B WITH REGARI |
|---|---|---|---|---|---|
| 0.25 | PR1ba = Preference of good 1 in country B made in country A | 0.05 | PR1bc = Preference of good 1 in country B made in country C | 0 | AR1b = Tariff (Arancel) of good 1 in country B |
| 0.18 | PR2ba = Preference of good 2 in country B made in country A | 0.16 | PR2bc = Preference of good 2 in country B made in country C | 0 | AR2b = Tariff (Arancel) of good 2 in country B |
| 0.14 | PR3ba = Preference of good 3 in country B made in country A | 0.21 | PR3bc = Preference of good 3 in country B made in country C | 0 | AR3b = Tariff (Arancel) of good 3 in country B |
| 0.22 | PR4ba = Preference of good 4 in country B made in country A | 0.17 | PR4bc = Preference of good 4 in country B made in country C | 0 | AR4b = Tariff (Arancel) of good 4 in country B |

**Country C = CANADA**

| | PREFERENCES OF CONSUMERS OF COUNTRY C WITH REGARD TO GOODS OF COUNTRY A | | PREFERENCES OF CONSUMERS OF COUNTRY C WITH REGARD TO GOODS OF COUNTRY B | | TARIFF (ARANCEL) OF COUNTRY C WITH REGARI |
|---|---|---|---|---|---|
| 0.15 | PR1ca = Preference of good 1 in country C made in country A | 0.19 | PR1cb= Preference of good 1 in country C made in country B | 0 | AR1c = Tariff (Arancel) of good 1 in country C |
| 0.25 | PR2ca = Preference of good 2 in country C made in country A | 0.15 | PR2cb= Preference of good 2 in country C made in country B | 0 | AR2c = Tariff (Arancel) of good 2 in country C |
| 0.15 | PR3ca = Preference of good 3 in country C made in country A | 0.14 | PR3cb= Preference of good 3 in country C made in country B | 0 | AR3c = Tariff (Arancel) of good 3 in country C |
| 0.18 | PR4ca = Preference of good 4 in country C made in country A | 0.14 | PR4cb= Preference of good 4 in country C made in country B | 0 | AR4c = Tariff (Arancel) of good 4 in country C |

# Appendix Spreadsheet 7

## Image Section 7.1
## COMMERCIAL BALANCES.

| | | | | | | | | | | | | | | | | |
|---|---|---|---|---|---|---|---|---|---|---|---|---|---|---|---|---|
| Spreadsheet 7 - MODEL SIMULATION INTERNATIONAL TRADE OF 3 COUNTRIES: | | | | | | | | | | | | | | | | |
| | | | | | | | | | SECTION 7.1: COMMERCIAL BALANCES. MODEL EQUATIONS | | | | | | | |
| | | Equation 8.7 Defining the Trade Balance of Country A (United States) with All Other Countries (BCa.bc) | | | | | | | | | | | | | | |
| | | | | BCa.bc = BCab1 + BCab2 + BCab3 + BCab4 + BCac1 + BCac2 + BCac3 + BCac4 | | | | | | | | | | | | |
| BCa.bc | = | BCab1 | + | BCab2 | + | BCab3 | + | BCab4 | + | BCac1 | + | BCac2 | + | BCac3 | + | BCac4 |
| -381.35 | = | -64.04 | + | (96.06) | + | (96.06) | + | (64.04) | + | (12.23) | + | (18.35) | + | (18.35) | + | (12.23) |
| | | Equation 8.8 Defining the Trade Balance of Country B (Mexico) with All Other Countries (BCb.ac) | | | | | | | | | | | | | | |
| | | | | BCb.ac = BCba1 + BCba2 + BCba3 + BCba4 + BCbc1 + BCbc2 + BCbc3 + BCbca4 | | | | | | | | | | | | |
| BCb.ac | = | BCba1 | + | BCba2 | + | BCba3 | + | BCba4 | + | BCbc1 | + | BCbc2 | + | BCbc3 | + | BCbc4 |
| 366.19 | = | 64.04 | | 96.06 | | 96.06 | | 64.04 | | 9.20 | | 13.80 | | 13.80 | | 9.20 |
| | | Equation 8.9 Defining the Trade Balance of Country C (Canada) with All Other Countries (BCc.ab) | | | | | | | | | | | | | | |
| | | | | BCc.ab = BCca1 + BCca2 + BCca3 + BCca4 + BCcb1 + BCcb2 + BCcb3 + BCcb4 | | | | | | | | | | | | |
| BCc.ab | = | BCca1 | + | BCca2 | + | BCca3 | + | BCca4 | + | BCcb1 | + | BCcb2 | + | BCcb3 | + | BCcb4 |
| 15.17 | = | 12.23 | + | 18.35 | + | 18.35 | + | 12.23 | + | (9.20) | + | (13.80) | + | (13.80) | + | (9.20) |
| | | | | | COMMERCIAL BALANCES | | | | | | | | | | | |
| | | Country A - United States | | | | | Country B - Mexico | | | | | Country C - Canada | | | | |
| | | | | Deficit or SURPLUS US | | | | | Déficit o Superávit Méx | | | | | Déficit o Superávit Cand | | |
| | | EXa EU | IMa EU | US$ BCa.bc = (Exa - Ima) | | | EXb Mex | IMb Mex | US$ BCb.ac = (EXb - IMb) | | | EXc Cand | IMc Cand | US$ BCc.ab = (EXc - IMc) | | |
| | | - | 381.35 | (381.35) | | | 366.19 | - | 366.19 | | | 61.16 | 46.00 | 15.17 | | |
| | | EXab | IMab | | | | EXba | IMba | | | | EXca | IMca | | | |
| | | - | 320.19 | (320.19) | | | 320.19 | - | 320.19 | | | 61.16 | - | 61.16 | | |
| | | EXac | IMac | | | | EXbc | IMbc | | | | EXcb | IMcb | | | |
| | | - | 61.16 | (61.16) | | | 46.00 | - | 46.00 | | | - | 46.00 | (46.00) | | |

# Appendix Spreadsheet 7
# Image Section 7.5
# MODEL PARAMETERS

**SECTION 7.5: MODEL PARAMETERS**

| | PRODUCTIVITY OF WORK IN THE COUNTRY A | | PRODUCTIVITY OF WORK IN COUNTRY B | | PRODUCTIVITY OF WORK IN COUNTRY C |
|---|---|---|---|---|---|
| 3.00 | Za1 | 3.00 | Zb1 | 3.00 | Zc1 |
| 2.00 | Za2 | 2.00 | Zb2 | 2.00 | Zc2 |
| 1.00 | Za3 | 1.00 | Zb3 | 1.00 | Zc3 |
| 0.50 | Za4 | 0.50 | Zb4 | 0.50 | Zc4 |

**MONETARY OFFER OMi, MARGIN OF GAIN MGi, POPULATION Ni, SALARY Wi, PROFITS Gij, WORK FORCE FT, CONSUMER PREFERENCES PR AND TARIFFS AR**

| | | | | | |
|---|---|---|---|---|---|
| 2,500 | OMa = Money Offer Country A US dollar | 0.3650 | MGa = Average Margin of Gain or Profit in country A | 325.00 | Na = Population of country A United States |
| 25,000 | OMb = Money Offer Country B Mexico peso | 0.5450 | MGb = Average Margin of Gain or Profit in country B | 137.00 | Nb = Population of country B México |
| 500 | OMc = Money Offer Country C Canda dollar | 0.4100 | MGc = Average Margin of Gain or Profit in country C | 36.00 | Nc = Population of country C Canada |

| | | | |
|---|---|---|---|
| 4.88 | Wa = Salary in country A for each unit of work hired = (OMa(1-MGa))/Na | 8.42 | Ga1 = Gain or Profit in country A corresponds to good 1 for each unit produced = (OMa(MGa)(FTa1))/Qa1 |
| 83.03 | Wb = Salary in country B for each unit of work hired = (OMb(1-MGb))/Nb | 5.62 | Ga2 = Gain or Profit in country A corresponds to good 2 for each unit produced = (OMa(MGa)(FTa2))/Qa2 |
| 8.19 | Wc = Salary in country C for each unit of work hired = (OMc(1-MGc))/Nc | 2.81 | Ga3 = Gain or Profit in country A corresponds to good 3 for each unit produced = (OMa(MGa)(FTa3))/Qa3 |
| | | 1.40 | Ga4 = Gain or Profit in country A corresponds to good 4 for each unit produced = (OMa(MGa)(FTa4))/Qa4 |
| | | 298.36 | Gb1 = Gain or Profit in country B corresponds to good 1 for each unit produced = (OMb(MGb)(FTb1))/Qb1 |
| | | 198.91 | Gb2 = Gain or Profit in country B corresponds to good 2 for each unit produced = (OMb(MGb)(FTb2))/Qb2 |
| | | 99.45 | Gb3 = Gain or Profit in country B corresponds to good 3 for each unit produced = (OMb(MGb)(FTb3))/Qb3 |
| | | 49.73 | Gb4 = Gain or Profit in country B corresponds to good 3 for each unit produced = (OMb(MGb)(FTb4))/Qb4 |
| | | 17.08 | Gc1 = Gain or Profit in country C corresponds to good 1 for each unit produced = (OMc(MGc)(FTc1))/Qc1 |
| | | 11.39 | Gc2 = Gain or Profit in country C corresponds to good 2 for each unit produced = (OMc(MGc)(FTc2))/Qc2 |
| | | 5.69 | Gc3 = Gain or Profit in country C corresponds to good 3 for each unit produced = (OMc(MGc)(FTc3))/Qc3 |
| | | 2.85 | Gc4 = Gain or Profit in country C corresponds to good 4 for each unit produced = (OMc(MGc)(FTc4))/Qc4 |

**DISTRIBUTION OF THE FORCE OF WORK**

| | DISTRIBUTION OF THE COUNTRY A FORCE OF WORK | | DISTRIBUTION OF THE COUNTRY B FORCE OF WORK | | DISTRIBUTION OF THE COUNTRY C FORCE OF WORK |
|---|---|---|---|---|---|
| 0.20 | FT1a = Proportion force of work producing good 1 in country A | 0.20 | FT1b = Proportion force of work producing good 1 in country B | 0.20 | FT1c = Proportion force of work producing good 1 in country C |
| 0.30 | FT2a = Proportion force of work producing good 2 in country A | 0.30 | FT2b = Proportion force of work producing good 2 in country B | 0.30 | FT2c = Proportion force of work producing good 2 in country C |
| 0.30 | FT3a = Proportion force of work producing good 3 in country A | 0.30 | FT3b = Proportion force of work producing good 3 in country B | 0.30 | FT3c = Proportion force of work producing good 3 in country C |
| 0.20 | FT4a = Proportion force of work producing good 4 in country A | 0.20 | FT4b = Proportion force of work producing good 4 in country B | 0.20 | FT4c = Proportion force of work producing good 4 in country C |

**CONSUMER PREFERENCES AND TARIFF RATES**

**Country A = UNITED STATES**

| | PREFERENCES OF CONSUMERS OF COUNTRY A WITH REGARD TO GOODS OF COUNTRY B | | PREFERENCES OF CONSUMERS OF COUNTRY A WITH REGARD TO GOODS OF COUNTRY C | | TARIFF (ARANCEL) OF COUNTRY A WITH REGARI |
|---|---|---|---|---|---|
| 0 | PR1ab = Preference of good 1 in country A made in country B | 0 | PR1ac = Preference of good 1 in country A made in country C | 0 | AR1a = Tariff (Arancel) of good 1 in country A |
| 0 | PR2ab = Preference of good 2 in country A made in country B | 0 | PR2ac = Preference of good 2 in country A made in country C | 0 | AR2a = Tariff (Arancel) of good 2 in country A |
| 0 | PR3ab = Preference of good 3 in country A made in country B | 0 | PR3ac = Preference of good 3 in country A made in country C | 0 | AR3a = Tariff (Arancel) of good 3 in country A |
| 0 | PR4ab = Preference of good 4 in country A made in country B | 0 | PR4ac = Preference of good 4 in country A made in country C | 0 | AR4a = Tariff (Arancel) of good 4 in country A |

**Country B = MEXICO**

| | PREFERENCES OF CONSUMERS OF COUNTRY B WITH REGARD TO GOODS OF COUNTRY A | | PREFERENCES OF CONSUMERS OF COUNTRY B WITH REGARD TO GOODS OF COUNTRY C | | TARIFF (ARANCEL) OF COUNTRY B WITH REGARI |
|---|---|---|---|---|---|
| 0 | PR1ba = Preference of good 1 in country B made in country A | 0 | PR1bc = Preference of good 1 in country B made in country C | 0 | AR1b = Tariff (Arancel) of good 1 in country B |
| 0 | PR2ba = Preference of good 2 in country B made in country A | 0 | PR2bc = Preference of good 2 in country B made in country C | 0 | AR2b = Tariff (Arancel) of good 2 in country B |
| 0 | PR3ba = Preference of good 3 in country B made in country A | 0 | PR3bc = Preference of good 3 in country B made in country C | 0 | AR3b = Tariff (Arancel) of good 3 in country B |
| 0 | PR4ba = Preference of good 4 in country B made in country A | 0 | PR4bc = Preference of good 4 in country B made in country C | 0 | AR4b = Tariff (Arancel) of good 4 in country B |

**Country C = CANADA**

| | PREFERENCES OF CONSUMERS OF COUNTRY C WITH REGARD TO GOODS OF COUNTRY A | | PREFERENCES OF CONSUMERS OF COUNTRY C WITH REGARD TO GOODS OF COUNTRY B | | TARIFF (ARANCEL) OF COUNTRY C WITH REGARI |
|---|---|---|---|---|---|
| 0 | PR1ca = Preference of good 1 in country C made in country A | 0 | PR1cb= Preference of good 1 in country C made in country B | 0 | AR1c = Tariff (Arancel) of good 1 in country C |
| 0 | PR2ca = Preference of good 2 in country C made in country A | 0 | PR2cb= Preference of good 2 in country C made in country B | 0 | AR2c = Tariff (Arancel) of good 2 in country C |
| 0 | PR3ca = Preference of good 3 in country C made in country A | 0 | PR3cb= Preference of good 3 in country C made in country B | 0 | AR3c = Tariff (Arancel) of good 3 in country C |
| 0 | PR4ca = Preference of good 4 in country C made in country A | 0 | PR4cb= Preference of good 4 in country C made in country B | 0 | AR4c = Tariff (Arancel) of good 4 in country C |

# Appendix Spreadsheet 8

## Image Section 8.1

| | | | | | | | | | | | | | | | | |
|---|---|---|---|---|---|---|---|---|---|---|---|---|---|---|---|---|
| **Spreadsheet 8 - MODEL SIMULATION INTERNATIONAL TRADE OF 3 COUNTRIES:** | | | | | | | | | | | | | | | | |
| | | | | | | | | | **SECTION 8.1: COMMERCIAL BALANCES. MODEL EQUATIONS** | | | | | | | |
| | | **Equation 8.7 Defining the Trade Balance of Country A (United States) with All Other Countries (BCa.bc)** | | | | | | | | | | | | | | |
| | | | | **BCa.bc = BCab1 + BCab2 + BCab3 + BCab4 + BCac1 + BCac2 + BCac3 + BCac4** | | | | | | | | | | | | |
| **BCa.bc** | **=** | **BCab1** | **+** | **BCab2** | **+** | **BCab3** | **+** | **BCab4** | **+** | **BCac1** | **+** | **BCac2** | **+** | **BCac3** | **+** | **BCac4** |
| **0.00** | **=** | **0.00** | **+** | **(0.00)** | **+** | **(0.00)** | **+** | **(0.00)** | **+** | **0.00** | **+** | **0.00** | **+** | **0.00** | **+** | **0.00** |
| | | **Equation 8.8 Defining the Trade Balance of Country B (Mexico) with All Other Countries (BCb.ac)** | | | | | | | | | | | | | | |
| | | | | **BCb.ac = BCba1 + BCba2 + BCba3 + BCba4 + BCbc1 + BCbc2 + BCbc3 + BCbca4** | | | | | | | | | | | | |
| **BCb.ac** | **=** | **BCba1** | **+** | **BCba2** | **+** | **BCba3** | **+** | **BCba4** | **+** | **BCbc1** | **+** | **BCbc2** | **+** | **BCbc3** | **+** | **BCbc4** |
| **0.00** | **=** | **0.00** | | **0.00** | | **0.00** | | **0.00** | | **0.00** | | **0.00** | | **0.00** | | **0.00** |
| | | **Equation 8.9 Defining the Trade Balance of Country C (Canada) with All Other Countries (BCc.ab)** | | | | | | | | | | | | | | |
| | | | | **BCc.ab = BCca1 + BCca2 + BCca3 + BCca4 + BCcb1 + BCcb2 + BCcb3 + BCcb4** | | | | | | | | | | | | |
| **BCc.ab** | **=** | **BCca1** | **+** | **BCca2** | **+** | **BCca3** | **+** | **BCca4** | **+** | **BCcb1** | **+** | **BCcb2** | **+** | **BCcb3** | **+** | **BCcb4** |
| **0.00** | **=** | **0.00** | **+** | **(0.00)** | **+** | **(0.00)** | **+** | **(0.00)** | **+** | **(0.00)** | **+** | **(0.00)** | **+** | **(0.00)** | **+** | **(0.00)** |
| | | | | | **COMMERCIAL BALANCES** | | | | | | | | | | | |
| | | **Country A - United States** | | | | | **Country B - Mexico** | | | | | **Country C - Canada** | | | | |
| | | | | **Deficit or SURPLUS US** | | | | | **Déficit o Superávit Méx** | | | | | **Déficit o Superávit Cand** | | |
| | | **EXa EU** | **IMa EU** | **US$ BCa.bc = (Exa - Ima)** | | | **EXb Mex** | **IMb Mex** | **US$ BCb.ac = (EXb - IMb)** | | | **EXc Cand** | **IMc Cand** | **US$ BCc.ab = (EXc - IMc)** | | |
| | | **0.00** | **0.00** | **(0.00)** | | | **0.00** | **-** | **0.00** | | | **-** | **0.00** | **(0.00)** | | |
| | | **EXab** | **IMab** | | | | **EXba** | **IMba** | | | | **EXca** | **IMca** | | | |
| | | **-** | **0.00** | **(0.00)** | | | **0.00** | **-** | **0.00** | | | **-** | **0.00** | **(0.00)** | | |
| | | **EXac** | **IMac** | | | | **EXbc** | **IMbc** | | | | **EXcb** | **IMcb** | | | |
| | | **0.00** | **-** | **0.00** | | | **0.00** | **-** | **0.00** | | | **-** | **0.00** | **(0.00)** | | |

# Apendix Spreadsheed 9

## Image Section 9.1

**Spreadsheet 9 - MODEL SIMULATION INTERNATIONAL TRADE OF 3 COUNTRIES:**

**SECTION 9.1: COMMERCIAL BALANCES. MODEL EQUATIONS**

**Equation 8.7 Defining the Trade Balance of Country A (United States) with All Other Countries (BCa.bc)**

**BCa.bc = BCab1 + BCab2 + BCab3 + BCab4 + BCac1 + BCac2 + BCac3 + BCac4**

| | | | | | | | | | | | | | | | | |
|---|---|---|---|---|---|---|---|---|---|---|---|---|---|---|---|---|
| BCa.bc | = | BCab1 | + | BCab2 | + | BCab3 | + | BCab4 | + | BCac1 | + | BCac2 | + | BCac3 | + | BCac4 |
| -503.44 | = | 0.00 | + | (148.23) | + | (331.79) | + | (46.78) | + | 0.00 | + | 19.48 | + | 0.00 | + | 3.89 |

**Equation 8.8 Defining the Trade Balance of Country B (Mexico) with All Other Countries (BCb.ac)**

**BCb.ac = BCba1 + BCba2 + BCba3 + BCba4 + BCbc1 + BCbc2 + BCbc3 + BCbca4**

| | | | | | | | | | | | | | | | | |
|---|---|---|---|---|---|---|---|---|---|---|---|---|---|---|---|---|
| BCb.ac | = | BCba1 | + | BCba2 | + | BCba3 | + | BCba4 | + | BCbc1 | + | BCbc2 | + | BCbc3 | + | BCbc4 |
| 525.68 | = | 0.00 | | 148.23 | | 331.79 | | 46.78 | | 8.79 | | (0.97) | | (6.99) | | (1.95) |

**Equation 8.9 Defining the Trade Balance of Country C (Canada) with All Other Countries (BCc.ab)**

**BCc.ab = BCca1 + BCca2 + BCca3 + BCca4 + BCcb1 + BCcb2 + BCcb3 + BCcb4**

| | | | | | | | | | | | | | | | | |
|---|---|---|---|---|---|---|---|---|---|---|---|---|---|---|---|---|
| BCc.ab | = | BCca1 | + | BCca2 | + | BCca3 | + | BCca4 | + | BCcb1 | + | BCcb2 | + | BCcb3 | + | BCcb4 |
| -22.24 | = | 0.00 | + | (19.48) | + | (0.00) | + | (3.89) | + | (8.79) | + | 0.97 | + | 6.99 | + | 1.95 |

**COMMERCIAL BALANCES**

| Country A - United States | | | Country B - Mexico | | | Country C - Canada | | |
|---|---|---|---|---|---|---|---|---|
| | | Deficit or SURPLUS US | | | Déficit o Superávit Méx | | | Déficit o Superávit Cand |
| EXa EU | IMa EU | US$ BCa.bc = (Exa - Ima) | EXb Mex | IMb Mex | US$ BCb.ac = (EXb - IMb) | EXc Cand | IMc Cand | US$ BCc.ab = (EXc - IMc) |
| 279.59 | 783.04 | (503.44) | 795.20 | 269.52 | 525.68 | 80.85 | 103.09 | (22.24) |
| EXab | IMab | | EXba | IMba | | EXca | IMca | |
| 222.45 | 749.25 | (526.81) | 749.25 | 222.45 | 526.81 | 33.78 | 57.15 | (23.36) |
| EXac | IMac | | EXbc | IMbc | | EXcb | IMcb | |
| 57.15 | 33.78 | 23.36 | 45.95 | 47.07 | (1.12) | 47.07 | 45.95 | 1.12 |

# Appendix Spreadsheet 9
# Image Section 9.5

**SECTION 9.5: MODEL PARAMETERS**

| PRODUCTIVITY OF WORK IN THE COUNTRY A | | PRODUCTIVITY OF WORK IN COUNTRY B | | PRODUCTIVITY OF WORK IN COUNTRY C | |
|---|---|---|---|---|---|
| 3.00 | Za1 | 3.00 | Zb1 | 3.00 | Zc1 |
| 2.00 | Za2 | 2.00 | Zb2 | 2.00 | Zc2 |
| 1.00 | Za3 | 1.00 | Zb3 | 1.00 | Zc3 |
| 0.50 | Za4 | 0.50 | Zb4 | 0.50 | Zc4 |

**MONETARY OFFER OMi, MARGIN OF GAIN MGi, POPULATION Ni, SALARY Wi, PROFITS Gij, WORK FORCE FT, CONSUMER PREFERENCES PR AND TARIFFS AR**

| | | | | | |
|---|---|---|---|---|---|
| 2,500 | OMa = Money Offer Country A US dollar | 0.3650 | MGa = Average Margin of Gain or Profit in country A | 325.00 | Na = Population of country A United States |
| 25,000 | OMb = Money Offer Country B Mexico peso | 0.5450 | MGb = Average Margin of Gain or Profit in country B | 137.00 | Nb = Population of country B México |
| 500 | OMc = Money Offer Country C Canda dollar | 0.4100 | MGc = Average Margin of Gain or Profit in country C | 36.00 | Nc = Population of country C Canada |

| | | | |
|---|---|---|---|
| 4.88 | Wa = Salary in country A for each unit of work hired = (OMa(1-MGa))/Na | 8.42 | Ga1 = Gain or Profit in country A corresponds to good 1 for each unit produced = (OMa(MGa)(FTa1))/Qa1 |
| 83.03 | Wb = Salary in country B for each unit of work hired = (OMb(1-MGb))/Nb | 5.62 | Ga2 = Gain or Profit in country A corresponds to good 2 for each unit produced = (OMa(MGa)(FTa2))/Qa2 |
| 8.19 | Wc = Salary in country C for each unit of work hired = (OMc(1-MGc))/Nc | 2.81 | Ga3 = Gain or Profit in country A corresponds to good 3 for each unit produced = (OMa(MGa)(FTa3))/Qa3 |
| | | 1.40 | Ga4 = Gain or Profit in country A corresponds to good 4 for each unit produced = (OMa(MGa)(FTa4))/Qa4 |
| | | | |
| | | 298.36 | Gb1 = Gain or Profit in country B corresponds to good 1 for each unit produced = (OMb(MGb)(FTb1))/Qb1 |
| | | 198.91 | Gb2 = Gain or Profit in country B corresponds to good 2 for each unit produced = (OMb(MGb)(FTb2))/Qb2 |
| | | 99.45 | Gb3 = Gain or Profit in country B corresponds to good 3 for each unit produced = (OMb(MGb)(FTb3))/Qb3 |
| | | 49.73 | Gb4 = Gain or Profit in country B corresponds to good 3 for each unit produced = (OMb(MGb)(FTb4))/Qb4 |
| | | | |
| | | 17.08 | Gc1 = Gain or Profit in country C corresponds to good 1 for each unit produced = (OMc(MGc)(FTc1))/Qc1 |
| | | 11.39 | Gc2 = Gain or Profit in country C corresponds to good 2 for each unit produced = (OMc(MGc)(FTc2))/Qc2 |
| | | 5.69 | Gc3 = Gain or Profit in country C corresponds to good 3 for each unit produced = (OMc(MGc)(FTc3))/Qc3 |
| | | 2.85 | Gc4 = Gain or Profit in country C corresponds to good 4 for each unit produced = (OMc(MGc)(FTc4))/Qc4 |

**DISTRIBUTION OF THE FORCE OF WORK**

| DISTRIBUTION OF THE COUNTRY A FORCE OF WORK | | DISTRIBUTION OF THE COUNTRY B FORCE OF WORK | | DISTRIBUTION OF THE COUNTRY C FORCE OF WORK | |
|---|---|---|---|---|---|
| 0.20 | FT1a = Proportion force of work producing good 1 in country A | 0.20 | FT1b = Proportion force of work producing good 1 in country B | 0.20 | FT1c = Proportion force of work producing good 1 in country C |
| 0.30 | FT2a = Proportion force of work producing good 2 in country A | 0.30 | FT2b = Proportion force of work producing good 2 in country B | 0.30 | FT2c = Proportion force of work producing good 2 in country C |
| 0.30 | FT3a = Proportion force of work producing good 3 in country A | 0.30 | FT3b = Proportion force of work producing good 3 in country B | 0.30 | FT3c = Proportion force of work producing good 3 in country C |
| 0.20 | FT4a = Proportion force of work producing good 4 in country A | 0.20 | FT4b = Proportion force of work producing good 4 in country B | 0.20 | FT4c = Proportion force of work producing good 4 in country C |

**CONSUMER PREFERENCES AND TARIFF RATES**

**Country A = UNITED STATES**

| PREFERENCES OF CONSUMERS OF COUNTRY A WITH REGARD TO GOODS OF COUNTRY B | | PREFERENCES OF CONSUMERS OF COUNTRY A WITH REGARD TO GOODS OF COUNTRY C | | TARIFF (ARANCEL) OF COUNTRY A WITH REGARI | |
|---|---|---|---|---|---|
| 0.25 | PR1ab = Preference of good 1 in country A made in country B | 0.15 | PR1ac = Preference of good 1 in country A made in country C | 0 | AR1a = Tariff (Arancel) of good 1 in country A |
| 0.50 | PR2ab = Preference of good 2 in country A made in country B | 0.05 | PR2ac = Preference of good 2 in country A made in country C | 0 | AR2a = Tariff (Arancel) of good 2 in country A |
| 0.75 | PR3ab = Preference of good 3 in country A made in country B | 0.15 | PR3ac = Preference of good 3 in country A made in country C | 0 | AR3a = Tariff (Arancel) of good 3 in country A |
| 0.38 | PR4ab = Preference of good 4 in country A made in country B | 0.12 | PR4ac = Preference of good 4 in country A made in country C | 0 | AR4a = Tariff (Arancel) of good 4 in country A |

**Country B = MEXICO**

| PREFERENCES OF CONSUMERS OF COUNTRY B WITH REGARD TO GOODS OF COUNTRY A | | PREFERENCES OF CONSUMERS OF COUNTRY B WITH REGARD TO GOODS OF COUNTRY C | | TARIFF (ARANCEL) OF COUNTRY B WITH REGARI | |
|---|---|---|---|---|---|
| 0.25 | PR1ba = Preference of good 1 in country B made in country A | 0.05 | PR1bc = Preference of good 1 in country B made in country C | 0 | AR1b = Tariff (Arancel) of good 1 in country B |
| 0.18 | PR2ba = Preference of good 2 in country B made in country A | 0.16 | PR2bc = Preference of good 2 in country B made in country C | 0 | AR2b = Tariff (Arancel) of good 2 in country B |
| 0.14 | PR3ba = Preference of good 3 in country B made in country A | 0.21 | PR3bc = Preference of good 3 in country B made in country C | 0 | AR3b = Tariff (Arancel) of good 3 in country B |
| 0.22 | PR4ba = Preference of good 4 in country B made in country A | 0.17 | PR4bc = Preference of good 4 in country B made in country C | 0 | AR4b = Tariff (Arancel) of good 4 in country B |

**Country C = CANADA**

| PREFERENCES OF CONSUMERS OF COUNTRY C WITH REGARD TO GOODS OF COUNTRY A | | PREFERENCES OF CONSUMERS OF COUNTRY C WITH REGARD TO GOODS OF COUNTRY B | | TARIFF (ARANCEL) OF COUNTRY C WITH REGARI | |
|---|---|---|---|---|---|
| 0.15 | PR1ca = Preference of good 1 in country C made in country A | 0.19 | PR1cb= Preference of good 1 in country C made in country B | 0 | AR1c = Tariff (Arancel) of good 1 in country C |
| 0.25 | PR2ca = Preference of good 2 in country C made in country A | 0.15 | PR2cb= Preference of good 2 in country C made in country B | 0 | AR2c = Tariff (Arancel) of good 2 in country C |
| 0.15 | PR3ca = Preference of good 3 in country C made in country A | 0.14 | PR3cb= Preference of good 3 in country C made in country B | 0 | AR3c = Tariff (Arancel) of good 3 in country C |
| 0.18 | PR4ca = Preference of good 4 in country C made in country A | 0.14 | PR4cb= Preference of good 4 in country C made in country B | 0 | AR4c = Tariff (Arancel) of good 4 in country C |

# Apendix Spreadsheed 10

## Image Section 10.1.

**Spreadsheet 10 - MODEL SIMULATION INTERNATIONAL TRADE OF 3 COUNTRIES:**

**SECTION 10.1: COMMERCIAL BALANCES. MODEL EQUATIONS**

**Equation 8.7 Defining the Trade Balance of Country A (United States) with All Other Countries (BCa.bc)**

BCa.bc = BCab1 + BCab2 + BCab3 + BCab4 + BCac1 + BCac2 + BCac3 + BCac4

| BCa.bc | = | BCab1 | + | BCab2 | + | BCab3 | + | BCab4 | + | BCac1 | + | BCac2 | + | BCac3 | + | BCac4 |
|---|---|---|---|---|---|---|---|---|---|---|---|---|---|---|---|---|
| 0.00 | = | 123.57 | + | (7.81) | + | (262.21) | + | 63.18 | + | 12.44 | + | 35.43 | + | 18.66 | + | 16.73 |

**Equation 8.8 Defining the Trade Balance of Country B (Mexico) with All Other Countries (BCb.ac)**

BCb.ac = BCba1 + BCba2 + BCba3 + BCba4 + BCbc1 + BCbc2 + BCbc3 + BCbca4

| BCb.ac | = | BCba1 | + | BCba2 | + | BCba3 | + | BCba4 | + | BCbc1 | + | BCbc2 | + | BCbc3 | + | BCbc4 |
|---|---|---|---|---|---|---|---|---|---|---|---|---|---|---|---|---|
| 0.00 | = | -123.57 | | 7.81 | | 262.21 | | (63.18) | | (6.74) | | (25.70) | | (32.26) | | (18.57) |

**Equation 8.9 Defining the Trade Balance of Country C (Canada) with All Other Countries (BCc.ab)**

BCc.ab = BCca1 + BCca2 + BCca3 + BCca4 + BCcb1 + BCcb2 + BCcb3 + BCcb4

| BCc.ab | = | BCca1 | + | BCca2 | + | BCca3 | + | BCca4 | + | BCcb1 | + | BCcb2 | + | BCcb3 | + | BCcb4 |
|---|---|---|---|---|---|---|---|---|---|---|---|---|---|---|---|---|
| 0.00 | = | -12.44 | + | (35.43) | + | (18.66) | + | (16.73) | + | 6.74 | + | 25.70 | + | 32.26 | + | 18.57 |

**COMMERCIAL BALANCES**

| Country A - United States | | Déficit o Superávit EU | Country B - Mexico | | Déficit o Superávit Méx | Country C - Canada | | Déficit o Superávit Cand |
|---|---|---|---|---|---|---|---|---|
| EXa EU | IMa EU | US$ BCa.bc = (Exa - Ima) | EXb Mex | IMb Mex | US$ BCb.ac = (EXb - IMb) | EXc Cand | IMc Cand | US$ BCc.ab = (EXc - IMc) |
| 683.80 | 683.80 | (0.00) | 682.59 | 682.60 | (0.00) | 103.71 | 103.71 | 0.00 |
| EXab | IMab | | EXba | IMba | | EXca | IMca | |
| 589.71 | 672.98 | (83.26) | 672.98 | 589.71 | 83.26 | 10.82 | 94.09 | (83.26) |
| EXac | IMac | | EXbc | IMbc | | EXcb | IMcb | |
| 94.09 | 10.82 | 83.26 | 9.62 | 92.89 | (83.27) | 92.89 | 9.62 | 83.27 |

# Apendix Spreadsheed 11

## Image Section 11.1

| | | | | | | | | | | | | | | | | |
|---|---|---|---|---|---|---|---|---|---|---|---|---|---|---|---|---|
| Spreadsheet 11 - MODEL SIMULATION INTERNATIONAL TRADE OF 3 COUNTRIES: | | | | | | | | | | | | | | | | |
| | | | | | | | | | SECTION 11.1: COMMERCIAL BALANCES. MODEL EQUATIONS | | | | | | | |
| | | Equation 8.7 Defining the Trade Balance of Country A (United States) with All Other Countries (BCa.bc) | | | | | | | | | | | | | | |
| | | | | BCa.bc = BCab1 + BCab2 + BCab3 + BCab4 + BCac1 + BCac2 + BCac3 + BCac4 | | | | | | | | | | | | |
| BCa.bc | = | BCab1 | + | BCab2 | + | BCab3 | + | BCab4 | + | BCac1 | + | BCac2 | + | BCac3 | + | BCac4 |
| -94.54 | = | -18.38 | + | (50.85) | + | (60.22) | + | 18.72 | + | (2.73) | + | 5.01 | + | 8.01 | + | 5.91 |
| | | Equation 8.8 Defining the Trade Balance of Country B (Mexico) with All Other Countries (BCb.ac) | | | | | | | | | | | | | | |
| | | | | BCb.ac = BCba1 + BCba2 + BCba3 + BCba4 + BCbc1 + BCbc2 + BCbc3 + BCbca4 | | | | | | | | | | | | |
| BCb.ac | = | BCba1 | + | BCba2 | + | BCba3 | + | BCba4 | + | BCbc1 | + | BCbc2 | + | BCbc3 | + | BCbc4 |
| 104.96 | = | 18.38 | | 50.85 | | 60.22 | | (18.72) | | 4.38 | | 0.20 | | (6.68) | | (3.66) |
| | | Equation 8.9 Defining the Trade Balance of Country C (Canada) with All Other Countries (BCc.ab) | | | | | | | | | | | | | | |
| | | | | BCc.ab = BCca1 + BCca2 + BCca3 + BCca4 + BCcb1 + BCcb2 + BCcb3 + BCcb4 | | | | | | | | | | | | |
| BCc.ab | = | BCca1 | + | BCca2 | + | BCca3 | + | BCca4 | + | BCcb1 | + | BCcb2 | + | BCcb3 | + | BCcb4 |
| -10.42 | = | 2.73 | + | (5.01) | + | (8.01) | + | (5.91) | + | (4.38) | + | (0.20) | + | 6.68 | + | 3.66 |
| | | | | | COMMERCIAL BALANCES | | | | | | | | | | | |
| | | Country A - United States | | | | | Country B - Mexico | | | | | Country C - Canada | | | | |
| | | | | Déficit o Superávit EU | | | | | Déficit o Superávit Méx | | | | | Déficit o Superávit Cand | | |
| | | EXa EU | IMa EU | US$ BCa.bc = (Exa - Ima) | | | EXb Mex | IMb Mex | US$ BCb.ac = (EXb - IMb) | | | EXc Cand | IMc Cand | US$ BCc.ab = (EXc - IMc) | | |
| | | 121.26 | 215.80 | (94.54) | | | 221.02 | 116.06 | 104.96 | | | 22.27 | 32.69 | (10.42) | | |
| | | EXab | IMab | | | | EXba | IMba | | | | EXca | IMca | | | |
| | | 99.43 | 210.16 | (110.73) | | | 210.16 | 99.43 | 110.73 | | | 5.64 | 21.83 | (16.18) | | |
| | | EXac | IMac | | | | EXbc | IMbc | | | | EXcb | IMcb | | | |
| | | 21.83 | 5.64 | 16.18 | | | 10.86 | 16.63 | (5.77) | | | 16.63 | 10.86 | 5.77 | | |

# Appendix Spreadsheet 11
# Sección 11.5

## SECTION 11.5: MODEL PARAMETERS

| | | | | | |
|---|---|---|---|---|---|
| | PRODUCTIVITY OF WORK IN THE COUNTRY A | | PRODUCTIVITY OF WORK IN COUNTRY B | | PRODUCTIVITY OF WORK IN COUNTRY C |
| 1.042169 | Za1 = ka1(1 - (Na1/(Na1 + Nb1 + Nc1))) | 2.8996 | Zb1 = kb1(1 - (Nb1/(Na1 + Nb1 + Nc1))) | 3.2470 | Zc1 = kc1(1 - (Nc1/(Na1 + Nb1 + Nc1))) |
| 0.694779 | Za2 = ka2(1 - (Na2/(Na2 + Nb2 + Nc2))) | 2.1747 | Zb2 = kb2(1 - (Nb2/(Na2 + Nb2 + Nc2))) | 2.3193 | Zc2 = kc2(1 - (Nc2/(Na2 + Nb2 + Nc2))) |
| 0.34739 | Za3= ka3(1 - (Na3/(Na3 + Nb3 + Nc3))) | 1.4498 | Zb3= kb3(1 - (Nb3/(Na3 + Nb3 + Nc3))) | 1.3916 | Zc3= kc3(1 - (Nc3/(Na3 + Nb3 + Nc3))) |
| 0.173695 | Za4 = ka4(1 - (Na4/(Na4 + Nb4 + Nc4))) | 0.7249 | Zb4 = kb4(1 - (Nb4/(Na4 + Nb4 + Nc4))) | 0.6958 | Zc4 = kc4(1 - (Nc4/(Na4 + Nb4 + Nc4))) |

## MONETARY OFFER OMi, MARGIN OF GAIN MGi, POPULATION Ni, SALARY Wi, PROFITS Gij, WORK FORCE FT, CONSUMER PREFERENCES PR AND TARIFFS AR

| | | | | | |
|---|---|---|---|---|---|
| 2,500 | OMa = Money Offer Country A US dollar | 0.365 | MGa = Average Margin of Gain or Profit in country A | 325.00 | Na = Population of the country A United States |
| 25,000 | OMb = Money Offer Country B Mexico | 0.545 | MGb = Average Margin of Gain or Profit in country B | 137.00 | Nb = Population of the country B México |
| 500 | OMc = Money Offer Country C Canda dollar | 0.410 | MGc = Average Margin of Gain or Profit in country C | 36.00 | Nc = Population of the country C Canadá |

| | |
|---|---|
| 4.88 | Wa = Salary in country A for each unit of work hired = (OMa(1-MGa))/Na |
| 83.03 | Wb = Salary in country B for each unit of work hired = (OMb(1-MGb))/Nb |
| 8.19 | Wc = Salary in country C for each unit of work hired = (OMc(1-MGc))/Nc |
| 2.93 | Ga1 = Gain or Profit in country A corresponds to good 1 for each unit produced = (OMa(MGa)(FTa1))/Qa1 |
| 1.95 | Ga2 = Gain or Profit in country A corresponds to good 2 for each unit produced = (OMa(MGa)(FTa2))/Qa2 |
| 0.98 | Ga3 = Gain or Profit in country A corresponds to good 3 for each unit produced = (OMa(MGa)(FTa3))/Qa3 |
| 0.49 | Ga4 = Gain or Profit in country A corresponds to good 4 for each unit produced = (OMa(MGa)(FTa4))/Qa4 |
| | |
| 288.37 | Gb1 = Gain or Profit in country B corresponds to good 1 for each unit produced = (OMb(MGb)(FTb1))/Qb1 |
| 216.28 | Gb2 = Gain or Profit in country B corresponds to good 2 for each unit produced = (OMb(MGb)(FTb2))/Qb2 |
| 144.19 | Gb3 = Gain or Profit in country B corresponds to good 3 for each unit produced = (OMb(MGb)(FTb3))/Qb3 |
| 72.09 | Gb4 = Gain or Profit in country B corresponds to good 3 for each unit produced = (OMb(MGb)(FTb4))/Qb4 |
| | |
| 18.49 | Gc1 = Gain or Profit in country C corresponds to good 1 for each unit produced = (OMc(MGc)(FTc1))/Qc1 |
| 13.21 | Gc2 = Gain or Profit in country C corresponds to good 2 for each unit produced = (OMc(MGc)(FTc2))/Qc2 |
| 7.92 | Gc3 = Gain or Profit in country C corresponds to good 3 for each unit produced = (OMc(MGc)(FTc3))/Qc3 |
| 3.96 | Gc4 = Gain or Profit in country C corresponds to good 4 for each unit produced = (OMc(MGc)(FTc4))/Qc4 |

### DISTRIBUTION OF THE FORCE OF WORK

| | | | | | |
|---|---|---|---|---|---|
| | DISTRIBUTION OF THE COUNTRY A FORCE OF WORK | | DISTRIBUTION OF THE COUNTRY B FORCE OF WORK | | DISTRIBUTION OF THE COUNTRY C FORCE OF WORK |
| 0.20 | FT1a = Proportion force of work producing good 1 in country A | 0.20 | FT1b = Proportion force of work producing good 1 in country B | 0.20 | FT1c = Proportion force of work producing good 1 in country C |
| 0.30 | FT2a = Proportion force of work producing good 2 in country A | 0.30 | FT2b = Proportion force of work producing good 2 in country B | 0.30 | FT2c = Proportion force of work producing good 2 in country C |
| 0.30 | FT3a = Proportion force of work producing good 3 in country A | 0.30 | FT3b = Proportion force of work producing good 3 in country B | 0.30 | FT3c = Proportion force of work producing good 3 in country C |
| 0.20 | FT4a = Proportion force of work producing good 4 in country A | 0.20 | FT4b = Proportion force of work producing good 4 in country B | 0.20 | FT4c = Proportion force of work producing good 4 in country C |

### DISTRIBUTION OF THE FORCE OF WORK

| | | | | | |
|---|---|---|---|---|---|
| | DISTRIBUTION OF THE COUNTRY A FORCE OF WORK | | DISTRIBUTION OF THE COUNTRY B FORCE OF WORK | | DISTRIBUTION OF THE COUNTRY C FORCE OF WORK |
| 65.00 | Na1 = FT1a(Na) = Number of workers producing good 1 in country A | 27.4 | Nb1 = FT1b(Nb) = Number of workers producing good 1 in country B | 7.2 | Nc1 = FT1c(Nc) = Number of workers producing good 1 in country C |
| 97.50 | Na2 = FT2a(Na) = Number of workers producing good 2 in country A | 41.1 | Nb2 = FT2b(Nb) = Number of workers producing good 2 in country B | 10.8 | Nc2 = FT2c(Nc) = Number of workers producing good 2 in country C |
| 97.50 | Na3 = FT3a(Na) = Number of workers producing good 3 in country A | 41.1 | Nb3 = FT3b(Nb) = Number of workers producing good 3 in country B | 10.8 | Nc3 = FT3c(Nc) = Number of workers producing good 3 in country C |
| 65.00 | Na4 = FT4a(Na) = Number of workers producing good 4 in country A | 27.4 | Nb4 = FT4b(Nb) = Number of workers producing good 4 in country B | 7.2 | Nc4 = FT4c(Nc) = Number of workers producing good 4 in country C |

### ARBITRARY CONSTANTS OF PRODUCTIVITY OF WORK

| | | | | | |
|---|---|---|---|---|---|
| | CONSTANT PRODUCTIVITY BY INDUSTRIAL SECTOR IN THE COUNTRY A | | CONSTANT PRODUCTIVITY BY INDUSTRIAL SECTOR IN THE COUNTRY B | | CONSTANT PRODUCTIVITY BY INDUSTRIAL SECTOR IN THE COUNTRY C |
| 3.00 | ka1 = Arbitrary constant of labor productivity in good 1 of country A | 4.00 | kb1 = Arbitrary constant of labor productivity in good 1 of country B | 3.50 | kc1 = Arbitrary constant of labor productivity in good 1 of country C |
| 2.00 | ka2 = Arbitrary constant of labor productivity in good 2 of country A | 3.00 | kb2 = Arbitrary constant of labor productivity in good 2 of country B | 2.50 | kc2 = Arbitrary constant of labor productivity in good 2 of country C |
| 1.00 | ka3 = Arbitrary constant of labor productivity in good 3 of country A | 2.00 | kb3 = Arbitrary constant of labor productivity in good 3 of country B | 1.50 | kc3 = Arbitrary constant of labor productivity in good 3 of country C |
| 0.50 | ka4 = Arbitrary constant of labor productivity in good 4 of country A | 1.00 | kb4 = Arbitrary constant of labor productivity in good 4 of country B | 0.75 | kc4 = Arbitrary constant of labor productivity in good 4 of country C |

### CONSUMER PREFERENCES AND TARIFF RATES

#### Country A = UNITED STETES

| | | | | | |
|---|---|---|---|---|---|
| | PREFERENCES OF CONSUMERS OF COUNTRY A WITH REGARD TO GOODS OF COUNTRY B | | PREFERENCES OF CONSUMERS OF COUNTRY A WITH REGARD TO GOODS OF COUNTRY C | | TARIFF (ARANCEL) OF COUNTRY A WITH REGARD TO GOODS |
| 0.25 | PR1ab = Preference of good 1 in country A made in country B | 0.15 | PR1ac = Preference of good 1 in country A made in country C | 0 | AR1a = Tariff (Arancel) of good 1 in country A |
| 0.50 | PR2ab = Preference of good 2 in country A made in country B | 0.05 | PR2ac = Preference of good 2 in country A made in country C | 0 | AR2a = Tariff (Arancel) of good 2 in country A |
| 0.75 | PR3ab = Preference of good 3 in country A made in country B | 0.15 | PR3ac = Preference of good 3 in country A made in country C | 0 | AR3a = Tariff (Arancel) of good 3 in country A |
| 0.38 | PR4ab = Preference of good 4 in country A made in country B | 0.12 | PR4ac = Preference of good 4 in country A made in country C | 0 | AR4a = Tariff (Arancel) of good 4 in country A |

#### Country B = MEXICO

| | | | | | |
|---|---|---|---|---|---|
| | PREFERENCES OF CONSUMERS OF COUNTRY B WITH REGARD TO GOODS OF COUNTRY A | | PREFERENCES OF CONSUMERS OF COUNTRY B WITH REGARD TO GOODS OF COUNTRY C | | TARIFF (ARANCEL) OF COUNTRY B WITH REGARD TO GOODS |
| 0.25 | PR1ba = Preference of good 1 in country B made in country A | 0.05 | PR1bc = Preference of good 1 in country B made in country C | 0 | AR1b = Tariff (Arancel) of good 1 in country B |
| 0.18 | PR2ba = Preference of good 2 in country B made in country A | 0.16 | PR2bc = Preference of good 2 in country B made in country C | 0 | AR2b = Tariff (Arancel) of good 2 in country B |
| 0.14 | PR3ba = Preference of good 3 in country B made in country A | 0.21 | PR3bc = Preference of good 3 in country B made in country C | 0 | AR3b = Tariff (Arancel) of good 3 in country B |
| 0.22 | PR4ba = Preference of good 4 in country B made in country A | 0.17 | PR4bc = Preference of good 4 in country B made in country C | 0 | AR4b = Tariff (Arancel) of good 4 in country B |

#### Country C = CANADA

| | | | | | |
|---|---|---|---|---|---|
| | PREFERENCES OF CONSUMERS OF COUNTRY C WITH REGARD TO GOODS OF COUNTRY A | | PREFERENCES OF CONSUMERS OF COUNTRY C WITH REGARD TO GOODS OF COUNTRY B | | TARIFF (ARANCEL) OF COUNTRY C WITH REGARD TO GOODS |
| 0.15 | PR1ca = Preference of good 1 in country C made in country A | 0.19 | PR1cb= Preference of good 1 in country C made in country B | 0 | AR1c = Tariff (Arancel) of good 1 in country C |
| 0.25 | PR2ca = Preference of good 2 in country C made in country A | 0.15 | PR2cb= Preference of good 2 in country C made in country B | 0 | AR2c = Tariff (Arancel) of good 2 in country C |
| 0.15 | PR3ca = Preference of good 3 in country C made in country A | 0.14 | PR3cb= Preference of good 3 in country C made in country B | 0 | AR3c = Tariff (Arancel) of good 3 in country C |
| 0.18 | PR4ca = Preference of good 4 in country C made in country A | 0.14 | PR4cb= Preference of good 4 in country C made in country B | 0 | AR4c = Tariff (Arancel) of good 4 in country C |

# Apendix Spreadsheet 12

## Image Section 12.1

| | | | | | | | | | | | | | | | | |
|---|---|---|---|---|---|---|---|---|---|---|---|---|---|---|---|---|
| **Spreadsheet 12 - MODEL SIMULATION INTERNATIONAL TRADE OF 3 COUNTRIES:** | | | | | | | | | | | | | | | | |
| | | | | | | | | | **SECTION 12.1: COMMERCIAL BALANCES. MODEL EQUATIONS** | | | | | | | |
| | | **Equation 8.7 Defining the Trade Balance of Country A (United States) with All Other Countries (BCa.bc)** | | | | | | | | | | | | | | |
| | | | | **BCa.bc = BCab1 + BCab2 + BCab3 + BCab4 + BCac1 + BCac2 + BCac3 + BCac4** | | | | | | | | | | | | |
| **BCa.bc** | **=** | **BCab1** | **+** | **BCab2** | **+** | **BCab3** | **+** | **BCab4** | **+** | **BCac1** | **+** | **BCac2** | **+** | **BCac3** | **+** | **BCac4** |
| **0.00** | **=** | **1.16** | **+** | **(26.53)** | **+** | **(40.47)** | **+** | **43.80** | **+** | **(1.46)** | **+** | **7.30** | **+** | **9.36** | **+** | **6.83** |
| | | **Equation 8.8 Defining the Trade Balance of Country B (Mexico) with All Other Countries (BCb.ac)** | | | | | | | | | | | | | | |
| | | | | **BCb.ac = BCba1 + BCba2 + BCba3 + BCba4 + BCbc1 + BCbc2 + BCbc3 + BCbca4** | | | | | | | | | | | | |
| **BCb.ac** | **=** | **BCba1** | **+** | **BCba2** | **+** | **BCba3** | **+** | **BCba4** | **+** | **BCbc1** | **+** | **BCbc2** | **+** | **BCbc3** | **+** | **BCbc4** |
| **0.00** | **=** | **-1.16** | | **26.53** | | **40.47** | | **(43.80)** | | **0.98** | | **(5.35)** | | **(11.08)** | | **(6.58)** |
| | | **Equation 8.9 Defining the Trade Balance of Country C (Canada) with All Other Countries (BCc.ab)** | | | | | | | | | | | | | | |
| | | | | **BCc.ab = BCca1 + BCca2 + BCca3 + BCca4 + BCcb1 + BCcb2 + BCcb3 + BCcb4** | | | | | | | | | | | | |
| **BCc.ab** | **=** | **BCca1** | **+** | **BCca2** | **+** | **BCca3** | **+** | **BCca4** | **+** | **BCcb1** | **+** | **BCcb2** | **+** | **BCcb3** | **+** | **BCcb4** |
| **0.00** | **=** | **1.46** | **+** | **(7.30)** | **+** | **(9.36)** | **+** | **(6.83)** | **+** | **(0.98)** | **+** | **5.35** | **+** | **11.08** | **+** | **6.58** |
| | | | | | **COMMERCIAL BALANCES** | | | | | | | | | | | |
| | | **Country A - United States** | | | | | **Country B - Mexico** | | | | | **Country C - Canada** | | | | |
| | | | | **Déficit o Superávit EU** | | | | | **Déficit o Superávit Méx** | | | | | **Déficit o Superávit Cand** | | |
| | | **EXa EU** | **IMa EU** | **US$ BCa.bc = (Exa - Ima)** | | | **EXb Mex** | **IMb Mex** | **US$ BCb.ac = (EXb - IMb)** | | | **EXc Cand** | **IMc Cand** | **US$ BCc.ab = (EXc - IMc)** | | |
| | | **197.02** | **197.02** | **0.00** | | | **197.21** | **197.21** | **0.00** | | | **30.49** | **30.49** | **(0.00)** | | |
| | | **EXab** | **IMab** | | | | **EXba** | **IMba** | | | | **EXca** | **IMca** | | | |
| | | **170.86** | **192.89** | **(22.03)** | | | **192.89** | **170.86** | **22.03** | | | **4.14** | **26.17** | **(22.03)** | | |
| | | **EXac** | **IMac** | | | | **EXbc** | **IMbc** | | | | **EXcb** | **IMcb** | | | |
| | | **26.17** | **4.14** | **22.03** | | | **4.32** | **26.35** | **(22.03)** | | | **26.35** | **4.32** | **22.03** | | |

# Appendix Spreadsheet 12

## Image Section 12.5

**SECTION 12.5: MODEL PARAMETERS**

| | | | | | |
|---|---|---|---|---|---|
| | PRODUCTIVITY OF WORK IN THE COUNTRY A | | PRODUCTIVITY OF WORK IN COUNTRY B | | PRODUCTIVITY OF WORK IN COUNTRY C |
| 1.042169 | Za1 = ka1(1 - (Na1/(Na1 + Nb1 + Nc1))) | 2.8996 | Zb1 = kb1(1 - (Nb1/(Na1 + Nb1 + Nc1))) | 3.2470 | Zc1 = kc1(1 - (Nc1/(Na1 + Nb1 + Nc1))) |
| 0.694779 | Za2 = ka2(1 - (Na2/(Na2 + Nb2 + Nc2))) | 2.1747 | Zb2 = kb2(1 - (Nb2/(Na2 + Nb2 + Nc2))) | 2.3193 | Zc2 = kc2(1 - (Nc2/(Na2 + Nb2 + Nc2))) |
| 0.34739 | Za3= ka3(1 - (Na3/(Na3 + Nb3 + Nc3))) | 1.4498 | Zb3= kb3(1 - (Nb3/(Na3 + Nb3 + Nc3))) | 1.3916 | Zc3= kc3(1 - (Nc3/(Na3 + Nb3 + Nc3))) |
| 0.173695 | Za4 = ka4(1 - (Na4/(Na4 + Nb4 + Nc4))) | 0.7249 | Zb4 = kb4(1 - (Nb4/(Na4 + Nb4 + Nc4))) | 0.6958 | Zc4 = kc4(1 - (Nc4/(Na4 + Nb4 + Nc4))) |

**MONETARY OFFER OMi, MARGIN OF GAIN MGi, POPULATION Ni, SALARY Wi, PROFITS Gij, WORK FORCE FT, CONSUMER PREFERENCES PR AND TARIFFS AR**

| | | | | | |
|---|---|---|---|---|---|
| 2,500 | OMa = Money Offer Country A US dollar | 0.365 | MGa = Average Margin of Gain or Profit in country A | 325.00 | Na = Population of the country A United States |
| 25,000 | OMb = Money Offer Country B Mexico | 0.545 | MGb = Average Margin of Gain or Profit in country B | 137.00 | Nb = Population of the country B México |
| 500 | OMc = Money Offer Country C Canda dollar | 0.410 | MGc = Average Margin of Gain or Profit in country C | 36.00 | Nc = Population of the country C Canadá |

| | | | |
|---|---|---|---|
| 4.88 | Wa = Salary in country A for each unit of work hired = (OMa(1-MGa))/Na | 2.93 | Ga1 = Gain or Profit in country A corresponds to good 1 for each unit produced = (OMa(MGa)(FTa1))/Qa1 |
| 83.03 | Wb = Salary in country B for each unit of work hired = (OMb(1-MGb))/Nb | 1.95 | Ga2 = Gain or Profit in country A corresponds to good 2 for each unit produced = (OMa(MGa)(FTa2))/Qa2 |
| 8.19 | Wc = Salary in country C for each unit of work hired = (OMc(1-MGc))/Nc | 0.98 | Ga3 = Gain or Profit in country A corresponds to good 3 for each unit produced = (OMa(MGa)(FTa3))/Qa3 |
| | | 0.49 | Ga4 = Gain or Profit in country A corresponds to good 4 for each unit produced = (OMa(MGa)(FTa4))/Qa4 |
| | | 288.37 | Gb1 = Gain or Profit in country B corresponds to good 1 for each unit produced = (OMb(MGb)(FTb1))/Qb1 |
| | | 216.28 | Gb2 = Gain or Profit in country B corresponds to good 2 for each unit produced = (OMb(MGb)(FTb2))/Qb2 |
| | | 144.19 | Gb3 = Gain or Profit in country B corresponds to good 3 for each unit produced = (OMb(MGb)(FTb3))/Qb3 |
| | | 72.09 | Gb4 = Gain or Profit in country B corresponds to good 3 for each unit produced = (OMb(MGb)(FTb4))/Qb4 |
| | | 18.49 | Gc1 = Gain or Profit in country C corresponds to good 1 for each unit produced = (OMc(MGc)(FTc1))/Qc1 |
| | | 13.21 | Gc2 = Gain or Profit in country C corresponds to good 2 for each unit produced = (OMc(MGc)(FTc2))/Qc2 |
| | | 7.92 | Gc3 = Gain or Profit in country C corresponds to good 3 for each unit produced = (OMc(MGc)(FTc3))/Qc3 |
| | | 3.96 | Gc4 = Gain or Profit in country C corresponds to good 4 for each unit produced = (OMc(MGc)(FTc4))/Qc4 |

**DISTRIBUTION OF THE FORCE OF WORK**

| | | | | | |
|---|---|---|---|---|---|
| | DISTRIBUTION OF THE COUNTRY A FORCE OF WORK | | DISTRIBUTION OF THE COUNTRY B FORCE OF WORK | | DISTRIBUTION OF THE COUNTRY C FORCE OF WORK |
| 0.20 | FT1a = Proportion force of work producing good 1 in country A | 0.20 | FT1b = Proportion force of work producing good 1 in country B | 0.20 | FT1c = Proportion force of work producing good 1 in country C |
| 0.30 | FT2a = Proportion force of work producing good 2 in country A | 0.30 | FT2b = Proportion force of work producing good 2 in country B | 0.30 | FT2c = Proportion force of work producing good 2 in country C |
| 0.30 | FT3a = Proportion force of work producing good 3 in country A | 0.30 | FT3b = Proportion force of work producing good 3 in country B | 0.30 | FT3c = Proportion force of work producing good 3 in country C |
| 0.20 | FT4a = Proportion force of work producing good 4 in country A | 0.20 | FT4b = Proportion force of work producing good 4 in country B | 0.20 | FT4c = Proportion force of work producing good 4 in country C |

**DISTRIBUTION OF THE FORCE OF WORK**

| | | | | | |
|---|---|---|---|---|---|
| | DISTRIBUTION OF THE COUNTRY A FORCE OF WORK | | DISTRIBUTION OF THE COUNTRY B FORCE OF WORK | | DISTRIBUTION OF THE COUNTRY C FORCE OF WORK |
| 65.00 | Na1 = FT1a(Na) = Number of workers producing good 1 in country A | 27.4 | Nb1 = FT1b(Nb) = Number of workers producing good 1 in country B | 7.2 | Nc1 = FT1c(Nc) = Number of workers producing good 1 in co |
| 97.50 | Na2 = FT2a(Na) = Number of workers producing good 2 in country A | 41.1 | Nb2 = FT2b(Nb) = Number of workers producing good 2 in country B | 10.8 | Nc2 = FT2c(Nc) = Number of workers producing good 2 in co |
| 97.50 | Na3 = FT3a(Na) = Number of workers producing good 3 in country A | 41.1 | Nb3 = FT3b(Nb) = Number of workers producing good 3 in country B | 10.8 | Nc3 = FT3c(Nc) = Number of workers producing good 3 in co |
| 65.00 | Na4 = FT4a(Na) = Number of workers producing good 4 in country A | 27.4 | Nb4 = FT4b(Nb) = Number of workers producing good 4 in country B | 7.2 | Nc4 = FT4c(Nc) = Number of workers producing good 4 in co |

**ARBITRARY CONSTANTS OF PRODUCTIVITY OF WORK**

| | | | | | |
|---|---|---|---|---|---|
| | CONSTANT PRODUCTIVITY BY INDUSTRIAL SECTOR IN THE COUNTRY A | | CONSTANT PRODUCTIVITY BY INDUSTRIAL SECTOR IN THE COUNTRY B | | CONSTANT PRODUCTIVITY BY INDUSTRIAL SECTOR IN THE C |
| 3.00 | ka1 = Arbitrary constant of labor productivity in good 1 of country A | 4.00 | kb1 = Arbitrary constant of labor productivity in good 1 of country B | 3.50 | kc1 = Arbitrary constant of labor productivity in good 1 of c |
| 2.00 | ka2 = Arbitrary constant of labor productivity in good 2 of country A | 3.00 | kb2 = Arbitrary constant of labor productivity in good 2 of country B | 2.50 | kc2 = Arbitrary constant of labor productivity in good 2 of c |
| 1.00 | ka3 = Arbitrary constant of labor productivity in good 3 of country A | 2.00 | kb3 = Arbitrary constant of labor productivity in good 3 of country B | 1.50 | kc3 = Arbitrary constant of labor productivity in good 3 of c |
| 0.50 | ka4 = Arbitrary constant of labor productivity in good 4 of country A | 1.00 | kb4 = Arbitrary constant of labor productivity in good 4 of country B | 0.75 | kc4 = Arbitrary constant of labor productivity in good 4 of c |

**CONSUMER PREFERENCES AND TARIFF RATES**

**Country A = UNITED STETES**

| | | | | | |
|---|---|---|---|---|---|
| | PREFERENCES OF CONSUMERS OF COUNTRY A WITH REGARD TO GOODS OF COUNTRY B | | PREFERENCES OF CONSUMERS OF COUNTRY A WITH REGARD TO GOODS OF COUNTRY C | | TARIFF (ARANCEL) OF COUNTRY A WITH REGARI |
| 0.25 | PR1ab = Preference of good 1 in country A made in country B | 0.15 | PR1ac = Preference of good 1 in country A made in country C | 0 | AR1a = Tariff (Arancel) of good 1 in country A |
| 0.50 | PR2ab = Preference of good 2 in country A made in country B | 0.05 | PR2ac = Preference of good 2 in country A made in country C | 0 | AR2a = Tariff (Arancel) of good 2 in country A |
| 0.75 | PR3ab = Preference of good 3 in country A made in country B | 0.15 | PR3ac = Preference of good 3 in country A made in country C | 0 | AR3a = Tariff (Arancel) of good 3 in country A |
| 0.38 | PR4ab = Preference of good 4 in country A made in country B | 0.12 | PR4ac = Preference of good 4 in country A made in country C | 0 | AR4a = Tariff (Arancel) of good 4 in country A |

**Country B = MEXICO**

| | | | | | |
|---|---|---|---|---|---|
| | PREFERENCES OF CONSUMERS OF COUNTRY B WITH REGARD TO GOODS OF COUNTRY A | | PREFERENCES OF CONSUMERS OF COUNTRY B WITH REGARD TO GOODS OF COUNTRY C | | TARIFF (ARANCEL) OF COUNTRY B WITH REGARI |
| 0.25 | PR1ba = Preference of good 1 in country B made in country A | 0.05 | PR1bc = Preference of good 1 in country B made in country C | 0 | AR1b = Tariff (Arancel) of good 1 in country B |
| 0.18 | PR2ba = Preference of good 2 in country B made in country A | 0.16 | PR2bc = Preference of good 2 in country B made in country C | 0 | AR2b = Tariff (Arancel) of good 2 in country B |
| 0.14 | PR3ba = Preference of good 3 in country B made in country A | 0.21 | PR3bc = Preference of good 3 in country B made in country C | 0 | AR3b = Tariff (Arancel) of good 3 in country B |
| 0.22 | PR4ba = Preference of good 4 in country B made in country A | 0.17 | PR4bc = Preference of good 4 in country B made in country C | 0 | AR4b = Tariff (Arancel) of good 4 in country B |

**Country C = CANADA**

| | | | | | |
|---|---|---|---|---|---|
| | PREFERENCES OF CONSUMERS OF COUNTRY C WITH REGARD TO GOODS OF COUNTRY A | | PREFERENCES OF CONSUMERS OF COUNTRY C WITH REGARD TO GOODS OF COUNTRY B | | TARIFF (ARANCEL) OF COUNTRY C WITH REGARI |
| 0.15 | PR1ca = Preference of good 1 in country C made in country A | 0.19 | PR1cb= Preference of good 1 in country C made in country B | 0 | AR1c = Tariff (Arancel) of good 1 in country C |
| 0.25 | PR2ca = Preference of good 2 in country C made in country A | 0.15 | PR2cb= Preference of good 2 in country C made in country B | 0 | AR2c = Tariff (Arancel) of good 2 in country C |
| 0.15 | PR3ca = Preference of good 3 in country C made in country A | 0.14 | PR3cb= Preference of good 3 in country C made in country B | 0 | AR3c = Tariff (Arancel) of good 3 in country C |
| 0.18 | PR4ca = Preference of good 4 in country C made in country A | 0.14 | PR4cb= Preference of good 4 in country C made in country B | 0 | AR4c = Tariff (Arancel) of good 4 in country C |

# Apendix Spreadsheed 13

## Image Section 13.1

| | | | | | | | | | | | | | | | | |
|---|---|---|---|---|---|---|---|---|---|---|---|---|---|---|---|---|
| Spreadsheet 13 - MODEL SIMULATION INTERNATIONAL TRADE OF 3 COUNTRIES: | | | | | | | | | | | | | | | | |
| | | | | | | | | | SECTION 13.1: COMMERCIAL BALANCES. MODEL EQUATIONS | | | | | | | |
| | | Equation 8.7 Defining the Trade Balance of Country A (United States) with All Other Countries (BCa.bc) | | | | | | | | | | | | | | |
| | | | | BCa.bc = BCab1 + BCab2 + BCab3 + BCab4 + BCac1 + BCac2 + BCac3 + BCac4 | | | | | | | | | | | | |
| BCa.bc | = | BCab1 | + | BCab2 | + | BCab3 | + | BCab4 | + | BCac1 | + | BCac2 | + | BCac3 | + | BCac4 |
| -127.80 | = | -41.48 | + | (31.62) | + | (42.78) | + | 26.13 | + | (45.81) | + | 1.86 | + | 2.08 | + | 3.81 |
| | | Equation 8.8 Defining the Trade Balance of Country B (Mexico) with All Other Countries (BCb.ac) | | | | | | | | | | | | | | |
| | | | | BCb.ac = BCba1 + BCba2 + BCba3 + BCba4 + BCbc1 + BCbc2 + BCbc3 + BCbca4 | | | | | | | | | | | | |
| BCb.ac | = | BCba1 | + | BCba2 | + | BCba3 | + | BCba4 | + | BCbc1 | + | BCbc2 | + | BCbc3 | + | BCbc4 |
| 73.05 | = | 41.48 | | 31.62 | | 42.78 | | (26.13) | | (16.01) | | 0.74 | | (0.41) | | (1.00) |
| | | Equation 8.9 Defining the Trade Balance of Country C (Canada) with All Other Countries (BCc.ab) | | | | | | | | | | | | | | |
| | | | | BCc.ab = BCca1 + BCca2 + BCca3 + BCca4 + BCcb1 + BCcb2 + BCcb3 + BCcb4 | | | | | | | | | | | | |
| BCc.ab | = | BCca1 | + | BCca2 | + | BCca3 | + | BCca4 | + | BCcb1 | + | BCcb2 | + | BCcb3 | + | BCcb4 |
| 54.75 | = | 45.81 | + | (1.86) | + | (2.08) | + | (3.81) | + | 16.01 | + | (0.74) | + | 0.41 | + | 1.00 |
| | | | | | COMMERCIAL BALANCES | | | | | | | | | | | |
| | | Country A - United States | | | | | Country B - Mexico | | | | | Country C - Canada | | | | |
| | | | | Déficit o Superávit EU | | | | | Déficit o Superávit Méx | | | | | Déficit o Superávit Cand | | |
| | | EXa EU | IMa EU | US$ BCa.bc = (Exa - Ima) | | | EXb Mex | IMb Mex | US$ BCb.ac = (EXb - IMb) | | | EXc Cand | IMc Cand | US$ BCc.ab = (EXc - IMc) | | |
| | | 121.05 | 248.85 | (127.80) | | | 206.06 | 133.01 | 73.05 | | | 65.52 | 10.76 | 54.75 | | |
| | | EXab | IMab | | | | EXba | IMba | | | | EXca | IMca | | | |
| | | 113.30 | 203.05 | (89.75) | | | 203.05 | 113.30 | 89.75 | | | 45.81 | 7.75 | 38.06 | | |
| | | EXac | IMac | | | | EXbc | IMbc | | | | EXcb | IMcb | | | |
| | | 7.75 | 45.81 | (38.06) | | | 3.02 | 19.71 | (16.69) | | | 19.71 | 3.02 | 16.69 | | |

# Apendix Spreadsheed 14

## Image Section 14.1

| | | | | | | | | | | | | | | | | |
|---|---|---|---|---|---|---|---|---|---|---|---|---|---|---|---|---|
| **Spreadsheet 14 - MODEL SIMULATION INTERNATIONAL TRADE OF 3 COUNTRIES:** | | | | | | | | | | | | | | | | |
| | | | | | | | | | **SECTION 14.1: COMMERCIAL BALANCES. MODEL EQUATIONS** | | | | | | | |
| | | **Equation 8.7 Defining the Trade Balance of Country A (United States) with All Other Countries (BCa.bc)** | | | | | | | | | | | | | | |
| | | | | **BCa.bc = BCab1 + BCab2 + BCab3 + BCab4 + BCac1 + BCac2 + BCac3 + BCac4** | | | | | | | | | | | | |
| **BCa.bc** | **=** | **BCab1** | **+** | **BCab2** | **+** | **BCab3** | **+** | **BCab4** | **+** | **BCac1** | **+** | **BCac2** | **+** | **BCac3** | **+** | **BCac4** |
| **0.00** | **=** | **-28.59** | **+** | **2.55** | **+** | **(14.57)** | **+** | **56.34** | **+** | **(34.40)** | **+** | **4.57** | **+** | **4.85** | **+** | **9.26** |
| | | **Equation 8.8 Defining the Trade Balance of Country B (Mexico) with All Other Countries (BCb.ac)** | | | | | | | | | | | | | | |
| | | | | **BCb.ac = BCba1 + BCba2 + BCba3 + BCba4 + BCbc1 + BCbc2 + BCbc3 + BCbca4** | | | | | | | | | | | | |
| **BCb.ac** | **=** | **BCba1** | **+** | **BCba2** | **+** | **BCba3** | **+** | **BCba4** | **+** | **BCbc1** | **+** | **BCbc2** | **+** | **BCbc3** | **+** | **BCbc4** |
| **0.00** | **=** | **28.59** | | **(2.55)** | | **14.57** | | **(56.34)** | | **8.94** | | **2.53** | | **1.50** | | **2.75** |
| | | **Equation 8.9 Defining the Trade Balance of Country C (Canada) with All Other Countries (BCc.ab)** | | | | | | | | | | | | | | |
| | | | | **BCc.ab = BCca1 + BCca2 + BCca3 + BCca4 + BCcb1 + BCcb2 + BCcb3 + BCcb4** | | | | | | | | | | | | |
| **BCc.ab** | **=** | **BCca1** | **+** | **BCca2** | **+** | **BCca3** | **+** | **BCca4** | **+** | **BCcb1** | **+** | **BCcb2** | **+** | **BCcb3** | **+** | **BCcb4** |
| **0.00** | **=** | **34.40** | **+** | **(4.57)** | **+** | **(4.85)** | **+** | **(9.26)** | **+** | **(8.94)** | **+** | **(2.53)** | **+** | **(1.50)** | **+** | **(2.75)** |
| | | | | | **COMMERCIAL BALANCES** | | | | | | | | | | | |
| | | **Country A - United States** | | | | | **Country B - Mexico** | | | | | **Country C - Canada** | | | | |
| | | | | **Déficit o Superávit EU** | | | | | **Déficit o Superávit Méx** | | | | | **Déficit o Superávit Cand** | | |
| | | **EXa EU** | **IMa EU** | **US$ BCa.bc = (Exa - Ima)** | | | **EXb Mex** | **IMb Mex** | **US$ BCb.ac = (EXb - IMb)** | | | **EXc Cand** | **IMc Cand** | **US$ BCc.ab = (EXc - IMc)** | | |
| | | **215.05** | **215.05** | **0.00** | | | **203.14** | **203.14** | **(0.00)** | | | **41.17** | **41.18** | **(0.00)** | | |
| | | **EXab** | **IMab** | | | | **EXba** | **IMba** | | | | **EXca** | **IMca** | | | |
| | | **196.37** | **180.64** | **15.73** | | | **180.64** | **196.37** | **(15.73)** | | | **34.40** | **18.68** | **15.72** | | |
| | | **EXac** | **IMac** | | | | **EXbc** | **IMbc** | | | | **EXcb** | **IMcb** | | | |
| | | **18.68** | **34.40** | **(15.72)** | | | **22.50** | **6.77** | **15.73** | | | **6.77** | **22.50** | **(15.73)** | | |

# Apendix Spreadsheed 15

## Image Section 15.1

| | | | | | | | | | | | | | | | | |
|---|---|---|---|---|---|---|---|---|---|---|---|---|---|---|---|---|
| Spreadsheet 15 - MODEL SIMULATION INTERNATIONAL TRADE OF 3 COUNTRIES: | | | | | | | | | | | | | | | | |
| | | | | | | | | | SECTION 15.1: COMMERCIAL BALANCES. MODEL EQUATIONS | | | | | | | |
| | | Equation 8.7 Defining the Trade Balance of Country A (United States) with All Other Countries (BCa.bc) | | | | | | | | | | | | | | |
| | | | | BCa.bc = BCab1 + BCab2 + BCab3 + BCab4 + BCac1 + BCac2 + BCac3 + BCac4 | | | | | | | | | | | | |
| BCa.bc | = | BCab1 | + | BCab2 | + | BCab3 | + | BCab4 | + | BCac1 | + | BCac2 | + | BCac3 | + | BCac4 |
| 0.00 | = | 14.48 | + | (32.19) | + | (54.28) | + | 39.07 | + | - | + | 11.56 | + | 15.18 | + | 6.20 |
| | | Equation 8.8 Defining the Trade Balance of Country B (Mexico) with All Other Countries (BCb.ac) | | | | | | | | | | | | | | |
| | | | | BCb.ac = BCba1 + BCba2 + BCba3 + BCba4 + BCbc1 + BCbc2 + BCbc3 + BCbca4 | | | | | | | | | | | | |
| BCb.ac | = | BCba1 | + | BCba2 | + | BCba3 | + | BCba4 | + | BCbc1 | + | BCbc2 | + | BCbc3 | + | BCbc4 |
| 0.00 | = | -14.48 | | 32.19 | | 54.28 | | (39.07) | | 6.42 | | (5.68) | | (21.47) | | (12.21) |
| | | Equation 8.9 Defining the Trade Balance of Country C (Canada) with All Other Countries (BCc.ab) | | | | | | | | | | | | | | |
| | | | | BCc.ab = BCca1 + BCca2 + BCca3 + BCca4 + BCcb1 + BCcb2 + BCcb3 + BCcb4 | | | | | | | | | | | | |
| BCc.ab | = | BCca1 | + | BCca2 | + | BCca3 | + | BCca4 | + | BCcb1 | + | BCcb2 | + | BCcb3 | + | BCcb4 |
| 0.00 | = | 0.00 | + | (11.56) | + | (15.18) | + | (6.20) | + | (6.42) | + | 5.68 | + | 21.47 | + | 12.21 |
| | | | | | COMMERCIAL BALANCES | | | | | | | | | | | |
| | | Country A - United States | | | | | Country B - Mexico | | | | | Country C - Canada | | | | |
| | | | | Deficit or Surplus US | | | | | Deficit or Surplus Méx | | | | | Deficit or Surplus Cand | | |
| | | EXa EU | IMa EU | US$ BCa.bc = (Exa - Ima) | | | EXb Mex | IMb Mex | US$ BCb.ac = (EXb - IMb) | | | EXc Cand | IMc Cand | US$ BCc.ab = (EXc - IMc) | | |
| | | 240.43 | 240.42 | 0.00 | | | 251.89 | 251.90 | (0.00) | | | 44.41 | 44.41 | (0.00) | | |
| | | EXab | IMab | | | | EXba | IMba | | | | EXca | IMca | | | |
| | | 207.49 | 240.42 | (32.93) | | | 240.42 | 207.49 | 32.93 | | | - | 32.94 | (32.94) | | |
| | | EXac | IMac | | | | EXbc | IMbc | | | | EXcb | IMcb | | | |
| | | 32.94 | - | 32.94 | | | 11.47 | 44.41 | (32.94) | | | 44.41 | 11.47 | 32.94 | | |

# Appendix Spreadsheet 15

## Image Section 15.5

**SECTION 15.5: MODEL PARAMETERS**

| | | | | | |
|---|---|---|---|---|---|
| | PRODUCTIVITY OF WORK IN THE COUNTRY A | | PRODUCTIVITY OF WORK IN COUNTRY B | | PRODUCTIVITY OF WORK IN COUNTRY C |
| 3.00 | Za1 | 4.00 | Zb1 | 3.50 | Zc1 |
| 2.00 | Za2 | 3.00 | Zb2 | 2.50 | Zc2 |
| 1.00 | Za3 | 2.00 | Zb3 | 1.50 | Zc3 |
| 0.50 | Za4 | 1.00 | Zb4 | 0.75 | Zc4 |

**MONETARY OFFER OMi, MARGIN OF GAIN MGi, POPULATION Ni, SALARY Wi, PROFITS Gij, WORK FORCE FT, CONSUMER PREFERENCES PR AND TARIFFS AR**

| | | | | | |
|---|---|---|---|---|---|
| 2,500 | OMa = Money Offer Country A US dollar | 0.3650 | MGa = Average Margin of Gain or Profit in country A | 325.00 | Na = Population of country A United States |
| 25,000 | OMb = Money Offer Country B Mexico peso | 0.5450 | MGb = Average Margin of Gain or Profit in country B | 137.00 | Nb = Population of country B México |
| 500 | OMc = Money Offer Country C Canda dollar | 0.4100 | MGc = Average Margin of Gain or Profit in country C | 36.00 | Nc = Population of country C Canada |

| | | | |
|---|---|---|---|
| 4.88 | Wa = Salary in country A for each unit of work hired = (OMa(1-MGa))/Na | 8.42 | Ga1 = Gain or Profit in country A corresponds to good 1 for each unit produced = (OMa(MGa)(FTa1))/Qa1 |
| 83.03 | Wb = Salary in country B for each unit of work hired = (OMb(1-MGb))/Nb | 5.62 | Ga2 = Gain or Profit in country A corresponds to good 2 for each unit produced = (OMa(MGa)(FTa2))/Qa2 |
| 8.19 | Wc = Salary in country C for each unit of work hired = (OMc(1-MGc))/Nc | 2.81 | Ga3 = Gain or Profit in country A corresponds to good 3 for each unit produced = (OMa(MGa)(FTa3))/Qa3 |
| | | 1.40 | Ga4 = Gain or Profit in country A corresponds to good 4 for each unit produced = (OMa(MGa)(FTa4))/Qa4 |
| | | | |
| | | 397.81 | Gb1 = Gain or Profit in country B corresponds to good 1 for each unit produced = (OMb(MGb)(FTb1))/Qb1 |
| | | 298.36 | Gb2 = Gain or Profit in country B corresponds to good 2 for each unit produced = (OMb(MGb)(FTb2))/Qb2 |
| | | 198.91 | Gb3 = Gain or Profit in country B corresponds to good 3 for each unit produced = (OMb(MGb)(FTb3))/Qb3 |
| | | 99.45 | Gb4 = Gain or Profit in country B corresponds to good 3 for each unit produced = (OMb(MGb)(FTb4))/Qb4 |
| | | | |
| | | 19.93 | Gc1 = Gain or Profit in country C corresponds to good 1 for each unit produced = (OMc(MGc)(FTc1))/Qc1 |
| | | 14.24 | Gc2 = Gain or Profit in country C corresponds to good 2 for each unit produced = (OMc(MGc)(FTc2))/Qc2 |
| | | 8.54 | Gc3 = Gain or Profit in country C corresponds to good 3 for each unit produced = (OMc(MGc)(FTc3))/Qc3 |
| | | 4.27 | Gc4 = Gain or Profit in country C corresponds to good 4 for each unit produced = (OMc(MGc)(FTc4))/Qc4 |

**DISTRIBUTION OF THE FORCE OF WORK**

| | | | | | |
|---|---|---|---|---|---|
| | DISTRIBUTION OF THE COUNTRY A FORCE OF WORK | | DISTRIBUTION OF THE COUNTRY B FORCE OF WORK | | DISTRIBUTION OF THE COUNTRY C FORCE OF WORK |
| 0.20 | FT1a = Proportion force of work producing good 1 in country A | 0.20 | FT1b = Proportion force of work producing good 1 in country B | 0.20 | FT1c = Proportion force of work producing good 1 in country C |
| 0.30 | FT2a = Proportion force of work producing good 2 in country A | 0.30 | FT2b = Proportion force of work producing good 2 in country B | 0.30 | FT2c = Proportion force of work producing good 2 in country C |
| 0.30 | FT3a = Proportion force of work producing good 3 in country A | 0.30 | FT3b = Proportion force of work producing good 3 in country B | 0.30 | FT3c = Proportion force of work producing good 3 in country C |
| 0.20 | FT4a = Proportion force of work producing good 4 in country A | 0.20 | FT4b = Proportion force of work producing good 4 in country B | 0.20 | FT4c = Proportion force of work producing good 4 in country C |

**CONSUMER PREFERENCES AND TARIFF RATES**

**Country A = UNITED STATES**

| | | | | | |
|---|---|---|---|---|---|
| | PREFERENCES OF CONSUMERS OF COUNTRY A WITH REGARD TO GOODS OF COUNTRY B | | PREFERENCES OF CONSUMERS OF COUNTRY A WITH REGARD TO GOODS OF COUNTRY C | | TARIFF (ARANCEL) OF COUNTRY A WITH REGARI |
| 0.25 | PR1ab = Preference of good 1 in country A made in country B | 0.15 | PR1ac = Preference of good 1 in country A made in country C | 0.4801 | AR1a = Tariff (Arancel) of good 1 in country A |
| 0.50 | PR2ab = Preference of good 2 in country A made in country B | 0.05 | PR2ac = Preference of good 2 in country A made in country C | 0.4712 | AR2a = Tariff (Arancel) of good 2 in country A |
| 0.75 | PR3ab = Preference of good 3 in country A made in country B | 0.15 | PR3ac = Preference of good 3 in country A made in country C | 0.5300 | AR3a = Tariff (Arancel) of good 3 in country A |
| 0.38 | PR4ab = Preference of good 4 in country A made in country B | 0.12 | PR4ac = Preference of good 4 in country A made in country C | 0.0000 | AR4a = Tariff (Arancel) of good 4 in country A |

**Country B = MEXICO**

| | | | | | |
|---|---|---|---|---|---|
| | PREFERENCES OF CONSUMERS OF COUNTRY B WITH REGARD TO GOODS OF COUNTRY A | | PREFERENCES OF CONSUMERS OF COUNTRY B WITH REGARD TO GOODS OF COUNTRY C | | TARIFF (ARANCEL) OF COUNTRY B WITH REGARI |
| 0.25 | PR1ba = Preference of good 1 in country B made in country A | 0.05 | PR1bc = Preference of good 1 in country B made in country C | 0.0000 | AR1b = Tariff (Arancel) of good 1 in country B |
| 0.18 | PR2ba = Preference of good 2 in country B made in country A | 0.16 | PR2bc = Preference of good 2 in country B made in country C | 0.0000 | AR2b = Tariff (Arancel) of good 2 in country B |
| 0.14 | PR3ba = Preference of good 3 in country B made in country A | 0.21 | PR3bc = Preference of good 3 in country B made in country C | 0.0000 | AR3b = Tariff (Arancel) of good 3 in country B |
| 0.22 | PR4ba = Preference of good 4 in country B made in country A | 0.17 | PR4bc = Preference of good 4 in country B made in country C | 0.0000 | AR4b = Tariff (Arancel) of good 4 in country B |

**Country C = CANADA**

| | | | | | |
|---|---|---|---|---|---|
| | PREFERENCES OF CONSUMERS OF COUNTRY C WITH REGARD TO GOODS OF COUNTRY A | | PREFERENCES OF CONSUMERS OF COUNTRY C WITH REGARD TO GOODS OF COUNTRY B | | TARIFF (ARANCEL) OF COUNTRY C WITH REGARI |
| 0.15 | PR1ca = Preference of good 1 in country C made in country A | 0.19 | PR1cb= Preference of good 1 in country C made in country B | 0.1300 | AR1c = Tariff (Arancel) of good 1 in country C |
| 0.25 | PR2ca = Preference of good 2 in country C made in country A | 0.15 | PR2cb= Preference of good 2 in country C made in country B | 0.1000 | AR2c = Tariff (Arancel) of good 2 in country C |
| 0.15 | PR3ca = Preference of good 3 in country C made in country A | 0.14 | PR3cb= Preference of good 3 in country C made in country B | 0.1015 | AR3c = Tariff (Arancel) of good 3 in country C |
| 0.18 | PR4ca = Preference of good 4 in country C made in country A | 0.14 | PR4cb= Preference of good 4 in country C made in country B | 0.2500 | AR4c = Tariff (Arancel) of good 4 in country C |

# Apendix Spreadsheed 16

## Image Section 16.1

**Spreadsheet 16 - MODEL SIMULATION INTERNATIONAL TRADE OF 4 COUNTRIES:**

**SECTION 16.1: COMMERCIAL BALANCES. MODEL EQUATIONS**

**Equation 9.1 Defining the Trade Balance of Country A (United States) with All Other Countries (BCa.bcd)**

**BCa.bcd = BCab1 + BCab2 + BCab3 + BCab4 + BCac1 + BCac2 + BCac3 + BCac4 + BCad1 + BCad2 + BCad3 + BCad4**

| BCa.bcd | = | BCab1 | + | BCab2 | + | BCab3 | + | BCab4 | + | BCac1 | + | BCac2 | + | BCac3 | + | BCac4 | + | BCad1 | + | BCad2 | + | BCad3 | + | BCad4 |
|---|---|---|---|---|---|---|---|---|---|---|---|---|---|---|---|---|---|---|---|---|---|---|---|---|
| -188.61 | = | -39.67 | + | -107.73 | + | -126.95 | + | 37.31 | + | -7.92 | + | 12.25 | + | 20.50 | + | 15.32 | + | -3.68 | + | 0.89 | + | 6.63 | + | 4.42 |

**Equation 9.2 Defining the Trade Balance of Country B (Mexico) with All Other Countries (BCb.acd)**

**BCb.acd = BCba1 + BCba2 + BCba3 + BCba4 + BCbc1 + BCbc2 + BCbc3 + BCbc4 + BCbd1 + BCbd2 + BCbd3 + BCbd4**

| BCb.acd | = | BCba1 | + | BCba2 | + | BCba3 | + | BCba4 | + | BCbc1 | + | BCbc2 | + | BCbc3 | + | BCbc4 | + | BCbd1 | + | BCbd2 | + | BCbd3 | + | BCbd4 |
|---|---|---|---|---|---|---|---|---|---|---|---|---|---|---|---|---|---|---|---|---|---|---|---|---|
| 225.78 | = | 39.67 | + | 107.73 | + | 126.95 | + | -37.31 | + | 11.51 | + | 0.27 | + | -17.96 | + | -9.88 | + | -1.11 | + | 1.86 | + | 2.96 | + | 1.10 |

**Equation 9.3 Defining the Trade Balance of Country C (Canada) with All Other Countries (BCc.abd)**

**BCc.abd = BCca1 + BCca2 + BCca3 + BCca4 + BCcb1 + BCcb2 + BCcb3 + BCcb4 + BCcd1 + BCcd2 + BCcd3 + BCcd4**

| BCc.abd | = | BCca1 | + | BCca2 | + | BCca3 | + | BCca4 | + | BCcb1 | + | BCcb2 | + | BCcb3 | + | BCcb4 | + | BCcd1 | + | BCcd2 | + | BCcd3 | + | BCcd4 |
|---|---|---|---|---|---|---|---|---|---|---|---|---|---|---|---|---|---|---|---|---|---|---|---|---|
| -18.43 | = | 7.92 | + | -12.25 | + | -20.50 | + | -15.32 | + | -11.51 | + | -0.27 | + | 17.96 | + | 9.88 | + | -2.93 | + | 1.06 | + | 4.18 | + | 3.37 |

**Equation 9.4 Defining the Trade Balance of Country D (Dominican Republic) with All Other Countries (BCd.abc)**

**BCd.abc = BCda1 + BCda2 + BCda3 + BCda4 + BCdb1 + BCdb2 + BCdb3 + BCdb4 + BCdc1 + BCdc2 + BCdc3 + BCdc4**

| BCd.abc | = | BCda1 | + | BCda2 | + | BCda3 | + | BCda4 | + | BCdb1 | + | BCdb2 | + | BCdb3 | + | BCdb4 | + | BCdc1 | + | BCdc2 | + | BCdc3 | + | BCdc4 |
|---|---|---|---|---|---|---|---|---|---|---|---|---|---|---|---|---|---|---|---|---|---|---|---|---|
| -18.75 | = | 3.68 | + | -0.89 | + | -6.63 | + | -4.42 | + | 1.11 | + | -1.86 | + | -2.96 | + | -1.10 | + | 2.93 | + | -1.06 | + | -4.18 | + | -3.37 |

**COMMERCIAL BALANCES**

| | | | | | | | | | | | | | | | | | | |
|---|---|---|---|---|---|---|---|---|---|---|---|---|---|---|---|---|---|---|
| **Country A - United States** | | | | | **Country B - Mexico** | | | | | **Country C - Canada** | | | | | **Country D - Dominican Republic** | | | |
| | | **Deficit or Surplus US** | | | | | **Deficit or Surplus Mex** | | | | | **Deficit or Surplus Cand** | | | | | **Deficit or Surplus Dominican Republic** | |
| EXa EU | IMa EU | US$ Bca.bcd | | | EXb Mex | IMb Mex | US$ BCb.acd | | | EXc Cand | IMc Cand | US$ BCc.abd | | | EXd RD | IMd RD | US$ BCd.abc | |
| 272 | 460.87 | (188.61) | | (2.01) | 479.20 | 253.42 | 225.78 | | | 71.49 | 89.92 | (18.43) | | | 17.38 | 36.13 | (18.75) | |
| EXab | IMab | | | | EXba | IMba | | | | EXca | IMca | | | | EXda | IMda | | |
| 202.43 | 439.46 | ############ | | 6.15 | 439.46 | 202.43 | 237.03 | 6.46 | | 16.01 | 56.17 | (40.16) | (17.39) | | 5.40 | 13.66 | (8.26) | (0.08) |
| EXac | IMac | | | | EXbc | IMbc | | | | EXcb | IMcb | | | | EXdb | IMdb | | |
| 56.17 | 16.01 | 40.160592 | | (7.98) | 28.35 | 44.41 | (16.05) | (0.44) | | 44.41 | 28.35 | 16.05 | 6.95 | | 6.58 | 11.39 | (4.81) | (0.05) |
| EXad | IMad | | | | EXbd | IMbd | | | | EXcd | IMcd | | | | EXdc | IMdc | | |
| 14 | 5.40 | 8.26255 | | (0.19) | 11.39 | 6.58 | 4.81 | 0.13 | | 11.08 | 5.40 | 5.68 | 2.46 | | 5.40 | 11.08 | (5.68) | (0.06) |

# Appendix Spreadsheet 16

## Image Section 16.5

## Model Parameter

**SECTION 16.5: MODEL PARAMETERS**

| | PRODUCTIVITY OF WORK IN THE COUNTRY A | | PRODUCTIVITY OF WORK IN COUNTRY B | | PRODUCTIVITY OF WORK IN COUNTRY C | | PRODUCTIVITY OF WORK IN COUNTRY D |
|---|---|---|---|---|---|---|---|
| 2.40 | Za1 = ka1(1 - (Na1/(Na1 + Nb1 + Nc1 + Nd1))) | 3.20 | Zb1 = kb1(1 - (Nb1/(Na1 + Nb1 + Nc1 + Nd1))) | 2.80 | Zc1 = kc1(1 - (Nc1/(Na1 + Nb1 + Nc1 + Nd1))) | 3.60 | Zd1 = kd1(1 - (Nd1/(Na1 + Nb1 + Nc1 + Nd1))) |
| 1.40 | Za2 = ka2(1 - (Na2/(Na2 + Nb2 + Nc2 + Nd2))) | 2.10 | Zb2 = kb2(1 - (Nb2/(Na2 + Nb2 + Nc2 + Nd2))) | 1.75 | Zc2 = kc2(1 - (Nc2/(Na2 + Nb2 + Nc2 + Nd2))) | 2.63 | Zd2 = kd2(1 - (Nc2/(Na2 + Nb2 + Nc2 + Nd2))) |
| 0.70 | Za3= ka3(1 - (Na3/(Na3 + Nb3 + Nc3 + Nd3))) | 1.40 | Zb3= kb3(1 - (Nb3/(Na3 + Nb3 + Nc3 + Nd3))) | 1.05 | Zc3= kc3(1 - (Nc3/(Na3 + Nb3 + Nc3 + Nd3))) | 1.75 | Zd3= kd3(1 - (Nc3/(Na3 + Nb3 + Nc3 + Nd3))) |
| 0.40 | Za4 = ka4(1 - (Na4/(Na4 + Nb4 + Nc4 + Nd4))) | 0.80 | Zb4 = kb4(1 - (Nb4/(Na4 + Nb4 + Nc4 + Nd4))) | 0.60 | Zc4 = kc4(1 - (Nc4/(Na4 + Nb4 + Nc4 + Nd4))) | 1.00 | Zd4 = kd4(1 - (Nc4/(Na4 + Nb4 + Nc4 + Nd4))) |

**MONETARY OFFER OMi, MARGIN OF GAIN MGi, POPULATION Ni, SALARY Wi, PROFITS Gij, WORK FORCE FT, CONSUMER PREFERENCES PR AND TARIFFS AR**

| | | | | | |
|---|---|---|---|---|---|
| 2,500 | OMa = Money Offer Country A US dollar | 0.365 | MGa = Average Margin of Gain or Profit in country A | 325 | Na = Population of country A United States |
| 25,000 | OMb = Money Offer Country B Mexico peso | 0.545 | MGb = Average Margin of Gain or Profit in country B | 137 | Nb = Population of country B México |
| 500 | OMc = Money Offer Country C Canda dollar | 0.410 | MGc = Average Margin of Gain or Profit in country C | 36 | Nc = Population of country C Canada |
| 5,000 | OMdr = Money Offer Country D Dominican Re | 0.410 | MGd = Average Margin of Gain or Profit in country D | 12 | Nd = Population of country D Dominican Republic |

| | | | |
|---|---|---|---|
| 4.88 | Wa = Salary in country A for each unit of work hired = (OMa(1-MGa))/Na | 6.74 | Ga1 = Gain or Profit in country A corresponds to good 1 for each unit produced = (OMa(MGa)(FTa1))/Qa1 |
| 83.03 | Wb = Salary in country B for each unit of work hired = (OMb(1-MGb))/Nb | 3.93 | Ga2 = Gain or Profit in country A corresponds to good 2 for each unit produced = (OMa(MGa)(FTa2))/Qa2 |
| 8.19 | Wc = Salary in country C for each unit of work hired = (OMc(1-MGc))/Nc | 1.97 | Ga3 = Gain or Profit in country A corresponds to good 3 for each unit produced = (OMa(MGa)(FTa3))/Qa3 |
| 245.83 | Wd = Salary in country D for each unit of work hired = (OMd(1-MGd))/Nd | 1.12 | Ga4 = Gain or Profit in country A corresponds to good 4 for each unit produced = (OMa(MGa)(FTa4))/Qa4 |
| | | | |
| | | 318.25 | Gb1 = Gain or Profit in country B corresponds to good 1 for each unit produced = (OMb(MGb)(FTb1))/Qb1 |
| | | 208.85 | Gb2 = Gain or Profit in country B corresponds to good 2 for each unit produced = (OMb(MGb)(FTb2))/Qb2 |
| | | 139.23 | Gb3 = Gain or Profit in country B corresponds to good 3 for each unit produced = (OMb(MGb)(FTb3))/Qb3 |
| | | 79.56 | Gb4 = Gain or Profit in country B corresponds to good 3 for each unit produced = (OMb(MGb)(FTb4))/Qb4 |
| | | | |
| | | 15.94 | Gc1 = Gain or Profit in country C corresponds to good 1 for each unit produced = (OMc(MGc)(FTc1))/Qc1 |
| | | 9.97 | Gc2 = Gain or Profit in country C corresponds to good 2 for each unit produced = (OMc(MGc)(FTc2))/Qc2 |
| | | 5.98 | Gc3 = Gain or Profit in country C corresponds to good 3 for each unit produced = (OMc(MGc)(FTc3))/Qc3 |
| | | 3.42 | Gc4 = Gain or Profit in country C corresponds to good 4 for each unit produced = (OMc(MGc)(FTc4))/Qc4 |
| | | | |
| | | 615.00 | Gd1 = Gain or Profit in country D corresponds to good 1 for each unit produced = (OMd(MGd)(FTd1))/Qd1 |
| | | 448.44 | Gc2 = Gain or Profit in country D corresponds to good 2 for each unit produced = (OMc(MGc)(FTc2))/Qc2 |
| | | 298.96 | Gc3 = Gain or Profit in country D corresponds to good 3 for each unit produced = (OMc(MGc)(FTc3))/Qc3 |
| | | 170.83 | Gc4 = Gain or Profit in country D corresponds to good 4 for each unit produced = (OMc(MGc)(FTc4))/Qc4 |

**DISTRIBUTION OF THE FORCE OF WORK**

| | DISTRIBUTION OF THE COUNTRY A FORCE OF WORK | | DISTRIBUTION OF THE COUNTRY B FORCE OF WORK | | DISTRIBUTION OF THE COUNTRY C FORCE OF WORK | | DISTRIBUTION OF THE COUNTRY D FORCE OF WOR |
|---|---|---|---|---|---|---|---|
| 0.2 | FT1a = Proportion force of work producing good 1 in country A | 0.2 | FT1b = Proportion force of work producing good 1 in country B | 0.2 | FT1c = Proportion force of work producing good 1 in country C | 0.2 | FT1d = Proportion force of work producing good |
| 0.3 | FT2a = Proportion force of work producing good 2 in country A | 0.3 | FT2b = Proportion force of work producing good 2 in country B | 0.3 | FT2c = Proportion force of work producing good 2 in country C | 0.3 | FT2d = Proportion force of work producing good |
| 0.3 | FT3a = Proportion force of work producing good 3 in country A | 0.3 | FT3b = Proportion force of work producing good 3 in country B | 0.3 | FT3c = Proportion force of work producing good 3 in country C | 0.3 | FT3d = Proportion force of work producing good |
| 0.2 | FT4a = Proportion force of work producing good 4 in country A | 0.2 | FT4b = Proportion force of work producing good 4 in country B | 0.2 | FT4c = Proportion force of work producing good 4 in country C | 0.2 | FT4d = Proportion force of work producing good |

**DISTRIBUTION OF THE NUMBER OF WORKERS**

| | NUMBER OF WORKERS BY INDUSTRIAL SECTOR IN THE COUNTRY A | | NUMBER OF WORKERS BY INDUSTRIAL SECTOR IN THE COUNTRY B | | NUMBER OF WORKERS BY INDUSTRIAL SECTOR IN THE COUNTRY C | | NUMBER C |
|---|---|---|---|---|---|---|---|
| 65.00 | Na1 = FT1a(Na) = Number of workers producing good 1 in country A | 27.4 | Nb1 = FT1b(Nb) = Number of workers producing good 1 in country B | 7.2 | Nc1 = FT1c(Nc) = Number of workers producing good 1 in country C | 2.4 | Nd1 = FT1d |
| 97.50 | Na2 = FT2a(Na) = Number of workers producing good 2 in country A | 41.1 | Nb2 = FT2b(Nb) = Number of workers producing good 2 in country B | 10.8 | Nc2 = FT2c(Nc) = Number of workers producing good 2 in country C | 3.6 | Nd2 = FT2d |
| 97.50 | Na3 = FT3a(Na) = Number of workers producing good 3 in country A | 41.1 | Nb3 = FT3b(Nb) = Number of workers producing good 3 in country B | 10.8 | Nc3 = FT3c(Nc) = Number of workers producing good 3 in country C | 3.6 | Nd3 = FT3d |
| 65.00 | Na4 = FT4a(Na) = Number of workers producing good 4 in country A | 27.4 | Nb4 = FT4b(Nb) = Number of workers producing good 4 in country B | 7.2 | Nc4 = FT4c(Nc) = Number of workers producing good 4 in country C | 2.4 | Nd4 = FT4d |

**ARBITRARY CONSTANTS OF PRODUCTIVITY OF WORK**

| | CONSTANT PRODUCTIVITY BY INDUSTRIAL SECTOR IN THE COUNTRY A | | CONSTANT PRODUCTIVITY BY INDUSTRIAL SECTOR IN THE COUNTRY B | | CONSTANT PRODUCTIVITY BY INDUSTRIAL SECTOR IN THE COUNTRY C | | CONSTANT |
|---|---|---|---|---|---|---|---|
| 3.00 | ka1 = Arbitrary constant of labor productivity in good 1 of country A | 4.00 | kb1 = Arbitrary constant of labor productivity in good 1 of country B | 3.50 | kc1 = Arbitrary constant of labor productivity in good 1 of country C | 4.50 | kd1 = Arbit |
| 2.00 | ka2 = Arbitrary constant of labor productivity in good 2 of country A | 3.00 | kb2 = Arbitrary constant of labor productivity in good 2 of country B | 2.50 | kc2 = Arbitrary constant of labor productivity in good 2 of country C | 3.75 | kd2 = Arbit |
| 1.00 | ka3 = Arbitrary constant of labor productivity in good 3 of country A | 2.00 | kb3 = Arbitrary constant of labor productivity in good 3 of country B | 1.50 | kc3 = Arbitrary constant of labor productivity in good 3 of country C | 2.50 | kd3 = Arbit |
| 0.50 | ka4 = Arbitrary constant of labor productivity in good 4 of country A | 1.00 | kb4 = Arbitrary constant of labor productivity in good 4 of country B | 0.75 | kc4 = Arbitrary constant of labor productivity in good 4 of country C | 1.25 | kd4 = Arbit |

**CONSUMER PREFERENCES AND TARIFF RATES**

Country A = UNITED STATES

| | PREFERENCES OF CONSUMERS OF COUNTRY A WITH REGARD TO GOODS OF COUNTRY B | | PREFERENCES OF CONSUMERS OF COUNTRY A WITH REGARD TO GOODS OF COUNTRY C | | PREFERENCES OF CONSUMERS OF COUNTRY A WITH REGARD TO GOODS OF COUNTRY |
|---|---|---|---|---|---|
| 0.250 | PR1ab = Preference of good 1 in country A made in country B | 0.15 | PR1ac = Preference of good 1 in country A made in country C | 0.150 | PR1ad = Preference of good 1 in country A made in country D |
| 0.500 | PR2ab = Preference of good 2 in country A made in country B | 0.05 | PR2ac = Preference of good 2 in country A made in country C | 0.120 | PR2ad = Preference of good 2 in country A made in country D |
| 0.750 | PR3ab = Preference of good 3 in country A made in country B | 0.15 | PR3ac = Preference of good 3 in country A made in country C | 0.150 | PR3ad = Preference of good 3 in country A made in country D |
| 0.380 | PR4ab = Preference of good 4 in country A made in country B | 0.12 | PR4ac = Preference of good 4 in country A made in country C | 0.050 | PR4ad = Preference of good 4 in country A made in country D |

Country B = MEXICO

| | PREFERENCES OF CONSUMERS OF COUNTRY B WITH REGARD TO GOODS OF COUNTRY A | | PREFERENCES OF CONSUMERS OF COUNTRY B WITH REGARD TO GOODS OF COUNTRY C | | PREFERENCES OF CONSUMERS OF COUNTRY B WITH REGARD TO GOODS OF COUNTRY |
|---|---|---|---|---|---|
| 0.250 | PR1ba = Preference of good 1 in country B made in country A | 0.050 | PR1bc = Preference of good 1 in country B made in country C | 0.140 | PR1bd= Preference of good 1 in country B made in country D |
| 0.180 | PR2ba = Preference of good 2 in country B made in country A | 0.160 | PR2bc = Preference of good 2 in country B made in country C | 0.220 | PR2bd = Preference of good 2 in country B made in country D |
| 0.140 | PR3ba = Preference of good 3 in country B made in country A | 0.210 | PR3bc = Preference of good 3 in country B made in country C | 0.050 | PR3bd = Preference of good 3 in country B made in country D |
| 0.220 | PR4ba = Preference of good 4 in country B made in country A | 0.170 | PR4bc = Preference of good 4 in country B made in country C | 0.160 | PR4bd = Preference of good 4 in country B made in country D |

Country C = CANADA

| | PREFERENCES OF CONSUMERS OF COUNTRY C WITH REGARD TO GOODS OF COUNTRY A | | PREFERENCES OF CONSUMERS OF COUNTRY C WITH REGARD TO GOODS OF COUNTRY B | | PREFERENCES OF CONSUMERS OF COUNTRY C WITH REGARD TO GOODS OF COUNTRY |
|---|---|---|---|---|---|
| 0.150 | PR1ca = Preference of good 1 in country C made in country A | 0.190 | PR1cb= Preference of good 1 in country C made in country B | 0.190 | PR1cd= Preference of good 1 in country C made in country D |
| 0.250 | PR2ca = Preference of good 2 in country C made in country A | 0.150 | PR2cb= Preference of good 2 in country C made in country B | 0.150 | PR2cd= Preference of good 2 in country C made in country D |
| 0.150 | PR3ca = Preference of good 3 in country C made in country A | 0.140 | PR3cb= Preference of good 3 in country C made in country B | 0.140 | PR3cd= Preference of good 3 in country C made in country D |
| 0.180 | PR4ca = Preference of good 4 in country C made in country A | 0.140 | PR4cb= Preference of good 4 in country C made in country B | 0.140 | PR4cd= Preference of good 4 in country C made in country D |

Country D = DOMINICAN REPUBLIC

| | PREFERENCES OF CONSUMERS OF COUNTRY D WITH REGARD TO GOODS OF COUNTRY A | | PREFERENCES OF CONSUMERS OF COUNTRY D WITH REGARD TO GOODS OF COUNTRY B | | PREFERENCES OF CONSUMERS OF COUNTRY D WITH REGARD TO GOODS OF COUNTRY |
|---|---|---|---|---|---|
| 0.190 | PR1da = Preference of good 1 in country D made in country A | 0.150 | PR1db = Preference of good 1 in country D made in country B | 0.140 | PR1dc = Preference of good 1 in country D made in country C |
| 0.150 | PR2da = Preference of good 2 in country D made in country A | 0.250 | PR2db = Preference of good 2 in country D made in country B | 0.150 | PR2dc = Preference of good 2 in country D made in country C |
| 0.140 | PR3ca = Preference of good 3 in country D made in country A | 0.150 | PR3db = Preference of good 3 in country D made in country B | 0.140 | PR3dc = Preference of good 3 in country D made in country C |
| 0.140 | PR4da = Preference of good 4 in country D made in country A | 0.180 | PR4db = Preference of good 4 in country D made in country B | 0.190 | PR4dc = Preference of good 4 in country D made in country C |

| | TARIFF (ARANCEL) OF COUNTRY A WITH REGARD TO GOODS 1, 2... 4 | | TARIFF (ARANCEL) OF COUNTRY B WITH REGARD TO GOODS 1, 2... 4 | | TARIFF (ARANCEL) OF COUNTRY C WITH REGARD TO GOODS 1, 2... 4 | | TARIFF (ARANCEL) OF COUNTRY D WITH REGARD T |
|---|---|---|---|---|---|---|---|
| 0.000 | AR1a = Tariff (Arancel) of good 1 in country A | 0.000 | AR1b = Tariff (Arancel) of good 1 in country B | 0.000 | AR1c = Tariff (Arancel) of good 1 in country C | 0.000 | AR1d = Tariff (Arancel) of good 1 in country D |
| 0.000 | AR2a = Tariff (Arancel) of good 2 in country A | 0.000 | AR2b = Tariff (Arancel) of good 2 in country B | 0.000 | AR2c = Tariff (Arancel) of good 2 in country C | 0.000 | AR2d = Tariff (Arancel) of good 2 in country D |
| 0.000 | AR3a = Tariff (Arancel) of good 3 in country A | 0.000 | AR3b = Tariff (Arancel) of good 3 in country B | 0.000 | AR3c = Tariff (Arancel) of good 3 in country C | 0.000 | AR3d = Tariff (Arancel) of good 3 in country D |
| 0.000 | AR4a = Tariff (Arancel) of good 4 in country A | 0.000 | AR4b = Tariff (Arancel) of good 4 in country B | 0.000 | AR4c = Tariff (Arancel) of good 4 in country C | 0.000 | AR4d = Tariff (Arancel) of good 4 in country D |

# Appendix Spreadsheed 17

## Image Section 17.1

**Spreadsheet 17 - MODEL SIMULATION INTERNATIONAL TRADE OF 4 COUNTRIES:**

**SECTION 17.1: COMMERCIAL BALANCES. MODEL EQUATIONS**

**Equation 9.1 Defining the Trade Balance of Country A (United States) with All Other Countries (BCa.bcd)**

**BCa.bcd = BCab1 + BCab2 + BCab3 + BCab4 + BCac1 + BCac2 + BCac3 + BCac4 + BCad1 + BCad2 + BCad3 + BCad4**

| BCa.bcd | = | BCab1 | + | BCab2 | + | BCab3 | + | BCab4 | + | BCac1 | + | BCac2 | + | BCac3 | + | BCac4 | + | BCad1 | + | BCad2 | + | BCad3 | + | BCad4 |
|---|---|---|---|---|---|---|---|---|---|---|---|---|---|---|---|---|---|---|---|---|---|---|---|---|
| 0.00 | = | -1.74 | + | -60.61 | + | -88.75 | + | 86.19 | + | -4.32 | + | 18.72 | + | 24.55 | + | 17.93 | + | -3.70 | + | 0.78 | + | 6.56 | + | 4.37 |

**Equation 9.2 Defining the Trade Balance of Country B (Mexico) with All Other Countries (BCb.acd)**

**BCb.acd = BCba1 + BCba2 + BCba3 + BCba4 + BCbc1 + BCbc2 + BCbc3 + BCbc4 + BCbd1 + BCbd2 + BCbd3 + BCbd4**

| BCb.acd | = | BCba1 | + | BCba2 | + | BCba3 | + | BCba4 | + | BCbc1 | + | BCbc2 | + | BCbc3 | + | BCbc4 | + | BCbd1 | + | BCbd2 | + | BCbd3 | + | BCbd4 |
|---|---|---|---|---|---|---|---|---|---|---|---|---|---|---|---|---|---|---|---|---|---|---|---|---|
| 0.00 | = | 1.74 | + | 60.61 | + | 88.75 | + | -86.19 | + | 3.36 | + | -13.08 | + | -28.89 | + | -17.12 | + | -3.53 | + | -2.44 | + | -1.41 | + | -1.79 |

**Equation 9.3 Defining the Trade Balance of Country C (Canada) with All Other Countries (BCc.abd)**

**BCc.abd = BCca1 + BCca2 + BCca3 + BCca4 + BCcb1 + BCcb2 + BCcb3 + BCcb4 + BCcd1 + BCcd2 + BCcd3 + BCcd4**

| BCc.abd | = | BCca1 | + | BCca2 | + | BCca3 | + | BCca4 | + | BCcb1 | + | BCcb2 | + | BCcb3 | + | BCcb4 | + | BCcd1 | + | BCcd2 | + | BCcd3 | + | BCcd4 |
|---|---|---|---|---|---|---|---|---|---|---|---|---|---|---|---|---|---|---|---|---|---|---|---|---|
| 0.00 | = | 4.32 | + | -18.72 | + | -24.55 | + | -17.93 | + | -3.36 | + | 13.08 | + | 28.89 | + | 17.12 | + | -3.70 | + | -0.35 | + | 2.77 | + | 2.43 |

**Equation 9.4 Defining the Trade Balance of Country D (Dominican Republic) with All Other Countries (BCd.abc)**

**BCd.abc = BCda1 + BCda2 + BCda3 + BCda4 + BCdb1 + BCdb2 + BCdb3 + BCdb4 + BCdc1 + BCdc2 + BCdc3 + BCdc4**

| BCd.abc | = | BCda1 | + | BCda2 | + | BCda3 | + | BCda4 | + | BCdb1 | + | BCdb2 | + | BCdb3 | + | BCdb4 | + | BCdc1 | + | BCdc2 | + | BCdc3 | + | BCdc4 |
|---|---|---|---|---|---|---|---|---|---|---|---|---|---|---|---|---|---|---|---|---|---|---|---|---|
| 0.00 | = | 3.70 | + | -0.78 | + | -6.56 | + | -4.37 | + | 3.53 | + | 2.44 | + | 1.41 | + | 1.79 | + | 3.70 | + | 0.35 | + | -2.77 | + | -2.43 |

**COMMERCIAL BALANCES**

| | | | | | | | | | | | | | | | | | | |
|---|---|---|---|---|---|---|---|---|---|---|---|---|---|---|---|---|---|---|
| **Country A - United States** | | | | | **Country B - Mexico** | | | | | **Country C - Canada** | | | | | **Country D - Dominican Republic** | | | |
| | | **Deficit or Surplus US** | | | | | **Deficit or Surplus Mex** | | | | | **Deficit or Surplus Cand** | | | | | **Deficit or Surplus Dominican Republic** | |
| EXa EU | IMa EU | US$ Bca.bcd | | | EXb Mex | IMb Mex | US$ BCb.acd | | | EXc Cand | IMc Cand | US$ BCc.abd | | | EXd RD | IMd RD | US$ BCd.abc | |
| 424 | 423.75 | 0.00 | | (0.00) | 423.74 | 423.74 | 0.00 | | | 88.58 | 88.58 | (0.00) | | | 26.93 | 26.93 | (0.00) | |
| EXab | IMab | | | | EXba | IMba | | | | EXca | IMca | | | | EXda | IMda | | |
| 341.82 | 406.73 | (64.906798) | | 0.00 | 406.73 | 341.82 | 64.91 | 2.04 | | 11.55 | 68.44 | (56.88) | (25.74) | | 5.47 | 13.49 | (8.02) | (0.08) |
| EXac | IMac | | | | EXbc | IMbc | | | | EXcb | IMcb | | | | EXdb | IMdb | | |
| 68.44 | 11.55 | 56.883252 | | (0.00) | 12.43 | 68.16 | (55.73) | (1.75) | | 68.16 | 12.43 | 55.73 | 25.21 | | 13.75 | 4.58 | 9.17 | 0.09 |
| EXad | IMad | | | | EXbd | IMbd | | | | EXcd | IMcd | | | | EXdc | IMdc | | |
| 13 | 5.47 | 8.02372 | | (0.00) | 4.58 | 13.75 | (9.17) | (0.29) | | 8.87 | 7.71 | 1.15 | 0.52 | | 7.71 | 8.87 | (1.15) | (0.01) |

# Appendix Spreadsheed 18

## Image Section 18.1

**Spreadsheet 18 - MODEL SIMULATION INTERNATIONAL TRADE OF 4 COUNTRIES:**

**SECTION 18.1: COMMERCIAL BALANCES. MODEL EQUATIONS**

**Equation 9.1 Defining the Trade Balance of Country A (United States) with All Other Countries (BCa.bcd)**

**BCa.bcd = BCab1 + BCab2 + BCab3 + BCab4 + BCac1 + BCac2 + BCac3 + BCac4 + BCad1 + BCad2 + BCad3 + BCad4**

| BCa.bcd | = | BCab1 | + | BCab2 | + | BCab3 | + | BCab4 | + | BCac1 | + | BCac2 | + | BCac3 | + | BCac4 | + | BCad1 | + | BCad2 | + | BCad3 | + | BCad4 |
|---|---|---|---|---|---|---|---|---|---|---|---|---|---|---|---|---|---|---|---|---|---|---|---|---|
| 0.00 | = | -29.53 | + | -22.46 | + | 31.20 | + | 20.80 | + | -5.99 | + | -4.76 | + | 6.45 | + | 4.30 | + | -2.85 | + | -1.33 | + | 2.51 | + | 1.67 |

**Equation 9.2 Defining the Trade Balance of Country B (Mexico) with All Other Countries (BCb.acd)**

**BCb.acd = BCba1 + BCba2 + BCba3 + BCba4 + BCbc1 + BCbc2 + BCbc3 + BCbc4 + BCbd1 + BCbd2 + BCbd3 + BCbd4**

| BCb.acd | = | BCba1 | + | BCba2 | + | BCba3 | + | BCba4 | + | BCbc1 | + | BCbc2 | + | BCbc3 | + | BCbc4 | + | BCbd1 | + | BCbd2 | + | BCbd3 | + | BCbd4 |
|---|---|---|---|---|---|---|---|---|---|---|---|---|---|---|---|---|---|---|---|---|---|---|---|---|
| 0.00 | = | 29.53 | + | 22.46 | + | -31.20 | + | -20.80 | + | 3.76 | + | 2.64 | + | -3.84 | + | -2.56 | + | -0.75 | + | 0.28 | + | 0.28 | + | 0.19 |

**Equation 9.3 Defining the Trade Balance of Country C (Canada) with All Other Countries (BCc.abd)**

**BCc.abd = BCca1 + BCca2 + BCca3 + BCca4 + BCcb1 + BCcb2 + BCcb3 + BCcb4 + BCcd1 + BCcd2 + BCcd3 + BCcd4**

| BCc.abd | = | BCca1 | + | BCca2 | + | BCca3 | + | BCca4 | + | BCcb1 | + | BCcb2 | + | BCcb3 | + | BCcb4 | + | BCcd1 | + | BCcd2 | + | BCcd3 | + | BCcd4 |
|---|---|---|---|---|---|---|---|---|---|---|---|---|---|---|---|---|---|---|---|---|---|---|---|---|
| -0.02 | = | 5.99 | + | 4.76 | + | -6.45 | + | -4.30 | + | -3.76 | + | -2.64 | + | 3.84 | + | 2.56 | + | -1.57 | + | -0.30 | + | 1.11 | + | 0.74 |

**Equation 9.4 Defining the Trade Balance of Country D (Dominican Republic) with All Other Countries (BCd.abc)**

**BCd.abc = BCda1 + BCda2 + BCda3 + BCda4 + BCdb1 + BCdb2 + BCdb3 + BCdb4 + BCdc1 + BCdc2 + BCdc3 + BCdc4**

| BCd.abc | = | BCda1 | + | BCda2 | + | BCda3 | + | BCda4 | + | BCdb1 | + | BCdb2 | + | BCdb3 | + | BCdb4 | + | BCdc1 | + | BCdc2 | + | BCdc3 | + | BCdc4 |
|---|---|---|---|---|---|---|---|---|---|---|---|---|---|---|---|---|---|---|---|---|---|---|---|---|
| 0.02 | = | 2.85 | + | 1.33 | + | -2.51 | + | -1.67 | + | 0.75 | + | -0.28 | + | -0.28 | + | -0.19 | + | 1.57 | + | 0.30 | + | -1.11 | + | -0.74 |

**COMMERCIAL BALANCES**

| | | | | | | | | | | | | | | | | | | |
|---|---|---|---|---|---|---|---|---|---|---|---|---|---|---|---|---|---|---|
| **Country A - United States** | | | | | **Country B - Mexico** | | | | | **Country C - Canada** | | | | | **Country D - Dominican Republic** | | | |
| | | **Deficit or Surplus US** | | | | | **Deficit or Surplus Mex** | | | | | **Deficit or Surplus Cand** | | | | | **Deficit or Surplus Dominican Republic** | |
| EXa EU | IMa EU | US$ Bca.bcd | | | EXb Mex | IMb Mex | US$ BCb.acd | | | EXc Cand | IMc Cand | US$ BCc.abd | | | EXd RD | IMd RD | US$ BCd.abc | |
| 67 | 66.92 | 0.00 | | (0.01) | 59.14 | 59.14 | 0.00 | | | 19.00 | 19.02 | (0.02) | | | 6.80 | 6.78 | 0.02 | |
| EXab | IMab | | | | EXba | IMba | | | | EXca | IMca | | | | EXda | IMda | | |
| 51.99 | 51.99 | 0.000000 | | 0.00 | 51.99 | 51.99 | (0.00) | (0.00) | | 10.75 | 10.75 | (0.00) | (0.00) | | 4.18 | 4.18 | (0.00) | (0.00) |
| EXac | IMac | | | | EXbc | IMbc | | | | EXcb | IMcb | | | | EXdb | IMdb | | |
| 10.75 | 10.75 | 0.000000 | | (0.01) | 6.40 | 6.39 | 0.01 | 0.00 | | 6.39 | 6.40 | (0.01) | (0.00) | | 0.75 | 0.74 | 0.01 | 0.00 |
| EXad | IMad | | | | EXbd | IMbd | | | | EXcd | IMcd | | | | EXdc | IMdc | | |
| 4 | 4.18 | 0.00000 | | 0.00 | 0.74 | 0.75 | (0.01) | (0.00) | | 1.85 | 1.86 | -0.01 | (0.00) | | 1.86 | 1.85 | 0.01 | 0.00 |

# Appendix Spreadsheed 19

## Page 1
## Image Section 19.1

**Spreadsheet 16 - MODEL SIMULATION INTERNATIONAL TRADE OF 4 COUNTRIES:**

**SECTION 16.1: COMMERCIAL BALANCES. MODEL EQUATIONS**

**Equation 9.1 Defining the Trade Balance of Country A (United States) with All Other Countries (BCa.bcd)**

**BCa.bcd = BCab1 + BCab2 + BCab3 + BCab4 + BCac1 + BCac2 + BCac3 + BCac4 + BCad1 + BCad2 + BCad3 + BCad4**

| BCa.bcd | = | BCab1 | + | BCab2 | + | BCab3 | + | BCab4 | + | BCac1 | + | BCac2 | + | BCac3 | + | BCac4 | + | BCad1 | + | BCad2 | + | BCad3 | + | BCad4 |
|---|---|---|---|---|---|---|---|---|---|---|---|---|---|---|---|---|---|---|---|---|---|---|---|---|
| -172.20 | = | -37.17 | + | -104.67 | + | -124.50 | + | 40.63 | + | -7.05 | + | 13.81 | + | 21.64 | + | 15.95 | + | -3.50 | + | 1.22 | + | 6.86 | + | 4.57 |

**Equation 9.2 Defining the Trade Balance of Country B (Mexico) with All Other Countries (BCb.acd)**

**BCb.acd = BCba1 + BCba2 + BCba3 + BCba4 + BCbc1 + BCbc2 + BCbc3 + BCbc4 + BCbd1 + BCbd2 + BCbd3 + BCbd4**

| BCb.acd | = | BCba1 | + | BCba2 | + | BCba3 | + | BCba4 | + | BCbc1 | + | BCbc2 | + | BCbc3 | + | BCbc4 | + | BCbd1 | + | BCbd2 | + | BCbd3 | + | BCbd4 |
|---|---|---|---|---|---|---|---|---|---|---|---|---|---|---|---|---|---|---|---|---|---|---|---|---|
| 214.33 | = | 37.17 | + | 104.67 | + | 124.50 | + | -40.63 | + | 11.64 | + | 0.27 | + | -18.16 | + | -9.99 | + | -1.12 | + | 1.88 | + | 2.99 | + | 1.11 |

**Equation 9.3 Defining the Trade Balance of Country C (Canada) with All Other Countries (BCc.abd)**

**BCc.abd = BCca1 + BCca2 + BCca3 + BCca4 + BCcb1 + BCcb2 + BCcb3 + BCcb4 + BCcd1 + BCcd2 + BCcd3 + BCcd4**

| BCc.abd | = | BCca1 | + | BCca2 | + | BCca3 | + | BCca4 | + | BCcb1 | + | BCcb2 | + | BCcb3 | + | BCcb4 | + | BCcd1 | + | BCcd2 | + | BCcd3 | + | BCcd4 |
|---|---|---|---|---|---|---|---|---|---|---|---|---|---|---|---|---|---|---|---|---|---|---|---|---|
| -22.37 | = | 7.05 | + | -13.81 | + | -21.64 | + | -15.95 | + | -11.64 | + | -0.27 | + | 18.16 | + | 9.99 | + | -2.96 | + | 1.07 | + | 4.23 | + | 3.41 |

**Equation 9.4 Defining the Trade Balance of Country D (Dominican Republic) with All Other Countries (BCd.abc)**

**BCd.abc = BCda1 + BCda2 + BCda3 + BCda4 + BCdb1 + BCdb2 + BCdb3 + BCdb4 + BCdc1 + BCdc2 + BCdc3 + BCdc4**

| BCd.abc | = | BCda1 | + | BCda2 | + | BCda3 | + | BCda4 | + | BCdb1 | + | BCdb2 | + | BCdb3 | + | BCdb4 | + | BCdc1 | + | BCdc2 | + | BCdc3 | + | BCdc4 |
|---|---|---|---|---|---|---|---|---|---|---|---|---|---|---|---|---|---|---|---|---|---|---|---|---|
| -19.75 | = | 3.50 | + | -1.22 | + | -6.86 | + | -4.57 | + | 1.12 | + | -1.88 | + | -2.99 | + | -1.11 | + | 2.96 | + | -1.07 | + | -4.23 | + | -3.41 |

**COMMERCIAL BALANCES**

| Country A - United States | | | | Country B - Mexico | | | Country C - Canada | | | Country D - Dominican Republic | | |
|---|---|---|---|---|---|---|---|---|---|---|---|---|
| | | Deficit or Surplus US | | | | Deficit or Surplus Mex | | | Deficit or Surplus Cand | | | Deficit or Surplus Dominican Republic |
| EXa EU | IMa EU | US$ Bca.bcd | | EXb Mex | IMb Mex | US$ BCb.acd | EXc Cand | IMc Cand | US$ BCc.abd | EXd RD | IMd RD | US$ BCd.abc |
| 285 | 457.67 | (172.20) | (4.09) | 477.87 | 263.54 | 214.33 | 70.88 | 93.25 | (22.37) | 17.30 | 37.05 | (19.75) |
| EXab | IMab | | | EXba | IMba | | EXca | IMca | | EXda | IMda | |
| 211.99 | 437.70 | ############ | 5.90 | 437.70 | 211.99 | 225.71 | 14.78 | 59.13 | (44.35) | 5.19 | 14.35 | (9.16) |
| EXac | IMac | | | EXbc | IMbc | | EXcb | IMcb | | EXdb | IMdb | |
| 59.13 | 14.78 | 44.352629 | (9.79) | 28.66 | 44.90 | (16.24) | 44.90 | 28.66 | 16.24 | 6.66 | 11.51 | (4.85) |
| EXad | IMad | | | EXbd | IMbd | | EXcd | IMcd | | EXdc | IMdc | |
| 14 | 5.19 | 9.15837 | (0.20) | 11.51 | 6.66 | 4.85 | 11.20 | 5.46 | 5.74 | 5.46 | 11.20 | (5.74) |

# Appendix Spreadsheed 19

## Page 1
## Image Section 19.7 and 19.8

**SECTION 16.7: PRIMARY, CONTRA PRIMARY AND NON-PRIMARY CHANGE RATES**

| | |
|---|---|
| 0.027541 | **TCab = Primary and equivalent exchange rate of country A with country B** |
| 0.437717 | **TCac = Primary and equivalent exchange rate of country A with country C** |
| 0.010079 | TCad = Tasa de cambio primaria y equivalente del país D con el país A |
| 36.309502 | **TCba = 1/TCab = Contra-Primary and equivalent exchange rate of country B with country A** |
| 2.284581 | **TCca = 1/TCac = Contra-Primary and equivalent exchange rate of country C with country A** |
| 99.21619 | TCda = 1/TCad = Tasa de cambio equivalente del país D con el país A |
| 15.893286 | **TCbc = TCac/TCab = Non-Primary and equivalent exchange rate of country B with country C** |
| 0.062920 | **TCcb = TCab/TCac = Non-Primary and equivalent exchange rate of country C with country B** |
| 0.365963 | TCbd = TCad/TCab = Tasa de cambio equivalente del país B con el país D |
| 2.732513 | TCdb = TCab/TCad = Tasa de cambio equivalente del país D con el país B |
| 0.023026 | TCcd = TCad/TCac = Tasa de cambio equivalente del país C con el país D |
| 43.428614 | TCdc = TCac/TCad = Tasa de cambio equivalente del país D con el país C |

| 1 | 2 | 3 |
|---|---|---|
| BC | BC/100 | BC-(BC/100) |
| 214.33 | 2.1433 | 212.18 |
| (22.37) | (0.2237) | (22.15) |
| (19.75) | (0.1975) | (19.55) |

**SECTION 16.7: INITIAL CHANGE RATES**

| | |
|---|---|
| 0.027541 | **TCab In = (Pa1+Pa2+Pa3+Pa4)/(Pb1+Pb2+Pb3+Pb4) = Initial exchange rate of country A with country B** |
| 0.437717 | **TCac In = (Pa1+Pa2+Pa3+Pa4)/(Pc1+Pc2+Pc3+Pc4) = Initial exchange rate of country A with country C** |
| 0.010079 | TCad Inicial = (Pa1+Pa2+Pa3+Pa4)/(Pd1+Pd2+Pd3+Pd4) = Tasa de cambio Injcial del país A con el país D |
| 36.310144 | **TCba In = (Pb1+Pb2+Pb3+Pb4)/(Pa1+Pa2+Pa3+Pa4) = 1/Tcab = Initial exchange rate of country B with country A** |
| 2.284580 | **TCca In = (Pc1+Pc2+Pc3+Pc4)/(Pa1+Pa2+Pa3+Pa4) = 1/TCac = Initial exchange rate of country C with country A** |
| 99.213435 | TCda Inicial = (Pd1+Pd2+Pd3+Pd4)/(Pa1+Pa2+Pa3+Pa4) = Tasa de cambio Injcial del país D con el país A |
| 15.893572 | **TCbc In = (Pb1+Pb2+Pb3+Pb4)/(Pc1+Pc2+Pc3+Pc4) = TCba/TCca =Initial exchange rate of country B with country C** |
| 0.062919 | **TCcb In = (Pc1+Pc2+Pc3+Pc4)/(Pb1+Pb2+Pb3+Pb4) = TCab/TCac = Initial exchange rate of country C with country B** |
| 0.365980 | TCbd Inicial = (Pb1+Pb2+Pb3+Pb4)/(Pd1+Pd2+Pd3+Pd4) = TCba/TCda = Tasa de cambio equivalente del país B con el país D |
| 2.732389 | TCdb Inicial = (Pd1+Pd2+Pd3+Pd4)/(Pb1+Pb2+Pb3+Pb4) = TCda/TCba = Tasa de cambio equivalente del país D con el país B |
| 0.023027 | TCcd Inicial = (Pc1+Pc2+Pc3+Pc4)/(Pd1+Pd2+Pd3+Pd4) = TCca/TCda = Tasa de cambio equivalente del país C con el país D |
| 43.427419 | TCdc Inicial = (Pd1+Pd2+Pd3+Pd4)/(Pc1+Pc2+Pc3+Pc4) = TCda/TCca = Tasa de cambio equivalente del país D con el país C |

# Appendix Spreadsheed 19

## Page 2
## Image Section 19.1

**Spreadsheet 16 - MODEL SIMULATION INTERNATIONAL TRADE OF 4 COUNTRIES:**

**SECTION 16.1: COMMERCIAL BALANCES. MODEL EQUATIONS**

**Equation 9.1 Defining the Trade Balance of Country A (United States) with All Other Countries (BCa.bcd)**

**BCa.bcd = BCab1 + BCab2 + BCab3 + BCab4 + BCac1 + BCac2 + BCac3 + BCac4 + BCad1 + BCad2 + BCad3 + BCad4**

| | | | | | | | | | | | | | | | | | | | | | | | | |
|---|---|---|---|---|---|---|---|---|---|---|---|---|---|---|---|---|---|---|---|---|---|---|---|---|
| BCa.bcd | = | BCab1 | + | BCab2 | + | BCab3 | + | BCab4 | + | BCac1 | + | BCac2 | + | BCac3 | + | BCac4 | + | BCad1 | + | BCad2 | + | BCad3 | + | BCad4 |
| -171.82 | = | -37.08 | + | -104.57 | + | -124.42 | + | 40.74 | + | -7.04 | + | 13.84 | + | 21.66 | + | 15.97 | + | -3.52 | + | 1.19 | + | 6.84 | + | 4.56 |

**Equation 9.2 Defining the Trade Balance of Country B (Mexico) with All Other Countries (BCb.acd)**

**BCb.acd = BCba1 + BCba2 + BCba3 + BCba4 + BCbc1 + BCbc2 + BCbc3 + BCbc4 + BCbd1 + BCbd2 + BCbd3 + BCbd4**

| | | | | | | | | | | | | | | | | | | | | | | | | |
|---|---|---|---|---|---|---|---|---|---|---|---|---|---|---|---|---|---|---|---|---|---|---|---|---|
| BCb.acd | = | BCba1 | + | BCba2 | + | BCba3 | + | BCba4 | + | BCbc1 | + | BCbc2 | + | BCbc3 | + | BCbc4 | + | BCbd1 | + | BCbd2 | + | BCbd3 | + | BCbd4 |
| 213.75 | = | 37.08 | + | 104.57 | + | 124.42 | + | -40.74 | + | 11.64 | + | 0.26 | + | -18.18 | + | -10.01 | + | -1.15 | + | 1.83 | + | 2.94 | + | 1.08 |

**Equation 9.3 Defining the Trade Balance of Country C (Canada) with All Other Countries (BCc.abd)**

**BCc.abd = BCca1 + BCca2 + BCca3 + BCca4 + BCcb1 + BCcb2 + BCcb3 + BCcb4 + BCcd1 + BCcd2 + BCcd3 + BCcd4**

| | | | | | | | | | | | | | | | | | | | | | | | | |
|---|---|---|---|---|---|---|---|---|---|---|---|---|---|---|---|---|---|---|---|---|---|---|---|---|
| BCc.abd | = | BCca1 | + | BCca2 | + | BCca3 | + | BCca4 | + | BCcb1 | + | BCcb2 | + | BCcb3 | + | BCcb4 | + | BCcd1 | + | BCcd2 | + | BCcd3 | + | BCcd4 |
| -22.53 | = | 7.04 | + | -13.84 | + | -21.66 | + | -15.97 | + | -11.64 | + | -0.26 | + | 18.18 | + | 10.01 | + | -2.99 | + | 1.03 | + | 4.19 | + | 3.38 |

**Equation 9.4 Defining the Trade Balance of Country D (Dominican Republic) with All Other Countries (BCd.abc)**

**BCd.abc = BCda1 + BCda2 + BCda3 + BCda4 + BCdb1 + BCdb2 + BCdb3 + BCdb4 + BCdc1 + BCdc2 + BCdc3 + BCdc4**

| | | | | | | | | | | | | | | | | | | | | | | | | |
|---|---|---|---|---|---|---|---|---|---|---|---|---|---|---|---|---|---|---|---|---|---|---|---|---|
| BCd.abc | = | BCda1 | + | BCda2 | + | BCda3 | + | BCda4 | + | BCdb1 | + | BCdb2 | + | BCdb3 | + | BCdb4 | + | BCdc1 | + | BCdc2 | + | BCdc3 | + | BCdc4 |
| -19.40 | = | 3.52 | + | -1.19 | + | -6.84 | + | -4.56 | + | 1.15 | + | -1.83 | + | -2.94 | + | -1.08 | + | 2.99 | + | -1.03 | + | -4.19 | + | -3.38 |

**COMMERCIAL BALANCES**

| | | | | | | | | | | | | | | | | |
|---|---|---|---|---|---|---|---|---|---|---|---|---|---|---|---|---|
| **Country A - United States** | | | | **Country B - Mexico** | | | | | **Country C - Canada** | | | | **Country D - Dominican Republic** | | | |
| | | **Deficit or Surplus US** | | | | **Deficit or Surplus Mex** | | | | | **Deficit or Surplus Cand** | | | | **Deficit or Surplus Dominican Republic** | |
| EXa EU | IMa EU | US$ Bca.bcd | | EXb Mex | IMb Mex | US$ BCb.acd | | | EXc Cand | IMc Cand | US$ BCc.abd | | EXd RD | IMd RD | US$ BCd.abc | |
| 286 | 457.60 | (171.82) | (4.17) | 477.71 | 263.96 | 213.75 | 212.18 | | 70.82 | 93.35 | (22.53) | (22.15) | 17.44 | 36.84 | (19.40) | (19.55) |
| EXab | IMab | | | EXba | IMba | | | | EXca | IMca | | | EXda | IMda | | |
| 212.31 | 437.64 | (225.331424) | 5.89 | 437.64 | 212.31 | 225.33 | | | 14.76 | 59.19 | (44.44) | | 5.21 | 14.29 | (9.08) | |
| EXac | IMac | | | EXbc | IMbc | | | | EXcb | IMcb | | | EXdb | IMdb | | |
| 59.19 | 14.76 | 44.436570 | (9.86) | 28.64 | 44.93 | (16.29) | | | 44.93 | 28.64 | 16.29 | | 6.72 | 11.43 | (4.71) | |
| EXad | IMad | | | EXbd | IMbd | | | | EXcd | IMcd | | | EXdc | IMdc | | |
| 14 | 5.21 | 9.07610 | (0.20) | 11.43 | 6.72 | 4.71 | | | 11.13 | 5.51 | 5.61 | | 5.51 | 11.13 | (5.61) | |

# Appendix Spreadsheed 19

## Page 2
## Image Section 19.7 and 19.8

| | | | | |
|---|---|---|---|---|
| | **SECTION 16.7: PRIMARY, CONTRA PRIMARY AND NON-PRIMARY CHANGE RATES** | Set CUCTC(1/100) | | **SECTION 16.7: INITIAL CHANGE RATES** |
| 0.027551 | **TCab = Primary and equivalent exchange rate of country A with country B** | 0.00001000 | 0.027541 | **TCab In = (Pa1+Pa2+Pa3+Pa4)/(Pb1+Pb2+Pb3+Pb4) = Initial exchange rate of country A with country B** |
| 0.437817 | **TCac = Primary and equivalent exchange rate of country A with country C** | 0.00010000 | 0.437717 | **TCac In = (Pa1+Pa2+Pa3+Pa4)/(Pc1+Pc2+Pc3+Pc4) = Initial exchange rate of country A with country C** |
| 0.010069 | TCad = Tasa de cambio primaria y equivalente del país D con el país A | -0.00001000 | 0.010079 | TCad Inicial = (Pa1+Pa2+Pa3+Pa4)/(Pd1+Pd2+Pd3+Pd4) = Tasa de cambio Injcial del país A con el país D |
| 36.296323 | **TCba = 1/TCab = Contra-Primary and equivalent exchange rate of country B with country A** | | 36.310144 | **TCba In = (Pb1+Pb2+Pb3+Pb4)/(Pa1+Pa2+Pa3+Pa4) = 1/Tcab = Initial exchange rate of country B with country A** |
| 2.284059 | **TCca = 1/TCac = Contra-Primary and equivalent exchange rate of country C with country A** | 1 | 2.284580 | **TCca In = (Pc1+Pc2+Pc3+Pc4)/(Pa1+Pa2+Pa3+Pa4) = 1/TCac = Initial exchange rate of country C with country A** |
| 99.31473 | TCda = 1/TCad = Tasa de cambio equivalente del país D con el país A | | 99.213435 | TCda Inicial = (Pd1+Pd2+Pd3+Pd4)/(Pa1+Pa2+Pa3+Pa4) = Tasa de cambio Injcial del país D con el país A |
| 15.891147 | **TCbc = TCac/TCab = Non-Primary and equivalent exchange rate of country B with country C** | | 15.893572 | **TCbc In = (Pb1+Pb2+Pb3+Pb4)/(Pc1+Pc2+Pc3+Pc4) = TCba/TCca =Initial exchange rate of country B with country C** |
| 0.062928 | **TCcb = TCab/TCac = Non-Primary and equivalent exchange rate of country C with country B** | | 0.062919 | **TCcb In = (Pc1+Pc2+Pc3+Pc4)/(Pb1+Pb2+Pb3+Pb4) = TCab/TCac = Initial exchange rate of country C with country B** |
| 0.365468 | TCbd = TCad/TCab = Tasa de cambio equivalente del país B con el país D | | 0.365980 | TCbd Inicial = (Pb1+Pb2+Pb3+Pb4)/(Pd1+Pd2+Pd3+Pd4) = TCba/TCda = Tasa de cambio equivalente del país B con el país D |
| 2.736220 | TCdb = TCab/TCad = Tasa de cambio equivalente del país D con el país B | | 2.732389 | TCdb Inicial = (Pd1+Pd2+Pd3+Pd4)/(Pb1+Pb2+Pb3+Pb4) = TCda/TCba = Tasa de cambio equivalente del país D con el país B |
| 0.022998 | TCcd = TCad/TCac = Tasa de cambio equivalente del país C con el país D | | 0.023027 | TCcd Inicial = (Pc1+Pc2+Pc3+Pc4)/(Pd1+Pd2+Pd3+Pd4) = TCca/TCda = Tasa de cambio equivalente del país C con el país D |
| 43.481676 | TCdc = TCac/TCad = Tasa de cambio equivalente del país D con el país C | | 43.427419 | TCdc Inicial = (Pd1+Pd2+Pd3+Pd4)/(Pc1+Pc2+Pc3+Pc4) = TCda/TCca = Tasa de cambio equivalente del país D con el país C |

# Appendix Spreadsheed 19

## Page 3
## Image Section 19.1

**Spreadsheet 16 - MODEL SIMULATION INTERNATIONAL TRADE OF 4 COUNTRIES:**

**SECTION 16.1: COMMERCIAL BALANCES. MODEL EQUATIONS**

**Equation 9.1 Defining the Trade Balance of Country A (United States) with All Other Countries (BCa.bcd)**

**BCa.bcd = BCab1 + BCab2 + BCab3 + BCab4 + BCac1 + BCac2 + BCac3 + BCac4 + BCad1 + BCad2 + BCad3 + BCad4**

| | | | | | | | | | | | | | | | | | | | | | | | | |
|---|---|---|---|---|---|---|---|---|---|---|---|---|---|---|---|---|---|---|---|---|---|---|---|---|
| BCa.bcd | = | BCab1 | + | BCab2 | + | BCab3 | + | BCab4 | + | BCac1 | + | BCac2 | + | BCac3 | + | BCac4 | + | BCad1 | + | BCad2 | + | BCad3 | + | BCad4 |
| -170.48 | = | -36.82 | + | -104.25 | + | -124.16 | + | 41.09 | + | -7.02 | + | 13.87 | + | 21.68 | + | 15.98 | + | -3.50 | + | 1.22 | + | 6.86 | + | 4.57 |

**Equation 9.2 Defining the Trade Balance of Country B (Mexico) with All Other Countries (BCb.acd)**

**BCb.acd = BCba1 + BCba2 + BCba3 + BCba4 + BCbc1 + BCbc2 + BCbc3 + BCbc4 + BCbd1 + BCbd2 + BCbd3 + BCbd4**

| | | | | | | | | | | | | | | | | | | | | | | | | |
|---|---|---|---|---|---|---|---|---|---|---|---|---|---|---|---|---|---|---|---|---|---|---|---|---|
| BCb.acd | = | BCba1 | + | BCba2 | + | BCba3 | + | BCba4 | + | BCbc1 | + | BCbc2 | + | BCbc3 | + | BCbc4 | + | BCbd1 | + | BCbd2 | + | BCbd3 | + | BCbd4 |
| 212.18 | = | 36.82 | + | 104.25 | + | 124.16 | + | -41.09 | + | 11.60 | + | 0.13 | + | -18.31 | + | -10.09 | + | -1.15 | + | 1.83 | + | 2.94 | + | 1.08 |

**Equation 9.3 Defining the Trade Balance of Country C (Canada) with All Other Countries (BCc.abd)**

**BCc.abd = BCca1 + BCca2 + BCca3 + BCca4 + BCcb1 + BCcb2 + BCcb3 + BCcb4 + BCcd1 + BCcd2 + BCcd3 + BCcd4**

| | | | | | | | | | | | | | | | | | | | | | | | | |
|---|---|---|---|---|---|---|---|---|---|---|---|---|---|---|---|---|---|---|---|---|---|---|---|---|
| BCc.abd | = | BCca1 | + | BCca2 | + | BCca3 | + | BCca4 | + | BCcb1 | + | BCcb2 | + | BCcb3 | + | BCcb4 | + | BCcd1 | + | BCcd2 | + | BCcd3 | + | BCcd4 |
| -22.15 | = | 7.02 | + | -13.87 | + | -21.68 | + | -15.98 | + | -11.60 | + | -0.13 | + | 18.31 | + | 10.09 | + | -2.97 | + | 1.05 | + | 4.22 | + | 3.40 |

**Equation 9.4 Defining the Trade Balance of Country D (Dominican Republic) with All Other Countries (BCd.abc)**

**BCd.abc = BCda1 + BCda2 + BCda3 + BCda4 + BCdb1 + BCdb2 + BCdb3 + BCdb4 + BCdc1 + BCdc2 + BCdc3 + BCdc4**

| | | | | | | | | | | | | | | | | | | | | | | | | |
|---|---|---|---|---|---|---|---|---|---|---|---|---|---|---|---|---|---|---|---|---|---|---|---|---|
| BCd.abc | = | BCda1 | + | BCda2 | + | BCda3 | + | BCda4 | + | BCdb1 | + | BCdb2 | + | BCdb3 | + | BCdb4 | + | BCdc1 | + | BCdc2 | + | BCdc3 | + | BCdc4 |
| -19.55 | = | 3.50 | + | -1.22 | + | -6.86 | + | -4.57 | + | 1.15 | + | -1.83 | + | -2.94 | + | -1.08 | + | 2.97 | + | -1.05 | + | -4.22 | + | -3.40 |

**COMMERCIAL BALANCES**

| | | | | | | | | | | | | | | | | | | |
|---|---|---|---|---|---|---|---|---|---|---|---|---|---|---|---|---|---|---|
| **Country A - United States** | | | | | **Country B - Mexico** | | | | | | **Country C - Canada** | | | | | **Country D - Dominican Republic** | | |
| | | **Deficit or Surplus US** | | | | | **Deficit or Surplus Mex** | | | | | **Deficit or Surplus Cand** | | | | | **Deficit or Surplus Dominican Republic** | |
| EXa EU | IMa EU | US$ Bca.bcd | | | EXb Mex | IMb Mex | US$ BCb.acd | | | EXc Cand | IMc Cand | US$ BCc.abd | | | EXd RD | IMd RD | US$ BCd.abc | |
| 287 | 457.38 | (170.48) | | (4.04) | 477.32 | 265.14 | 212.18 | 212.18 | | 71.02 | 93.16 | (22.15) | (22.15) | | 17.40 | 36.95 | (19.55) | (19.55) |
| EXab | IMab | | | | EXba | IMba | | | | EXca | IMca | | | | EXda | IMda | | |
| 213.30 | 437.45 | ############ | | 5.85 | 437.45 | 213.30 | 224.14 | | | 14.74 | 59.25 | (44.51) | | | 5.19 | 14.35 | (9.16) | |
| EXac | IMac | | | | EXbc | IMbc | | | | EXcb | IMcb | | | | EXdb | IMdb | | |
| 59.25 | 14.74 | 44.506253 | | (9.70) | 28.44 | 45.10 | (16.66) | | | 45.10 | 28.44 | 16.66 | | | 6.73 | 11.43 | (4.70) | |
| EXad | IMad | | | | EXbd | IMbd | | | | EXcd | IMcd | | | | EXdc | IMdc | | |
| 14 | 5.19 | 9.15590 | | (0.20) | 11.43 | 6.73 | 4.70 | | | 11.17 | 5.48 | 5.70 | | | 5.48 | 11.17 | (5.70) | |

# Appendix Spreadsheed 19

## Page 3
## Image Section 19.7 and 19.8

| | | |
|---|---|---|
| | **SECTION 16.7: PRIMARY, CONTRA PRIMARY AND NON-PRIMARY CHANGE RATES** | Set CUCTC(1/100) |
| 0.027582 | **TCab = Primary and equivalent exchange rate of country A with country B** | 0.00004120 |
| 0.437900 | **TCac = Primary and equivalent exchange rate of country A with country C** | 0.00018300 |
| 0.010079 | TCad = Tasa de cambio primaria y equivalente del país D con el país A | -0.00000030 |
| 36.255266 | **TCba = 1/TCab = Contra-Primary and equivalent exchange rate of country B with country A** | |
| 2.283626 | **TCca = 1/TCac = Contra-Primary and equivalent exchange rate of country C with country A** | 1 |
| 99.21915 | TCda = 1/TCad = Tasa de cambio equivalente del país D con el país A | |
| 15.876181 | **TCbc = TCac/TCab = Non-Primary and equivalent exchange rate of country B with country C** | |
| 0.062987 | **TCcb = TCab/TCac = Non-Primary and equivalent exchange rate of country C with country B** | |
| 0.365406 | TCbd = TCad/TCab = Tasa de cambio equivalente del país B con el país D | |
| 2.736682 | TCdb = TCab/TCad = Tasa de cambio equivalente del país D con el país B | |
| 0.023016 | TCcd = TCad/TCac = Tasa de cambio equivalente del país C con el país D | |
| 43.448064 | TCdc = TCac/TCad = Tasa de cambio equivalente del país D con el país C | |

| | |
|---|---|
| | **SECTION 16.7: INITIAL CHANGE RATES** |
| 0.027541 | **TCab In = (Pa1+Pa2+Pa3+Pa4)/(Pb1+Pb2+Pb3+Pb4) = Initial exchange rate of country A with country B** |
| 0.437717 | **TCac In = (Pa1+Pa2+Pa3+Pa4)/(Pc1+Pc2+Pc3+Pc4) = Initial exchange rate of country A with country C** |
| 0.010079 | TCad Inicial = (Pa1+Pa2+Pa3+Pa4)/(Pd1+Pd2+Pd3+Pd4) = Tasa de cambio Injcial del país A con el país D |
| 36.310144 | **TCba In = (Pb1+Pb2+Pb3+Pb4)/(Pa1+Pa2+Pa3+Pa4) = 1/Tcab = Initial exchange rate of country B with country A** |
| 2.284580 | **TCca In = (Pc1+Pc2+Pc3+Pc4)/(Pa1+Pa2+Pa3+Pa4) = 1/TCac = Initial exchange rate of country C with country A** |
| 99.213435 | TCda Inicial = (Pd1+Pd2+Pd3+Pd4)/(Pa1+Pa2+Pa3+Pa4) = Tasa de cambio Injcial del país D con el país A |
| 15.893572 | **TCbc In = (Pb1+Pb2+Pb3+Pb4)/(Pc1+Pc2+Pc3+Pc4) = TCba/TCca =Initial exchange rate of country B with country C** |
| 0.062919 | **TCcb In = (Pc1+Pc2+Pc3+Pc4)/(Pb1+Pb2+Pb3+Pb4) = TCab/TCac = Initial exchange rate of country C with country B** |
| 0.365980 | TCbd Inicial = (Pb1+Pb2+Pb3+Pb4)/(Pd1+Pd2+Pd3+Pd4) = TCba/TCda = Tasa de cambio equivalente del país B con el país D |
| 2.732389 | TCdb Inicial = (Pd1+Pd2+Pd3+Pd4)/(Pb1+Pb2+Pb3+Pb4) = TCda/TCba = Tasa de cambio equivalente del país D con el país B |
| 0.023027 | TCcd Inicial = (Pc1+Pc2+Pc3+Pc4)/(Pd1+Pd2+Pd3+Pd4) = TCca/TCda = Tasa de cambio equivalente del país C con el país D |
| 43.427419 | TCdc Inicial = (Pd1+Pd2+Pd3+Pd4)/(Pc1+Pc2+Pc3+Pc4) = TCda/TCca = Tasa de cambio equivalente del país D con el país C |

# Appendix Spreadsheed 19

## Page 4

## Image Section 19.1

**Spreadsheet 16 - MODEL SIMULATION INTERNATIONAL TRADE OF 4 COUNTRIES:**

**SECTION 16.1: COMMERCIAL BALANCES. MODEL EQUATIONS**

**Equation 9.1 Defining the Trade Balance of Country A (United States) with All Other Countries (BCa.bcd)**

**BCa.bcd = BCab1 + BCab2 + BCab3 + BCab4 + BCac1 + BCac2 + BCac3 + BCac4 + BCad1 + BCad2 + BCad3 + BCad4**

| | | | | | | | | | | | | | |
|---|---|---|---|---|---|---|---|---|---|---|---|---|---|
| **BCa.bcd** | **=** | **BCab1** | **+** | **BCab2** | **+** | **BCab3** | **+** | **BCab4** | **+** | **BCac1** | **+** | **BCac2** | **+** |
| **14.16** | **=** | **0.45** | **+** | **-57.85** | **+** | **-86.49** | **+** | **88.93** | **+** | **-3.64** | **+** | **19.93** | **+** |

| | | | | | | | | | | |
|---|---|---|---|---|---|---|---|---|---|---|
| **BCac3** | **+** | **BCac4** | **+** | **BCad1** | **+** | **BCad2** | **+** | **BCad3** | **+** | **BCad4** |
| **25.26** | **+** | **18.42** | **+** | **-3.50** | **+** | **1.22** | **+** | **6.86** | **+** | **4.57** |

**Equation 9.2 Defining the Trade Balance of Country B (Mexico) with All Other Countries (BCb.acd)**

**BCb.acd = BCba1 + BCba2 + BCba3 + BCba4 + BCbc1 + BCbc2 + BCbc3 + BCbc4 + BCbd1 + BCbd2 + BCbd3 + BCbd4**

| | | | | | | | | | | | | | |
|---|---|---|---|---|---|---|---|---|---|---|---|---|---|
| **BCb.acd** | **=** | **BCba1** | **+** | **BCba2** | **+** | **BCba3** | **+** | **BCba4** | **+** | **BCbc1** | **+** | **BCbc2** | **+** |
| **-9.61** | **=** | **-0.45** | **+** | **57.85** | **+** | **86.49** | **+** | **-88.93** | **+** | **3.46** | **+** | **-13.08** | **+** |

| | | | | | | | | | | |
|---|---|---|---|---|---|---|---|---|---|---|
| **BCbc3** | **+** | **BCbc4** | **+** | **BCbd1** | **+** | **BCbd2** | **+** | **BCbd3** | **+** | **BCbd4** |
| **-29.07** | **+** | **-17.22** | **+** | **-3.51** | **+** | **-2.26** | **+** | **-1.21** | **+** | **-1.67** |

**Equation 9.3 Defining the Trade Balance of Country C (Canada) with All Other Countries (BCc.abd)**

**BCc.abd = BCca1 + BCca2 + BCca3 + BCca4 + BCcb1 + BCcb2 + BCcb3 + BCcb4 + BCcd1 + BCcd2 + BCcd3 + BCcd4**

| | | | | | | | | | | | | | |
|---|---|---|---|---|---|---|---|---|---|---|---|---|---|
| **BCc.abd** | **=** | **BCca1** | **+** | **BCca2** | **+** | **BCca3** | **+** | **BCca4** | **+** | **BCcb1** | **+** | **BCcb2** | **+** |
| **-2.32** | **=** | **3.64** | **+** | **-19.93** | **+** | **-25.26** | **+** | **-18.42** | **+** | **-3.46** | **+** | **13.08** | **+** |

| | | | | | | | | | | |
|---|---|---|---|---|---|---|---|---|---|---|
| **BCcb3** | **+** | **BCcb4** | **+** | **BCcd1** | **+** | **BCcd2** | **+** | **BCcd3** | **+** | **BCcd4** |
| **29.07** | **+** | **17.22** | **+** | **-3.69** | **+** | **-0.16** | **+** | **3.00** | **+** | **2.59** |

**Equation 9.4 Defining the Trade Balance of Country D (Dominican Republic) with All Other Countries (BCd.abc)**

**BCd.abc = BCda1 + BCda2 + BCda3 + BCda4 + BCdb1 + BCdb2 + BCdb3 + BCdb4 + BCdc1 + BCdc2 + BCdc3 + BCdc4**

| | | | | | | | | | | | | | |
|---|---|---|---|---|---|---|---|---|---|---|---|---|---|
| **BCd.abc** | **=** | **BCda1** | **+** | **BCda2** | **+** | **BCda3** | **+** | **BCda4** | **+** | **BCdb1** | **+** | **BCdb2** | **+** |
| **-2.23** | **=** | **3.50** | **+** | **-1.22** | **+** | **-6.86** | **+** | **-4.57** | **+** | **3.51** | **+** | **2.26** | **+** |

| | | | | | | | | | | |
|---|---|---|---|---|---|---|---|---|---|---|
| **BCdb3** | **+** | **BCdb4** | **+** | **BCdc1** | **+** | **BCdc2** | **+** | **BCdc3** | **+** | **BCdc4** |
| **1.21** | **+** | **1.67** | **+** | **3.69** | **+** | **0.16** | **+** | **-3.00** | **+** | **-2.59** |

**COMMERCIAL BALANCES**

**Country A - United States**

| | | |
|---|---|---|
| | | **Deficit or Surplus US** |
| EXa EU | IMa EU | US$ Bca.bcd |
| 435 | 420.46 | 14.16 |
| EXab | IMab | |
| 349.56 | 404.52 | (54.965898) |
| EXac | IMac | |
| 70.72 | 10.75 | 59.970269 |
| EXad | IMad | |
| 14 | 5.19 | 9.15590 |

**Country B - Mexico**

| | | | | |
|---|---|---|---|---|
| | | | **Deficit or Surplus Mex** | |
| | EXb Mex | IMb Mex | US$ BCb.acd | |
| (1.39) | 422.07 | 431.68 | (9.61) | 212.18 |
| | EXba | IMba | | |
| (0.30) | 404.52 | 349.56 | 54.97 | |
| | EXbc | IMbc | | |
| (1.06) | 12.62 | 68.53 | (55.91) | |
| | EXbd | IMbd | | |
| (0.02) | 4.93 | 13.60 | (8.66) | |

**Country C - Canada**

| | | | |
|---|---|---|---|
| | | **Deficit or Surplus Cand** | |
| EXc Cand | IMc Cand | US$ BCc.abd | |
| 88.53 | 90.85 | (2.32) | (22.15) |
| EXca | IMca | | |
| 10.75 | 70.72 | (59.97) | |
| EXcb | IMcb | | |
| 68.53 | 12.62 | 55.91 | |
| EXcd | IMcd | | |
| 9.25 | 7.51 | 1.74 | |

**Country D - Dominican Republic**

| | | | |
|---|---|---|---|
| | | **Deficit or Surplus Dominican Republic** | |
| EXd RD | IMd RD | US$ BCd.abc | |
| 26.29 | 28.52 | (2.23) | (19.55) |
| EXda | IMda | | |
| 5.19 | 14.35 | (9.16) | |
| EXdb | IMdb | | |
| 13.60 | 4.93 | 8.66 | |
| EXdc | IMdc | | |
| 7.51 | 9.25 | (1.74) | |

# Appendix Spreadsheed 19

## Page 4

## Image Section 19.7 and 19.8

| | **SECTION 16.7: PRIMARY, CONTRA PRIMARY AND NON-PRIMARY CHANGE RATES** | Set CUCTC | | **SECTION 16.7: INITIAL CHANGE RATES** |
|---|---|---|---|---|
| 0.031661 | **TCab = Primary and equivalent exchange rate of country A with country B** | 0.00412000 | 0.027541 | **TCab In = (Pa1+Pa2+Pa3+Pa4)/(Pb1+Pb2+Pb3+Pb4) = Initial exchange rate of country A with country B** |
| 0.456017 | **TCac = Primary and equivalent exchange rate of country A with country C** | 0.01830000 | 0.437717 | **TCac In = (Pa1+Pa2+Pa3+Pa4)/(Pc1+Pc2+Pc3+Pc4) = Initial exchange rate of country A with country C** |
| 0.010079 | TCad = Tasa de cambio primaria y equivalente del país D con el país A | -0.00000030 | 0.010079 | TCad Inicial = (Pa1+Pa2+Pa3+Pa4)/(Pd1+Pd2+Pd3+Pd4) = Tasa de cambio Injcial del país A con el país D |
| 31.584599 | **TCba = 1/TCab = Contra-Primary and equivalent exchange rate of country B with country A** | | 36.310144 | **TCba In = (Pb1+Pb2+Pb3+Pb4)/(Pa1+Pa2+Pa3+Pa4) = 1/Tcab = Initial exchange rate of country B with country A** |
| 2.192901 | **TCca = 1/TCac = Contra-Primary and equivalent exchange rate of country C with country A** | 100 | 2.284580 | **TCca In = (Pc1+Pc2+Pc3+Pc4)/(Pa1+Pa2+Pa3+Pa4) = 1/TCac = Initial exchange rate of country C with country A** |
| 99.21915 | TCda = 1/TCad = Tasa de cambio equivalente del país D con el país A | | 99.213435 | TCda Inicial = (Pd1+Pd2+Pd3+Pd4)/(Pa1+Pa2+Pa3+Pa4) = Tasa de cambio Injcial del país D con el país A |
| 14.403114 | **TCbc = TCac/TCab = Non-Primary and equivalent exchange rate of country B with country C** | | 15.893572 | **TCbc In = (Pb1+Pb2+Pb3+Pb4)/(Pc1+Pc2+Pc3+Pc4) = TCba/TCca =Initial exchange rate of country B with country C** |
| 0.069429 | **TCcb = TCab/TCac = Non-Primary and equivalent exchange rate of country C with country B** | | 0.062919 | **TCcb In = (Pc1+Pc2+Pc3+Pc4)/(Pb1+Pb2+Pb3+Pb4) = TCab/TCac = Initial exchange rate of country C with country B** |
| 0.318332 | TCbd = TCad/TCab = Tasa de cambio equivalente del país B con el país D | | 0.365980 | TCbd Inicial = (Pb1+Pb2+Pb3+Pb4)/(Pd1+Pd2+Pd3+Pd4) = TCba/TCda = Tasa de cambio equivalente del país B con el país D |
| 3.141377 | TCdb = TCab/TCad = Tasa de cambio equivalente del país D con el país B | | 2.732389 | TCdb Inicial = (Pd1+Pd2+Pd3+Pd4)/(Pb1+Pb2+Pb3+Pb4) = TCda/TCba = Tasa de cambio equivalente del país D con el país B |
| 0.022102 | TCcd = TCad/TCac = Tasa de cambio equivalente del país C con el país D | | 0.023027 | TCcd Inicial = (Pc1+Pc2+Pc3+Pc4)/(Pd1+Pd2+Pd3+Pd4) = TCca/TCda = Tasa de cambio equivalente del país C con el país D |
| 45.245617 | TCdc = TCac/TCad = Tasa de cambio equivalente del país D con el país C | | 43.427419 | TCdc Inicial = (Pd1+Pd2+Pd3+Pd4)/(Pc1+Pc2+Pc3+Pc4) = TCda/TCca = Tasa de cambio equivalente del país D con el país C |

# Appendix Spreadsheed 19

## Page 5

## Image Section 19.1

**Spreadsheet 16 - MODEL SIMULATION INTERNATIONAL TRADE OF 4 COUNTRIES:**

**SECTION 16.1: COMMERCIAL BALANCES. MODEL EQUATIONS**

**Equation 9.1 Defining the Trade Balance of Country A (United States) with All Other Countries (BCa.bcd)**

**BCa.bcd = BCab1 + BCab2 + BCab3 + BCab4 + BCac1 + BCac2 + BCac3 + BCac4 + BCad1 + BCad2 + BCad3 + BCad4**

| BCa.bcd | = | BCab1 | + | BCab2 | + | BCab3 | + | BCab4 | + | BCac1 | + | BCac2 | + | BCac3 | + | BCac4 | + | BCad1 | + | BCad2 | + | BCad3 | + | BCad4 |
|---|---|---|---|---|---|---|---|---|---|---|---|---|---|---|---|---|---|---|---|---|---|---|---|---|
| 0.00 | = | -1.74 | + | -60.61 | + | -88.75 | + | 86.19 | + | -4.32 | + | 18.72 | + | 24.55 | + | 17.93 | + | -3.70 | + | 0.78 | + | 6.56 | + | 4.37 |

**Equation 9.2 Defining the Trade Balance of Country B (Mexico) with All Other Countries (BCb.acd)**

**BCb.acd = BCba1 + BCba2 + BCba3 + BCba4 + BCbc1 + BCbc2 + BCbc3 + BCbc4 + BCbd1 + BCbd2 + BCbd3 + BCbd4**

| BCb.acd | = | BCba1 | + | BCba2 | + | BCba3 | + | BCba4 | + | BCbc1 | + | BCbc2 | + | BCbc3 | + | BCbc4 | + | BCbd1 | + | BCbd2 | + | BCbd3 | + | BCbd4 |
|---|---|---|---|---|---|---|---|---|---|---|---|---|---|---|---|---|---|---|---|---|---|---|---|---|
| 0.00 | = | 1.74 | + | 60.61 | + | 88.75 | + | -86.19 | + | 3.36 | + | -13.08 | + | -28.89 | + | -17.12 | + | -3.53 | + | -2.44 | + | -1.41 | + | -1.79 |

**Equation 9.3 Defining the Trade Balance of Country C (Canada) with All Other Countries (BCc.abd)**

**BCc.abd = BCca1 + BCca2 + BCca3 + BCca4 + BCcb1 + BCcb2 + BCcb3 + BCcb4 + BCcd1 + BCcd2 + BCcd3 + BCcd4**

| BCc.abd | = | BCca1 | + | BCca2 | + | BCca3 | + | BCca4 | + | BCcb1 | + | BCcb2 | + | BCcb3 | + | BCcb4 | + | BCcd1 | + | BCcd2 | + | BCcd3 | + | BCcd4 |
|---|---|---|---|---|---|---|---|---|---|---|---|---|---|---|---|---|---|---|---|---|---|---|---|---|
| 0.00 | = | 4.32 | + | -18.72 | + | -24.55 | + | -17.93 | + | -3.36 | + | 13.08 | + | 28.89 | + | 17.12 | + | -3.70 | + | -0.35 | + | 2.77 | + | 2.43 |

**Equation 9.4 Defining the Trade Balance of Country D (Dominican Republic) with All Other Countries (BCd.abc)**

**BCd.abc = BCda1 + BCda2 + BCda3 + BCda4 + BCdb1 + BCdb2 + BCdb3 + BCdb4 + BCdc1 + BCdc2 + BCdc3 + BCdc4**

| BCd.abc | = | BCda1 | + | BCda2 | + | BCda3 | + | BCda4 | + | BCdb1 | + | BCdb2 | + | BCdb3 | + | BCdb4 | + | BCdc1 | + | BCdc2 | + | BCdc3 | + | BCdc4 |
|---|---|---|---|---|---|---|---|---|---|---|---|---|---|---|---|---|---|---|---|---|---|---|---|---|
| 0.00 | = | 3.70 | + | -0.78 | + | -6.56 | + | -4.37 | + | 3.53 | + | 2.44 | + | 1.41 | + | 1.79 | + | 3.70 | + | 0.35 | + | -2.77 | + | -2.43 |

**COMMERCIAL BALANCES**

| **Country A - United States** | | | | **Country B - Mexico** | | | **Country C - Canada** | | | **Country D - Dominican Republic** | | |
|---|---|---|---|---|---|---|---|---|---|---|---|---|
| | | **Deficit or Surplus US** | | | | **Deficit or Surplus Mex** | | | **Deficit or Surplus Cand** | | | **Deficit or Surplus Dominican Republic** |
| EXa EU | IMa EU | US$ Bca.bcd | | EXb Mex | IMb Mex | US$ BCb.acd | EXc Cand | IMc Cand | US$ BCc.abd | EXd RD | IMd RD | US$ BCd.abc |
| 424 | 423.75 | 0.00 | (0.00) | 423.74 | 423.74 | 0.00 | 88.58 | 88.58 | (0.00) | 26.93 | 26.93 | (0.00) |
| EXab | IMab | | | EXba | IMba | | EXca | IMca | | EXda | IMda | |
| 341.82 | 406.73 | (64.906798) | 0.00 | 406.73 | 341.82 | 64.91 | 11.55 | 68.44 | (56.88) | 5.47 | 13.49 | (8.02) |
| EXac | IMac | | | EXbc | IMbc | | EXcb | IMcb | | EXdb | IMdb | |
| 68.44 | 11.55 | 56.883252 | (0.00) | 12.43 | 68.16 | (55.73) | 68.16 | 12.43 | 55.73 | 13.75 | 4.58 | 9.17 |
| EXad | IMad | | | EXbd | IMbd | | EXcd | IMcd | | EXdc | IMdc | |
| 13 | 5.47 | 8.02372 | (0.00) | 4.58 | 13.75 | (9.17) | 8.87 | 7.71 | 1.15 | 7.71 | 8.87 | (1.15) |

# Appendix Spreadsheed 19

## Page 5

## Image Section 19.7 and 19.8

| | **SECTION 16.7: PRIMARY, CONTRA PRIMARY AND NON-PRIMARY CHANGE RATES** | Set CUCTC | | **SECTION 16.7: INITIAL CHANGE RATES** |
|---|---|---|---|---|
| 0.031438 | **TCab = Primary and equivalent exchange rate of country A with country B** | 0.003897 | 0.027541 | **TCab In = (Pa1+Pa2+Pa3+Pa4)/(Pb1+Pb2+Pb3+Pb4) = Initial exchange rate of country A with country B** |
| 0.452446 | **TCac = Primary and equivalent exchange rate of country A with country C** | 0.014729 | 0.437717 | **TCac In = (Pa1+Pa2+Pa3+Pa4)/(Pc1+Pc2+Pc3+Pc4) = Initial exchange rate of country A with country C** |
| 0.009936 | TCad = Tasa de cambio primaria y equivalente del país D con el país A | -0.000143 | 0.010079 | TCad Inicial = (Pa1+Pa2+Pa3+Pa4)/(Pd1+Pd2+Pd3+Pd4) = Tasa de cambio Injcial del país A con el país D |
| 31.808953 | **TCba = 1/TCab = Contra-Primary and equivalent exchange rate of country B with country A** | | 36.310144 | **TCba In = (Pb1+Pb2+Pb3+Pb4)/(Pa1+Pa2+Pa3+Pa4) = 1/Tcab = Initial exchange rate of country B with country A** |
| 2.210207 | **TCca = 1/TCac = Contra-Primary and equivalent exchange rate of country C with country A** | | 2.284580 | **TCca In = (Pc1+Pc2+Pc3+Pc4)/(Pa1+Pa2+Pa3+Pa4) = 1/TCac = Initial exchange rate of country C with country A** |
| 100.64108 | TCda = 1/TCad = Tasa de cambio equivalente del país D con el país A | | 99.213435 | TCda Inicial = (Pd1+Pd2+Pd3+Pd4)/(Pa1+Pa2+Pa3+Pa4) = Tasa de cambio Injcial del país D con el país A |
| 14.391846 | **TCbc = TCac/TCab = Non-Primary and equivalent exchange rate of country B with country C** | | 15.893572 | **TCbc In = (Pb1+Pb2+Pb3+Pb4)/(Pc1+Pc2+Pc3+Pc4) = TCba/TCca =Initial exchange rate of country B with country C** |
| 0.069484 | **TCcb = TCab/TCac = Non-Primary and equivalent exchange rate of country C with country B** | | 0.062919 | **TCcb In = (Pc1+Pc2+Pc3+Pc4)/(Pb1+Pb2+Pb3+Pb4) = TCab/TCac = Initial exchange rate of country C with country B** |
| 0.316063 | TCbd = TCad/TCab = Tasa de cambio equivalente del país B con el país D | | 0.365980 | TCbd Inicial = (Pb1+Pb2+Pb3+Pb4)/(Pd1+Pd2+Pd3+Pd4) = TCba/TCda = Tasa de cambio equivalente del país B con el país D |
| 3.163923 | TCdb = TCab/TCad = Tasa de cambio equivalente del país D con el país B | | 2.732389 | TCdb Inicial = (Pd1+Pd2+Pd3+Pd4)/(Pb1+Pb2+Pb3+Pb4) = TCda/TCba = Tasa de cambio equivalente del país D con el país B |
| 0.021961 | TCcd = TCad/TCac = Tasa de cambio equivalente del país C con el país D | | 0.023027 | TCcd Inicial = (Pc1+Pc2+Pc3+Pc4)/(Pd1+Pd2+Pd3+Pd4) = TCca/TCda = Tasa de cambio equivalente del país C con el país D |
| 45.534696 | TCdc = TCac/TCad = Tasa de cambio equivalente del país D con el país C | | 43.427419 | TCdc Inicial = (Pd1+Pd2+Pd3+Pd4)/(Pc1+Pc2+Pc3+Pc4) = TCda/TCca = Tasa de cambio equivalente del país D con el país C |

# Appendix Spreadsheed 20

## Image Section 20.1

**Spreadsheet 19 - MODEL SIMULATION INTERNATIONAL TRADE OF 4 COUNTRIES:**

**SECTION 19.1: COMMERCIAL BALANCES. MODEL EQUATIONS**

**Equation 9.1 Defining the Trade Balance of Country A (United States) with All Other Countries (BCa.bcd)**

**BCa.bcd = BCab1 + BCab2 + BCab3 + BCab4 + BCac1 + BCac2 + BCac3 + BCac4 + BCad1 + BCad2 + BCad3 + BCad4**

| | | | | | | | | | | | | | | | | | | | | | | | | |
|---|---|---|---|---|---|---|---|---|---|---|---|---|---|---|---|---|---|---|---|---|---|---|---|---|
| BCa.bcd | = | BCab1 | + | BCab2 | + | BCab3 | + | BCab4 | + | BCac1 | + | BCac2 | + | BCac3 | + | BCac4 | + | BCad1 | + | BCad2 | + | BCad3 | + | BCad4 |
| -6.82 | = | -1.74 | + | -60.61 | + | -88.75 | + | 86.19 | + | -4.32 | + | 18.72 | + | 24.55 | + | 17.93 | + | -4.23 | + | -2.12 | + | 4.53 | + | 3.02 |

**Equation 9.2 Defining the Trade Balance of Country B (Mexico) with All Other Countries (BCb.acd)**

**BCb.acd = BCba1 + BCba2 + BCba3 + BCba4 + BCbc1 + BCbc2 + BCbc3 + BCbc4 + BCbd1 + BCbd2 + BCbd3 + BCbd4**

| | | | | | | | | | | | | | | | | | | | | | | | | |
|---|---|---|---|---|---|---|---|---|---|---|---|---|---|---|---|---|---|---|---|---|---|---|---|---|
| BCb.acd | = | BCba1 | + | BCba2 | + | BCba3 | + | BCba4 | + | BCbc1 | + | BCbc2 | + | BCbc3 | + | BCbc4 | + | BCbd1 | + | BCbd2 | + | BCbd3 | + | BCbd4 |
| -6.74 | = | 1.74 | + | 60.61 | + | 88.75 | + | -86.19 | + | 3.36 | + | -13.08 | + | -28.89 | + | -17.12 | + | -4.08 | + | -4.98 | + | -3.47 | + | -3.39 |

**Equation 9.3 Defining the Trade Balance of Country C (Canada) with All Other Countries (BCc.abd)**

**BCc.abd = BCca1 + BCca2 + BCca3 + BCca4 + BCcb1 + BCcb2 + BCcb3 + BCcb4 + BCcd1 + BCcd2 + BCcd3 + BCcd4**

| | | | | | | | | | | | | | | | | | | | | | | | | |
|---|---|---|---|---|---|---|---|---|---|---|---|---|---|---|---|---|---|---|---|---|---|---|---|---|
| BCc.abd | = | BCca1 | + | BCca2 | + | BCca3 | + | BCca4 | + | BCcb1 | + | BCcb2 | + | BCcb3 | + | BCcb4 | + | BCcd1 | + | BCcd2 | + | BCcd3 | + | BCcd4 |
| -8.55 | = | 4.32 | + | -18.72 | + | -24.55 | + | -17.93 | + | -3.36 | + | 13.08 | + | 28.89 | + | 17.12 | + | -4.23 | + | -3.14 | + | -0.33 | + | 0.31 |

**Equation 9.4 Defining the Trade Balance of Country D (Dominican Republic) with All Other Countries (BCd.abc)**

**BCd.abc = BCda1 + BCda2 + BCda3 + BCda4 + BCdb1 + BCdb2 + BCdb3 + BCdb4 + BCdc1 + BCdc2 + BCdc3 + BCdc4**

| | | | | | | | | | | | | | | | | | | | | | | | | |
|---|---|---|---|---|---|---|---|---|---|---|---|---|---|---|---|---|---|---|---|---|---|---|---|---|
| BCd.abc | = | BCda1 | + | BCda2 | + | BCda3 | + | BCda4 | + | BCdb1 | + | BCdb2 | + | BCdb3 | + | BCdb4 | + | BCdc1 | + | BCdc2 | + | BCdc3 | + | BCdc4 |
| 22.11 | = | 4.23 | + | 2.12 | + | -4.53 | + | -3.02 | + | 4.08 | + | 4.98 | + | 3.47 | + | 3.39 | + | 4.23 | + | 3.14 | + | 0.33 | + | -0.31 |

**COMMERCIAL BALANCES**

| Country A - United States | | | | Country B - Mexico | | | Country C - Canada | | | Country D - Dominican Republic | | |
|---|---|---|---|---|---|---|---|---|---|---|---|---|
| | | Deficit or Surplus US | | | | Deficit or Surplus Mex | | | Deficit or Surplus Cand | | | Deficit or Surplus Dominican Republic |
| EXa EU | IMa EU | US$ Bca.bcd | | EXb Mex | IMb Mex | US$ BCb.acd | EXc Cand | IMc Cand | US$ BCc.abd | EXd RD | IMd RD | US$ BCd.abc |
| 419 | 425.51 | (6.82) | (3.88) | 420.21 | 426.95 | (6.74) | 83.93 | 92.48 | (8.55) | 35.81 | 13.70 | 22.11 |
| EXab | IMab | | | EXba | IMba | | EXca | IMca | | EXda | IMda | |
| 341.82 | 406.73 | (64.906798) | (0.21) | 406.73 | 341.82 | 64.91 | 11.55 | 68.44 | (56.88) | 7.23 | 8.43 | (1.20) |
| EXac | IMac | | | EXbc | IMbc | | EXcb | IMcb | | EXdb | IMdb | |
| 68.44 | 11.55 | 56.883252 | (3.87) | 12.43 | 68.16 | (55.73) | 68.16 | 12.43 | 55.73 | 16.97 | 1.05 | 15.92 |
| EXad | IMad | | | EXbd | IMbd | | EXcd | IMcd | | EXdc | IMdc | |
| 8 | 7.23 | 1.20196 | 0.20 | 1.05 | 16.97 | (15.92) | 4.21 | 11.61 | -7.40 | 11.61 | 4.21 | 7.40 |

# Appendix Spreadsheed 21

## Page 2
## Image Section 21.1

**Hoja de Cálculo 21 - MODELO SIMULACION COMERCIO INTERNACIONAL DE 4 PAISES:**

**SECCION 21.1: BALANZAS COMERCIALES. ECUACIONES DEL MODELO**

**Ecuación 9.1 que Define la Balanza Comercial Del País A (Estaos Unidos) con Todos los Demás Países (BCa.bcd)**

**BCa.bcd = BCab1 + BCab2 + BCab3 + BCab4 + BCac1 + BCac2 + BCac3 + BCac4 + BCad1 + BCad2 + BCad3 + BCad4**

| BCa.bcd | = | BCab1 | + | BCab2 | + | BCab3 | + | BCab4 | + | BCac1 | + | BCac2 | + | BCac3 | + | BCac4 | + | BCad1 | + | BCad2 | + | BCad3 | + | BCad4 |
|---|---|---|---|---|---|---|---|---|---|---|---|---|---|---|---|---|---|---|---|---|---|---|---|---|
| -191.67 | = | -37.17 | + | -104.68 | + | -124.50 | + | 40.62 | + | -9.59 | + | 9.22 | + | 17.30 | + | 14.09 | + | -4.09 | + | -1.34 | + | 5.09 | + | 3.39 |

**Ecuación 9.2 que Define la Balanza Comercial Del País B (México) con Todos los Demás Países (BCb.acd)**

**BCb.acd = BCba1 + BCba2 + BCba3 + BCba4 + BCbc1 + BCbc2 + BCbc3 + BCbc4 + BCbd1 + BCbd2 + BCbd3 + BCbd4**

| BCb.acd | = | BCba1 | + | BCba2 | + | BCba3 | + | BCba4 | + | BCbc1 | + | BCbc2 | + | BCbc3 | + | BCbc4 | + | BCbd1 | + | BCbd2 | + | BCbd3 | + | BCbd4 |
|---|---|---|---|---|---|---|---|---|---|---|---|---|---|---|---|---|---|---|---|---|---|---|---|---|
| 192.74 | = | 37.17 | + | 104.68 | + | 124.50 | + | -40.62 | + | 8.73 | + | -3.88 | + | -21.68 | + | -12.42 | + | -2.66 | + | -0.72 | + | 0.27 | + | -0.62 |

**Ecuación 9.3 que Define la Balanza Comercial Del País C (Canadá) con Todos los Demás Países (BCc.abd)**

**BCc.abd = BCca1 + BCca2 + BCca3 + BCca4 + BCcb1 + BCcb2 + BCcb3 + BCcb4 + BCcd1 + BCcd2 + BCcd3 + BCcd4**

| BCc.abd | = | BCca1 | + | BCca2 | + | BCca3 | + | BCca4 | + | BCcb1 | + | BCcb2 | + | BCcb3 | + | BCcb4 | + | BCcd1 | + | BCcd2 | + | BCcd3 | + | BCcd4 |
|---|---|---|---|---|---|---|---|---|---|---|---|---|---|---|---|---|---|---|---|---|---|---|---|---|
| -1.46 | = | 9.59 | + | -9.22 | + | -17.30 | + | -14.09 | + | -8.73 | + | 3.88 | + | 21.68 | + | 12.42 | + | -3.51 | + | -0.58 | + | 2.31 | + | 2.09 |

**Ecuación 9.4 que Define la Balanza Comercial Del País D (República Dominicana) con Todos los Demás Países (BCd.abc)**

**BCd.abc = BCda1 + BCda2 + BCda3 + BCda4 + BCdb1 + BCdb2 + BCdb3 + BCdb4 + BCdc1 + BCdc2 + BCdc3 + BCdc4**

| BCd.abc | = | BCda1 | + | BCda2 | + | BCda3 | + | BCda4 | + | BCdb1 | + | BCdb2 | + | BCdb3 | + | BCdb4 | + | BCdc1 | + | BCdc2 | + | BCdc3 | + | BCdc4 |
|---|---|---|---|---|---|---|---|---|---|---|---|---|---|---|---|---|---|---|---|---|---|---|---|---|
| 0.38 | = | 4.09 | + | 1.34 | + | -5.09 | + | -3.39 | + | 2.66 | + | 0.72 | + | -0.27 | + | 0.62 | + | 3.51 | + | 0.58 | + | -2.31 | + | -2.09 |

SECCION D : BALANZAS COMERCIALES

| País A = Estados Unidos | | | País B = México | | | País C = Canadá | | | País D = República Dominicana | | |
|---|---|---|---|---|---|---|---|---|---|---|---|
| | | Déficit o Superávit EU | | | Déficit o Superávit Méx | | | Déficit o Superávit Cand | | | Déficit o Superávit RD |
| EXa EU | IMa EU | US$ BCa.bcd | EXb Mex | IMb Mex | US$ BCb.acd | EXc Cand | IMc Cand | US$ BCc.abd | EXd RD | IMd RD | US$ BCd.abc |
| 272 | 463.83 | (191.67) | 464.95 | 272.21 | 192.74 | 77.31 | 78.77 | (1.46) | 24.52 | 24.13 | 0.38 |
| EXab | IMab | | EXba | IMba | | EXcb | IMcb | | EXdb | IMdb | |
| 211.97 | 437.70 | (225.729970) | 437.70 | 211.97 | 225.73 | 50.06 | 20.81 | 29.26 | 10.17 | 6.44 | 3.73 |
| EXac | IMac | | EXbc | IMbc | | EXca | IMca | | EXdc | IMdc | |
| 50.38 | 19.36 | 31.019615 | 20.81 | 50.06 | (29.26) | 19.36 | 50.38 | (31.02) | 7.58 | 7.88 | (0.31) |
| EXad | IMad | | EXbd | IMbd | | EXcd | IMcd | | EXda | IMda | |
| 10 | 6.77 | 3.04256 | 6.44 | 10.17 | (3.73) | 7.88 | 7.58 | 0.31 | 6.77 | 9.81 | (3.04) |

# Appendix Spreadsheed 21

## Page 2
## Image Sections 21.7 and 21.8

| | SECCION 21.7 : TASAS DE CAMBIO PRIMARIAS Y EQUIVALENTES | | Conjunto CUCTC | | SECCION 21.8 : TASAS DE CAMBIO INICIALES |
|---|---|---|---|---|---|
| 0.027541 | TCab = Tasa de cambio primaria y equivalente del país A con el país B | 0.031438 | 0.000000 | 0.027541 | TCab Inicial = (Pa1+Pa2+Pa3+Pa4)/(Pb1+Pb2+Pb3+Pb4) = Tasa de cambio Injcial del país A con el país B |
| 0.423553 | TCac = Tasa de cambio primaria y equivalente del país A con el país C | 0.452446 | -0.032360 | 0.437717 | TCac Inicial = (Pa1+Pa2+Pa3+Pa4)/(Pc1+Pc2+Pc3+Pc4) = Tasa de cambio Inicial del país A con el país C |
| 0.009216 | TCad = Tasa de cambio primaria y equivalente del país D con el país A | 0.009936 | -0.085690 | 0.010079 | TCad Inicial = (Pa1+Pa2+Pa3+Pa4)/(Pd1+Pd2+Pd3+Pd4) = Tasa de cambio Injcial del país A con el país D |
| 36.310144 | TCba = 1/TCab = Tasa de cambio equivalente del país B con el país A | | | 36.310144 | TCba Inicial = (Pb1+Pb2+Pb3+Pb4)/(Pa1+Pa2+Pa3+Pa4) = 1/Tcab = Tasa de cambio Inicial del país B con el país A |
| 2.360982 | TCca = 1/TCac = Tasa de cambio equivalente del país A con el país C | | | 2.284580 | TCca Inicial = (Pc1+Pc2+Pc3+Pc4)/(Pa1+Pa2+Pa3+Pa4) = 1/TCac = Tasa de cambio Inicial del país C con el país A |
| 108.51181 | TCda = 1/TCad = Tasa de cambio equivalente del país D con el país A | | | 99.213435 | TCda Inicial = (Pd1+Pd2+Pd3+Pd4)/(Pa1+Pa2+Pa3+Pa4) = Tasa de cambio Injcial del país D con el país A |
| 15.379256 | TCbc = TCac/TCab = Tasa de cambio equivalente del país B con el país C | | | 15.893572 | TCbc Inicial = (Pb1+Pb2+Pb3+Pb4)/(Pc1+Pc2+Pc3+Pc4) = TCba/TCca = Tasa de cambio Inicial del país B con el país C |
| 0.065023 | TCcb = TCab/TCac = Tasa de cambio equivalente del país C con el país B | | | 0.062919 | TCcb Inicial = (Pc1+Pc2+Pc3+Pc4)/(Pb1+Pb2+Pb3+Pb4) o TCab/TCac = Tasa de cambio Inicial del país C con el país B |
| 0.334619 | TCbd = TCad/TCab = Tasa de cambio equivalente del país B con el país D | | | 0.365980 | TCbd Inicial = (Pb1+Pb2+Pb3+Pb4)/(Pd1+Pd2+Pd3+Pd4) = TCba/TCda = Tasa de cambio equivalente del país B con el país D |
| 2.988471 | TCdb = TCab/TCad = Tasa de cambio equivalente del país D con el país B | | | 2.732389 | TCdb Inicial = (Pd1+Pd2+Pd3+Pd4)/(Pb1+Pb2+Pb3+Pb4) = TCda/TCba = Tasa de cambio equivalente del país D con el país B |
| 0.021758 | TCcd = TCad/TCac = Tasa de cambio equivalente del país C con el país D | | | 0.023027 | TCcd Inicial = (Pc1+Pc2+Pc3+Pc4)/(Pd1+Pd2+Pd3+Pd4) = TCca/TCda = Tasa de cambio equivalente del país C con el país D |
| 45.960460 | TCdc = TCac/TCad = Tasa de cambio equivalente del país D con el país C | | | 43.427419 | TCdc Inicial = (Pd1+Pd2+Pd3+Pd4)/(Pc1+Pc2+Pc3+Pc4) = TCda/TCca = Tasa de cambio equivalente del país D con el país C |

# Appendix Spreadsheed 21

## Page 3
## Image Section 21.1

**Hoja de Cálculo 21 - MODELO SIMULACION COMERCIO INTERNACIONAL DE 4 PAISES:**

**SECCION 21.1: BALANZAS COMERCIALES. ECUACIONES DEL MODELO**

**Ecuación 9.1 que Define la Balanza Comercial Del País A (Estaos Unidos) con Todos los Demás Países (BCa.bcd)**

**BCa.bcd = BCab1 + BCab2 + BCab3 + BCab4 + BCac1 + BCac2 + BCac3 + BCac4 + BCad1 + BCad2 + BCad3 + BCad4**

| BCa.bcd | = | BCab1 | + | BCab2 | + | BCab3 | + | BCab4 | + | BCac1 | + | BCac2 | + | BCac3 | + | BCac4 | + | BCad1 | + | BCad2 | + | BCad3 | + | BCad4 |
|---|---|---|---|---|---|---|---|---|---|---|---|---|---|---|---|---|---|---|---|---|---|---|---|---|
| -192.16 | = | -37.17 | + | -104.68 | + | -124.50 | + | 40.62 | + | -9.70 | + | 9.03 | + | 17.09 | + | 14.01 | + | -4.09 | + | -1.30 | + | 5.12 | + | 3.41 |

**Ecuación 9.2 que Define la Balanza Comercial Del País B (México) con Todos los Demás Países (BCb.acd)**

**BCb.acd = BCba1 + BCba2 + BCba3 + BCba4 + BCbc1 + BCbc2 + BCbc3 + BCbc4 + BCbd1 + BCbd2 + BCbd3 + BCbd4**

| BCb.acd | = | BCba1 | + | BCba2 | + | BCba3 | + | BCba4 | + | BCbc1 | + | BCbc2 | + | BCbc3 | + | BCbc4 | + | BCbd1 | + | BCbd2 | + | BCbd3 | + | BCbd4 |
|---|---|---|---|---|---|---|---|---|---|---|---|---|---|---|---|---|---|---|---|---|---|---|---|---|
| 192.31 | = | 37.17 | + | 104.68 | + | 124.50 | + | -40.62 | + | 8.60 | + | -4.06 | + | -21.83 | + | -12.53 | + | -2.64 | + | -0.68 | + | 0.31 | + | -0.59 |

**Ecuación 9.3 que Define la Balanza Comercial Del País C (Canadá) con Todos los Demás Países (BCc.abd)**

**BCc.abd = BCca1 + BCca2 + BCca3 + BCca4 + BCcb1 + BCcb2 + BCcb3 + BCcb4 + BCcd1 + BCcd2 + BCcd3 + BCcd4**

| BCc.abd | = | BCca1 | + | BCca2 | + | BCca3 | + | BCca4 | + | BCcb1 | + | BCcb2 | + | BCcb3 | + | BCcb4 | + | BCcd1 | + | BCcd2 | + | BCcd3 | + | BCcd4 |
|---|---|---|---|---|---|---|---|---|---|---|---|---|---|---|---|---|---|---|---|---|---|---|---|---|
| -0.06 | = | 9.70 | + | -9.03 | + | -17.09 | + | -14.01 | + | -8.60 | + | 4.06 | + | 21.83 | + | 12.53 | + | -3.49 | + | -0.50 | + | 2.40 | + | 2.14 |

**Ecuación 9.4 que Define la Balanza Comercial Del País D (República Dominicana) con Todos los Demás Países (BCd.abc)**

**BCd.abc = BCda1 + BCda2 + BCda3 + BCda4 + BCdb1 + BCdb2 + BCdb3 + BCdb4 + BCdc1 + BCdc2 + BCdc3 + BCdc4**

| BCd.abc | = | BCda1 | + | BCda2 | + | BCda3 | + | BCda4 | + | BCdb1 | + | BCdb2 | + | BCdb3 | + | BCdb4 | + | BCdc1 | + | BCdc2 | + | BCdc3 | + | BCdc4 |
|---|---|---|---|---|---|---|---|---|---|---|---|---|---|---|---|---|---|---|---|---|---|---|---|---|
| -0.10 | = | 4.09 | + | 1.30 | + | -5.12 | + | -3.41 | + | 2.64 | + | 0.68 | + | -0.31 | + | 0.59 | + | 3.49 | + | 0.50 | + | -2.40 | + | -2.14 |

SECCION D : BALANZAS COMERCIALES

| País A = Estados Unidos | | | País B = México | | | País C = Canadá | | | País D = República Dominicana | | |
|---|---|---|---|---|---|---|---|---|---|---|---|
| | | Déficit o Superávit EU | | | Déficit o Superávit Méx | | | Déficit o Superávit Cand | | | Déficit o Superávit RD |
| EXa EU | IMa EU | US$ BCa.bcd | EXb Mex | IMb Mex | US$ BCb.acd | EXc Cand | IMc Cand | US$ BCc.abd | EXd RD | IMd RD | US$ BCd.abc |
| 272 | 464.02 | (192.16) | 464.69 | 272.38 | 192.31 | 77.88 | 77.93 | (0.06) | 24.31 | 24.41 | (0.10) |
| EXab | IMab | | EXba | IMba | | EXcb | IMcb | | EXdb | IMdb | |
| 211.97 | 437.70 | (225.729970) | 437.70 | 211.97 | 225.73 | 50.29 | 20.47 | 29.82 | 10.12 | 6.52 | 3.60 |
| EXac | IMac | | EXbc | IMbc | | EXca | IMca | | EXdc | IMdc | |
| 50.01 | 19.58 | 30.430079 | 20.47 | 50.29 | (29.82) | 19.58 | 50.01 | (30.43) | 7.45 | 8.01 | (0.55) |
| EXad | IMad | | EXbd | IMbd | | EXcd | IMcd | | EXda | IMda | |
| 10 | 6.74 | 3.13822 | 6.52 | 10.12 | (3.60) | 8.01 | 7.45 | 0.55 | 6.74 | 9.88 | (3.14) |

# Appendix Spreadsheed 21

## Page 3
## Image Sections 21.7 and 21.8

| | SECCION 21.7 : TASAS DE CAMBIO PRIMARIAS Y EQUIVALENTES | | Conjunto CUCTC | | SECCION 21.8 : TASAS DE CAMBIO INICIALES |
|---|---|---|---|---|---|
| 0.027541 | TCab = Tasa de cambio primaria y equivalente del país A con el país B | 0.031438 | 0.000000 | 0.027541 | TCab Inicial = (Pa1+Pa2+Pa3+Pa4)/(Pb1+Pb2+Pb3+Pb4) = Tasa de cambio Injcial del país A con el país B |
| 0.422940 | TCac = Tasa de cambio primaria y equivalente del país A con el país C | 0.452446 | -0.033760 | 0.437717 | TCac Inicial = (Pa1+Pa2+Pa3+Pa4)/(Pc1+Pc2+Pc3+Pc4) = Tasa de cambio Inicial del país A con el país C |
| 0.009230 | TCad = Tasa de cambio primaria y equivalente del país D con el país A | 0.009936 | -0.084270 | 0.010079 | TCad Inicial = (Pa1+Pa2+Pa3+Pa4)/(Pd1+Pd2+Pd3+Pd4) = Tasa de cambio Injcial del país A con el país D |
| 36.310144 | TCba = 1/TCab = Tasa de cambio equivalente del país B con el país A | | | 36.310144 | TCba Inicial = (Pb1+Pb2+Pb3+Pb4)/(Pa1+Pa2+Pa3+Pa4) = 1/Tcab = Tasa de cambio Inicial del país B con el país A |
| 2.364403 | TCca = 1/TCac = Tasa de cambio equivalente del país A con el país C | | | 2.284580 | TCca Inicial = (Pc1+Pc2+Pc3+Pc4)/(Pa1+Pa2+Pa3+Pa4) = 1/TCac = Tasa de cambio Inicial del país C con el país A |
| 108.34355 | TCda = 1/TCad = Tasa de cambio equivalente del país D con el país A | | | 99.213435 | TCda Inicial = (Pd1+Pd2+Pd3+Pd4)/(Pa1+Pa2+Pa3+Pa4) = Tasa de cambio Injcial del país D con el país A |
| 15.357005 | TCbc = TCac/TCab = Tasa de cambio equivalente del país B con el país C | | | 15.893572 | TCbc Inicial = (Pb1+Pb2+Pb3+Pb4)/(Pc1+Pc2+Pc3+Pc4) = TCba/TCca = Tasa de cambio Inicial del país B con el país C |
| 0.065117 | TCcb = TCab/TCac = Tasa de cambio equivalente del país C con el país B | | | 0.062919 | TCcb Inicial = (Pc1+Pc2+Pc3+Pc4)/(Pb1+Pb2+Pb3+Pb4) o TCab/TCac = Tasa de cambio Inicial del país C con el país B |
| 0.335139 | TCbd = TCad/TCab = Tasa de cambio equivalente del país B con el país D | | | 0.365980 | TCbd Inicial = (Pb1+Pb2+Pb3+Pb4)/(Pd1+Pd2+Pd3+Pd4) = TCba/TCda = Tasa de cambio equivalente del país B con el país D |
| 2.983837 | TCdb = TCab/TCad = Tasa de cambio equivalente del país D con el país B | | | 2.732389 | TCdb Inicial = (Pd1+Pd2+Pd3+Pd4)/(Pb1+Pb2+Pb3+Pb4) = TCda/TCba = Tasa de cambio equivalente del país D con el país B |
| 0.021823 | TCcd = TCad/TCac = Tasa de cambio equivalente del país C con el país D | | | 0.023027 | TCcd Inicial = (Pc1+Pc2+Pc3+Pc4)/(Pd1+Pd2+Pd3+Pd4) = TCca/TCda = Tasa de cambio equivalente del país C con el país D |
| 45.822797 | TCdc = TCac/TCad = Tasa de cambio equivalente del país D con el país C | | | 43.427419 | TCdc Inicial = (Pd1+Pd2+Pd3+Pd4)/(Pc1+Pc2+Pc3+Pc4) = TCda/TCca = Tasa de cambio equivalente del país D con el país C |

# Appendix Spreadsheed 21

## Page 4
## Image Section 21.1

**Hoja de Cálculo 21 - MODELO SIMULACION COMERCIO INTERNACIONAL DE 4 PAISES:**

**SECCION 21.1: BALANZAS COMERCIALES. ECUACIONES DEL MODELO**

**Ecuación 9.1 que Define la Balanza Comercial Del País A (Estaos Unidos) con Todos los Demás Países (BCa.bcd)**

**BCa.bcd = BCab1 + BCab2 + BCab3 + BCab4 + BCac1 + BCac2 + BCac3 + BCac4 + BCad1 + BCad2 + BCad3 + BCad4**

| BCa.bcd | = | BCab1 | + | BCab2 | + | BCab3 | + | BCab4 | + | BCac1 | + | BCac2 | + | BCac3 | + | BCac4 | + | BCad1 | + | BCad2 | + | BCad3 | + | BCad4 |
|---|---|---|---|---|---|---|---|---|---|---|---|---|---|---|---|---|---|---|---|---|---|---|---|---|
| -192.23 | = | -37.17 | + | -104.68 | + | -124.50 | + | 40.62 | + | -9.70 | + | 9.01 | + | 17.08 | + | 14.00 | + | -4.09 | + | -1.31 | + | 5.11 | + | 3.41 |

**Ecuación 9.2 que Define la Balanza Comercial Del País B (México) con Todos los Demás Países (BCb.acd)**

**BCb.acd = BCba1 + BCba2 + BCba3 + BCba4 + BCbc1 + BCbc2 + BCbc3 + BCbc4 + BCbd1 + BCbd2 + BCbd3 + BCbd4**

| BCb.acd | = | BCba1 | + | BCba2 | + | BCba3 | + | BCba4 | + | BCbc1 | + | BCbc2 | + | BCbc3 | + | BCbc4 | + | BCbd1 | + | BCbd2 | + | BCbd3 | + | BCbd4 |
|---|---|---|---|---|---|---|---|---|---|---|---|---|---|---|---|---|---|---|---|---|---|---|---|---|
| 192.23 | = | 37.17 | + | 104.68 | + | 124.50 | + | -40.62 | + | 8.59 | + | -4.07 | + | -21.84 | + | -12.53 | + | -2.64 | + | -0.69 | + | 0.30 | + | -0.60 |

**Ecuación 9.3 que Define la Balanza Comercial Del País C (Canadá) con Todos los Demás Países (BCc.abd)**

**BCc.abd = BCca1 + BCca2 + BCca3 + BCca4 + BCcb1 + BCcb2 + BCcb3 + BCcb4 + BCcd1 + BCcd2 + BCcd3 + BCcd4**

| BCc.abd | = | BCca1 | + | BCca2 | + | BCca3 | + | BCca4 | + | BCcb1 | + | BCcb2 | + | BCcb3 | + | BCcb4 | + | BCcd1 | + | BCcd2 | + | BCcd3 | + | BCcd4 |
|---|---|---|---|---|---|---|---|---|---|---|---|---|---|---|---|---|---|---|---|---|---|---|---|---|
| 0.00 | = | 9.70 | + | -9.01 | + | -17.08 | + | -14.00 | + | -8.59 | + | 4.07 | + | 21.84 | + | 12.53 | + | -3.49 | + | -0.51 | + | 2.39 | + | 2.14 |

**Ecuación 9.4 que Define la Balanza Comercial Del País D (República Dominicana) con Todos los Demás Países (BCd.abc)**

**BCd.abc = BCda1 + BCda2 + BCda3 + BCda4 + BCdb1 + BCdb2 + BCdb3 + BCdb4 + BCdc1 + BCdc2 + BCdc3 + BCdc4**

| BCd.abc | = | BCda1 | + | BCda2 | + | BCda3 | + | BCda4 | + | BCdb1 | + | BCdb2 | + | BCdb3 | + | BCdb4 | + | BCdc1 | + | BCdc2 | + | BCdc3 | + | BCdc4 |
|---|---|---|---|---|---|---|---|---|---|---|---|---|---|---|---|---|---|---|---|---|---|---|---|---|
| 0.00 | = | 4.09 | + | 1.31 | + | -5.11 | + | -3.41 | + | 2.64 | + | 0.69 | + | -0.30 | + | 0.60 | + | 3.49 | + | 0.51 | + | -2.39 | + | -2.14 |

SECCION D : BALANZAS COMERCIALES

| País A = Estados Unidos | | | País B = México | | | País C = Canadá | | | País D = República Dominicana | | |
|---|---|---|---|---|---|---|---|---|---|---|---|
| | | Déficit o Superávit EU | | | Déficit o Superávit Méx | | | Déficit o Superávit Cand | | | Déficit o Superávit RD |
| EXa EU | IMa EU | US$ BCa.bcd | EXb Mex | IMb Mex | US$ BCb.acd | EXc Cand | IMc Cand | US$ BCc.abd | EXd RD | IMd RD | US$ BCd.abc |
| 272 | 464.04 | (192.23) | 464.65 | 272.41 | 192.23 | 77.89 | 77.89 | (0.00) | 24.35 | 24.35 | (0.00) |
| EXab | IMab | | EXba | IMba | | EXcb | IMcb | | EXdb | IMdb | |
| 211.97 | 437.70 | (225.729970) | 437.70 | 211.97 | 225.73 | 50.31 | 20.45 | 29.86 | 10.13 | 6.50 | 3.63 |
| EXac | IMac | | EXbc | IMbc | | EXca | IMca | | EXdc | IMdc | |
| 49.98 | 19.59 | 30.387999 | 20.45 | 50.31 | (29.86) | 19.59 | 49.98 | (30.39) | 7.47 | 7.99 | (0.53) |
| EXad | IMad | | EXbd | IMbd | | EXcd | IMcd | | EXda | IMda | |
| 10 | 6.75 | 3.11126 | 6.50 | 10.13 | (3.63) | 7.99 | 7.47 | 0.53 | 6.75 | 9.86 | (3.11) |

# Appendix Spreadsheed 21

## Page 4
## Image Sections 21.7 and 21.8

| | SECCION 21.7 : TASAS DE CAMBIO PRIMARIAS Y EQUIVALENTES | | Conjunto CUCTC | | SECCION 21.8 : TASAS DE CAMBIO INICIALES |
|---|---|---|---|---|---|
| 0.02754 | TCab = Tasa de cambio primaria y equivalente del país A con el país B | 0.031438 | 0.000000 | 0.027541 | TCab Inicial = (Pa1+Pa2+Pa3+Pa4)/(Pb1+Pb2+Pb3+Pb4) = Tasa de cambio Injcial del país A con el país B |
| 0.42290 | TCac = Tasa de cambio primaria y equivalente del país A con el país C | 0.452446 | -0.033860 | 0.437717 | TCac Inicial = (Pa1+Pa2+Pa3+Pa4)/(Pc1+Pc2+Pc3+Pc4) = Tasa de cambio Inicial del país A con el país C |
| 0.00923 | TCad = Tasa de cambio primaria y equivalente del país D con el país A | 0.009936 | -0.084670 | 0.010079 | TCad Inicial = (Pa1+Pa2+Pa3+Pa4)/(Pd1+Pd2+Pd3+Pd4) = Tasa de cambio Injcial del país A con el país D |
| 36.310144 | TCba = 1/TCab = Tasa de cambio equivalente del país B con el país A | | | 36.310144 | TCba Inicial = (Pb1+Pb2+Pb3+Pb4)/(Pa1+Pa2+Pa3+Pa4) = 1/Tcab = Tasa de cambio Inicial del país B con el país A |
| 2.364647 | TCca = 1/TCac = Tasa de cambio equivalente del país A con el país C | | | 2.284580 | TCca Inicial = (Pc1+Pc2+Pc3+Pc4)/(Pa1+Pa2+Pa3+Pa4) = 1/TCac = Tasa de cambio Inicial del país C con el país A |
| 108.39089 | TCda = 1/TCad = Tasa de cambio equivalente del país D con el país A | | | 99.213435 | TCda Inicial = (Pd1+Pd2+Pd3+Pd4)/(Pa1+Pa2+Pa3+Pa4) = Tasa de cambio Injcial del país D con el país A |
| 15.355416 | TCbc = TCac/TCab = Tasa de cambio equivalente del país B con el país C | | | 15.893572 | TCbc Inicial = (Pb1+Pb2+Pb3+Pb4)/(Pc1+Pc2+Pc3+Pc4) = TCba/TCca = Tasa de cambio Inicial del país B con el país C |
| 0.065124 | TCcb = TCab/TCac = Tasa de cambio equivalente del país C con el país B | | | 0.062919 | TCcb Inicial = (Pc1+Pc2+Pc3+Pc4)/(Pb1+Pb2+Pb3+Pb4) o TCab/TCac = Tasa de cambio Inicial del país C con el país B |
| 0.334993 | TCbd = TCad/TCab = Tasa de cambio equivalente del país B con el país D | | | 0.365980 | TCbd Inicial = (Pb1+Pb2+Pb3+Pb4)/(Pd1+Pd2+Pd3+Pd4) = TCba/TCda = Tasa de cambio equivalente del país B con el país D |
| 2.985141 | TCdb = TCab/TCad = Tasa de cambio equivalente del país D con el país B | | | 2.732389 | TCdb Inicial = (Pd1+Pd2+Pd3+Pd4)/(Pb1+Pb2+Pb3+Pb4) = TCda/TCba = Tasa de cambio equivalente del país D con el país B |
| 0.021816 | TCcd = TCad/TCac = Tasa de cambio equivalente del país C con el país D | | | 0.023027 | TCcd Inicial = (Pc1+Pc2+Pc3+Pc4)/(Pd1+Pd2+Pd3+Pd4) = TCca/TCda = Tasa de cambio equivalente del país C con el país D |
| 45.838077 | TCdc = TCac/TCad = Tasa de cambio equivalente del país D con el país C | | | 43.427419 | TCdc Inicial = (Pd1+Pd2+Pd3+Pd4)/(Pc1+Pc2+Pc3+Pc4) = TCda/TCca = Tasa de cambio equivalente del país D con el país C |